\documentclass[twoside,12pt]{report}

\usepackage{vmargin}

\setpapersize[portrait]{custom}{210mm}{297mm}

\setmarginsrb{24mm}{33mm}{24mm}{13mm}{5mm}{10mm}{5mm}{10mm}

\usepackage{fancyhdr}

\renewcommand{\chaptermark}[1]%
              {\markboth{#1}{}}
\renewcommand{\sectionmark}[1]%
              {\markright{\thesection\ #1}}

\usepackage[bf]{caption2}

\newcommand{\mychap}[1]{\chapter*{#1}\markboth{#1}{}\addcontentsline{toc}{chapter}{#1}}

\usepackage[psamsfonts]{amsfonts}

\usepackage[psamsfonts]{amssymb}

\usepackage{cmmib57}

\usepackage{times}

\normalfont

\usepackage[T1]{fontenc}

\usepackage{textcomp}

 \usepackage{amssymb,amsmath,slashed,epsfig,cite,bbm,multicol}
 \usepackage[dutch,english]{babel}
\newcommand{\bc}{boundary condition}

\newcommand{\dof}{degrees of freedom}

\newcommand{\NS}{Neveu-Schwarz}
\newcommand{\RR}{Ramond-Ramond}

\newcommand{\st}{space-time}
\newcommand{\ST}{string theory}

\newcommand{\sugra}{supergravity}

\newcommand{\wv}{world-volume}

\newcommand{\ft}{\tfrac}

\newcommand{\id}{\mathbbm{1}}

\newcommand{\SLTR}{SL(2,\mathbb{R})}
\newcommand{\SLTZ}{SL(2,\mathbb{Z})}

\newcommand{\wdg}{\ensuremath{{\scriptstyle \wedge}}}

\def\rmd{{\rm d}}

\newcommand{\C}[1]{C^{(#1)}}
\newcommand{\G}[1]{G^{(#1)}}

\newcommand{\intdep}{U}
\newcommand{\nephat}{}

\newcommand{\CL}{\ensuremath{\mathcal{L}}}

\begin{document}

 \selectlanguage{english}

 \rightline{\tt hep-th/0408175}
 \rightline{UG-04/05}

 \thispagestyle{empty}
 \vskip 1.5truecm

 \centerline{{\LARGE \bf M-theory and Gauged Supergravities}}
 \vskip 1truecm

 \centerline{\bf D.~Roest\footnote{Based on the author's Ph.D. thesis, defended {\it cum laude} on June 25, 2004.}}
 \bigskip
 \centerline{Centre for Theoretical Physics, University of Groningen}
 \centerline{Nijenborgh 4, 9747 AG Groningen, The Netherlands}
 \centerline{E-mail: {\tt d.roest@phys.rug.nl}}

 \vskip 1.8truecm

\centerline{ABSTRACT} \bigskip

We present a pedagogical discussion of the emergence of gauged supergravities from
M-theory. First, a review of maximal supergravity and its global symmetries and
supersymmetric solutions is given. Next, different procedures of dimensional reduction
are explained: reductions over a torus, a group manifold and a coset manifold and
reductions with a twist. Emphasis is placed on the consistency of the truncations, the
resulting gaugings and the possibility to generate field equations without an action.

Using these techniques, we construct a number of gauged maximal supergravities in diverse
dimensions with a string or M-theory origin. One class consists of the $CSO$ gaugings,
which comprise the analytic continuations and group contractions of $SO(n)$ gaugings. We
construct the corresponding half-supersymmetric domain walls and discuss their uplift to
D- and M-brane distributions. Furthermore, a number of gauged maximal supergravities are
constructed that do not have an action.

\vskip 2.8truecm

\centerline{}

\vfill \eject

 \tableofcontents

 \chapter{Introduction}

This review article\footnote{This article is based on the author's Ph.D. thesis, which
also includes a historical introduction to high-energy physics and a crash course on
perturbative string theory, while the other chapters are virtually identical to the
material presented here. The thesis can be found on {\it
http://www.ub.rug.nl/eldoc/dis/science/d.roest/}; if you are interested in a hard copy
version, please contact me.} deals with the construction of different gauged
supergravities that arise in the framework of string and M-theory. The latter are thought
to be consistent theories of quantum gravity, unifying the four different forces. One of
their particular features is their critical dimension: these theories necessarily live in
ten or eleven dimensions. For this reason one needs a procedure to obtain effective
four-dimensional descriptions, which goes under the name of Kaluza-Klein theory or
dimensional reduction. In this article we will discuss a number of possible dimensional
reductions and the resulting lower-dimensional descriptions. It will be useful to be
acquinted with perturbative string theory (see e.g.~\cite{Green:1987sp, Green:1987mn,
Polchinski:1998rq}) and the basic concepts of string dualities (see
e.g.~\cite{Schwarz:1997bh, Sen:1998kr, Polchinski:1998rr}); in this article, emphasis
will be placed on supergravity aspects.

Proposed in 1974 \cite{Scherk:1974ca}, the idea of string theory as a theory of quantum
gravity was not really picked up until the "first superstring revolution" in the mid
1980s. After this period, there were five different perturbative superstring theories:
four of closed strings (type IIA, IIB and heterotic with gauge group ${E_8 \times E_8}$
or ${SO(32)}$) and one of open and closed strings (type I). This situation changed with
the discovery of string dualities, culminating in the "second superstring revolution" in
the mid 1990s. It was found that the different string theories are related to each other
for different values of certain parameters; for example, the strong coupling limit of one
theory yields another theory at weak coupling (S-duality) and string theories on
different backgrounds are equivalent (T-duality). The upshot was that the five string
theories could be unified in a single eleven-dimensional theory, which was named M-theory
\cite{Witten:1995ex}. The different string theories are thus understood as perturbative
expansions in different limits of the parameter space of M-theory. This appreciation is
known as U-duality \cite{Hull:1995ys} and has spectacularly changed our understanding of
string theory and the distinction between perturbative and non-perturbative effects.

Of central importance for the different dualities are D$p$-branes \cite{Dai:1989ua,
Polchinski:1995mt}, which are extended objects of $p$ spatial dimensions. These branes
are required to fill out the multiplets of string dualities, e.g.~the fundamental string
is mapped onto the D$1$-brane under S-duality. In addition, different descriptions of
D-branes play a crucial role in the string theory calculation \cite{Strominger:1996sh} of
the Bekenstein-Hawking entropy of a black hole and in the AdS/CFT correspondence
\cite{Maldacena:1998re}, relating a string theory in a particular background (IIB on
AdS$_5$ $\times$ S$^5$) to a particular and supersymmetric QFT ($N=4$ SYM in $D=4$).

The low-energy limit of string theory, supergravity, has proven to be an important tool
to study the different phenomena in string theory. Many features of string and M-theory
are also present in its supergravity limit, such as D-branes and U-duality, and it is
therefore interesting to study this effective description. In particular, one can extract
effective lower-dimensional descriptions by considering string or M-theory on a compact
internal manifold, which is taken to be very small (i.e.~dimensional reduction).
Different reductions give rise to different lower-dimensional supergravities. Thus it is
clearly very desirable to have a proper understanding of the different reduction
procedures and their resulting lower-dimensional descriptions. In particular, we will be
interested in gauged supergravities as the lower-dimensional theories.

Ungauged supergravities have a global symmetry group $G$, which is a consequence of the
U-duality of M-theory. In gauged supergravities a subgroup of this global group is
elevated to a gauge symmetry by the introduction of mass parameters. The combination of a
gauge group and local supersymmetry implies the appearance of a scalar potential, which
is quadratic in the mass parameters.

It is the scalar potential which makes gauged supergravities interesting since it
generically breaks the Minkowski vacuum to solutions like (Anti-)de Sitter space-time
(AdS or dS), domain walls or cosmological solutions. These play important roles in the
AdS/CFT correspondence\footnote{Indeed, the effective description of IIB string theory on
the particular background AdS$_5$ $\times$ S$^5$ is a gauged supergravity: the $N=4$
$SO(5)$ theory in $D=5$, see also section~\ref{sec:CSO-gaugings}.} and its
generalisation, the DW/QFT correspondence \cite{Itzhaki:1998dd, Boonstra:1998mp},
brane-world scenarios \cite{Randall:1999ee, Randall:1999vf} and accelerating cosmologies
\cite{Townsend:2001ea, Kallosh:2001gr}. From various points of view, it would therefore
be highly advantageous to have a classification of gauged supergravities in the different
dimensions.

We are only interested, however, in gauged supergravities with a higher-dimensional
origin in string or M-theory: the lower-dimensional theory must be obtainable via
dimensional reduction. Our approach consists of the dimensional reduction of eleven- and
ten-dimensional maximal supergravities and the investigation of the resulting gauged
supergravity. We have applied two reduction methods, both preserving supersymmetry:
reduction with a twist and reduction on a group manifold. In the twisted reduction one
employs a global symmetry of the parent theory to induce a gauging of one of its
subgroups in the lower dimension. In the group manifold reduction one reduces over a
number of isometries that do not commute and form the algebra of a Lie group. This
results in the gauging of this group in the lower dimension. The consistency of both
reductions is guaranteed by symmetry, as proven by Scherk and Schwarz in 1979
\cite{Scherk:1979ta, Scherk:1979zr} and as opposed to reduction on a coset manifold,
whose consistency remains to be understood in generality.

The outline of this review article is as follows. Chapter~\ref{ch:supergravity} is
devoted to supergravity, the low-energy limit of string and M-theory. In particular, we
focus on the maximal supergravities, their global symmetries and their supersymmetric
solutions. In chapter~\ref{ch:reductions} we describe a number of techniques to generate
lower-dimensional gauged supergravities. Reduction over a torus, with a twist, over a
group manifold and over a coset manifold are explained, with proper attention to the
consistency of the truncation and the resulting gauging. In the last section of this
chapter we discuss a subtlety which can arise for certain dimensional reductions,
yielding gauged supergravities without an action. This concludes the more general part of
this review.

One finds the application of the different dimensional reductions in
chapter~\ref{ch:gauged}, where different gauged theories are constructed. By applying
reductions with a twist and over a group manifold, we generate a number of gaugings in
ten, nine and eight dimensions. We also discuss the class of $CSO$ gaugings in lower
dimensions, which are obtainable by reduction over coset or other manifolds. Finally, in
chapter~\ref{ch:domain-walls} we construct and discuss half-supersymmetric domain wall
solutions for the different gauged supergravities. The topic of the first section is the
D8-brane. Next, we treat the lower-dimensional domain walls and their relation to
higher-dimensional branes, with a special treatment of the 9D and 8D cases. We end with a
discussion of 1/4-supersymmetric intersections of domain walls and strings.

 \chapter{Supergravity} \label{ch:supergravity}

As mentioned in the introduction, supergravities in ten and eleven dimensions emerge as
the effective low-energy description of string and M-theory. In this chapter we will
discuss supersymmetry and supergravity in various dimensions, some supersymmetric
solutions and their relations.

\section{Supersymmetry} \label{sec:susy}

\subsection{Superalgebra and Supercharges} \label{sec:susy-1}

The symmetry of supergravity theories is the super-Poincar\'{e} symmetry, which is an
extension of the usual Poincar\'{e} symmetry of gravity theories with the generators of
supersymmetry. Thus, it contains the Lorentz generators, the generators of translations
(a vector under the Lorentz symmetry) and the supersymmetry generators (spinors under the
Lorentz symmetry). In addition, the super-Poincar\'{e} algebra, or superalgebra in short,
can be extended with a number of gauge generators, which are bosonic generators whose
parameter is a $p$-form.

Due to the intertwining of the fermionic generators of supersymmetry and the bosonic
generators of translations and gauge symmetries in the superalgebra\footnote{The
Poincar\'{e} symmetry and gauge symmetries always form a direct product in a bosonic
group~\cite{Coleman:1967ad}. A non-trivial intertwining of these symmetries is only
possible when including fermionic generators~\cite{Haag:1975qh}.}, the requirement of
local supersymmetry \cite{Freedman:1976xh} has profound implications. In particular, it
leads to the inclusion of gravity, due to the presence of translations in the
superalgebra. Thus any locally supersymmetric theory contains gravity and is usually
called a supergravity.

For the discussion of supersymmetry in $D$ dimensions we will now consider fermionic
representations of the Lorentz group $SO(1,D-1)$. This is the Dirac representation and its
generators are given by $[ \Gamma_\mu, \Gamma_\nu$], where the $\Gamma$-matrices $\Gamma_\mu$
satisfy the Clifford algebra
\begin{align}
 \{ \Gamma_\mu, \Gamma_\nu \} = 2 \eta_{\mu \nu} \,.
\end{align}
The dimension\footnote{We always refer to the {\it real} dimension.} of this
representation of the Clifford algebra is $2^{[D/2]+1}$, where the notation $[D/2]$ means
the integer part of $D/2$.

Since spinors transform under the fermionic representation of the Lorentz group, their
number of components in principle equals the dimension of the Dirac representation. These
are called Dirac spinors. However, in certain dimensions Dirac spinors are reducible,
allowing one to impose conditions that are preserved under Lorentz symmetry. For example,
in even dimensions one can impose a chirality condition: spinors are required to have
eigenvalue $\pm 1$ under the chirality operator
 \begin{align}
  \Gamma_c = i^{D/2-1} \Gamma_{\underline{01 \cdots D-1}} \,,
 \end{align}
giving rise to Weyl spinors. In other cases it is possible to impose a reality condition,
leading to Majorana spinors. In addition it is possible that both these conditions can be
imposed, leading to Majorana-Weyl spinors. In table~\ref{tab:spinors} we give the minimal
spinors in different dimensions and their number of components $q$, where minimal spinors
have the smallest number of components, i.e.~all possible and mutually consistent
conditions are imposed. A more detailed account can be found in
e.g.~\cite{VanProeyen:1999ni, deWit:2002vz}.

\begin{table}[ht]
 \begin{center}
 \begin{tabular}{||c||c|c||}
  \hline \rule[-1mm]{0mm}{6mm}
 Dimension & Spinors & Components ($q$) \\
  \hline
  \hline \rule[-1mm]{0mm}{6mm}
 2 mod 8 & Maj.-Weyl & $2^{D/2-1}$ \\
  \hline \rule[-1mm]{0mm}{6mm}
 3,9 mod 8 & Majorana & $2^{(D-1)/2}$ \\
  \hline \rule[-1mm]{0mm}{6mm}
 4,8 mod 8 & Maj. / Weyl & $2^{D/2}$ \\
  \hline \rule[-1mm]{0mm}{6mm}
 5,7 mod 8 & Dirac & $2^{(D+1)/2}$ \\
  \hline \rule[-1mm]{0mm}{6mm}
 6 mod 8 & Weyl & $2^{D/2}$ \\
 \hline
\end{tabular}
 \caption{\it The different minimal spinors in different space-time dimensions and their number of
 components. Note that one can define either Majorana or Weyl but not Majorana-Weyl spinors in $D=4,8$ mod $8$.}
 \label{tab:spinors}
\end{center}
\end{table}

The parameter of supersymmetry is a spinor and thus the number of supercharges $Q$, associated to
supersymmetry generators, is always a multiple $N$ of the dimension of the irreducible
representation:
 \begin{align}
  Q = N q \,.
 \end{align}
However, there is a bound on the number of supercharges \cite{Nahm:1978tg}. For theories with
global supersymmetry, thus not containing gravity, the bound is 16 supercharges. Theories with
local supersymmetry, therefore including gravity, can have up to 32 supercharges. Superalgebras
with more than 32 supercharges will only have representations that include states of helicity
higher than two. When coupling these to other fields one breaks the associated gauge symmetry, thus
rendering the interaction inconsistent. For this reason these higher-spin theories are usually
discarded, although there are attempts to remedy the problems \cite{Vasiliev:1996dn}. Theories with
exactly 32 supercharges are called maximal supergravities.

\begin{table}[ht]
 \begin{center}
 \begin{tabular}{||c|c||}
  \hline \rule[-1mm]{0mm}{6mm}
 Dimension &  Supergravity ($N$) \\
  \hline \hline \rule[-1mm]{0mm}{6mm}
 11 & 1 \\
  \hline \rule[-1mm]{0mm}{6mm}
 10 & 1, IIA, IIB \\
  \hline \rule[-1mm]{0mm}{6mm}
 7,8,9 & 1,2 \\
  \hline \rule[-1mm]{0mm}{6mm}
 6 & 1, iia, iib, 4 \\
 \hline
\end{tabular}
 \caption{\it Supergravity in different space-time dimensions, labelled by their number of
 supersymmetry generators.}
 \label{tab:sugras}
\end{center}
\end{table}

\subsection{Possible Supergravity Theories} \label{sec:susy-2}

When combining the bound on the number of supercharges with the dimension of the minimal
spinor in the different dimensions, we can survey the different possibilities for $N$ in
different dimensions\footnote{We will always restrict ourselves to $D > 2$, since
theories in two dimensions are special in many respects.}, as summarised in
table~\ref{tab:sugras}. One dramatic conclusion is that in dimensions twelve or higher
there are no supergravity theories\footnote{At least with Lorentzian signature, as is our
assumption here.} since the dimension of the minimum spinor is 64. Thus $D=11$ is the tip
of the pyramid of supergravities, where one can only have maximal supergravity with 32
supercharges. We will discuss 11D supergravity in subsection~\ref{sec:parent-sugra-1}.

In ten dimensions one can have either $N=1$ or $N=2$ supersymmetry, corresponding to 16
or 32 supercharges, respectively. Only the first of these cases does not necessarily
contain gravity. The second case contains two Majorana-Weyl spinors of certain
chiralities and thus allows for two different theories with spinors of either the
opposite or the same chirality: type IIA and IIB supergravity with $(1,1)$ or $(2,0)$
supersymmetry, respectively (in this notation the first and second entries denote the
number of supersymmetry generators with positive and negative chirality, respectively).
In fact, $D=10$ is the only dimension which has two inequivalent maximal supergravities;
it is unique in all other dimensions. We will discuss IIA and IIB supergravity in
subsection~\ref{sec:parent-sugra-3}.

The structure of maximal supergravities in ten dimensions nicely dovetails with the
possible string theories with maximal supersymmetry. In ten dimensions, one has IIA and
IIB string theory, whose low-energy effective actions are provided by the corresponding
supergravities. For a long time, it was somewhat of a mystery what eleven-dimensional
supergravity should correspond to (i.e.~of which underlying theory it should be the
effective action). This was clarified by the appearance of eleven-dimensional M-theory in
the strong-coupling limit of IIA string theory \cite{Townsend:1995kk, Witten:1995ex,
Schwarz:1996jq}, see also subsection~\ref{sec:parent-sugra-4}.

Another interesting phenomenon occurs in six dimensions, where there are Weyl spinors
with eight components. Of the maximal superalgebras with $N=4$, only the $(2,2)$ case
gives rise to a supergravity theory; other choices contain states with higher helicity.
When considering 16 supercharges, there are two choices: one finds $(1,1)$ and $(2,0)$
supersymmetry as well, leading to two distinct $Q=16$ supergravities in six dimensions,
labelled iia and iib. In all other dimensions than six, the superalgebra with $Q=16$
supercharges is unique.

We would like to make a few remarks about the explicit supergravity realisation of the
superalgebras. The supergravity fields form massless multiplets under supersymmetry,
called supermultiplets. These are usually christened after the field with the highest
helicity. The best-known example is the graviton multiplet, which includes the graviton
(spin 2), the gravitino (spin 3/2) and fields with lower spin. All supergravity theories
contain this multiplet. Maximal supersymmetry only allows for this supermultiplet while a
smaller amount of supersymmetry allows for other multiplets without gravity as well.
Examples are the gravitino and the vector multiplet with highest spins $3/2$ and $1$,
respectively; see subsection~\ref{sec:parent-sugra-2}.

\begin{table}[ht]
 \begin{center}
 \begin{tabular}{||c|c|c|c||}
  \hline \rule[-1mm]{0mm}{6mm}
 Name & Symbol & Spin & On-shell d.o.f. \\
  \hline \hline \rule[-1mm]{0mm}{6mm}
 Graviton & $g_{\mu \nu}$ & $2$ & $(D-2)(D-1)/2 -1$ \\
  \hline \rule[-1mm]{0mm}{6mm}
 Gravitino & $\psi_\mu$ & $3/2$ & $(D-3) \cdot q /2$ \\
  \hline \rule[-1mm]{0mm}{6mm}
 Rank-$d$ potential & $C^{(d)}_{\mu_1 \cdots \mu_d}$ & $1$ & $\left( \; \begin{array}{c} D-2 \\ d \end{array} \; \right)$ \\
  \hline \rule[-1mm]{0mm}{6mm}
 Dilatino & $\lambda$ & $1/2$ & $q / 2$ \\
  \hline \rule[-1mm]{0mm}{6mm}
 Scalar & $\phi$ or $\chi$ & $0$ & $1$ \\
 \hline
\end{tabular}
 \caption{\it On-shell degrees of freedom of D-dimensional supergravity fields.}
 \label{tab:dof}
\end{center}
\end{table}

For supersymmetry to be a consistent symmetry, all supermultiplets must have an on-shell
matching of bosonic and fermionic degrees of freedom. The on-shell degrees of freedom are
multiplets of the little group $SO(D-2)$ for massless fields and are given in
table~\ref{tab:dof} for generic supergravity fields\footnote{We distinguish between two
types of scalars: dilatons $\phi$ and axions $\chi$. Loosely speaking, the difference
between these is that axions only appear with a derivative whereas the dilatons also
occur without it. A stricter definition of this distinction will be discussed in
subsection~\ref{sec:global-symmetries-1}.}${}^,$\footnote{In the case $d = (D-2)/2$ one
can impose a self-duality constraint on the $(d+1)$-form field strength. The potential
would then give rise to half the degrees of freedom as listed in table~\ref{tab:dof}.}.
Note that a $d$-form potential $C^{(d)}$ carries the same amount of degrees of freedom as
a $\tilde{d}$-form potential with $\tilde{d} = D-2-d$. What corresponds to an electric
charge in one potential is a magnetic charge in its dual potential and vice versa. This
equivalence between two potentials is called Hodge duality and is a generalisation of the
well-known electric-magnetic duality in 4D to higher ranks $d$ and $\tilde{d}$ and
dimension $D$.

\section{Maximal Supergravities in 11D and 10D} \label{sec:parent-sugra}

\subsection{Supergravity in 11D} \label{sec:parent-sugra-1}

In eleven dimensions one has maximal supersymmetry. The superalgebra allows for the
inclusion of a rank-2 and a rank-5 gauge symmetry. As is always the case with maximal
supersymmetry, there is only one massless supermultiplet, the graviton multiplet. It
consists of the on-shell degrees of freedom
 \begin{align}
   \text{D=11:} \qquad ({\bf 44} + {\bf 84})_\text{B} + ({\bf 128})_\text{F} \,,
   \label{SO(9)dof}
 \end{align}
which are multiplets of $SO(9)$. In 11D supergravity theory the graviton multiplet is usually
represented by the fields
 \begin{align}
  \text{D=11:} \qquad
  \{ { e}_{{\mu}}{}^{{ a}} , { C}_{{\mu}{\nu}{\rho}} ;
    { \psi}_{ { \mu}}  \} \,.
 \end{align}
These are the Vielbein, a three-form gauge potential and a Majorana gravitino,
respectively. The bosonic part of the corresponding Lagrangian \cite{Cremmer:1978km}
reads
\begin{align}
  \mathcal{L} = \sqrt{-{g}}
    [ {R} -\tfrac{1}{2} G \cdot G - \tfrac{1}{6} \star (G \wdg G \wdg C) ] \,,
 \label{11Daction}
\end{align}
where ${G} = d C$. Note that it consists of the Einstein-Hilbert term, a kinetic term for
the rank-three potential and a Chern-Simons term. The latter only depends on the
rank-three potential and is independent of the metric; for this reason it is also called
a topological term.

The 11D supergravity theory has an $\mathbb{R}^+$ symmetry which acts as
 \begin{align}
  g_{\mu \nu} \rightarrow \lambda^2 g_{\mu \nu} \,, \qquad
  C_{\mu \nu \rho} \rightarrow \lambda^3 C_{\mu \nu \rho} \,, \qquad
  \psi_\mu \rightarrow \lambda^{1/2} \psi_\mu \,,
 \label{11Dtrombone}
 \end{align}
with $\lambda \in \mathbb{R}^+$. Two remarks are in order here. The above symmetry acts
covariantly on the field equations (as all symmetries) but does not leave the Lagrangian
invariant: it transforms as ${\cal L} \rightarrow \lambda^{9} {\cal L}$. All terms in
$\cal L$ scale with the same weight: for this reason it is called a trombone
symmetry\footnote{Alternatively, such trombone symmetries can be seen as a scaling of the
only length scale of the theory, i.e.~Newton's constant $G_N$ or the string length
$\alpha^\prime$, see e.g.~\cite{deWit:2002vz}. We thank Bernard de Wit for pointing this
out.} \cite{Cremmer:1998xj}. Secondly, the covariant scaling of $\cal L$ only holds at
lowest order. Higher-derivative corrections will scale with different weights and thus
break the symmetry \eqref{11Dtrombone} of the field equations.

The occurrence of trombone symmetries will be a generic feature in ungauged or massless
supergravities. The weights of the fields are always determined by a simple rule: for the
bosonic fields the weights equal the number of Lorentz indices while for the fermions it
is one-half less. The Lagrangian will scale as ${\cal L} \rightarrow \lambda^{D-2} {\cal
L}$ under such symmetries. The scaling of bosonic terms is easily understood from the two
derivatives they contain. Thus this symmetry is broken by terms with less (as in scalar
potentials, to be encountered in chapter~\ref{ch:gauged}) or more (as in higher-order
corrections) than two derivatives.

\subsection{Minimal Supergravity in 10D} \label{sec:parent-sugra-2}

In 10D the minimal spinor is a 16-component Majorana-Weyl spinor. Minimal $N=1$
supersymmetry in 10D therefore has 16 supercharges. Its superalgebra allows for the
inclusion of a rank-one and a self-dual rank-five gauge symmetry. Being non-maximal
supersymmetry, one finds different supermultiplets \cite{Nahm:1978tg}:
 \begin{align}
  \text{N=1:} \qquad
   \begin{cases}
    \text{vector:} \qquad &
     ({\bf 8}_\text{v})_\text{B} + ({\bf 8}_\text{c})_\text{F} \,, \\
    \text{graviton:} \qquad &
     [ {\bf 8}_\text{v} + {\bf 8}_\text{c} ] \times {\bf 8}_\text{v}
     = ({\bf 35}_\text{v} + {\bf 28} + {\bf 1})_\text{B} + ({\bf 56}_\text{s} + {\bf 8}_\text{s})_\text{F} \,, \\
    \text{gravitino A:} \qquad &
     [ {\bf 8}_\text{v} + {\bf 8}_\text{c} ] \times {\bf 8}_\text{s}
     = ({\bf 56}_\text{v} + {\bf 8}_\text{v})_\text{B} + ({\bf 56}_\text{s} + {\bf 8}_\text{s})_\text{F} \,, \\
    \text{gravitino B:} \qquad &
     [ {\bf 8}_\text{v} + {\bf 8}_\text{c} ] \times {\bf 8}_\text{c}
     = ({\bf 35}_\text{c} + {\bf 28} + {\bf 1})_\text{B} + ({\bf 56}_\text{s} + {\bf 8}_\text{s})_\text{F}
    \,,
   \end{cases}
 \label{N=1supermultiplets}
 \end{align}
Note that the little group $SO(8)$ has three 8-dimensional representations: one bosonic, the vector
${\bf 8}_\text{v}$, and two fermionic, spinors of opposite chirality ${\bf 8}_\text{c}$ and ${\bf
8}_\text{s}$. This special property of $SO(8)$ is known as triality.

Due to the appearance of several supermultiplets, non-maximal supergravity is not unique.
It always contains the graviton multiplet, which can be coupled in various ways to vector
multiplets, leading to different Yang-Mills sectors. An example is provided by the
low-energy limit of the three $N=1$ string theories, which consist of the graviton
multiplet plus 496 vector multiplets to obtain the $SO(32)$ or $E_8 \times E_8$ gauge
groups \cite{Green:1984sg}.

\subsection{IIA and IIB Supergravity} \label{sec:parent-sugra-3}

Turning to maximal $N=2$ supersymmetry in 10D, one has two possibilities: one can choose
Majorana-Weyl spinors of either opposite or equal chirality, leading to the non-chiral
IIA or the chiral IIB supergravity theories with $(1,1)$ and $(2,0)$ supersymmetry,
respectively. The IIA superalgebra can be extended with gauge symmetries of rank 0,1,2,4
and 5, while IIB allows for 1,1,3,5$^+$,5$^+$ and 5$^+$, where all five-form gauge
parameters  5$^+$ are self-dual. In fact, the IIB superalgebra has an additional $SO(2)$
R-symmetry, rotating the two supersymmetry spinors of equal chirality. Under this
R-symmetry, the central charges form doublets (for rank 1 and 5$^+$) and singlets (for
rank 3 and 5$^+$). We will discuss R-symmetries of lower-dimensional superalgebras in
section~\ref{sec:global-symmetries}.

As always, maximal supersymmetry allows for only one massless multiplet, whose on-shell degrees of
freedom are given by
 \begin{align}
   \text{IIA:} \qquad
    [ {\bf 8}_\text{v} + {\bf 8}_\text{c} ] \times [ {\bf 8}_\text{v} + {\bf 8}_\text{s} ] = \;
    & [ ({\bf 35}_\text{v} + {\bf 28} + {\bf 1})_\text{NS-NS} + ( {\bf 56}_\text{v} + {\bf 8}_\text{v} )_\text{R-R}
    ]_\text{B} \notag \\
    & + [ ({\bf 56}_\text{s} + {\bf 8}_\text{s})_\text{NS-R} + ({\bf 56}_\text{c} + {\bf
    8}_\text{c})_\text{R-NS}]_\text{F} \,, \notag \\
   \text{IIB:} \qquad
    [ {\bf 8}_\text{v} + {\bf 8}_\text{c} ] \times [ {\bf 8}_\text{v} + {\bf 8}_\text{c} ] = \;
    & [ ({\bf 35}_\text{v} + {\bf 28} + {\bf 1})_\text{NS-NS} + ( {\bf 35}_\text{c} + {\bf 28} + {\bf 1} )_\text{R-R}
    ]_\text{B} \notag \\
    & + [ ({\bf 56}_\text{s} + {\bf 8}_\text{s})_\text{NS-R} + ({\bf 56}_\text{s} + {\bf
    8}_\text{s})_\text{R-NS}]_\text{F} \,,
 \label{IIABonshell}
 \end{align}
Note that these $N=2$ supermultiplets are constructed from the $N=1$ supermultiplets: both $N=2$
graviton multiplets consist of the $N=1$ graviton and a gravitino multiplet. This is possible in
10D due to triality, which yields $N=1$ graviton and gravitino multiplets of equal size.

We will now consider the field-theoretic realisation of the graviton multiplet. The
common bosonic subsector, which is called the NS-NS subsector, contains gravity, a
rank-two potential and a dilaton. The remaining bosonic part is called the \RR\ subsector
and will only contain R-R rank-$d$ potentials where $d$ is odd in IIA and even in IIB.
The standard forms of the theories have $d=1,3$ for IIA and $d=0,2,4$ for IIB:
 \begin{align}
  & \text{IIA:} \qquad \{ {g}_{{\mu} {\nu}}, {B}_{\mu\nu}, {\phi},  \C{1}_{\mu},
   \C{3}_{\mu\nu\rho}; {\psi}_{\mu}, {\lambda} \} \, , \notag \\
  & \text{IIB:} \qquad \{ {g}_{{\mu} {\nu}}, {B}_{\mu\nu}, {\phi}, \C{0},
   {C}^{(2)}_{\mu\nu}, {C^{(4)+}_{\mu\nu\rho\sigma}}; {\psi}_{\mu}, {\lambda}
     \}\,.
 \label{IIABfieldcontentstandard}
 \end{align}
In the IIA case the fermions are real and contain two minimal spinors of both
chiralities, while in the IIB case they are complex and contain two minimal spinors of
the same chirality. The field strength of the IIB rank-four potential $C^{(4)+}$
satisfies a self-duality constraint, halving the number of degrees of freedom.

We would also like to present a special formulation of IIA and IIB supergravity which emphasises
the equivalence of dual R-R potentials, based on \cite{Bergshoeff:2001pv}, and introduces an extra
feature of IIA supergravity. To this end we will enlarge the field content by including all odd or
even R-R potentials, thus allowing for the ranges $d=1,3,5,7,9$ and $d=0,2,4,6,8$. The field
contents of IIA and IIB supergravity read in the double formulation
 \begin{align}
  \text{IIA}: & \qquad \{
  g_{\mu \nu},
  B_{\mu \nu},
  \phi,
  C^{(1)}_{\mu},
  C^{(3)}_{\mu \nu \rho},
  C^{(5)}_{\mu \cdots \rho},
  C^{(7)}_{\mu \cdots \rho};
  \psi_\mu,
  \lambda
  \} \,, \notag \\
  \text{IIB}: & \qquad \{
  g_{\mu \nu},
  B_{\mu \nu},
  \phi,
  C^{(0)},
  C^{(2)}_{\mu \nu},
  C^{(4)}_{\mu \cdots \rho},
  C^{(6)}_{\mu \cdots \rho},
  C^{(8)}_{\mu \cdots \rho};
  \psi_\mu,
  \lambda \}
  \, .
 \label{IIABfieldcontentdemocratic}
 \end{align}
To get the correct number of degrees of freedom, one must by hand impose duality
relations between the field strengths of rank-$d$ and rank-$(8-d)$ potentials, which read
\cite{Bergshoeff:2001pv}
 \begin{align}
  G^{(d+1)} = (-)^{[(d+1)/2]} \, e^{(d-4) \phi /2} \, \star G^{(9-d)} \,, \qquad
  G^{(d+1)} = d C^{(d)} - H \wdg C^{(d-2)} \,, \label{RR-duality}
 \end{align}
for vanishing fermions and where $H = d B$. The (bosonic part of the) field equations for
$\C{d}$ can be derived from the action \cite{Bergshoeff:2001pv}
 \begin{align}
  L = \sqrt{-g} [ & R - \tfrac{1}{2} (\partial{\phi})^2 - \tfrac{1}{2} e^{- \phi} H \cdot H
    - \sum_{d} \tfrac{1}{4} e^{(4-d) \phi /2} G^{(d+1)} \cdot G^{(d+1)} ] \,,
 \label{democratic-action}
 \end{align}
subject to the duality relations \eqref{RR-duality}. Due to these constraints, the above
is called a pseudo-action \cite{Bergshoeff:1996sq}. Note that the doubling of \RR\
potentials has two effects: the kinetic terms have coefficients $1/4$ instead of the
canonical $1/2$ and there are no explicit Chern-Simons terms in the action.

We would like to make the following two remarks. Note that the duality constraint on the
five-form field strength of IIB can not be eliminated, in contrast to the other duality
relations; it is a constraint on one field strength $G^{(5)}$ while the others relate two
different field strengths $G^{(d+1)}$ and $G^{(9-d)}$ for $d \neq 4$.

Secondly, one can include a nine-form potential $\C{9}$ in
\eqref{IIABfieldcontentdemocratic}, which carries no degrees of freedom (and thus is
consistent with \eqref{IIABonshell}) but is very natural from the point of view of R-R
equivalence \cite{Polchinski:1995mt}. The corresponding field strength trivially
satisfies the Bianchi identity. Its Hodge dual is a rank-zero field strength, which has
no corresponding potential nor a field equation. Its Bianchi identity implies it to be
constant. Thus we have effectively introduced a mass parameter in the theory, given by
 \begin{align}
   G^{(0)} = e^{-5 \phi/2} \, \star G^{(10)} \,. \label{RR-duality-mass}
 \end{align}
The corresponding action is given by \eqref{democratic-action} with $d= -1,1,\ldots,9$
\cite{Bergshoeff:2001pv} and the field strengths \cite{Green:1996bh}
 \begin{align}
  G^{(d+1)} = d C^{(d)} - H \wdg C^{(d-2)} + \frac{1}{(d+1)/2 \,!} \, \G{0} B \wdg \ldots \wdg B
  \,.
 \end{align}
Due to the equivalence of the different formulations, one should expect this mass
parameter to appear in the normal formulation as well. Indeed this deformation to massive
IIA supergravity has been found \cite{Romans:1986tz}, shortly after the inception of its
massless counterpart \cite{Giani:1984wc, Campbell:1984zc}. In this chapter, we
concentrate on the massless part and we will come back to the massive deformations in
sections~\ref{sec:massive-gauged-IIA} and \ref{sec:D8-brane}. Also, we leave the
formulation with R-R equivalence here and return to the standard formulation
\eqref{IIABfieldcontentstandard}.

The (bosonic part of the massless) IIA Lagrangian is given by
\begin{align}
  L_{IIA} = \sqrt{-g} [ & R - \tfrac{1}{2} (\partial{\phi})^2 - \tfrac{1}{2} e^{- \phi} H \cdot H
    - \sum_{d=1,3} \tfrac{1}{2} e^{(4-d) \phi /2} G^{(d+1)} \cdot G^{(d+1)} + \notag \\
  & - \tfrac{1}{2} \star \, ( d C^{(3)} \wdg d C^{(3)} \wdg B )] \,,
\label{IIAaction}
\end{align}
The IIA theory has two $\mathbb{R}^+$ symmetries. The first is a symmetry of the
Lagrangian \eqref{IIAaction} and is given by
 \begin{align}
  e^\phi \rightarrow \lambda e^\phi \,, \quad
  B_{\mu \nu}  \rightarrow \lambda^{1/2} B_{\mu \nu} \,, \quad
  C^{(1)}_\mu \rightarrow \lambda^{-3/4} C^{(1)}_\mu \,, \quad
  C^{(3)}_{\mu \nu \rho} \rightarrow \lambda^{-1/4} C^{(3)}_{\mu \nu \rho} \,,
 \label{IIAscale}
 \end{align}
with $\lambda \in \mathbb{R}^+$ and other fields invariant. The other is the 10D analog of the 11D
trombone symmetry \eqref{11Dtrombone} with weights as explained below the 11D weights.

The (bosonic part of the) field equations for IIB supergravity \cite{Schwarz:1983qr,
Howe:1984sr} can be derived from the Lagrangian
 \begin{align}
  L_{IIB} = \sqrt{-g} [ & R - \tfrac{1}{2} (\partial{\phi})^2 - \tfrac{1}{2} e^{- \phi} H \cdot H
    - \sum_{d=0,2,4} \tfrac{1}{2} e^{(4-d) \phi /2} G^{(d+1)} \cdot G^{(d+1)} + \notag \\
  & + \tfrac{1}{2} \star \, ( C^{(4)} \wdg d C^{(2)} \wdg H ) ] \,,
  \label{IIBaction}
 \end{align}
which has to be supplemented\footnote{An action without extra constraints can only be
constructed when including auxiliary fields \cite{Pasti:1997vs}.} with the self-duality
relation \eqref{RR-duality} for $d=4$ (for this reason it is called a pseudo-action
\cite{Bergshoeff:1996sq}). The IIB supergravity theory has a global $SL(2,\mathbb{R})$
symmetry \cite{Schwarz:1983wa}
\begin{align}
  {\tau} & \rightarrow \frac{a {\tau} +b}{c{\tau} +d} \,, \quad
  {B}^i  \rightarrow (\Omega^{-1})_j{}^i {B}^j \,, \quad
   C^{(4)} \rightarrow  C^{(4)} \,, \quad
  \Omega_i{}^j =
    \left( \begin{array}{cc} a&b\\c&d \end{array} \right) \in SL(2,\mathbb{R}) \,,
    \notag \\
  {\psi}_{{\mu}} & \rightarrow
    \left( \frac{c \, {\tau}^*+d}{c\, {\tau}+d} \right)^{1/4}
    {\psi}_{{\mu}} \,, \quad
  {\lambda}  \rightarrow \left( \frac{c \, {\tau}^*+d}{c \, {\tau}+d}
    \right)^{3/4} {\lambda} \,, \quad
  {\epsilon}  \rightarrow \left( \frac{c \, {\tau}^*+d}{c \, {\tau}+d}
    \right)^{1/4} {\epsilon} \,,
  \label{IIBSL2R}
\end{align}
where we have defined the doublet ${B}^i = (-B,C^{(2)})$ and the complex scalar $\tau =
\chi + i e^{-\phi}$ with the axion $\chi = C^{(0)}$. In terms of the real and imaginary
parts of $\tau$ the action of $SL(2,\mathbb{R})$ reads
\begin{align}
  e^\phi \rightarrow (c \chi +d)^2 e^{\phi} + c^2 e^{-\phi} \,, \qquad
  \chi \rightarrow \frac{ac + e^{2 \phi}(a\chi+b)(c\chi+d)}{c^2 + e^{2 \phi}(c \chi +d)^2} \,,
\end{align}
Note that the scalars transform non-linearly. We will discuss a more covariant way to
view this $SL(2,\mathbb{R})$ symmetry in section~\ref{sec:global-symmetries}. The
$SL(2,\mathbb{R})$ symmetry of IIB \sugra\ is broken to $\SLTZ$ in IIB \ST
\cite{Hull:1995ys}. The element $(a,b;c,d)=(0,1;-1,0)$ corresponds to the transformation
$\phi \rightarrow - \phi$ (for vanishing axion background), which relates the strong and
weak string coupling. For this reason this transformation is called S-duality. In
addition the IIB symmetry also has a trombone symmetry.

\subsection{Supergravity Relations and Dualities} \label{sec:parent-sugra-4}

As we will now show, the eleven- and ten-dimensional maximal supergravity theories are
not unrelated but rather can be connected via dimensional reduction. These relations can
be understood from the different dualities between the different string theories and
M-theory.

Ten-dimensional IIA supergravity can be obtained as a reduction of the unique
supergravity theory in $D=11$. This amounts to dimensionally reducing the 11D
supergravity, a procedure which is being elaborated upon in
section~\ref{sec:reduction-intro}, while only retaining the massless modes. The relations
between the supergravity fields are given in \eqref{11Dred}. Indeed, the full 11D
Lagrangian \eqref{11Daction} and supersymmetry transformations \eqref{11Dsusy} in this
way give rise to the IIA counterparts \eqref{IIAaction} and \eqref{IIAsusy}. In terms of
on-shell degrees of freedom, the 11D representations of $SO(9)$ \eqref{SO(9)dof} can be
decomposed into the IIA representations of $SO(8)$ \eqref{IIABonshell} via
 \begin{align}
   & \text{B}: \qquad {\bf 44} \rightarrow {\bf 35}_\text{v} + {\bf 8}_\text{v} + {\bf 1} \,, \qquad
   {\bf 84} \rightarrow {\bf 56}_\text{v} + {\bf 28} \,, \notag \\
   & \text{F}: \qquad {\bf 128} \rightarrow {\bf 56}_\text{s} + {\bf 8}_\text{s} + {\bf 56}_\text{c} + {\bf
   8}_\text{c} \,,
  \label{SO(9)decomposition}
 \end{align}
which reduces the 11D graviton multiplet to the IIA graviton multiplet.

As a side remark, from the relation \eqref{11Dred} between the supergravity fields one
can read off the following relations between the parameters of IIA and 11D on a circle:
 \begin{align}
  l_s{}^2 =  \frac{l_p{}^3}{R} \,, \qquad g_s = \Big( \frac{R}{l_p} \Big)^{3/2} \,,
 \end{align}
where $l_p$ is the 11D Planck length and $R$ the radius of the internal circle. This
supports the idea that strong coupling in IIA string theory corresponds to a large
radius, in which eleven-dimensional M-theory emerges. Though the appearance of
eleven-dimensional Lorentz covariance can not be proven in perturbative IIA string theory
(since its size is proportional to $e^\phi$), a lot of evidence for the existence of
M-theory has been put forward \cite{Townsend:1995kk, Witten:1995ex, Schwarz:1996jq}. For
example, the massive Kaluza-Klein states of 11D supergravity are interpreted as the
D$0$-brane states of IIA \ST\ \cite{Schwarz:1996jq, Sethi:1998pa}.

Similarly, IIA and IIB supergravity both reduce to the unique nine-dimensional maximal
supergravity. The corresponding reduction Ans\"{a}tze for the IIA and IIB supergravity
fields are given in \eqref{IIAred} and \eqref{IIBred}, respectively. These reduce the IIA
and IIB supersymmetry transformations and field equations to their 9D counterparts. Also
the IIA Lagrangian \eqref{IIAaction} can be reduced to the correct 9D action. The IIB
case requires a bit more discussion due to the self-duality constraint on the 5-form
field strength. Upon reduction it gives rise to a 4-form and a 5-form field strength and
a duality relation between the two. The latter can be used to eliminate either of the
field strengths, which is usually the 5-form. If properly treated the IIB
pseudo-Lagrangian \eqref{IIBaction} can also be reduced to the 9D Lagrangian. In terms of
on-shell degrees of freedom, the decompositions of the IIA and IIB representations of
$SO(8)$ \eqref{IIABonshell} under $SO(7)$ coincide, as can be read off explicitly:
 \begin{alignat}{2}
  & \text{IIA:}
   \begin{cases}
    \text{NS-NS:} \qquad & {\bf 35}_\text{v} \rightarrow {\bf 27} + {\bf 7} + {\bf 1} \,, \qquad
      {\bf 28} \rightarrow {\bf 21} + {\bf 7} \,, \qquad
     {\bf 1} \rightarrow {\bf 1} \,,  \\
    \text{R-R}: \qquad & {\bf 56}_\text{v} \rightarrow {\bf 35} + {\bf 21} \,, \qquad
     \quad \; {\bf 8}_\text{v} \rightarrow {\bf 7} + {\bf 1} \,,  \\
    \text{NS-R:} \qquad & {\bf 56}_\text{s} \rightarrow {\bf 48} + {\bf 8} \,, \qquad
     \quad \quad {\bf 8}_\text{s} \rightarrow {\bf 8} \,,  \\
    \text{R-NS:} \qquad & {\bf 56}_\text{c} \rightarrow {\bf 48} + {\bf 8} \,, \qquad
     \quad \quad {\bf 8}_\text{c} \rightarrow {\bf 8} \,, \\
   \end{cases} \notag \displaybreak[2] \\
  & \text{IIB:}
   \begin{cases}
    \text{NS-NS:} \qquad & {\bf 35}_\text{v} \rightarrow {\bf 27} + {\bf 7} + {\bf 1} \,, \qquad
      {\bf 28} \rightarrow {\bf 21} + {\bf 7} \,, \qquad
     {\bf 1} \rightarrow {\bf 1} \,,  \\
    \text{R-R}: \qquad & {\bf 35}_\text{c} \rightarrow {\bf 35} \,, \qquad \qquad \;
      \quad \; {\bf 28} \rightarrow {\bf 21} + {\bf 7} \,, \qquad {\bf 1} \rightarrow {\bf 1} \,,  \\
    \text{NS-R:} \qquad & {\bf 56}_\text{s} \rightarrow {\bf 48} + {\bf 8} \,, \qquad
      \quad \quad {\bf 8}_\text{s} \rightarrow {\bf 8} \,,  \\
    \text{R-NS:} \qquad & {\bf 56}_\text{s} \rightarrow {\bf 48} + {\bf 8} \,, \qquad
      \quad \quad {\bf 8}_\text{s} \rightarrow {\bf 8} \,,
   \end{cases}
 \label{SO(8)decomposition}
 \end{alignat}
Thus the massless modes of IIA and IIB supergravity on $S^1$ are equivalent and indeed
are described by the same effective theory, the unique $D=9$ maximal supergravity.

However, the massive modes of IIA and IIB supergravity on $S^1$, sometimes called
momentum modes, are distinct. For this reason, IIA and IIB supergravity are only
equivalent on very small circles, where such modes become infinitely massive (for more
detail, see section~\ref{sec:reduction-intro}). String theory modifies this situation in
the following way. Due to the fact closed strings can wind around the internal direction,
there is an entire tower of massive winding multiplets. Note that this phenomenon is
intrinsic to string theory and does not have a counterpart in field theory. It turns out
that the combination of massive momentum states and massive winding states yields the
same result for IIA and IIB \ST; to be precise, IIA on a circle with radius $R$ is
equivalent to IIB on a circle with radius $\tilde{R}$ with the relation $\tilde{R} =
\alpha' / R$ \cite{Dai:1989ua, Dine:1989vu}. Such a relation between theories on
different compactification manifold is generically called T-duality \cite{Giveon:1994fu}.
The towers of momentum and winding states are interchanged under the T-duality
transformation\footnote{A first confirmation can be found in the gauge vectors of 9D
supergravity that couple to these momentum and winding states. These are $A^{1}$ and $A$,
respectively, for the IIA theory and interchanged for the IIB theory, see \eqref{IIAred}
and \eqref{IIBred}. For a more extensive discussion of the inclusion of these massive
states in 9D supergravity, see \cite{Abou-Zeid:1999fv}.} on $S^1$. In accordance with
their accompanying string theories, the map between the (dimensionally reduced) IIA and
IIB supergravities is usually called T-duality.

The strong coupling limit of IIB \ST\ can be understood from its conjectured
$SL(2,\mathbb{Z})$ symmetry \cite{Hull:1995ys}. Indeed, this symmetry is shared by its
low-energy approximation and one of its generators acts on the IIB supergravity fields as
$\phi \rightarrow - \phi$ (for vanishing axion background). This corresponds to a
strong-weak coupling transformation due to the interpretation of the dilaton and is
called S-duality. For this reason, IIB \ST\ is understood to be self-dual\footnote{This
is very similar to the conjectured $\SLTZ$ duality of $N=4$ super-Yang Mills theory in 4D
\cite{Montonen:1977sn}.}. At weak coupling, strings are the fundamental, perturbative
\dof\ while at strong coupling, this role is played by the D$p$-branes with $p=1$.

\section{Scalar Cosets and Global Symmetries in $D \leq 9$} \label{sec:global-symmetries}

We now turn to the remaining maximal supergravities in $D \leq 9$. Being unique these can
all be obtained by dimensional reduction of any of the higher-dimensional theories, in
the same way that IIA supergravity can be obtained from 11 dimensions. Their construction
is rather straightforward and we will not consider it in great detail. One aspects
deserves proper discussion however: the scalar sector and its transformation under the
global symmetries of the theory. See \cite{Pope} for a clear discussion.

\subsection{Scalar Cosets} \label{sec:global-symmetries-1}

The field content of any $D \leq 9$-dimensional maximal supergravity is easily obtained
by dimensional reduction; its bosonic subsector consisting of gauge potentials is given
in table~\ref{tab:bosons}. The same holds for the Lagrangians and general formulae for
maximal supergravity in any dimension have been obtained \cite{Lu:1996yn}. The bosonic
part generically reads
 \begin{align}
  {\cal L}_D  = \sqrt{-g} [ R - \tfrac{1}{2} (\partial \vec{\phi})^2 - \sum_{d,i}
  \tfrac{1}{2} e^{\vec{\alpha}_d^i \cdot \vec{\phi}} G^{(d+1)}_i \cdot G^{(d+1)}_i ] + {\cal L}_{CS} \,,
 \end{align}
where the $G^{(d+1)}_i$ are rank-$(d+1)$ field strengths of gauge potentials $C^{(d)}_i$
with $d=0,\ldots,3$. The index $i$ denotes the different $d$-form potentials; its range
can be inferred from table~\ref{tab:bosons}. The number of dilatons $\vec{\phi}$ always
equals $11-D$ since all reduced dimensions will give rise to one dilaton. The length of
the vectors $\vec{\alpha}_d^i$ will always be given by
 \begin{align}
  \vec{\alpha}_d^i \cdot \vec{\alpha}_d^i = 4 - \frac{2 d (D-d-2)}{D-2} \,,
 \end{align}
in maximal supergravity.

Of special interest in this Lagrangian is the scalar sector, which we rewrite as
 \begin{align}
  {\cal L}_{\text{scalars}} = \sqrt{-g} [ - \tfrac{1}{2} (\partial \vec{\phi})^2 - \sum_{i}
  \tfrac{1}{2} e^{\vec{\alpha}^i \cdot \vec{\phi}} G^{(1)}_i \cdot G^{(1)}_i  ] \,.
  \label{scalars}
 \end{align}
where $\G{1}_i$ are the one-form field strengths of the axions $\chi^i$ and where we have
dropped the subscript $0$ on the vectors $\vec{\alpha}^i$. The vectors $\vec{\alpha}^i$
can be interpreted as positive root vectors of a simple Lie algebra. In the Cartan-Weyl
basis, the generators of this algebra are the Cartan generators $\vec{H}$, the positive
root generators $E_{\vec{\alpha}^i}$ and the negative root generators
$E_{-\vec{\alpha}^i}$ with commutation relations
 \begin{align}
  [ \vec{H} , \vec{H} ] = 0 \,, \qquad
  [ \vec{H} , E_{\vec{\alpha}^i} ] = \vec{\alpha}^i E_{\vec{\alpha}^i} \,, \qquad
  [ E_{\vec{\alpha}^i}, E_{\vec{\alpha}^j} ] = N(\vec{\alpha}^i, \vec{\alpha}^j) E_{\vec{\alpha}^i +
  \vec{\alpha}^j} \,,
 \label{Lie-algebra}
 \end{align}
and similarly for the negative root generators (replacing $\vec{\alpha}^i \rightarrow -
\vec{\alpha}^i$). The coefficients $N(\vec{\alpha}^i, \vec{\alpha}^j)$ are constants (possibly
zero) and characterise the algebra. We will now show that the scalar sector \eqref{scalars} is
invariant under the action of the corresponding semi-simple group $G$.

To this end we construct a particular representative of $G$, defined by\footnote{Other choices for
this representative are related by field redefinitions.}
 \begin{align}
  L = \text{exp} ( \sum_i \chi^i E_{\vec{\alpha}^i} ) \text{exp} ( -\vec{\phi} \cdot \vec{H}/2 ) \,,
 \label{L-representative}
 \end{align}
with parameters $\vec{\phi}$ corresponding to the Cartan generators and $\chi^i$ to the
positive root generators. This parameterises the coset $G/H$ with $H$ the maximal compact
subgroup of $G$. The group $H$ will turn out to be the R-symmetry group of the
superalgebra. Upon acting with a group element $g \in G$ from the left, the element $g L$
will generically no longer have the form of the $G/H$ representative
\eqref{L-representative}, i.e.~this can in general not be expressed as a transformation
$\vec{\phi} \rightarrow \vec{\phi}'$ and $\chi^i \rightarrow {\chi^i}'$. However, one can
employ the Iwasawa decomposition, which states that
 \begin{align}
  L \rightarrow g L = L' h \,,
 \label{Iwasawa}
 \end{align}
i.e.~the resulting matrix can be decomposed as $L'$ of the form \eqref{L-representative}
and a remainder $h \in H$. The latter will be dependent on $\vec{\phi}$ and $\chi^i$ in
general. Due to the Iwasawa decomposition we have defined a transformation
 \begin{align}
  L(\vec{\phi},\chi^i) \rightarrow L' = g L(\vec{\phi},\chi^i) h^{-1} = L (\vec{\phi}', {\chi^i}') \,,
 \label{scalartransformation}
 \end{align}
consisting of a left-acting $G$ element and a compensating right-acting $H$ element. Note
that for global $G$ transformations, the action of $H$ will be local due to the field
dependence via $\vec{\phi}$ and $\chi^i$. The $H$ transformation is called compensating
since it compensates for the $G$ transformation that does not preserve the $G/H$
representative \eqref{L-representative}.

The relevance of the transformation properties of $L$ stems from the fact that the scalar
kinetic terms \eqref{scalars} can be written as
 \begin{align}
  {\cal L}_{\text{scalars}} = \sqrt{-g} [ \tfrac{1}{4}\text{Tr}(\partial M \partial M^{-1}) ] \,,
  \label{scalarsLM}
 \end{align}
where we have defined $M = L L^T$. Note that $M$ does not see the compensating $H$
transformation: it transforms as $M \rightarrow g M g^T$ under
\eqref{scalartransformation}. Thus the scalar sector is by construction invariant under
global $G$ transformations. It turns out that this group is a symmetry not only of the
scalar subsector but of the entire theory\footnote{In many cases, however, the group $G$
is a symmetry of the equations of motion rather than the Lagrangian, since it requires
e.g.~the dualisation of some gauge potentials.}, i.e.~when also including the potentials
of higher rank and the fermions.

Let us take a step back and consider the significance of the compensating transformation
$H$. We have shown that the scalar kinetic terms \eqref{scalars} are invariant under the
global symmetry $G$ by constructing a particular $G/H$ representative $L$. Every $G$
transformation is accompanied by a compensating $H$ transformation to keep $L$ of the
same form. This can be seen as the gauge fixed version (with gauge choice
\eqref{L-representative}) of a more covariant system with global $G$ and local $H$
symmetry. The covariant system has kinetic term \eqref{scalarsLM} for arbitrary $L \in
G$. The extra degrees of freedom that are introduces in $L$ are cancelled by the extra
gauge degrees of freedom $L \rightarrow L h$ with $h \in H$ local. This is a completely
equivalent formulation of the scalar sector with advantages due to its covariance.

\subsection{Example: $SL(2,\mathbb{R})$ Symmetry of IIB} \label{sec:global-symmetries-2}

To make matters more concrete let us discuss the scalar sector of IIB supergravity as an example.
From its Lagrangian one reads off that it has one dilaton and one axion with positive root vector
$\alpha = 2$. This corresponds to the simple Lie algebra $sl(2)$ with generators (in the
fundamental representation)
 \begin{align}
  H = \left( \begin{array}{cc} 1 & 0 \\ 0 & -1 \end{array} \right) \,, \qquad
  E_{+2} = \left( \begin{array}{cc} 0 & 1 \\ 0 & 0 \end{array} \right) \,, \qquad
  E_{-2} = \left( \begin{array}{cc} 0 & 0 \\ 1 & 0 \end{array} \right) \,,
 \end{align}
satisfying the algebra \eqref{Lie-algebra}. Next we define the $SL(2,\mathbb{R})/SO(2)$
representative
 \begin{align}
  L = e^{\chi E_{+2}} e^{-\phi H/2} =
   \left( \begin{array}{cc} e^{-\phi/2} & e^{\phi/2} \chi \\ 0 & e^{\phi/2} \end{array}
   \right) \,.
  \end{align}
Any left-acting $SL(2,\mathbb{R})$ transformation on $L$ can be compensated by a
right-acting field-dependent $SO(2)$ transformation. Indeed one can easily identify these
in the explicit $SL(2,\mathbb{R})$ transformations \eqref{IIBSL2R} of IIB supergravity.
The two-form potentials transform linearly under $G$ while the fermions only transform
under the compensating $SO(2)$ transformations. Without gauge fixing the $G$
transformations would read (omitting $SL(2,\mathbb{R})$ indices)
\begin{alignat}{3}
  L & \rightarrow g \, L \, h_{SO(2)}^{-1} \,, \qquad &
  {B} & \rightarrow (g^{-1})^T {{B}} \,, \qquad &
   \C{4} & \rightarrow  \C{4} \,, \notag \\
  {\psi}_{{\mu}} & \rightarrow h_{U(1)}^{1/2} {\psi}_{{\mu}} \,, \qquad &
  {\lambda} & \rightarrow h_{U(1)}^{3/2} {\lambda} \,, \qquad &
  {\epsilon} & \rightarrow h_{U(1)}^{1/2} {\epsilon} \,.
  \label{IIBSL2Rcov}
\end{alignat}
where $g$ and $h$ are given by
 \begin{align}
  g = \left( \begin{array}{cc} a&b\\c&d \end{array} \right) \in SL(2,\mathbb{R}) \,, \quad
  h_{SO(2)} = \text{exp} \left( \begin{array}{cc} 0 & \theta(x) \\ -\theta(x) & 0 \end{array} \right) \,, \quad
  h_{U(1)} = \text{exp} ( i \theta(x) ) \,. \notag
 \end{align}
This clearly shows the two different symmetries that act independently in the covariant
formulation. The gauge fixing condition translates in the role of $H$ as compensating
transformation with
 \begin{align}
  \theta = - \text{arccos} \left( \frac{e^\phi (c \chi +d)}{\sqrt{c^2 + e^{2 \phi}(c \chi +d)^2}} \right) \,.
 \label{compensating}
 \end{align}
Indeed, the transformations $\eqref{IIBSL2Rcov}$ with constraint \eqref{compensating}
reduce to the non-linear transformations \eqref{IIBSL2R}.

\subsection{Global Symmetries of Maximal Supergravities} \label{sec:global-symmetries-3}

Having dealt with the simplest example in $D=10$, we now turn to lower-dimensional scalar
cosets. In table~\ref{tab:bosons} we give the groups $G$ and $H$ that one encounters. The
groups $G$ are symmetries of 11D supergravity on a torus; it is expected that $G$ is
broken to an arithmetic subgroup $G(\mathbb{Z})$ for the full M-theory on a torus
\cite{Hull:1995ys}.

The dimension of the scalar coset $G/H$ equals the number of scalars; the number of
axions is given by the number of positive roots of the algebra corresponding to $G$ while
the number of dilatons equals $11-D$ (one for every reduced dimension). In
table~\ref{tab:bosons} we also give the bosonic potentials of higher rank and their
transformation under the $G$ groups. The potentials form linear representations of $G$
while they are invariant under $H$. We do not give the fermionic field content; see
e.g.~\cite{deWit:2002vz}. In contrast to the bosons, the fermions are invariant under $G$
but transform under $H$. One can check these statements in the example of
$SL(2,\mathbb{R})$ symmetry in IIB supergravity, see \eqref{IIBSL2R} and
\eqref{IIBSL2Rcov}.

\begin{table}[ht]
\begin{center}
 \begin{tabular}{||c||c|c|c||c|c|c|c||}
  \hline \rule[-1mm]{0mm}{6mm}
 $D$ & G & H & Dim$[G/H]$ & $d=1$ & $d=2$ & $d=3$ & $d=4$\\
  \hline \hline \rule[-1mm]{0mm}{6mm}
 $11$ & $1$ & $1$ & $-$ & $-$ & $-$ & $\bf 1$ & $-$ \\
  \hline \rule[-1mm]{0mm}{6mm}
 IIA & $\mathbb{R}^+$ & $1$ & $1$ & $\bf 1$ & $\bf 1$ & $\bf 1$ & $-$ \\
  \hline \rule[-1mm]{0mm}{6mm}
 IIB & $SL(2,\mathbb{R})$ & $SO(2)$ & $2$ & $-$ & $\bf 2$ & $-$ & ${\bf 1}^+$ \\
  \hline \rule[-1mm]{0mm}{6mm}
 $9$ & $SL(2,\mathbb{R}) \times \mathbb{R}^+$ & $SO(2)$ & $3$ & ${\bf 2} + {\bf 1}$ & $\bf 2$ & $\bf1$ & $-$ \\
  \hline \rule[-1mm]{0mm}{6mm}
 $8$ & $\begin{array}{c} SL(3,\mathbb{R}) \times \\ \times SL(2,\mathbb{R}) \end{array}$
     & $\begin{array}{c} SO(3) \times \\ \times SO(2) \end{array}$
     & $7$ & $\bf (3,2)$ & $\bf (3,1)$ & ${\bf (1,2)}^+$ & $-$ \\
  \hline \rule[-1mm]{0mm}{6mm}
 $7$ & $SL(5,\mathbb{R})$ & $SO(5)$ & $14$ & $\bf 10$ & $\bf 5$ & $-$ & $-$ \\
  \hline \rule[-1mm]{0mm}{6mm}
 $6$ & $SO(5,5)$ & $SO(5) \times SO(5)$ & $25$ & $\bf 16$ & ${\bf 10}^+$ & $-$ & $-$ \\
  \hline \rule[-1mm]{0mm}{6mm}
 $5$ & $E_{6(+6)}$ & $USp(8)$ & $42$ & $\bf 27$ & $-$ & $-$ & $-$ \\
  \hline \rule[-1mm]{0mm}{6mm}
 $4$ & $E_{7(+7)}$ & $SU(8)$ & $70$ & ${\bf 56}^+$ & $-$ & $-$ & $-$ \\
 \hline
\end{tabular}
\caption{\it The groups $G$ and $H$ and the $d$-form gauge potentials as representations
of $G$ of maximal supergravity in $6 \leq D \leq 11$. The $+$ denotes self-dual
representations: there are duality constraints that halve the number of degrees of
freedom. In all but the IIB case the constraints can be eliminated at the cost of
breaking manifest $G$ covariance.}
 \label{tab:bosons}
\end{center}
\end{table}

Note that the global symmetry group $G$ in $D$ dimensions is often larger than the
$SL(11-D,\mathbb{R})$ that is expected from the connection with eleven-dimensional
supergravity (as will be explained in subsection~\ref{sec:toroidal-2}). For this reason,
the group $G$ is known as a hidden symmetry \cite{Cremmer:1978ds, Julia:1980gr,
Cremmer:1980gs}. Another important feature in even dimensions is that they are only
symmetries of the equations of motion and not of the Lagrangian. For example, this can
come about when the symmetry transformation involves a Hodge dualisation of a gauge
potential, which can only be performed straightforwardly on the field equations and not
on the Lagrangian. This is the origin of the self-dual representations in
table~\ref{tab:bosons}.

A number of complications turn up in $D \leq 5$, as can be inferred from
table~\ref{tab:bosons}. First of all, the exceptional groups of the A-D-E-classification
(of simply-laced simple Lie algebras) appear. Secondly, one needs to dualise potentials
of higher rank to axions to realise the symmetry group $G$. Also the $H$ groups are no
longer orthogonal and one needs a generalised notion of orthogonality. Some details can
be found in \cite{Cremmer:1998ct}.

\begin{figure}[h]
\centerline{\epsfig{file=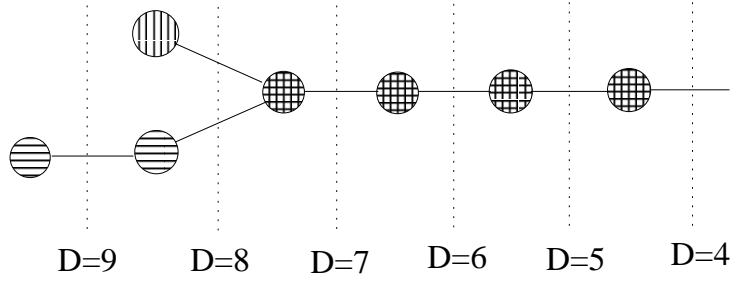,width=0.6\textwidth}}
 \caption{\it The Dynkin diagrams of the symmetry groups $G$ of maximal supergravity in different dimensions
 summarised in one picture.
 Given $D$, the part to the left of the corresponding split is relevant.
 The horizonal and vertical fillings correspond to the 11D and IIB origin, respectively.}
\label{fig:Dynkin}
\end{figure}

The appearing symmetry groups $G$ can be represented by Dynkin diagrams. Here each node
represents a simple root (spanning the space of positive roots) and the number of lines
(zero, one, two or three) between two nodes corresponds to an angle of 90, 120, 135 or
150 degrees between the associated simple roots. In the algebras that we encounter all
simple root vectors have the same length (simply laced algebras) and angles of 120
degrees with respect to each other. The Dynkin diagrams of maximal supergravity are
distilled into figure \ref{fig:Dynkin}. Indeed, continuation to $D < 6$ brings one to the
exceptional Lie algebras.

Note that the Dynkin diagram is very reminiscent of the possible maximal supergravities;
with a highest node in 11D, two possibilities in 10D and unique possibilities in $D \leq
9$. Indeed, one can view the symmetry group $G$ as coming from the higher-dimensional
origin: reduction over a $d$-torus gives rise to an $SL(d,\mathbb{R})$ symmetry (as
explained in subsection~\ref{sec:toroidal-2}). Thus, one can understand the horizontally
filled nodes as coming from 11D while the vertical fillings come from IIB. Together,
these two subgroups generate the full duality group $G$ in any dimension $D \leq 9$
\cite{Lavrinenko:1998hf}.

As an amusing note we would like to mention that the same phenomenon occurs in $Q=16$
supergravity. As discussed in subsection~\ref{sec:susy-2}, these are unique in all
dimensions but six, where one encounters non-chiral iia and chiral iib, similar to IIA
and IIB in $D=10$. Again, the existence of this extra supergravity in six dimensions
gives rise to an extra $SL(2,\mathbb{R})$ symmetry in four dimensions.

However, despite many similarities, the above discussion does not directly carry over to
theories with less supersymmetry. For example, the global symmetry group $G$ is always
maximally non-compact for the case of maximal supersymmetry. In less supersymmetric cases
this is not necessarily true, in which case one should not exponentiate all Cartan
generators but only the non-compact ones. Some of these issues are discussed in
\cite{Pope}.

\section{Supersymmetric Solutions} \label{sec:susy-solutions}

\subsection{Generic Brane Solutions} \label{sec:susy-solutions-1}

In the previous sections we have seen that supergravity theories generically contain
bosonic fields of spin 0,1 and 2, corresponding to a scalar, a rank-$d$ potential and the
graviton. In this subsection we will take a step back and discuss generic solutions to
this system called $p$-brane solutions. These are generalisations of the extremal
Reissner-Nordstr\"{o}m charged black hole to $d \neq 1$ (the rank of the gauge potential)
and $D \neq 4$ (the dimension of space-time). These will occur frequently as
supersymmetric solutions of supergravities, as we will find below. For reviews see
e.g.~\cite{Duff:1995an, Stelle:1998xg}.

The starting point is the D-dimensional toy model Lagrangian
 \begin{align}
  \mathcal{L} = \sqrt{-g} [R - \tfrac{1}{2} (\partial \phi)^2
   - \tfrac{1}{2} e^{a \phi} \G{d+1} \cdot \G{d+1} ] \,,
  \label{braneLagrangian}
 \end{align}
with the rank-$(d+1)$ field strength $\G{d+1} = d \C{d}$. It consists of an
Einstein-Hilbert term, a dilaton kinetic term and a kinetic term for a rank-$d$ potential
with arbitrary dilaton coupling, parameterised by $a$. For future use we define the
constants \cite{Lu:1995cs}
 \begin{align}
  \Delta = a^2 + \frac{2 d \tilde{d}}{D-2} \,, \qquad \tilde{d}=D-d-2 \,.
  \label{Delta}
 \end{align}
The constant $\Delta$ will play an important role in the characterisation of solutions.
In particular, in many supergravities it will be given by $4/n$ with $n$ a positive
integer and the corresponding $p$-brane solutions will preserve a fraction $1/2^n$ of the
supersymmetry.

Due to the presence of the gauge potential, solutions to this system can carry electric and
magnetic charge, defined by
 \begin{align}
  Q_e = \int_{S^{\tilde{d}+1}} e^{a \phi} \star \G{d+1} \,, \qquad
  Q_m = \int_{S^{d+1}} \G{d+1} \,.
  \label{charges}
 \end{align}
These are conserved due to the field equation of $\C{d}$ and the Bianchi identity of
$\G{d+1}$, respectively, and can be seen as generalisations of the Maxwell charges in 4D.
Hodge dualisation interchanges the electric and magnetic charges since the dual field
strengths are related by (in analogy to \eqref{RR-duality})
 \begin{align}
  e^{a \phi} \G{d+1} = \star {G}^{(\tilde{d}+1)} \,,
 \end{align}
where ${G}^{(\tilde{d}+1)} = d {C}^{(\tilde{d})}$. Under this dualisation the field
equations for $\C{d}$ is transformed to the Bianchi identity for the dual field strength
${G}^{(\tilde{d}+1)}$ while the Bianchi identity for $\G{d+1}$ corresponds to the field
equation for the dual potential ${C}^{(\tilde{d})}$. Also $\Delta$ is invariant under
Hodge dualisation since this interchanges $d$ and $\tilde{d}$ and flips the sign of $a$.

The system \eqref{braneLagrangian} allows for two $p$-brane solutions, where $p$ refers
to the dimensionality of the spatial extension of the brane, that carry one of the
charges \eqref{charges}:
 \begin{align}
  \text{electric~$p$-brane:} & \qquad p=d-1 \,, \notag \\
  \text{magnetic~$p$-brane:} & \qquad p=\tilde{d}-1 \,. \notag
 \end{align}
The dimension of the world-volume equals $p+1$ while the remainder $D-p-1$ is the
dimension of the transverse space and is called the codimension.

We will discuss the electric and magnetic $p$-brane solutions at the same time. To this
end, we split up the coordinates in the world-volume $t,x^i$ with $i=1,\ldots,p$ and the
transverse space $x^m$ with $m=p+1,\ldots,D-1$. The metric and dilaton are given by
 \begin{align}
  ds^2 & = H^{-4 \tilde{d}/(\Delta (D-2))} (-dt^2 + dx_i{}^2) + H^{4 d /(\Delta (D-2))} dx_m{}^2 \,,
  \quad
  e^\phi = H^{\pm 2 a/\Delta} \,,
  \label{p-brane-metric}
 \end{align}
where the electric and magnetic solutions have a $+$ and a $-$ sign, respectively. The
corresponding field strengths are given by\footnote{We give only the so-called brane
solutions with positive charge; anti-brane solutions carry negative charge and have an
extra $-$ sign in \eqref{p-brane-fieldstrengths}.}${}^,$\footnote{An additional
possibility for $d=D/2-1$ is the dyonic brane carrying both electric and magnetic charge.
In such cases, both lines of \eqref{p-brane-fieldstrengths} are valid, with an extra
factor of $1/2$ on the right-hand sides.}
 \begin{align}
  \G{d+1}_e = \frac{2}{\sqrt{\Delta}} \, dt \wdg dx^1 \wdg \cdots \wdg dx^p \wdg d H^{-1} \,, \quad
  \G{d+1}_m = \frac{2}{\sqrt{\Delta}} \star (dt \wdg dx^1 \wdg \cdots \wdg dx^p \wdg d H) \,.
  \label{p-brane-fieldstrengths}
 \end{align}
The $p$-branes are characterised by the function $H(x^m)$, which is given by (for the
moment we assume $p < D-3$; we will discuss the other cases later)
 \begin{align}
   H = c + \frac{Q}{r^{D-p-3}} \,,
 \label{single-p-brane-harmonic}
 \end{align}
with $r = \| x^m \|$. The integration constants $c$ and $Q$ are taken both positive to
avoid naked singularities at finite $r$. All such $p$-brane solutions have $ISO(1,p)
\times SO(D-p-1)$ isometry. For branes with $a \neq 0$ the constant $c$ can be related to
the asymptotic value of $\phi$ via $g_s = \exp(\phi)_\infty = c^{\pm 2a / \Delta}$.

The $p$-brane solutions have a horizon at $r=0$. Depending on $D$, $p$ and $\Delta$ the
horizon may coincide with a singularity or it may be possible to find a geodetically
complete extension of the solution. We will not pursue the solution behind the horizon
and will content ourselves with the description of the $0 < r < \infty$ part of
space-time, thus avoiding the possible necessity for a source term. This part
interpolates between two different vacua of the theory \cite{Gibbons:1993sv}: one finds
D-dimensional Minkowski space for $r \rightarrow \infty$ and a metric which is conformal
to a product of Anti-de Sitter space\footnote{An exception is the case $a^2 = 2
\tilde{d}^2 / (D-2)$: in this case the radius of the Anti-de Sitter space-time becomes
infinite and the AdS-part reduces to (p+2)-dimensional Minkowski space-time
\cite{Gibbons:1993sv, Behrndt:1999mk}.} and a higher-dimensional sphere:
 \begin{align}
   ds^2 = H^{2 a^2 / (\Delta (D-p-3))} (ds^2(AdS_{p+2}) + ds^2(S^{D-p-2})) \,,
 \label{confAdSxS}
 \end{align}
for $r \rightarrow 0$, which is called the near-horizon limit.

The $p$-brane solutions carry mass and charge density. The ADM mass per unit $p$-brane volume is
given by
 \begin{align}
  M = \frac{4 Q (D-p-3) g_s^{-a/2} \Omega_{D-p-2}}{\sqrt{\Delta}} \,,
  \label{ADMmass}
 \end{align}
where $\Omega_{D-p-2}$ is the volume of the unit $(D-p-2)$-sphere that surrounds the
$p+1$-dimensional world-volume. Computing the charge densities from \eqref{charges}, one finds that
there is an equality between the mass and (the absolute value of) the charge density: $M = | Q_e |$
for the electric solution and $M = | Q_m |$ for the magnetic solution. In supergravity theories
this will generically lead to an amount of preserved supersymmetry.

There are several generalisations of the prime examples \eqref{p-brane-metric},
\eqref{p-brane-fieldstrengths} of $p$-brane solutions. For instance, one can replace the
function $H=H(x^m)$ by any solution to the Laplace equation
 \begin{align}
  \Box H(x^m) = \partial_n \partial^n H(x^m) = 0 \,,
 \end{align}
in $(D-p-1)$-dimensional flat transverse space. Examples are
 \begin{itemize}
  \item
  the multi-center $p$-brane solution with
  \begin{align}
     H = c + \sum_i \frac{Q_i}{\|x^m - x_i^m\|^{D-p-3}} \,,
  \label{multi-p-brane-harmonic}
  \end{align}
  with all $Q_i$ positive to avoid naked singularities at finite $x^m$. Its interpretation
  consists of a number of $p$-branes located at $x_i^m$. Physically, this solution is possible
  since all separate $p$-branes have equal mass and charge; for this reason their attractive force (due to
  gravity and the scalar) cancels their repulsive force (due to the rank-$d$ potential).
  \item
  the smeared $p$-brane solution with $H=H(x^m)$ a harmonic function in a subspace of the full transverse
  space. An example is the following function for $p < D-4$:
   \begin{align}
    H = c + \frac{Q}{\| x^{\tilde{m}} \|^{D-p-4}} \,,
    \label{smeared-harmonic}
   \end{align}
  where $\tilde{m} = p+1, \ldots , D-2$, i.e.~the harmonic function does not depend on $x^{D-1}$.
  This can be interpreted as the configuration of a smooth distribution of $p$-brane in the $x^{D-1}$-direction.
  The smeared solutions will be very relevant later for the relation between the different solutions.
 \end{itemize}
These generalisations break part of the isometry group. However, since the mass and charge of these
solutions are still equal, they will preserve supersymmetry in a supergravity theory.

It is also possible to add mass to the $p$-brane solution without affecting its charge:
$Q_{e,m} < M$. This generically breaks the supersymmetry and (part of the) isometry of
the solutions. For example, one can construct non-supersymmetric solutions with isometry
group $\mathbb{R} \times ISO(p) \times SO(D-p-1)$ \cite{Horowitz:1991cd, Duff:1994ye}.
Such deformations are only possible for the single-center solution
\eqref{single-p-brane-harmonic} and not for its multi-center generalisation
\eqref{multi-p-brane-harmonic}, as can physically be understood from the inequality of
mass and charge: the attractive and repulsive forces between different constituents no
longer cancel.

\subsection{Branes with Little Transverse Space} \label{sec:susy-solutions-2}

Let us now discuss branes with $p \geq D-3$, starting with the case that saturates this
bound. Such branes are sometimes called vortex branes and have a two-dimensional
transverse space. The most symmetric harmonic function reads (with $r = \| x^m \|$)
 \begin{align}
  H = c + Q \log( r ) \,,
 \end{align}
giving rise to $ISO(1,D-3) \times SO(2)$ isometry. The limit $r \rightarrow \infty$ in
this case does not yield D-dimensional Minkowski but an asymptotically locally flat
space-time; locally this is Minkowski but a global difference occurs in the form of a
deficit angle in the 2D transverse space, stemming from the mass density of the
$(D-3)$-brane solution. The other limit, $r \rightarrow 0$, is not well-defined since the
harmonic function becomes negative at finite $r$, thus rendering this solution valid only
for $r$ large enough. However, there are modifications of this solution with the same
large-$r$ behaviour and a well-defined interior \cite{Greene:1990ya}.

The next case concerns $(D-2)$-branes which are usually referred to as domain walls. Their
transverse space is one-dimensional, on which the most general harmonic function reads (where $y =
x^{D-1}$)
 \begin{align}
  H = c + Q y \,,
 \end{align}
where we take $Q$ positive. Note that a potential of rank $D-1$, corresponding to an
electric domain wall, carries no degrees of freedom (see table~\ref{tab:dof}). Its Hodge
dual
 \begin{align}
  {G}^{(0)} = e^{a \phi} \star \G{D} = 2 Q / \sqrt{\Delta} \,,
 \end{align}
is a constant zero-form field strength and can be interpreted as a mass parameter. We
thus find that mass parameters can support domain walls. A necessary condition for this
is the quadratic term in \eqref{braneLagrangian} with $d=0$. Rather than a kinetic term
it is called a scalar potential (due to the coupling to the dilaton) and its form
determines the possible properties of domain wall solutions. We will encounter many
examples of scalar potentials in gauged supergravities, see chapter \ref{ch:gauged}.

Again, one might wonder if the domain wall solution interpolates between different vacua.
Due to the one-dimensional transverse space, the domain walls differ in this respect from
the other $p$-branes. One can always do a reparameterisation of the transverse coordinate
\cite{Bergshoeff:1998bs} to obtain the metric of either conformal Anti-de Sitter
space-time or of conformal Minkowski space-time. However, the domain wall as it stands is
certainly not a globally well-defined solution\footnote{Except for the case $a=0$, in
which the domain wall solution yields Anti-de Sitter space-time (without conformal
factor). Indeed, the scalar potential becomes a pure cosmological constant in this
limit.}: one finds that the harmonic function vanishes for finite $y$. To remedy the
resulting singularity, one has to patch solutions with different values for the mass
parameters. This requires the presence of source terms, whose charge is related to the
difference between the values of the mass parameters on both sides of the domain wall. We
will discuss an example of such a source term in section~\ref{sec:D8-brane}.

Domain walls of the above type are usually called thin domain walls: the source term
corresponds to a object of infinitesimal thickness in the transverse direction. Such
source terms are always necessary with potentials of the form \eqref{braneLagrangian}
with $p=D-2$, which have only one asymptotic minimum (with $\phi \rightarrow \pm
\infty$). In contrast, potentials with more than one (local) minima allow for solutions
interpolating between two minima. Such smooth configurations are called thick domain
walls. We will mostly encounter the thin version in this article, however.

Taking the $p$-brane classification one step further by considering $p=D-1$ brings us to
space-time-filling branes. All of space-time is world-volume and there is no transverse space.
Though not very interesting from a supergravity point of view there is an appreciation of
space-time filling branes in string theory \cite{Polchinski:1995mt, Bergshoeff:1998re}.

\subsection{Maximally Supersymmetric Solutions} \label{sec:susy-solutions-3}

In section~\ref{sec:parent-sugra} we have encountered different supergravity theories in
eleven and ten dimensions. In the next two subsections we will discuss solutions of these
theories that preserve a fraction of supersymmetry.

From the supersymmetry transformations one can deduce which solutions can preserve
supersymmetry. We will only consider bosonic solutions. For these to preserve
supersymmetry, the right-hand side of the supersymmetry transformations of the fermions
must vanish. These conditions are the Killing spinor equations. Here one distinguishes
two possibilities: either all terms in the variation of the fermions vanish separately,
leading to maximally supersymmetric solutions, or there is a cancellation between
non-zero terms. The latter case will involve a condition on the supersymmetry parameter
$\epsilon$ due to the different $\Gamma$-structures. The supersymmetry parameter subject
to this condition is called the Killing spinor. Since it is constrained this will lead to
solutions preserving only fractions of supersymmetry.

All maximally supersymmetric solutions to maximal supergravity in eleven and ten
dimensions have been classified \cite{Figueroa-O'Farrill:2002ft}. Minkowski space-time
without field strengths is a maximally supersymmetric solution to 11D, IIA and IIB
supergravity. In addition to this trivial vacuum, there are so-called AdS $\times$ S and
plane wave solutions that preserve all supersymmetry. The AdS $\times$ S metric consists
of a product of a $(d+1)$-dimensional Anti-de Sitter space-time and an
$(D-d-1)$-dimensional sphere, whose isometry group is $SO(1,d+1) \times SO(D-d)$ (which
is considerably larger than that of the brane solutions with rank-$(d+1)$ field
strengths). In addition, there is a flux of the rank-$(d+1)$ field strength though the
sphere. In eleven dimensions one has such solutions with $d=3$ and $d=6$
\cite{Freund:1980xh, Pilch:1984xy} while IIB allows for the $d=4$ case. The plane wave
solution, found in 11D \cite{Kowalski-Glikman:1984wv} and in IIB \cite{Blau:2001ne}, has
the metric of a gravitational plane wave and a constant null flux of the rank-four and
self-dual rank-five field strength, respectively. Only recently has it been appreciated
\cite{Blau:2002dy} that the maximally supersymmetric plane wave is the Penrose limit
\cite{Penrose:1984, Gueven:2000ru} of the AdS $\times$ S solutions.

\subsection{Half-supersymmetric Solutions} \label{sec:susy-solutions-4}

The solutions preserving half supersymmetry have also received a lot of attention. Here
the Killing spinor is subject to a projection:
 \begin{align}
  \tfrac{1}{2}(1 \pm O) \epsilon = \epsilon \,, \qquad O^2 = 1 \,.
 \end{align}
The possible projectors of 11D, IIA and IIB supergravity are given in
table~\ref{tab:projectors}. Each theory has a number of $p$-brane solutions while they
have the plane wave and Kaluza-Klein monopole in common.

\begin{table}[ht]
 \begin{center}
 \begin{tabular}{||c|c|||c|c||}
  \hline \rule[-1mm]{0mm}{6mm}
   $O$ & 11D, IIA, IIB solution & $O$ & 11D solution \\
  \hline \hline \rule[-1mm]{0mm}{6mm}
  $\Gamma_{\underline{01}}$ & pp-wave & $\Gamma_{\underline{012}}$ & M2-brane \\
  \hline \rule[-1mm]{0mm}{6mm}
  $\Gamma_{\underline{1234}}$ & Kaluza-Klein monopole & $\Gamma_{\underline{12345}}$ & M5-brane \\
  \hline \hline
  \hline \rule[-1mm]{0mm}{6mm}
  $O$ & IIA solution & $O$ & IIB solution \\
  \hline \hline \rule[-1mm]{0mm}{6mm}
  $\Gamma_{\underline{0}} \Gamma_{11}$ & D0-brane & $ \text{i} \Gamma_{\underline{01}} \star$ & D1-brane \\
  \hline \rule[-1mm]{0mm}{6mm}
  $\Gamma_{\underline{01}} \Gamma_{11}$ & F1-brane & $\Gamma_{\underline{01}} \star$ & F1-brane \\
  \hline \rule[-1mm]{0mm}{6mm}
  $\Gamma_{\underline{012}}$ & D2-brane & $ \Gamma_{\underline{01234}} \star$ & D3-brane \\
  \hline \rule[-1mm]{0mm}{6mm}
  $\Gamma_{\underline{12345}}$ & D4-brane & $ \text{i} \Gamma_{\underline{1234}} \star $ & D5-brane \\
  \hline \rule[-1mm]{0mm}{6mm}
  $\Gamma_{\underline{1234}} \Gamma_{11}$ & NS5-brane & $\Gamma_{\underline{1234}} \star$ & NS5-brane \\
  \hline \rule[-1mm]{0mm}{6mm}
  $\Gamma_{\underline{123}} \Gamma_{11}$ & D6-brane & $ \text{i} \Gamma_{\underline{12}} $ & D7-brane \\
  \hline
\end{tabular}
 \caption{\it Possible projection operators of the supersymmetry transformations
 of 11D, IIA and IIB supergravity and the corresponding half-supersymmetric solutions.}
 \label{tab:projectors}
\end{center}
\end{table}

The branes of table~\ref{tab:projectors} are labelled by their value of $p$, which equals
$d-1$ for the electric solution and $\tilde{d}-1 = D - d -3$ for the magnetic solution.
Their metric, dilaton and field strength are given in \eqref{p-brane-metric} and
\eqref{p-brane-fieldstrengths}. In addition, their values of $a$ (the dilaton coupling to
the field strength kinetic term in the electric formulation) can be read off from
\eqref{11Daction}, \eqref{IIAaction} and \eqref{IIBaction}:
\begin{itemize}
 \item $a=0$ for the M-branes \cite{Duff:1991xz, Gueven:1992hh},
 \item $a=-1$ for the F1-brane \cite{Dabholkar:1990yf},
 \item $a=\tfrac{1}{2}(3-p)$ for the D-branes \cite{Horowitz:1991cd},
 \item $a=+1$ for the NS5-brane \cite{Strominger:1990et}.
 \end{itemize}
From \eqref{Delta} it follows that these branes all have $\Delta = 4$. Such branes
preserve half of supersymmetry. Note that $a$ vanishes for the M2-, M5- and
D3-brane\footnote{In fact, the D3-brane carries both electric and magnetic charge (it is
dyonic), due to the self-duality condition on its five-form field strength. For this
reason, in contrast to all other branes, both lines of \eqref{p-brane-fieldstrengths} are
valid, but with an extra factor of $1/2$ on the right-hand sides.}. This has an important
consequence: their near-horizon limits \eqref{confAdSxS} are of the form $AdS^4 \times
S^7$, $AdS^7 \times S^4$ and $AdS_5 \times S^5$ (without conformal factor), respectively,
which are maximally supersymmetric vacua of 11D and IIB supergravity. Thus one finds
isometry and supersymmetry enhancement in the near-horizon limit for these branes.

We can now interpret the brane solutions of IIA and IIB supergravity in the context of string
theory. An important tool will be the dependence of the mass on the coupling constant $g_s$, which
is given by\footnote{The difference with \eqref{ADMmass} is due to the field redefinition $g_{\mu
\nu} \rightarrow e^{\phi/2} g_{\mu \nu}$ between Einstein and string frame.}
 \begin{align}
  M \sim g_s^{ - (2a+p+1) /4} \,. \label{gs-scaling}
 \end{align}
The F1-solution corresponds to the fundamental string, which is charged with respect to
the NS-NS 2-form $B$. Its mass scales like $g_s{}^0$. The D$p$-brane solutions are
interpreted as the $p+1$-dimensional hyperplanes on which open strings can end
\cite{Polchinski:1996df}, due to imposition of so-called Dirichlet boundary conditions.
These carry charge of the corresponding R-R potential $C^{(p+1)}$ and their masses scale
as $1/g_s$, which is in between fundamental and solitonic behaviour. The microscopic
understanding of D-branes in terms of open strings with Dirichlet \bc s\ was one of the
key insights that led to the second superstring revolution. Note that the remaining brane
solution, the NS5-brane, has a mass that scales like $1/g_s{}^2$ and can thus be
considered truly solitonic.

In addition to the brane solutions one has so-called pp-wave solutions\footnote{Here {\it pp}
stands for {\it plane fronted with parallel rays}. The former refers to the planar nature of the
wave fronts while the latter denotes the existence of a covariantly constant null vector.}. Its
metric and field strength read (in light-cone coordinates $x^\pm = t \pm x^{1}$ and $x^m$ with
$m=2,\ldots,D-1$):
 \begin{align}
  & ds^2 = 2 dx^+ dx^- + H(x^m,x^-) (dx^-)^2 + (dx^m)^2 \,, \qquad \G{d+1} = dx^- \wdg \xi^{(d)} \,,
  \label{planewaves}
 \end{align}
where $H$ and $\xi^{(d)}$ satisfy the requirements
 \begin{align}
  \Box H = - \tfrac{1}{4} \| \xi^{(d)} \|^2 \,, \qquad d \xi^{(d)} = d \star \xi^{(d)} = 0 \,,
 \end{align}
which are all defined on the transverse Euclidean space with coordinates $x^m$ (for all
$x^-$). The field strength $\G{d+1}$ can be the four-form field strength of 11D or
several field strengths of IIA and IIB. This pp-wave solution generically preserves half
supersymmetry (with the projector as given in table~\ref{tab:projectors}) but special
choices of $H$ and $\xi^{(d)}$ give rise to more supersymmetry \cite{Hull:1984vh,
Cvetic:2002si, Gauntlett:2002cs, Bena:2002kq}. For 11D and IIB one obtains maximal
supersymmetry for the truncation to the plane wave
 \begin{align}
  \text{11D:} \qquad &
  \begin{cases}
   \xi^{(3)} = \mu dx^2 \wdg dx^3 \wdg dx^4 \,, \\
   H(x^m,x^-)= -\tfrac19 \mu^2 ((x^2)^2 + (x^3)^2 + (x^4)^2) -\tfrac1{36} \mu^2 ((x^5)^2 + \cdots +(x^{10})^2) \,,
  \end{cases} \notag \\
  \text{IIB:} \qquad &
  \begin{cases}
   \xi^{(4)} = \mu dx^2 \wdg dx^3 \wdg dx^4 \wdg dx^5 + \mu dx^6 \wdg dx^7 \wdg dx^8 \wdg dx^ 9 \,,
   \\
   H(x^m,x^-)= -4 \mu^2 ((x^2)^2 + \cdots + (x^9)^2) \,.
  \end{cases}
 \end{align}
Another special case is the Brinkmann wave \cite{Brinkmann:1923}, a purely gravitational
solution with $\xi^{(d)} = 0$. It is described in terms of one harmonic function, i.e.~a
function satisfying $\Box H = 0$. There is in general no supersymmetry enhancement for
this case.

Another purely gravitational solution of 11D, IIA and IIB is provided by the Kaluza-Klein
monopole \cite{Gross:1983hb, Sorkin:1983ns} ($m=1,2,3$ and $i=5,\ldots,D-1$):
 \begin{align}
   ds^2 = - dt^2 + dx_i^2 + H^{-1}(dx^4 + A_m dx_m)^2 + H dx_m^2 \,,
 \end{align}
where the functions $H=H(x^m)$ and $A_m=A_m(x^n)$ are subject to the condition
 \begin{align}
  F_{mn} = \tfrac{1}{2}(\partial_m A_n - \partial_n A_m) =  \varepsilon_{mnp} \partial_p H
  \,. \label{KK-monopole-vector}
 \end{align}
This metric is the product of a Minkowski space-time and the 4D Euclidean Taub-NUT space
with isometry direction $x^4$. The $SO(3)$ isometric case is given by (where $r=\| x^m
\|$)
 \begin{align}
  H = c + \frac{Q}{r} \,.
 \end{align}
This gives rise to a regular geometry if the isometry direction $x^4$ is compact with
period $4 \pi Q$ \cite{Hawking:1977jb}. Its near-horizon limit $r \rightarrow 0$ gives
rise to flat space-time and thus indeed gives rise to both isometry and supersymmetry
enhancement. In addition to the $SO(3)$ isometric case, one can take also take
multi-centered solutions or smeared versions, as discussed in \ref{sec:susy-solutions-1}.
The Kaluza-Klein monopole also preserves half of supersymmetry for generic choices of the
harmonic function.

\subsection{Relations between Half-Supersymmetric Solutions} \label{sec:susy-solutions-5}

The above solutions constitute all known maximally and half-supersymmetric solutions of
eleven- and ten-dimensional maximal supergravity. Since the theories in 10D and 11D are
related to each other upon dimensional reduction, as we found in
subsection~\ref{sec:parent-sugra-4}, one can also relate their solutions. One provision
is that the solution must have the correct isometry to allow for this reduction.
Reduction in a transverse direction is therefore only possible for smeared solutions with
harmonic functions that have an extra isometry. Reduction in a world-volume direction is
always possible. Thus, reduction of the two M-branes gives rise to four different brane
solutions of IIA supergravity. Similar remarks hold for the relations between IIA and IIB
solutions.

\begin{figure}[h]
\centerline{\epsfig{file=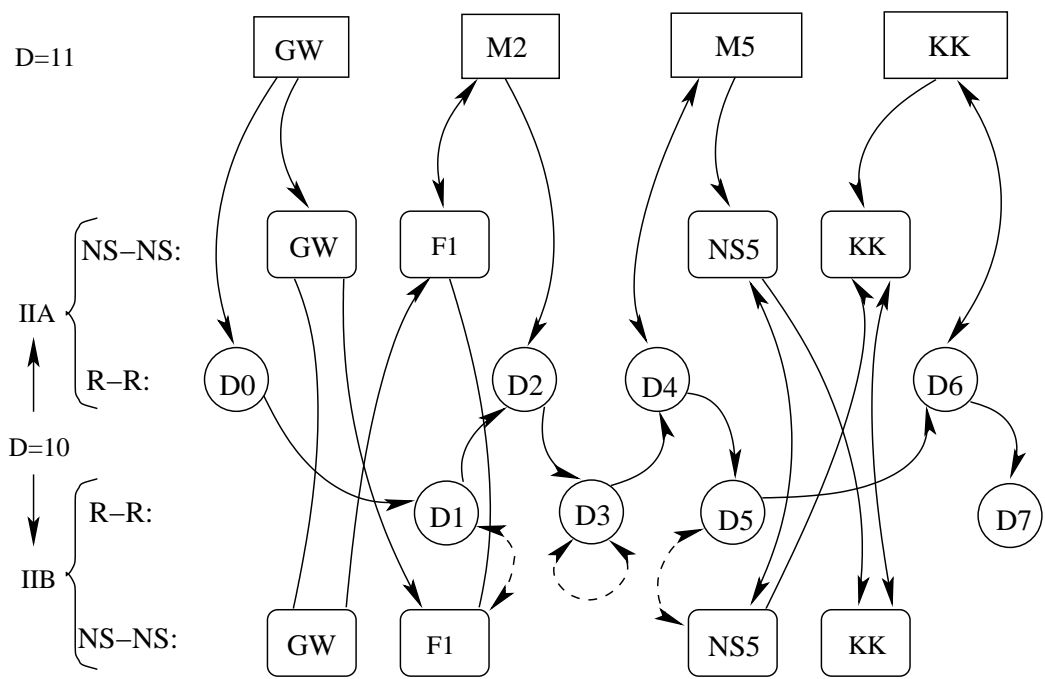,width=0.85\textwidth}}
 \caption{\it The web of half-supersymmetric solutions and their relations in D=10 and D=11 maximal supergravities.
  Solid lines correspond to dimensional reduction or T-duality, the dashed lines correspond to S-duality. If an arrow ends with a
  head, the operation leads to the maximally isometric solution; if not, one obtains a smeared version. Adapted from \cite{Bergshoeff:1998re}.}
 \label{fig:solutions-web}
\end{figure}

In figure \ref{fig:solutions-web} we show the relations between the different solutions
that preserve half of supersymmetry. Note that the solutions in the NS-NS sectors of IIA
and IIB transform into each other; the same holds for the D-branes\footnote{Indeed,
T-duality interchanges Neumann and Dirichlet \bc s; for this reason it also relates the
different D-branes in string theory.} of the R-R sectors. As for the pp-wave solutions,
we have only considered their purely gravitational limit (the gravitational wave) since
the solution is then expressible in terms of a harmonic function, which greatly
simplifies the T-duality discussion.

Furthermore, solutions with less than 1/2 supersymmetry have been studied extensively.
For example, it has been known for long that 11D supergravity allows for solutions
preserving 1/4 or 1/8 supersymmetry \cite{Gueven:1992hh}. Only later these were
understood as intersections of different solutions preserving 1/2 supersymmetry
\cite{Papadopoulos:1996uq}. A lot of intersections have been studied since, see
\cite{Gauntlett:1997cv} for a review.

 \chapter{Dimensional Reduction} \label{ch:reductions}

As discussed in the introduction, the most promising candidates for quantum gravity are
M- and string theory. It is of interest to investigate which four-dimensional effective
descriptions can be obtained from these ten- and eleven-dimensional theories. As a first
step, in this chapter we will discuss the techniques of extracting different effective
descriptions from a higher-dimensional field theory.

\section{Introduction} \label{sec:reduction-intro}

\subsection{Scalar Field and Kaluza-Klein States} \label{sec:reduction-intro-1}

Consider a complex scalar field\footnote{For our conventions concerning dimensional
reduction, see appendix~\ref{app:conventions}.} $\hat{\phi}$ in $\hat{D}$ dimensions,
depending on the coordinates $x^{\hat{\mu}} = (x^\mu , z)$. One can expand the dependence
on one of the coordinates via the Fourier decomposition:
 \begin{align}
  \hat{\phi} (x,z) = \int dk e^{i k z} \phi_k (x) \,,
 \end{align}
in terms of components $\phi_k$ with momentum $k$. If, in addition, the $z$ direction is
taken to be compact of length $2 \pi R$ and we impose the boundary condition
$\hat{\phi}(x,0) = \hat{\phi}(x,2 \pi R)$, the integral becomes the sum
 \begin{align}
  \hat{\phi} (x,z) = \sum_n e^{i n z / R} \phi_n (x) \,,
 \label{Fourier-summation}
 \end{align}
over a discrete spectrum of fields $\phi_n$ with momentum $k = n / R$ in the compact
direction.

Suppose the complex scalar $\hat{\phi}$ is subject to the Klein-Gordon equation
$\hat{\Box} \hat{\phi} = 0$ where $\hat{\Box} = \partial_\mu \partial^\mu + \partial_z
\partial^z$. Upon inserting the Fourier transform in this equation, one obtains separate
equations for components with different momentum:
 \begin{align}
  \Box \phi_k - k^2 \phi_k = \Box \phi_n - (n/R)^2 \phi_n = 0 \,,
 \label{mass-scalar}
 \end{align}
where $\Box = \partial_\mu \partial^\mu$. This is the equation for a scalar of (mass)$^2$
$k^2$ or $(n/R)^2$. Thus a massless scalar in $\hat{D}$ dimensions splits up in an
infinite number of scalar fields in $D=\hat{D}-1$ dimensions. In the context of
dimensional reduction, these are called Kaluza-Klein states. Only one of these (the
component $\phi_0$) is massless, while the other ones are massive. The spectrum of
Kaluza-Klein states is continuous for a non-compact internal direction and discrete for
$z$ compact. The latter spectrum therefore has a mass gap, which is an important
ingredient when considering compactifications.

\subsection{Consistency of Truncations} \label{sec:reduction-intro-2}

The fact that one obtains separate equations for the different Fourier components lies at
the heart of dimensional reduction. First one expresses a higher-dimensional field in an
infinite tower of lower-dimensional fields by expanding the dependence on the internal
coordinates into harmonics on the internal manifold. Next, one observes that one can
consistently truncate to a finite number of fields and set the rest of the spectrum equal
to zero. Here, a consistent truncation refers to the origin in the higher-dimensional
theory: every lower-dimensional solution should uplift to a higher-dimensional one.

Usually, one truncates to only the massless sector for the following reason. In
dimensional reduction the masses are inversely proportional to the size of the internal
manifold (as can be seen on dimensional grounds and in the example \eqref{mass-scalar}).
Since we live in an effectively four-dimensional world, any internal directions must be
very small. This means that the mass of states with non-zero momentum becomes very large.
Therefore, these modes are too massive to be physically interesting and are usually
discarded. In the above example, this would correspond to keeping only $\phi_0$ and
truncating the other components.

Note however that one does not need to take a very small size of the internal manifold
for the massive modes to decouple; in many cases it is always a consistent truncation to
retain only the massless modes, irrespective of whether the internal manifold is small or
large or indeed, whether it is compact or non-compact. Again, the scalar field serves as
an example: the Klein-Gordon equation for $\hat{\phi}$ splits up in many
lower-dimensional equations, which are all solved by $\Box \phi_0 = 0$ and $\phi_k = 0$
(in the non-compact case) or $\phi_n = 0$ (in the compact case). Thus any solution to the
equation for $\phi_0$ will also solve the higher-dimensional Klein-Gordon equation for
$\hat{\phi}$.

Another important point is that the lower-dimensional degrees of freedom are not always
massless. In such cases, the Fourier expansion of a field over the internal manifold does
not comprise any massless fields. A consistent truncation then only keeps the lightest
modes of a field. The set of lower-dimensional fields then do not have the same mass:
some may be massless (such as gravity and gauge vectors) while others are massive (such
as scalars). In the above discussion of consistent truncation, this corresponds to
replacing massless with lightest.

In the reduction procedures that we consider in this chapter, the number of \dof\ is
unchanged by the dimensional reduction: every higher-dimensional degree of freedom
corresponds, after the expansion and truncation, to one lower-dimensional degree of
freedom. These lower-dimensional fields fall in multiplets of the isometry group of the
internal manifold. In particular, when expanding a theory including gravity over a
manifold with isometry group $G$, one expects non-Abelian gauge vectors of $G$ to be
among the massless lower-dimensional modes, see e.g.~\cite{Duff:1986hr}. This will be an
essential feature in sections~\ref{sec:group} and \ref{sec:coset}.

Thus dimensional reduction consists of an expansion over an internal manifold and a
subsequent truncation to the lightest subsector. However, this is usually not what is
done in practice. Rather, a reduction Ansatz is constructed, relating higher-dimensional
fields to a set of lower-dimensional fields. This is the result of the expansion and
truncation: the lower-dimensional fields are the lightest modes of the expansion.
Dimensional reduction then consists of substituting the reduction Ansatz in the field
equations or Lagrangian. In many cases the reduction Ansatz contains a certain dependence
on the internal coordinates. To be able to interpret the resulting equations as a
lower-dimensional theory, this dependence should cancel at the end of the day. This
requirement is equivalent to the consistency of truncations to the finite number of
lower-dimensional fields.

In this chapter we will consider toroidal and twisted reductions and reductions over
group and coset manifolds, all of which are consistent reductions. In the case of
toroidal reduction, the reduction Ansatz is taken independent of the internal coordinates
$z^m$. Toroidal reduction is therefore obviously consistent. The other three reductions
require a certain $z^m$-dependence. For reductions with a twist and over a group
manifold, the cancellation of the internal coordinate dependence is guaranteed on
group-theoretical grounds, as will be explained in sections~\ref{sec:twist} and
\ref{sec:group}. In the remaining reduction over a coset manifold this cancellation is
quite miraculous and poorly understood; it has been proven only in a small number of
cases, see section~\ref{sec:coset}. Examples of reductions whose consistency (in the
above sense) has not been proven are Calabi-Yau compactifications\footnote{
  The metrics of CY spaces are not known in full generality
  so explicit reduction Ans\"{a}tze are not available.},
which we will not consider.

\section{Toroidal Reduction} \label{sec:toroidal}

In this section we will consider the reduction Ans\"{a}tze for toroidal reduction of
gravity, gauge potentials and fermions. As indicated above, for reduction over a torus
one does not include dependence on the internal coordinates and thus its consistency is
guaranteed. More information can be found in e.g.~\cite{Pope}.

\subsection{Gravity on a Circle} \label{sec:toroidal-1}

We will now consider the reduction of gravity in $\hat{D}$ dimensions over a circle to $D
= \hat{D}-1$ dimensions. To this end, the coordinates are split up according to
$x^{\hat{\mu}} = (x^\mu , z)$. We will use the following choice for the decomposition of
$\hat{D}$-dimensional gravity into $D$-dimensional fields:
 \begin{align}
  \hat{ds}^2 = e^{2 \alpha \phi} ds^2 + e^{2 \beta \phi} (dz + A_\mu dx^\mu)^2 \,,
 \label{Ansatz-gravity-circle}
 \end{align}
i.e.~gravity decomposes into a lower-dimensional gravity plus a vector $A_\mu$ and a
scalar $\phi$. The constants $\alpha$ and $\beta$ are in principle arbitrary. This Ansatz
gives rise to a lower-dimensional theory with the Lagrangian
 \begin{align}
  \CL = \sqrt{-\hat{g}} \hat{R} = \sqrt{-g} [ R -\tfrac{1}{2} (\partial \phi)^2
  - \tfrac{1}{2 \, 2!} e^{2(\beta - \alpha) \phi} F^2 ] \,,
 \label{action-gravity-circle}
 \end{align}
with $F = d A$. Here we have chosen the constants to the values
 \begin{align}
  \alpha^2 = \frac{1}{2(D-1)(D-2)} \,, \qquad \beta = - (D-2) \alpha \,,
 \label{constants-circle}
 \end{align}
to obtain the lower-dimensional Lagrangian in the conventional form
\eqref{action-gravity-circle}, i.e.~without dilaton coupling for the Ricci scalar and
with the factor $\tfrac{1}{2}$ in the dilaton kinetic term. Note that this a system of
the form that was considered in subsection~\ref{sec:susy-solutions-1} on brane
solutions\footnote{The corresponding electric and magnetic brane solutions will uplift to
the gravitational wave and Kaluza-Klein monopole in $\hat{D}$ dimensions, respectively,
as also seen in figure \ref{fig:solutions-web}.}, with parameter $\Delta$ as defined in
\eqref{Delta} equal to $4$ for all dimensions $D$.

The appearance of the Maxwell kinetic term was the reason for Kaluza \cite{Kaluza:1921tu}
and Klein \cite{Klein:1926tv} to consider such dimensional reductions: it seemed possible
to unify gravity and electromagnetism in 4D by the introduction of a fifth coordinate.
Note however that there is also an extra scalar, which can not be simply set equal to
zero: this would be inconsistent with the higher-dimensional field equations. Often these
extra fields are called the Kaluza-Klein scalar and vector. Also, the general procedure
of obtaining a lower-dimensional description from a higher-dimensional theory is
sometimes called Kaluza-Klein theory. We will not use this terminology, however, since we
need to make a distinction between the different possibilities within Kaluza-Klein
theory.

One can understand the lower-dimensional symmetries of the Lagrangian
\eqref{action-gravity-circle} by considering its higher-dimensional origin. In
particular, the $\hat{D}$-dimensional Einstein-Hilbert action is invariant under general
coordinate transformations
 \begin{align}
   \delta x^{\hat{\mu}} = - \hat{\xi}^{\hat{\mu}} \,, \qquad \Rightarrow
   \delta \hat{g}_{\hat{\mu} \hat{\nu}} =
   \hat{\xi}^{\hat{\rho}} \partial_{\hat{\rho}} \hat{g}_{\hat{\mu} \hat{\nu}} +
   \hat{g}_{\hat{\rho} \hat{\nu}} \partial_{\hat{\mu}} \hat{\xi}^{\hat{\rho}} +
   \hat{g}_{\hat{\mu} \hat{\rho}} \partial_{\hat{\nu}} \hat{\xi}^{\hat{\rho}} \,.
   \label{gct}
 \end{align}
In general, such a transformation will not preserve the form of the reduction Ansatz
\eqref{Ansatz-gravity-circle}, i.e.~the resulting metric will not be expressible as
\eqref{Ansatz-gravity-circle} with transformed fields. The Ansatz will only transform
covariantly under transformations with specific parameters. Such Ansatz-preserving
transformations and their effect on the lower-dimensional fields are the following:
 \begin{alignat}{2}
  & \delta x^\mu = - \xi^\mu (x) \,, \qquad && \Rightarrow
  \begin{cases}
    \delta g_{\mu \nu} =
    \xi^\rho \partial_\rho g_{\mu \nu} +
    g_{\rho \nu} \partial_\mu \xi^\rho + g_{\mu \rho} \partial_\nu \xi^\rho \,, \\
    \delta A_\mu = \xi^\rho \partial_\rho A_\mu + A_\rho \partial_\mu \xi^\rho \,, \\
    \delta \phi = \xi^\rho \partial_\rho \phi \,,
   \end{cases} \notag \displaybreak[2] \\
  & \delta z = - \lambda (x) \,, \qquad && \Rightarrow
  \begin{cases}
    \delta A_\mu = \partial_\mu \lambda \,,
   \end{cases} \notag \displaybreak[2] \\
  & \delta z = - c z \,, \qquad && \Rightarrow
  \begin{cases}
   \delta g_{\mu \nu} = - 2 \alpha c g_{\mu \nu} / \beta \,, \\
   \delta A_\mu = - c A_\mu \,, \\
   \delta \phi = c / \beta \,.
  \end{cases}
 \end{alignat}
These can respectively be understood as $D$-dimensional general coordinate
transformations, $U(1)$ gauge transformations and a global scale symmetry.

The latter can be integrated to give a finite, rather than infinitesimal, transformation.
In addition one has the higher-dimensional trombone symmetry $\hat{g}_{\hat{\mu}
\hat{\nu}} \rightarrow \lambda^2 \hat{g}_{\hat{\mu} \hat{\nu}}$, which also reduces to a
finite scale symmetry of the lower-dimensional theory. One can construct linear
combinations of these symmetries to obtain the following transformations
 \begin{align}
  g_{\mu \nu} \rightarrow \lambda_1{}^{2} g_{\mu \nu} \,, \qquad
  A_\mu \rightarrow \lambda_1 A_\mu \,,
  \label{trombone}
 \end{align}
where $\lambda_1 \in \mathbb{R}^+$. This is the lower-dimensional trombone symmetry (with
coefficients as explained in subsection~\ref{sec:parent-sugra-1}), which scales all terms
in the Lagrangian with the same factor, and is only a symmetry of the field equations.
The other combination reads
 \begin{align}
  A_{\mu} \rightarrow \lambda_2{}^{\alpha - \beta} A_{\mu} \,, \qquad
  e^\phi \rightarrow \lambda_2 e^\phi \,,
 \label{dilaton-shift}
 \end{align}
also with $\lambda_2 \in \mathbb{R}^+$. This corresponds to the only scale symmetry of
the Lagrangian. Indeed, this explains the two $\mathbb{R}^+$ symmetries of IIA
supergravity: they stem from combinations of the 11D trombone symmetry and internal
coordinate transformations.

\subsection{Gravity on a Torus} \label{sec:toroidal-2}

The reduction of gravity over a torus $T^n$ can be seen as successive reductions over $n$
circles. The reduction Ansatz of $\hat{D}$-dimensional gravity over an $n$-torus to $D =
\hat{D}-n$ dimensions reads (with a coordinate split $x^{\hat{\mu}} = (x^\mu , z^m)$
where $m=1,\ldots,n$)
 \begin{align}
  \hat{ds}^2 = e^{2 \alpha \phi} ds^2 + e^{2 \beta \phi} M_{mn} (d z^m + A^m_\mu dx^\mu) (d z^n + A^n_\mu dx^\mu)\,.
 \label{Ansatz-gravity-torus}
 \end{align}
The lower-dimensional field strength is a generalisation of the result of a torus
reduction: in addition to gravity one finds $n$ vectors $A^m_\mu$, a dilaton $\phi$ and a
scalar matrix $M_{mn}$ which parameterises a coset $SL(n,\mathbb{R}) / SO(n)$ (see
section~\ref{sec:global-symmetries} for scalar cosets). The latter corresponds to $n-1$
dilatons and $\tfrac{1}{2} n (n-1)$ axions. Again, one can obtain the lower-dimensional
Lagrangian by a reduction of the Einstein Hilbert term:
 \begin{align}
  \CL = \sqrt{-\hat{g}} \hat{R} = \sqrt{-g} [ R -\tfrac{1}{2} (\partial \phi)^2 + \tfrac{1}{4}
  \text{Tr}(\partial M \partial M^{-1}) - \tfrac{1}{2 \, 2!} e^{2(\beta - \alpha) \phi} M_{mn} F^m F^n ]
  \,,
 \label{action-gravity-torus}
 \end{align}
with $F^m = d A^m$. The convenient values for $\alpha$ and $\beta$ now read
 \begin{align}
  \alpha^2 = \frac{n}{2(D+n-2)(D-2)} \,, \qquad \beta = - \frac{(D-2) \alpha}{n} \,,
 \label{constants-torus}
 \end{align}
yielding the Lagrangian in the conventional form \eqref{action-gravity-torus}.

As in the reduction over the circle, one can wonder which general coordinate
transformations \eqref{gct} preserve the form of the reduction Ansatz and induce a
lower-dimensional transformation. For the torus reduction \eqref{Ansatz-gravity-torus}
these turn out to be
 \begin{align}
  \hat{\xi}^\mu = \xi^\mu (x) \,, \qquad \hat{\xi}^m = \lambda^m (x) + \Lambda^m{}_n z^n \,.
 \label{gct-reduction}
 \end{align}
These can respectively be understood as $D$-dimensional general coordinate
transformations, $U(1)^n$ gauge transformations and a global $GL(n,\mathbb{R})$ symmetry.
As in the torus case, the global transformations can be integrated to finite
transformations, where it is convenient to use a split into $SL(n,\mathbb{R})$ and
$\mathbb{R}^+$. The former acts in the obvious way on the $SL(n,\mathbb{R})$ indices
while the latter can again be combined with the reduced trombone symmetry to yield the
lower-dimensional trombone symmetry and the dilaton scale symmetry (formulae
\eqref{trombone} and \eqref{dilaton-shift} with an extra $m$ index for $A_\mu$). Thus, in
comparison with the circle case, the new features of the $n$-torus reduction are the $n$
Abelian gauge symmetries and the global $SL(n,\mathbb{R})$ symmetry.

\subsection{Inclusion of Gauge Potentials} \label{sec:toroidal-3}

We will now consider the reduction of a gauge potential of rank $d$ over a circle. The
dynamics of the higher-dimensional potential $\hat{C}^{(d)}$, coupled to gravity and
possibly a dilaton $\hat{\varphi}$, is determined by
 \begin{align}
  \CL = \sqrt{-\hat{g}} [ -\tfrac{1}{2} (\partial \hat{\varphi})^2
   - \tfrac{1}{2} e^{a \hat{\varphi}} \hat{G}^{(d+1)} \cdot \hat{G}^{(d+1)} ] \,.
 \end{align}
with $\hat{G}^{(d+1)} = d \hat{C}^{(d)}$, where we have included the dilaton kinetic
term. The parameter $a$ characterises the dilaton coupling. For gravity we will take the
reduction Ansatz \eqref{Ansatz-gravity-circle} while the rest of the reduction Ansatz
reads
 \begin{align}
   \hat{C}^{(d)} = C^{(d)} + (dz + A) \wdg C^{(d-1)} \,, \qquad \hat{\varphi} = \varphi \,.
 \label{p-form-Ansatz}
 \end{align}
where $A$ is the Kaluza-Klein vector field of the gravity Ansatz
\eqref{Ansatz-gravity-circle}. The resulting Lagrangian is described by
 \begin{align}
  \CL = \sqrt{-{g}} [ -\tfrac{1}{2} ({\partial} {\varphi})^2
   - \tfrac{1}{2} e^{a {\varphi} -2 d \alpha \phi} {G}^{(d+1)} \cdot {G}^{(d+1)}
   - \tfrac{1}{2} e^{a {\varphi} + 2 (D-d-1) \alpha \phi} G^{(d)} \cdot {G}^{(d)} ] \,,
 \end{align}
with field strengths ${G}^{(d+1)} = d {C}^{(d)} + F \wdg \C{d-1}$ and ${G}^{(d)} = d
{C}^{(d-1)}$. Note that $\Delta$, defined in \eqref{Delta}, is preserved under the
operation of dimensional reduction; the value associated to $\hat{G}^{(d+1)}$ is also
found for both ${G}^{(d+1)}$ and ${G}^{(d)}$:
 \begin{align}
  \Delta & = a^2 + \frac{2 d (\hat{D}-d-2)}{\hat{D}-2} \,, \notag \\
  & =a^2 + (2 d \alpha)^2 + \frac{2 d (D-d-2)}{D-2} \,, \notag \\
  & = a^2 + (2 (D-d-1) \alpha )^2 + \frac{2 (d-1)(D-d-1)}{D-2} \,.
 \end{align}
Indeed, this corresponds to the statement from subsection~\ref{sec:susy-solutions-1} that
the parameter $\Delta$ is invariant under toroidal reduction.

The reduction of a $d$-form gauge potential over a circle can be performed a number of
times. This corresponds to the reduction over a torus. We will not discuss the explicit
Ansatz here since it follows from \eqref{p-form-Ansatz} but clearly there are general
formulae for the reduction of a gauge potential over a torus, similar to
\eqref{Ansatz-gravity-torus}. However, it is useful to know the resulting field content.
From subsequent applications of \eqref{p-form-Ansatz} it can be seen that the reduction
of a $d$-form over an $n$-torus gives rise to an amount of
 \begin{align}
  \left( \begin{array}{c} n \\ d- \tilde{d} \end{array} \right) \,, \qquad
  \text{where~~} d-n \leq \tilde{d} \leq d \,,
 \label{d-form-torus}
 \end{align}
forms of rank $\tilde{d}$. For example, reduction of a $2$-form over a $2$-torus gives
rise to a $2$-form, two vectors and a scalar.

Upon reduction over a torus, the gauge symmetry $\delta \hat{C}^{(d)} = d
\hat{\lambda}^{(d-1)}$ splits up in different lower-dimensional gauge transformations,
corresponding to the different $\tilde{d}$-form potentials. In the case of a circle, for
example, the gauge transformations that act covariantly on the lower-dimensional
potentials are
 \begin{align}
  \hat{\lambda}^{(d-1)} = \lambda^{(d-1)} + (dz + A) \wdg \lambda^{(d-2)} \,,
 \end{align}
where $A$ is the Kaluza-Klein vector. The gauge parameters $\lambda^{(d-1)}$ and
$\lambda^{(d-2)}$ correspond to the potentials $C^{(d)}$ and $C^{(d-1)}$, respectively.

In addition, the higher-dimensional Lagrangian is of course invariant under the general
coordinate transformations \eqref{gct}. As in the case of gravity, the reduction Ansatz
for gauge potentials over a torus is only covariant for the restricted transformations
\eqref{gct-reduction}. The lower-dimensional potentials transform in the usual way under
the lower-dimensional coordinate transformations and they can also be assigned a weight
under the global scale symmetries. Moreover, the $\tilde{d}$-form potentials, the number
of which is given by \eqref{d-form-torus}, form linear representations of the global
$SL(n,\mathbb{R})$ symmetry.

\subsection{Global Symmetry Enhancement} \label{sec:toroidal-4}

However, this is not the full story of gravity and gauge potentials on tori. It turns out
that, in special cases, one obtains a larger symmetry group than the $SL(n,\mathbb{R})$
whose appearance was guaranteed by the higher-dimensional coordinate transformations. An
obvious example is provided by the Lagrangian \eqref{action-gravity-circle}, which is the
reduction of the Einstein-Hilbert action over a circle. Reduction of
\eqref{action-gravity-circle} over an $n$-torus will lead to the global symmetry
$SL(n+1,\mathbb{R})$ rather than $SL(n,\mathbb{R})$. In this case one can understand the
symmetry enhancement by the higher-dimensional origin of \eqref{action-gravity-circle}.
However, there are also examples where such an explanation is not available.

As an example, consider the bosonic string, whose low-energy limit consists of gravity, a
dilaton and a rank-two gauge potential. After appropriate field redefinitions, the action
takes the canonical form \eqref{braneLagrangian} of the gravity-dilaton-potential system
of subsection~\ref{sec:susy-solutions-1}, with a dilaton coupling corresponding to
$\Delta = 4$. Upon reduction over an $n$-torus, it turns out that the global symmetry
group is enhanced from $SL(n,\mathbb{R})$ to $SO(n,n)$, see e.g.~\cite{deWit:2002vz}. In
addition, the scalar coset is enhanced as well:
 \begin{align}
  \frac{SL(n,\mathbb{R})}{SO(n)} \qquad \Rightarrow \qquad
  \frac{SO(n,n)}{SO(n) \times SO(n)} \,.
 \end{align}
For this to be possible, there is a conspiracy between the scalars coming from the metric
(giving rise to the smaller coset) and those coming from the two-form, together giving
rise to the larger coset.

Another example is provided by the reduction of (the bosonic sector of)
eleven-dimensional supergravity, whose symmetry groups are given in
table~\ref{tab:bosons}. Again, the symmetry groups and scalar cosets are larger than the
naive $SL(n,\mathbb{R})$. In this case this requires a collaboration between the scalars
coming from the metric and those coming from the three-from gauge potential. Although
often appearing in the low-energy limits of string or M-theory, it should be stressed
that such symmetry enhancement is a miraculous phenomenon and strongly dependent on the
details of interactions.

\subsection{Fermionic Sector} \label{sec:toroidal-5}

If one wants to dimensionally reduce a supergravity theory, clearly a recipe is required
for the fermionic sector. Since this is rather strongly dependent on the dimensions of
the higher- and lower-dimensional theories, we will not present explicit formulae but
only discuss the conceptual aspects. In the explicit reduction of supergravities that we
will perform later, such explicit formulae are given while in this chapter however, we
will mainly consider the bosonic part. For more detail see e.g.~\cite{Duff:1986hr, Pope}.

The essential idea in fermionic dimensional reduction is to split up the spinors as a
tensor product of spinors in the lower-dimensional space and the internal space. For
toroidal reduction, the internal spinors are taken constant. Thus, the reduction Ansatz
for a dilatino sketchily reads
 \begin{align}
  \hat{\lambda} = \sum_i \lambda^i \otimes \eta^i \,,
 \end{align}
where $\lambda^i$ are the lower-dimensional spinors and $\eta^i$ the internal spinors.
The range of $i$ is equal to the number of independent spin-$1/2$ components on the
internal manifold and therefore strongly depends on $\hat{D}-D$. This range corresponds
to the quotient of the \dof\ of the minimal spinors in the higher- and lower-dimensional
theory. For example, reducing over a seven-torus, the $32$-component minimal spinor
$\hat{\lambda}$ splits up in $4$-component minimal spinors $\lambda^i$ and therefore $i$
ranges from $1$ to $8$. This corresponds for example to the reduction of $N=1$
supergravity in 11D to $N=8$ supergravity in 4D over the seven-torus, which indeed allows
for eight constant internal spinors.

In the case of spin-$3/2$ fermions, i.e.~if the fermions are carrying a space-time index
as well, the procedure is a combination of the bosonic and fermionic Ans\"{a}tze. Both
spinorial and \st\ indices are split up into the lower-dimensional ranges:
 \begin{align}
  \hat{\psi}_\mu = \sum_i \psi_\mu^i \otimes \eta^i \,, \qquad
  \hat{\psi}_m = \sum_j \lambda^j \otimes \eta_m^j \,,
 \end{align}
where $\eta^i$ and $\eta_m^j$ are constant fermions on the internal space of spin 1/2 and
3/2, respectively. Thus the resulting fermions are the gravitini $\psi_\mu^i$ and the
dilatini $\lambda^j$.

We will indicate the changes in the fermionic Ans\"{a}tze in the upcoming cases of
twisted reduction and reductions over group manifolds.

\section{Reduction with a Twist} \label{sec:twist}

We will now discuss a generalisation of toroidal reduction, leading to a different
lower-dimensional description including e.g.~a scalar potential. This generalisation is
possible whenever the higher-dimensional theory contains a global symmetry
\cite{Scherk:1979ta}.

\subsection{Boundary Conditions and Twisted Expansions} \label{sec:twist-1}

In subsection~\ref{sec:reduction-intro-1} we considered the expansion of a complex scalar
field over an internal dimension under the assumption $\hat{\phi}(x,2 \pi R) =
\hat{\phi}(x,0)$, i.e.~a periodic boundary condition. One can also impose the generalised
boundary condition
 \begin{align}
  \hat{\phi}(x,2 \pi R) = e^{2 \pi i m R} \hat{\phi}(x,0) \,, \label{generalised-bnd-cnd}
 \end{align}
for some constant $m$. We will call this the twisted boundary condition, giving rise to
reduction with a twist. It leads to the expansion
 \begin{align}
  \hat{\phi} (x,z) = \sum_n e^{i (m+n/R) z } \phi_n (x) \,,
 \label{twisted-expansion}
 \end{align}
with a discrete spectrum of fields $\phi_n$. Note that this twisted expansion is
invariant under the transformation
 \begin{align}
  m \rightarrow m + 1/R \,, \qquad \phi_n \rightarrow \phi_{n+1} \,.
 \label{KK-tower-symmetry}
 \end{align}
For this reason one can always take $|m| \leq \tfrac{1}{2} / R$ without loss of
generality. Substitution into the Klein-Gordon equation yields
 \begin{align}
  \Box \phi_n - (m+n/R)^2 \phi_n = 0 \,.
 \label{mass-scalar-twist}
 \end{align}
Again, the higher-dimensional equation decouples into separate equations for all
components $\phi_n$ of (mass)$^2$ $(m+n/R)^2$.

Again, we would like to truncate to the sector with the lowest mass; to which component
$\phi_n$ this corresponds to is determined by the parameter $m$. Adhering to the above
convention of taking $|m| \leq \tfrac{1}{2} / R$, the lowest sector corresponds to the
component $\phi_0$, as in the massless case. Note however that the lower-dimensional
description is different; the periodic boundary condition gave rise to a massless scalar
while the twisted boundary condition leads to a scalar of (mass)$^2$ $m^2$. However, both
reductions are consistent: the field equations for $\phi_n$ with $n \neq 0$ are satisfied
and, equivalently, the dependence on the internal coordinate $z$ has dropped out.

Note that one can take $m=n/R$, leaving the above convention, and truncate consistently
to the component $\phi_0$. However, this does not correspond to the lightest mode.
Indeed, due to the above symmetry \eqref{KK-tower-symmetry}, this corresponds to a
toroidal reduction with expansion \eqref{Fourier-summation} and subsequent truncation to
a heavier mode. The ambiguity in the lower-dimensional description (i.e.~a massless or
massive scalar) stems from the possibility to consistently truncate the Kaluza-Klein
tower \eqref{Fourier-summation} in infinitely many ways.

\subsection{Global Symmetries and Monodromy} \label{sec:twist-2}

One can extend the generalised boundary condition \eqref{generalised-bnd-cnd} for $U(1)$
to other groups if the theory is invariant under a global symmetry group $G$. Consider a
set of fields $\hat{\phi}$, which we take to be scalars for concreteness but the
discussion can easily be extended to other fields. The fields $\hat{\phi}$ are taken to
transform linearly under a global transformation: $\hat{\phi} \rightarrow g \hat{\phi}$
with $g \in G$, where we suppress group indices. This allows us to impose a more general
twisted boundary condition:
 \begin{align}
  \hat{\phi}(x,2 \pi R) = \mathcal{M}(g) \hat{\phi}(x,0) \,.
 \label{twisted-boundary-condition}
 \end{align}
Upon traversing the circle, the fields come back to themselves up to a symmetry
transformation: this transformation is called the monodromy. This boundary condition
leads to the twisted reduction Ansatz (i.e.~expansion and truncation to the lightest
modes)
 \begin{align}
  \hat{\phi}(x,z) = g(z) \phi(x) \,, \qquad \Rightarrow \qquad
  \mathcal{M}(g) = g(z=2 \pi R) g(z=0)^{-1} \,,
 \end{align}
with an element $g(z) \in G$ which depends on $z$. This is the generalisation to
arbitrary groups $G$ of the $U(1)$ twisted Ansatz \eqref{twisted-expansion} with $\phi_{n
\neq 0} =0$. For general groups $G$, the element $g(z)$ has to satisfy a consistency
criterium: the combination
 \begin{align}
  C = g(z)^{-1} \partial_z g(z) \,.
 \label{mass-matrix}
 \end{align}
must be a constant, which is required by the cancellation of the $z$-coordinate in the
lower-dimensional field equations and thus ensures consistency of the truncation to the
lightest modes $\phi$. Clearly, it can be solved by the $z$-dependence
 \begin{align}
  g(z) = \text{exp}(C z) \,, \qquad \text{with~~} \mathcal{M} = \text{exp}(2 \pi R C) \,.
 \end{align}
Thus the constants $C$ constitute an element of the Lie algebra of $G$. It determines
which linear combination of the generators of $G$ is employed in the twisted reduction.

This reduction Ansatz brings one from the higher-dimensional massless Klein-Gordon
equations to lower-dimensional massive Klein-Gordon equations:
 \begin{align}
  \hat{\Box} \hat{\phi} = 0 \qquad \Rightarrow \qquad
  \Box \phi + C^2 \phi = 0 \,.
 \end{align}
For this reason, the matrix $C$ is usually called the mass matrix. The eigenvalues of
$C^2$ are related to the (masses)$^2$ of the fields $\phi$: negative eigenvalues
correspond to positive (masses)$^2$ and vice versa. This depends on the compactness of
the subgroups of $G$ generated by $C$.

Note that the symmetry $G$ is generically broken upon twisted reduction: elements of $G$
do not preserve the field equations but rather transform the mass matrix by
 \begin{align}
  C \rightarrow g^{-1} C g \,.
 \end{align}
Only transformations for which the two mass matrices $C$ and $g^{-1} C g$ are equal
preserve the lower-dimensional field equations. This is in general only met by group
elements of the form as employed in the twisted reduction, i.e.~of the form
$\text{exp}(\lambda C)$. Note that $G$ is always preserved for $C = 0$, i.e.~under
toroidal reduction.

A special case consists of a mass matrix $C \neq 0$ that gives rise to a trivial
monodromy $\mathcal{M} = \mathbb{I}$. This is possible when $G$ contains a compact
subgroup and is the equivalent of the choice $m = 1/R$ considered in the previous
subsection: it corresponds to an expansion without twist (yielding trivial monodromy)
which is truncated to a massive mode (giving rise to the mass matrix), rather than the
massless mode \cite{Dabholkar:2002sy}. This situation will be encountered in
subsection~\ref{sec:D=9-gaugings-6}.

In this toy example, the group $G$ plays a central role. If $G$ is not a symmetry of the
theory, the reduction will not be consistent: one will (generally) not find cancellation
of all $z$-dependence in the lower-dimensional field equations. Thus, the existence of
$G$ allows for the twisted reduction Ansatz, as was first recognised by Scherk and
Schwarz\footnote{Their motivation was the spontaneous breaking of supersymmetry.}
\cite{Scherk:1979ta}.

Another important point is the fact that $G$ is only a global symmetry in the
higher-dimensional theory. It is impossible to perform twisted reductions of this kind
with local symmetries, as can easily be seen from our toy example. Suppose that the
higher-dimensional theory had a local symmetry $G$. Then the group element $g(z)$ in the
reduction Ansatz $\hat{\phi}(x,z) = g(z) \phi(x)$ can be brought to the left-hand side,
where it acts on $\hat{\phi}$. But this is just a symmetry transformation, which leaves
the higher-dimensional theory invariant. Thus the reduction Ans\"{a}tze $\hat{\phi}(x,z)
= \phi(x)$ and $\hat{\phi}(x,z) = g(z) \phi(x)$ will give the same result, a massless
lower-dimensional theory, if $g \in G$ is a local symmetry acting on $\hat{\phi}$.

\subsection{Gravity and Gaugings} \label{sec:twist-3}

We would like to apply our twisted reductions to supergravity in chapter \ref{ch:gauged}.
For this reason it is imperative to include gravity, which will bring in a number of new
features.

A useful subsector of supergravities to consider consists of only gravity and the
scalars. As in all maximal supergravities, the scalars parameterise a coset $G/H$,
denoted by $M$. Examples of $G$ and $H$ are given in table~\ref{tab:bosons}. The
Lagrangian reads
 \begin{align}
  \hat{\mathcal{L}} = \sqrt{-\hat{g}} [ \hat{R}
  + \tfrac{1}{4} \text{Tr}(\partial \hat{M} \partial \hat{M}^{-1}) ] \,.
 \end{align}
Toroidal reduction of this theory would correspond to the reduction Ansatz (in the case
of a circle)
 \begin{align}
  \hat{ds}^2 = e^{2 \alpha \phi} ds^2 + e^{2 \beta \phi} (dz + A_\mu dx^\mu)^2 \,, \qquad
  \hat{M} = M \,,
 \end{align}
with the constants $\alpha$ and $\beta$ given in \ref{constants-circle}. However, this
theory has a global symmetry\footnote{We restrict to symmetries of the action here. In
the case of symmetries that scale the action, e.g.~trombone symmetries, there is a
subtlety that is addressed in section~\ref{sec:no-action}.}, which acts as $M \rightarrow
\Omega M \Omega^T$ with $\Omega \in G$. Therefore it also allows for a twisted reduction,
parameterised by a mass matrix $C$ of the Lie algebra of $G$. The corresponding reduction
Ansatz reads
 \begin{align}
  \hat{ds}^2 = e^{2 \alpha \phi} ds^2 + e^{2 \beta \phi} (dz + A_\mu dx^\mu)^2 \,, \qquad
  \hat{M} = \intdep(z) M \intdep(z)^T \,,
 \label{twisted-reduction}
 \end{align}
for an element $\intdep(z) = \text{exp}(C z) \in G$. The resulting lower-dimensional
Lagrangian is given by
 \begin{align}
  \mathcal{L} = \sqrt{-g} & \, [ R - \tfrac{1}{2} (\partial \phi)^2 + \tfrac{1}{4}
  \text{Tr}(D M D M^{-1}) - \tfrac{1}{2 \, 2!} e^{2(\beta - \alpha) \phi} F^2 - V ] \,,
 \end{align}
where we have defined
 \begin{align}
  DM = d M + (C M + M C^T) A \,, \qquad
  V = \tfrac{1}{2} e^{2 (\alpha-\beta) \phi} \text{Tr}[C^2 + C^T M^{-1} C M] \,,
  \label{twisted-scalar-potential}
 \end{align}
where $DM$ and $V$ are the scalar field strength and the scalar potential, respectively.
These contain the deformations in terms of the mass matrix $C$.

In the previous discussion we have found the fate of the symmetry $G$ under twisted
reduction. Only a one-dimensional subgroup (with generator $C$) was preserved while the
remaining transformations were broken. When including gravity, it is interesting to
consider the action of the general coordinate transformations on the fields in the
twisted reduction. Recall the decomposition \eqref{gct-reduction} of the
higher-dimensional coordinate transformations $\delta x^{\hat{\mu}} = -
\hat{\xi}^{\hat{\mu}}$ into lower-dimensional coordinate transformations, a $U(1)$ gauge
symmetry and a global symmetry. The first of these transformations is unchanged,
i.e.~also the lower-dimensional theory has diffeomorphism invariance. The latter two are
modified due to the twist, however.

The $U(1)$ factor corresponds to the parameter $\hat{\xi}^{z} = \lambda (x)$. Note that
the scalar reduction Ansatz \eqref{twisted-reduction} is not invariant under this
coordinate transformation:
 \begin{align}
  \hat{M} = \intdep(z) M \intdep(z)^T \qquad \rightarrow \qquad
  \hat{M} & = \intdep(z-\lambda) M \intdep(z-\lambda)^T \,.
  \label{twisted-U(1)}
 \end{align}
Using $\intdep(z) = \text{exp}(C z) \in G$, an internal coordinate transformation
corresponds to the lower-dimensional transformation
 \begin{align}
  M \rightarrow \text{exp}(-C \lambda) M \text{exp}(- C^T \lambda) \,, \qquad
  A_\mu \rightarrow A_\mu + \partial_\mu \lambda \,.
 \end{align}
Indeed, the scalar field strength transforms covariantly under this local transformation.
Thus it turns out that the one-dimensional subgroup of $G$ generated by $C$ is in fact
gauged. This means that the global parameter of this transformation is elevated to a
local one. For this reason we say that twisted reduction leads to a non-trivial gauging
in the lower-dimensional theory.

The remaining parameter of the higher-dimensional diffeomorphisms, the constant $c$ in
the decomposition \eqref{gct-reduction}, acts as
 \begin{align}
  \hat{M} = \intdep(z) M \intdep(z)^T \qquad \rightarrow \qquad
  \hat{M} & = \intdep(z-cz) M \intdep(z-cz)^T \,.
 \end{align}
However, unlike the local $U(1)$ action \eqref{twisted-U(1)}, this can not be interpreted
as a lower-dimensional (i.e.~$z$-independent) transformation on $M$. For this reason the
extra scale symmetry is broken by the mass parameters $C$. Another way to see this stems
from the scale weight of the scalar potential under the global symmetry with parameter
$c$ given in \eqref{gct-reduction}. It is easily seen that the kinetic terms scale
differently than the scalar potential, which therefore breaks this symmetry.

In addition to the Ans\"{a}tze for gravity and scalars presented here, one can construct
similar formulae for the twisted reduction of e.g.~gauge potentials and fermions. The
guiding principle is the global symmetry: one modifies the toroidal Ans\"{a}tze by
inserting the transformation $\intdep(z)$ in the appropriate places, while the
consistency of such reductions is guaranteed by the global symmetry $G$. We will perform
twisted reductions of supergravities in section~\ref{sec:D=9-gaugings}.

\subsection{Enhanced Gaugings} \label{sec:twist-4}

However, one feature of enlarged field contents is noteworthy. In special cases, the
existence of extra gauge potentials in twisted reduction gives rise to an enhancement of
the gauging. This means that, in addition to the gauging of the twisted symmetry, one
finds other symmetries that have been elevated to local ones in the gauged theory.
Clearly, for this to be possible, one needs the global part of these symmetries to be
present in the ungauged theory. An additional requirement is the presence of the
corresponding gauge vectors, which are necessary to gauge the extra symmetries.

Rather than the most general possibility we will consider a specific example, which will
be important in section~\ref{sec:D=9-gaugings}. In addition to gravity and the scalar
coset $\hat{M}_{mn}$ of the previous subsection, we include a gauge vector $\hat{V}$. The
twist symmetry that we employ scales this gauge vector with a certain weight $\alpha$,
i.e.~$\hat{V} \rightarrow \Omega^{\alpha} \hat{V}$ with $\Omega \in \mathbb{R}^+$. In
addition, we have the gauge transformation $\delta \hat{V} = d \hat{\lambda}$.

As explained in the previous subsection, the transformation under the twist symmetry
determines the internal dependence of the reduction Ansatz, which therefore reads
 \begin{align}
  \hat{V} = \intdep^{\alpha} (V + \chi (dz + A)) \,, \qquad
  \hat{\lambda} = \intdep^{\alpha} \lambda
 \end{align}
where $\hat{V}$ splits up in a vector $V$ and an axion $\chi$, while the vector $A$ comes
from the metric. We have also included the reduction Ansatz for the gauge parameter
$\hat{\lambda}$. The internal dependence is inserted via the $\mathbb{R}^+$ group element
$\intdep = \text{exp}(m z)$.

Note that in the lower-dimensional theories there are two vectors: the Kaluza-Klein
vector $A$ coming from the metric and the vector $V$ coming from the higher-dimensional
vector. We will call the gauge parameters $\lambda_A$ and $\lambda_V$, respectively.
Their action on the axion $\chi$ reads
 \begin{align}
  \delta_A \chi = m \lambda_A \chi \,, \qquad \delta_V \chi = m \lambda_V \,.
 \end{align}
Thus one mass parameter yields two independent local transformations: we find gauge
symmetry enhancement. In fact, in this case the two gaugings are non-Abelian, since
 \begin{align}
  [ \delta_A , \delta_V ] = m^2 \lambda_A \lambda_V \,.
 \end{align}
These form the unique two-dimensional non-Abelian group, which we will denote by $A(1)$.

Though a general proof on the appearance of enhanced gaugings is lacking, the above
example seems to be typical for this phenomenon. The generic rule, applicable throughout
this article, is that any higher-dimensional gauge vector that transforms under the twist
symmetry will give rise to an extra gauging upon reduction. We will encounter different
examples of enhanced gaugings in section~\ref{sec:D=9-gaugings} and
\ref{sec:D=8-gaugings}, including the two-dimensional group $A(1)$.

\section{Reduction over a Group Manifold} \label{sec:group}

In the previous section we have seen how twisted reduction employs the global symmetries
of the higher-dimensional theory. In this section we will focus on the global symmetries
of the internal space instead, leading to group manifolds as internal spaces. For this
reason, the corresponding reduction procedure is only possible for theories which include
gravity.

\subsection{Group Manifolds} \label{sec:group-1}

A group manifold $G$ with coordinates $z^m$ consists of group elements $g=g(z^m) \in G$
(omitting group indices): points on the manifold correspond to elements of the group and
the dimension $n$ of the manifold equals $\text{dim}(G)$. Group multiplication, e.g.~$g
\rightarrow \Lambda_L g$ or $g \rightarrow g \Lambda_R$, corresponds to a coordinate
transformation. Both left and right multiplications correspond to transitively acting
coordinate transformations\footnote{Coordinate transformations are said to act
transitively if they relate all points on the manifold.} due to the group structure.

However, these coordinate transformations are not necessarily isometries of the metric.
To ensure that left multiplication gives rise to an isometry of the metric, we make the
choice
 \begin{align}
  ds_G^2 = g_{mn} \sigma^m \sigma^n \,, \qquad
  T_m \sigma^m = g^{-1} d g \,,
 \label{left-invariant-metric}
 \end{align}
with $g_{mn}$ arbitrary, $T_m$ generators and $g = g(z^m)$ elements of the group $G$. The
combinations $\sigma^m$ are called the Maurer-Cartan one-forms and can be written as
$\sigma^m = \intdep^m{}_n dz^n$ with $\intdep^m{}_n = \intdep^m{}_n (z^p)$. Since left
multiplication $g \rightarrow \Lambda_L g$ leaves $\sigma^m$ invariant it is an isometry
of the metric, which is therefore called the left-invariant metric. Note that the group
manifold with metric \eqref{left-invariant-metric} is homogeneous\footnote{We call a
manifold homogeneous if its metric allows for transitively acting isometries.} for all
values of $g_{mn}$ due to the transitively acting isometries of left multiplication.
These isometries are generated by the Killing vectors $L_n$, which by definition satisfy
the Maurer-Cartan equations
 \begin{align}
  [ L_m , L_n ] = f_{mn}{}^p L_p \,,
 \label{MC}
 \end{align}
where the $f_{mn}{}^{p}$ are given by
 \begin{align}
  f_{mn}{}^{p} = -2(\intdep^{-1})^{r}{}_{m} (\intdep^{-1})^{s}{}_{n}\,
  \partial_{[r} \intdep^{p}{}_{s]} \,.
 \label{structure-constants}
 \end{align}
Due to Lie's second theorem, these are always independent of $z^m$ and indeed are the
structure constants of the group $G$. Thus a group manifold with metric
\eqref{left-invariant-metric} has $n$ transitively acting isometries that span the group
$G$. Explicit examples of such Killing vectors are given in
subsection~\ref{sec:D=8-gaugings-2}.

With the choice of metric \eqref{left-invariant-metric}, right multiplication does not
give rise to an isometry for general $g_{mn}$: the transformation $g \rightarrow g
\Lambda_R$ is an isometry of the metric \eqref{left-invariant-metric} if and only if
$g_{mn}$ is given by the Cartan-Killing metric of the group $G$. Such a particular metric
is referred to as the bi-invariant metric since its isometry group is $G_L \times G_R$.

\subsection{Gravity on a Group Manifold} \label{sec:group-2}

To see how such group manifolds arise in reductions, we start out with the Ansatz for
toroidal reduction
 \begin{align}
  \hat{ds}^2 = e^{2 \alpha \phi} ds^2 + e^{2 \beta \phi} M_{mn} (d z^m + A^m_\mu dx^\mu) (d z^n + A^n_\mu dx^\mu) \,,
 \label{Ansatz-gravity-torus-2}
 \end{align}
with $\alpha$ and $\beta$ given in \eqref{constants-torus}. As noted before, this
reduction Ansatz transforms covariantly under general coordinate transformations of the
special form
 \begin{align}
   \hat{\xi}^\mu = \xi^\mu (x) \,, \qquad \hat{\xi}^m = \lambda^m (x) + \Lambda^m{}_n z^n \,.
 \label{gct-reduction-2}
 \end{align}
The latter term corresponds to $GL(n,\mathbb{R})$ transformations on the internal
coordinates $z^m$. These will reduce to global symmetries of the lower-dimensional
theory.

As is the case of global symmetries of the higher-dimensional theories, these internal
transformations can also be used for a generalised reduction procedure
\cite{Scherk:1979zr}. In complete analogy to the twisted reduction, one can take the
toroidal reduction Ansatz and perform a $GL(n,\mathbb{R})$ transformation on the
lower-dimensional fields, whose parameter we call $\intdep$. If this is a constant
parameter, the lower-dimensional theory is clearly unchanged due to its global symmetry.
However, we allow for a certain internal coordinate dependence of the transformation
parameter: $\intdep^m{}_n = \intdep^m{}_n (z^p)$. Thus, for reduction of gravity, the
Ansatz can be obtained by applying $\intdep^m{}_n$ transformations on all fields in the
toroidal Ansatz \eqref{Ansatz-gravity-torus-2} and reads
 \begin{align}
  \hat{ds}^2 & = e^{2 \alpha \phi} ds^2
   + e^{2 \beta \phi} \intdep^m{}_p \intdep^q{}_n M_{pq}
   (d z^m + (\intdep^{-1})^m{}_r A^r_\mu dx^\mu) (d z^n + (\intdep^{-1})^n{}_s A^s_\mu dx^\mu) \,,
   \notag \\
  & = e^{2 \alpha \phi} ds^2
  + e^{2 \beta \phi} M_{mn}(\sigma^m + A^m_\mu dx^\mu)(\sigma^n + A^n_\mu dx^\mu) \,,
 \label{Ansatz-gravity-group}
 \end{align}
with $\sigma^m = \intdep^m{}_n dz^n$. Thus the internal part of this metric, given by
 \begin{align}
  ds_{G}^2 = e^{2 \beta \phi} M_{mn} \sigma^m \sigma^n \,,
 \end{align}
corresponds to the left-invariant metric of a group manifold. Therefore this reduction
procedure corresponds to the reduction over a group manifold $G$, where one uses the
left-invariant metric on the group manifold \cite{DeWitt, Cho:1975sf, Scherk:1979zr}.

Upon reduction of the Einstein-Hilbert term, the $GL(n,\mathbb{R})$ transformation will
cancel in many places, due to the fact that it is a global symmetry of the
lower-dimensional theory. Only when the parameters $\intdep^m{}_n$ run into internal
derivatives, the cancellation of such terms is no longer guaranteed. However, it turns
out that the only combination of $\intdep^m{}_n$'s that survives upon reduction is
exactly the combination $f_{mn}{}^p$ of \eqref{structure-constants}. Therefore, to obtain
a lower-dimensional theory without $z^m$ dependence, one has to require that the
combinations $f_{mn}{}^{p}$ are $z^m$ independent. As we have seen in the previous
subsection, this is guaranteed if one takes the internal dependence of $\intdep^m{}_n$
such that
 \begin{align}
  T_m \intdep^m{}_n dz^n = g^{-1} d g \,,
 \end{align}
for group elements $g=g(z^m)$.

Explicitly, the lower-dimensional result of the reduction of the higher-dimensional
action with Einstein-Hilbert term is given by\footnote{Here we restrict to unimodular
groups, having structure constants with vanishing trace: $f_{mn}{}^n =0$. For
non-unimodular groups there is a number of complications which will be addressed in
section~\ref{sec:no-action}.}
 \begin{align}
  \mathcal{L} = \sqrt{-g} & \, \big[ R + \tfrac{1}{4} {\rm Tr}({D} {\cal M} {D}{\cal M}^{-1})
  - \tfrac{1}{2} (\partial \phi)^2 - \tfrac{1}{4} e^{2(\alpha-\beta)\phi} F^{m}{\cal M}_{mn} F^{n}
  - V \big] \,,
 \end{align}
where the field strengths are given by
 \begin{align}
  F^{m} = 2\partial A^{m} -f_{np}{}^{m}A^{n}A^{p} \,, \qquad
  D {\cal M}_{mn} = \partial {\cal M}_{mn} + 2 f_{q(m}{}^{p} A^{q} {\cal M}_{n)p}
  \,.
 \end{align}
In addition, one has a scalar potential
 \begin{align}
 & V = \tfrac{1}{4} e^{2(\beta-\alpha)\phi}\, [ 2{\cal M}^{nq}f_{mn}{}^{p} f_{pq}{}^{m}
                  + {\cal M}^{mq}{\cal M}^{nr}{\cal M}_{ps}
                  f_{mn}{}^{p}f_{qr}{}^{s} ] \,.
 \end{align}
Thus we find two differences when compared with toroidal reduction: the modification of
field strengths and the appearance of a scalar potential. These deformations of the
massless theory are linear and quadratic in the structure constants, respectively.

\subsection{Gaugings from Group Manifolds} \label{sec:group-3}

Again, it is natural to wonder about the lower-dimensional symmetries. The
higher-dimensional coordinate transformations that act covariantly on the reduction
Ansatz are
 \begin{align}
   \hat{\xi}^\mu = \xi^\mu (x) \,, \qquad \hat{\xi}^m = \intdep^m{}_n \lambda^n (x) \,,
 \label{gct-group-reduction}
 \end{align}
consisting of lower-dimensional coordinate transformations with parameter $\xi^\mu (x)$
and gauge transformations with parameter $\lambda^n (x)$. The effect of the latter on the
lower-dimensional fields is given by
 \begin{align}
  \delta A^m_\mu = \partial_\mu \lambda^m + f_{np}{}^m \lambda^n A_\mu^p \,, \qquad
  \delta M_{mn} = f_{mp}{}^q \lambda^p M_{qn} + f_{np}{}^q \lambda^p M_{mq} \,,
 \end{align}
while the metric is invariant. These are non-Abelian gauge transformations with gauge
vectors $A_\mu^m$ and structure constants $f_{mn}{}^p$.

As in the twisted reduction, the global symmetry employed in the reduction is generically
broken for the larger part. In the group manifold case, this symmetry is
$GL(n,\mathbb{R})$ and comes from the internal coordinate transformations with
$\hat{\xi}^m = \Lambda^m{}_n z^n$. In the gauged theory, the $GL(n,\mathbb{R})$ is in
general no longer a symmetry since it does not preserve the structure constants. The
unbroken part is exactly given by the automorphism group of the structure constants,
i.e.~the transformations satisfying
 \begin{align}
  f_{mn}{}^p = \Lambda_m{}^q \Lambda_n{}^r (\Lambda^{-1})_s{}^p f_{qr}{}^s \,.
 \end{align}
Of course it always includes the gauge group, which is embedded in the global symmetry
group $GL(n,\mathbb{R})$ via
\begin{equation}
  \Lambda_n{}^m = e^{\lambda^k f_{kn}{}^m} \,,
 \label{adjoint}
\end{equation}
where $\lambda^k$ are the local parameters of the gauge transformations. However, the
full automorphism group can be bigger; for instance, its dimension is $n^2$ in case of
$f_{mn}{}^p =0$. Of course this amounts to the fact that the ungauged theory has a
$GL(n,\mathbb{R})$ symmetry. All other cases have Dim(Aut) $< n^2$ and thus break the
$GL(n,\mathbb{R})$ symmetry to some extent.

Thus reduction over a group manifold leads to a gauging, where the adjoint representation
of the gauge group is embedded in the fundamental representation of the global symmetry
group \eqref{adjoint}. Therefore, reduction over a torus $T^n$ leads to a theory without
gauging, since the adjoint of $U(1)^n$ is trivial; we call this an ungauged theory. In
contrast, gauge groups with non-trivial adjoints lead to the gauging of a number of
global symmetries; these are called gauged theories.

Although we have only discussed gravity on a group manifold in this section, the same
reasoning can be applied to other fields, as was already done in \cite{Scherk:1979zr}.
The behaviour under the internal transformations \eqref{gct-reduction-2} determines the
reduction Ansatz and guarantees consistency of the reduction. In
section~\ref{sec:D=8-gaugings} we will apply group manifold reductions to $D=11$ maximal
supergravity.

\subsection{Consistency of Reduction over Group Manifolds} \label{sec:group-4}

The consistency of this procedures is guaranteed by group-theoretical arguments: there is
always an internal dependence such that only the structure constants appear in the
lower-dimensional theory. An equivalent statement is that the Kaluza-Klein tower of
fields, stemming from the expansion over the group manifold, is truncated to fields that
are singlets under $G_L$. Since singlets can never generate non-singlets, this guarantees
that the field equations for the non-singlets are automatically satisfied. In other
words, the consistency of this reduction can be understood from the presence of the
transitively acting isometries of $G_L$, over which one can reduce.

The metric \eqref{Ansatz-gravity-group} includes deformations from the bi-invariant
metric, which are parameterised by the lower-dimensional fields. Since the metric always
retains a set of transitively acting isometries, these are called homogeneous
deformations and reduce the isometry group from $G_L \times G_R$ to $G_L \times H_R$
where $H_R \subset G_R$. In the literature, such deformations are referred to as
squashings of the maximally symmetric metric \cite{Duff:1986hr}.

Another result from group theory is that the matrix $\intdep^m{}_n$, parameterising the
dependence on the internal coordinates, can be taken independent of a set of coordinates
that correspond to commuting isometries. Clearly, an extreme case is the torus, having
all isometries commuting and indeed allowing for a constant $\intdep^m{}_n$. The opposite
extreme has no two isometries that commute, in which case $\intdep^m{}_n$ depends on all
but one internal coordinates.

\subsection{Twisted vs.~Group Manifold Reductions} \label{sec:group-5}

Having treated both twisted and group manifold reductions, we would like to comment on
some similarities and differences.

An important common feature of the two reduction schemes is the reliance on global
symmetries in the reduction Ansatz. The twisted reduction employs a global symmetry of
the higher-dimensional theory while group manifold reduction makes use of the global
symmetries of the internal manifold. Due to these global symmetries, one can introduce a
certain dependence on the internal coordinate via $\intdep(z^p)$, which will either
cancel or appear in the specific combinations $C_m{}^n$ or $f_{mn}{}^p$ defined in
\eqref{mass-matrix} and \eqref{structure-constants}. Thus, to interpret the emerging
equations as lower-dimensional, one has to require these combinations to be
$z$-independent. For $C_m{}^n$ this implies that it is the Lie algebra element
corresponding to the twisted reduction while the $f_{mn}{}^p$'s are interpreted as the
structure constants of the isometry group of the internal manifold.

This brings us to an equally important difference: due to the different dependences on
the internal coordinates, the resulting deformations will be different as well. In the
twisted case, the mass matrix $C_m{}^n$ induces a gauging which is always one-dimensional
and therefore Abelian (in the generic cases without enhanced gaugings). On the contrary,
the structure constants $f_{mn}{}^p$ necessarily involve non-Abelian gaugings. Both
gaugings induce a scalar potential.

However, in certain cases there is a relation between the two reduction schemes. Consider
reduction over group manifolds with $n-1$ commuting isometries: these can be split up in
a toroidal reduction over $n-1$ dimensions followed by a twisted reduction over the
remaining dimension. In this scenario, the twist symmetry is a subgroup of the
$GL(n-1,\mathbb{R})$ global symmetry obtained from the toroidal reduction of gravity.
Thus, a twisted reduction with a symmetry that has a higher-dimensional origin can also
be interpreted as a group manifold reduction. Indeed, in such cases one must encounter
the phenomenon of gauging enhancement, as discussed in subsection~\ref{sec:twist-4}: the
twisted reduction must lead to a non-Abelian gauging. In this example, the extra gauge
vectors, transforming under the twist symmetry, are provided by the reduction of gravity
over the $T^{n-1}$. Explicit cases will be discussed in sections~\ref{sec:D=9-gaugings}
and \ref{sec:D=8-gaugings}.

\section{Reduction over a Coset Manifold} \label{sec:coset}

We turn to the most complicated reduction procedure that we will discuss, reduction over
a coset manifold. Unlike the preceding reductions, its consistency is not secured by
group-theoretical arguments and has been proven only in very special cases.

\subsection{Coset Manifolds} \label{sec:coset-1}

A coset manifold is defined as follows. Consider a group manifold $G$ with group elements
$g \in G$. A subgroup of $G$, denoted by $H$, can be used to construct a coset manifold
by identifying group elements that are related by a right-acting transformation of an
element of $H$:
 \begin{align}
  g \cong g h \,, \qquad \forall g \in G \,, \qquad \forall h \in H \subset G \,.
  \label{coset-identification}
 \end{align}
The corresponding coset manifold is denoted by $G/H$. Its dimension $n$ is equal to
$\text{dim}[G] - \text{dim}[H]$.

Remember that a group manifold has coordinate transformations corresponding to left- and
right-acting group multiplication. Indeed, the bi-invariant metric has isometry group
$G_L \times G_R$. For coset manifolds, only the left-acting group multiplication
corresponds to coordinate transformations, while right-acting multiplication takes one
outside of the coset manifold:
 \begin{align}
  g \rightarrow g \Lambda_R \ncong g h \Lambda_R \leftarrow g h \,,
 \end{align}
since $\Lambda_R^{-1} h \Lambda_R$ is not an element of $H$ in general. Therefore, the
most symmetric metric on a coset manifold $G/H$ will have isometry group $G$ (omitting
the subscript) rather than $G_L \times G_R$. This metric is usually called the round
metric. The subgroup $H$ is known as the isotropy group.

An important example of a coset manifold is the sphere $S^n$, which has isometry group $G
= SO(n+1)$ and isotropy group $H = SO(n)$. Indeed, for every point on the sphere, one can
perform $SO(n)$ rotations that leave this point invariant. This corresponds to the
identification \eqref{coset-identification}.

\subsection{Coset Reductions} \label{sec:coset-2}

The maximal isometry group of a coset manifold $G/H$ is $G$. However, for generic
metrics, the coset manifold has no isometries at all. Therefore, the deformations from
the maximally symmetric metric are called inhomogeneous: they break all isometries and
thus also homogene\"{\i}ty. The lower-dimensional fields parameterise these deformations
and fall in multiplets of the maximal isometry group $G$. In particular, one expects
massless gauge vectors corresponding to $G$.

The lack of isometries is an important issue in reductions over coset manifolds. Due to
this feature, reduction over a coset is a highly non-trivial procedure whose consistency
is not guaranteed by group-theoretical arguments. Only in very special cases the
consistency has been proven, though not at all understood. Most of these cases are
concerned with spheres $S^n$, resulting in massless $SO(n+1)$ gauge vectors upon
reduction. A necessary requirement for this to be possible is the presence of
$\tfrac{1}{2} n (n+1)$ gauge vectors in the lower dimensions. In addition, which is the
condition we will focus on, the ungauged theory must have a global symmetry that contains
$SO(n+1)$. This rules out coset reductions of pure gravity: we have seen in
subsection~\ref{sec:toroidal-2} that reduction of gravity over $T^n$ leads to an
$SL(n,\mathbb{R})$ global symmetry group, which does not contain $SO(n+1)$ and therefore
does not allow for such a gauge group.

Extending gravity with gauge fields and scalars, the situation looks more promising. We
already encountered examples of such global symmetry enhancement in
subsection~\ref{sec:toroidal-4}. Indeed, as we will discuss, these all allow for coset
reductions. We will consider gravity, a dilaton $\phi$ and an $n$-form field strength
$G^{(n)} = d C^{(n-1)}$ whose coupling to the dilaton is parameterised by the constant
$a$. Its Lagrangian reads
 \begin{align}
  \mathcal{L} = \sqrt{-g} [ R -\tfrac{1}{2} (\partial \phi)^2 - \tfrac{1}{2} e^{a \phi} \G{n} \cdot \G{n} ] \,,
 \label{coset-system}
 \end{align}
which is identical to the system giving rise to the brane solutions of
subsection~\ref{sec:susy-solutions-1} with $n =d+1$. Note that the dilaton decouples for
$a=0$ and can be consistently truncated away. Due to Hodge duality, the field strengths
$\G{n}$ and $\G{D-n}$ are equivalent and therefore we restrict to $n \leq \tfrac{1}{2}
D$. It turns out \cite{Cvetic:2000dm} that reduction of this system over $T^n$ gives rise
to an $SL(n+1,\mathbb{R})$ global symmetry (rather than just the $SL(n,\mathbb{R})$ that
follows from gravity) if the dilaton coupling is given by
 \begin{align}
  a^2 = \frac{8-2(n-3)(D-n-3)}{D-2} \,,
 \end{align}
corresponding to the value $\Delta = 4$. This is only a necessary and (in general) not a
sufficient condition. The following cases do allow for coset reductions:
 \begin{itemize}
 \item
 $n=1$: This is clearly not the most interesting of all cases since the manifold $S^1 \sim SO(2) / SO(1)$
 is not a coset since $SO(1)$ is trivial. A related point is that the necessary $SL(2,\mathbb{R})$ symmetry is
 already present in the higher-dimensional system \eqref{coset-system}. Therefore, in the
 discussion of coset reductions, we will not consider this case.
 \item
 $n=2$: In this case the system \eqref{coset-system} can be exactly obtained from the reduction of
 pure gravity over a circle \eqref{Ansatz-gravity-circle}. This higher-dimensional origin
 clearly explains the appearance of the $SL(3,\mathbb{R})$ symmetry rather than
 $SL(2,\mathbb{R})$, as noted in subsection~\ref{sec:toroidal-4}.
 The consistent reduction of this system over $S^2$ has been proven
 in \cite{Cvetic:2000dm}. An equivalent way to view this coset reduction of the
 Einstein-Maxwell-dilaton system is to perform an $SO(3)$ group manifold reduction on the
 higher-dimensional gravity \cite{Boonstra:1998mp}. We will
 encounter an example of this in section~\ref{sec:D=8-gaugings}.
 \item
 $n=3$: This is exactly the effective action of the bosonic string in $D$ dimensions.
 Indeed, the reduction of this effective action
 on an $n$-torus gives a global symmetry group $SO(n,n)$, as discussed in subsection~\ref{sec:toroidal-4}.
 The case $n=3$ then corresponds to $SO(3,3) \sim SL(4,\mathbb{R})$, which allows
 for a gauging of $SO(4)$. The consistency of the corresponding $S^3$ coset reduction was proven in
 \cite{Cvetic:2000dm}.
 \item
 $n=4$: Reality of $a$ implies $D \leq 11$. Let us first consider $D=11$, in which case $a$ vanishes.
 It has been proven that the reduction of the system \eqref{coset-system} is inconsistent:
 one needs an extra interaction term, which is called a Chern-Simons term and
 which is exactly present in (the bosonic sector of) 11D supergravity, see \eqref{11Daction}.
 The corresponding reduction Ans\"{a}tze on $S^4$
 \cite{Nastase:1999cb, Nastase:1999kf} and $S^7$ \cite{deWit:1987iy} have been proven to be
 consistent. Indeed, maximal supergravity in $D=7$ and $D=4$ include global symmetry groups
 $SL(5,\mathbb{R})$ and $SL(8,\mathbb{R})$.
 Other cases with $D<11$ and $a \neq 0$ correspond to toroidal reduction of 11D
 supergravity to $D$ dimensions, followed by the coset reductions.
 \item
 $n=5$: Reality of $a$ implies $D \leq 10$. Again, in the limiting case $D=10$ one finds
 a vanishing dilaton coupling $a$. Reduction of the system \eqref{coset-system} is not
 consistent, however: one needs to impose a self-duality constraint on the five-form field
 strength. The corresponding reduction Ansatz has been constructed in \cite{Cvetic:2000nc}.
 Note that one again encounters a (bosonic subsector of) supergravity, in this case IIB supergravity\footnote{
  However, the consistency of the $S^5$ reduction of the full IIB supergravity
  has not been proven so far.}. Indeed, 5D maximal supergravity includes a global
 symmetry group $SL(6,\mathbb{R})$. Lower-dimensional cases with $a \neq 0$ are obtainable by toroidal
 reduction of the prime example in $D=10$.
 \end{itemize}
This concludes all possible sphere reductions. Cases with $n>5$ and real $a$ are all
related to any of the above cases by Hodge duality.

Of these coset reductions, the first case with $n=2$ is readily understood from its
higher-dimensional origin. Indeed, one can always split up a reduction over the group $G$
into a reduction over the subgroup $H \subset G$ followed by a coset reduction $G/H$
\cite{Cvetic:2003jy}. Clearly, the consistency of such a coset reduction is implied by
its higher-dimensional origin. The above example corresponds to $G=SO(3)$ and $H=SO(2)$.

The next case, which has $n=3$, allows for a reduction over the coset manifold $S^3 =
SO(4) / SO(3)$, leading to an $SO(4)$ gauge group, of which three corresponding vectors
are provided by the metric while the remaining three are provided by the three-form field
strength. This can be contrasted to the reduction of the same theory over the group
manifold $SO(3)$. As discussed in subsection~\ref{sec:group-3}, this gives rise to the
gauge group $SO(3)$ of which the vectors are provided by the metric only. The peculiar
feature in this case is that the group and coset manifolds coincide for the maximally
symmetric case, having isometries $SO(4) \sim SO(3) \times SO(3)$. The two reduction
schemes differ in the deformations that are included in the reduction Ans\"{a}tze. In the
group manifold these only parameterise homogeneous deformations, keeping a transitively
acting $SO(3)$ group of isometries. In contrast, the coset manifold reduction includes
also inhomogeneous deformations, breaking all isometries.

Another noteworthy feature of the bosonic string effective action is the global symmetry
group $SO(n,n)$ that appears upon reduction over an $n$-torus. This has led to the
conjecture \cite{Duff:1986ya} that it allows for a consistent truncation over the coset
manifold $(G \times G)/G$, where $G$ has dimension $n$ and is a compact subgroup of
$SO(n)$. Though not proven in generality, such a truncation is believed to be consistent,
of which the above case $n=3$ (whose consistency has been proven \cite{Cvetic:2000dm})
provides an example.

It is remarkable that for the remaining cases $n=4$ and $n=5$, purely bosonic
considerations lead to subsectors of the highest-dimensional supergravities, while
consistency of the reduction requires exactly the interactions provided by supergravity.
These spherical reductions have been employed to generate lower-dimensional gauged
supergravities. We will discuss these in more detail in section~\ref{sec:CSO-gaugings}.

\begin{table}[ht]
 \begin{center}
 \begin{tabular}{||c||c|c|c||c|c||}
  \hline \rule[-1mm]{0mm}{6mm}
 Method & Requirement & Manifold & Gauging & Min. & Max. \\
  \hline \hline \rule[-1mm]{0mm}{6mm}
 Toroidal & $-$ & $U(1)^n$ & $-$ & $U(1)^n$ & $U(1)^n$ \\
  \hline \rule[-1mm]{0mm}{6mm}
 Twisted & Global symmetry & $U(1)$ & $U(1)$ & $U(1)$ & $U(1)$ \\
  \hline \rule[-1mm]{0mm}{6mm}
 Group manifold & Gravity & $G$ & $G_R$ & $G_L$ & $G_L \times G_R$ \\
   \hline \rule[-1mm]{0mm}{6mm}
 Coset manifold & Gravity and flux & $G/H$ & $G$ & $-$ & $G$ \\
 \hline
\end{tabular}
 \caption{\it The different reduction schemes with the requirements, the internal manifolds and the resulting gaugings
 of the lower-dimensional theories. We also give the minimum and maximum possible isometry groups
 of the internal manifold. Adapted from \cite{Cvetic:2003jy}.}
 \label{tab:reductions}
\end{center}
\end{table}

In addition to the aforementioned spherical reductions, one can also consider reductions
over hyperboloid spaces defined by a hypersurface
 \begin{align}
  \sum_{i=1}^n q_i \mu_i{}^2 = 1 \,,
 \end{align}
with parameters $q_i = \pm 1$. The case with all $q_i = +1$ is the only compact manifold,
corresponding to the sphere, while the other cases are non-compact. Despite its
non-compactness, one can still perform consistent reductions over such spaces, giving
rise to non-compact $SO(p,q)$ gaugings with $p+q = n$. The consistency of reductions over
such hyperboloids can be deduced from analytical continuation of the corresponding
spherical reduction \cite{Hull:1988jw}. We will encounter examples of such hyperboloids
in subsection~\ref{sec:CSO-gaugings-2}.

\section{Lagrangian vs.~Field Equations} \label{sec:no-action}

\begin{figure}[h]
\centerline{\epsfig{file=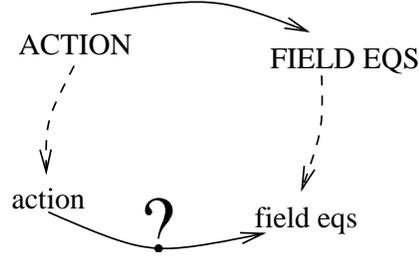,width=0.35\textwidth}}
 \caption{\it Reductions of the action or the field equations do not necessarily yield equivalent
 lower-dimensional field equations, i.e.~the operations of minimalisation (denoted by the solid arrows)
 and reduction (denoted by the dashed arrows) of the action do not necessarily commute.}
\label{fig:variation}
\end{figure}

In the preceding sections on toroidal, twisted and group manifold reductions, we have
often substituted the reduction Ansatz in the Lagrangian to obtain the lower-dimensional
Lagrangian. However, the reduction Ansatz comprises a truncation to the lightest modes,
and the consistency of truncations is determined by the field equations. In general,
there is no reason to assume that substitution in the Lagrangian yields the same result
as substitution in the field equations, as illustrated in figure \ref{fig:variation}. In
this section we will first discuss an explicit example in which this issue arises and
then discuss general conditions in which the two schemes yield the same result, i.e.~in
which the operations in figure \ref{fig:variation} do commute.

\subsection{Toy Example} \label{sec:no-action-1}

As a toy model in 10D, we start with the truncation of IIA and IIB supergravity to the
metric and the dilaton. The Lagrangian reads
\begin{align}
  \hat{\mathcal{L}} = \sqrt{-\hat{g}} [ \hat{R} - \tfrac{1}{2} (\partial \hat{\phi})^2 ] \,,
\end{align}
while the corresponding Euler-Lagrange equations are given by
\begin{align}
  [ \hat{g}^{\hat\mu \hat\nu} ]: \qquad
  & \hat{R}_{\hat\mu \hat\nu} - \tfrac{1}{2} \hat{R} \hat{g}_{\hat\mu \hat\nu}
    - \tfrac{1}{2} \partial_{\hat \mu} \hat \phi \partial_{\hat \nu} \hat \phi
    + \tfrac{1}{4} (\partial \hat{\phi})^2 \hat{g}_{\hat\mu \hat\nu} = 0 \,, \notag \\
  [ \hat\phi ]: \qquad
  & \Box \hat \phi = 0 \,.
\end{align}
This system has two global symmetries, as discussed in
subsection~\ref{sec:parent-sugra-3}: one can either scale the metric or one can shift the
dilaton, parameterised by $m_g$ and $m_\phi$, respectively:
\begin{align}
  \hat{g}_{\hat\mu \hat\nu} \rightarrow e^{2 m_g} \hat{g}_{\hat\mu \hat\nu} \,, \qquad
  \hat\phi \rightarrow \hat\phi + m_\phi \,.
\end{align}
The shift of the dilaton is a symmetry of the Lagrangian. The trombone symmetry, scaling
the metric, is a symmetry of the field equations only; it scales the Lagrangian. This
will prove an important difference when performing twisted reductions. We will show that
one has to reduce field equations, rather than the Lagrangian, when performing reductions
with twist symmetries of the field equations only.

Using an arbitrary linear combination of the two global symmetries we make the following
Ansatz for twisted reduction over $z$ to nine dimensions:
\begin{align}
  \hat{g}_{\hat{\mu} \hat{\nu}} & =
    e^{2 m_g z} \left(\begin{array}{cc}
    e^{\sqrt{7}\varphi/14} g_{\mu \nu} & 0 \\
    0 & e^{-\sqrt{7}\varphi/2}
  \end{array} \right) \,, \qquad
  \hat \phi = \phi + m_\phi z \,,
\end{align}
where we have omitted the Kaluza-Klein vector $A_\mu$ for simplicity. Using this Ansatz
the 10D field equations yield the following 9D equations:
\begin{align}
  [\hat{g}^{\mu \nu}]: \qquad
  & {R}_{\mu \nu} - \tfrac{1}{2} {R} {g}_{\mu \nu}
    - \tfrac{1}{2} \partial_{ \mu}  \phi \partial_{ \nu}  \phi
    + \tfrac{1}{4} (\partial {\phi})^2 {g}_{\mu \nu}
    - \tfrac{1}{2} \partial_{ \mu}  \varphi \partial_{ \nu}  \varphi
    + \tfrac{1}{4} (\partial {\varphi})^2 {g}_{\mu \nu} + \notag \\
  & + e^{4 \varphi/\sqrt{7}} (\tfrac{1}{4} m_\phi{}^2+28 m_g{}^2)
  g_{\mu\nu} = 0 \,, \notag \\
  [\hat\phi]: \qquad
  & \Box \phi + 8 m_g m_\phi e^{4 \varphi/\sqrt{7}} = 0 \,, \notag \\
  [\hat{g}^{zz}]: \qquad
  & \Box \varphi - \tfrac{2}{\sqrt{7}} m_\phi{}^2 e^{4 \varphi/\sqrt{7}} = 0 \,.
\label{redfieldeqs}
\end{align}
Note that the field equations of the metric and both scalars get bilinear massive
deformations. In addition one has the reduction of the $\hat{g}^{z \mu}$ field equation
 \begin{align}
  [\hat{g}^{z \mu}]: \qquad
  & 2 \sqrt{7} m_g \partial_\mu \varphi
    + \tfrac{1}{2} m_\phi \partial_\mu \phi = 0 \,,
 \label{redvectoreqs}
 \end{align}
which is the equation of motion for the Kaluza-Klein vector $A_\mu$.

We would like to discuss whether the field equations can be reproduced by a Lagrangian.
We will not consider the field equation for the vector \eqref{redvectoreqs} since it is
not important for our argument, and restrict to \eqref{redfieldeqs}. If one performs the
twisted reduction on the 10D Lagrangian, instead of on the field equations, the result
reads $\hat{\mathcal{L}} = e^{8 m_g z} \mathcal{L}$ with the 9D Lagrangian given by
\begin{align}
  \mathcal{L} = \sqrt{-g} [ {R} - \tfrac{1}{2} (\partial {\phi})^2
  - \tfrac{1}{2} (\partial {\varphi})^2
  - V(\phi,\varphi) ] \text{~~~with~~~}
   V(\phi,\varphi) = e^{4 \varphi/7} (\tfrac{1}{2} m_\phi{}^2 + 72 m_g{}^2) \,.
\label{redlagr}
\end{align}
The corresponding Euler-Lagrange equations read
\begin{align}
  [{g}^{\mu \nu}]: \qquad
  & {R}_{\mu \nu} - \tfrac{1}{2} {R} {g}_{\mu \nu}
    - \tfrac{1}{2} \partial_{ \mu}  \phi \partial_{ \nu}  \phi
    + \tfrac{1}{4} (\partial {\phi})^2 {g}_{\mu \nu}
    - \tfrac{1}{2} \partial_{ \mu}  \varphi \partial_{ \nu}  \varphi
    + \tfrac{1}{4} (\partial {\varphi})^2 {g}_{\mu \nu} + \notag \\
  & + e^{4 \varphi/\sqrt{7}} (\tfrac{1}{4} m_\phi{}^2+36 m_g{}^2)
  g_{\mu\nu} = 0 \,, \notag \\
  [\phi]: \qquad
  & \Box \phi  = 0 \,, \notag \\
  [\varphi]: \qquad
  & \Box \varphi - \tfrac{4}{\sqrt{7}} e^{4 \varphi/\sqrt{7}}
    (\tfrac{1}{2} m_\phi{}^2 + 72 m_g{}^2) = 0 \,.
\label{redEL}
\end{align}
These Euler-Lagrange equations only coincide with the reduction of the 10D Euler-Lagrange
equations \eqref{redfieldeqs} provided $m_g = 0$. Thus the twisted reduction of the
Lagrangian does not give the correct answer if the Lagrangian scales: the Euler-Lagrange
equations \eqref{redEL} are not equal to the field equations \eqref{redfieldeqs} for $m_g
\neq 0$. In fact, the situation is worse \cite{Howe:1998qt, Lavrinenko:1998qa}: for $m_g
\neq 0$ there is no Lagrangian $\mathcal{L}$ with potential $V(\phi,\varphi)$ whose
Euler-Lagrange equations are the correct field equations \eqref{redfieldeqs}. The metric
field equation would require
\begin{align}
  V(\phi,\varphi)= e^{4 \varphi/\sqrt{7}} (\tfrac{1}{2} m_\phi{}^2 + 56 m_g{}^2) \,,
\end{align}
but this is inconsistent with the $\phi$ and $\varphi$ field equations for $m_g \neq 0$.

Thus we conclude that twisted reduction of the Lagrangian is only legitimate when the
exploited symmetry leaves the Lagrangian invariant rather than covariant. For symmetries
that scale the Lagrangian one has to reduce the field equations. Including the full field
content, such as the Kaluza-Klein vector $A_\mu$, does not change this conclusion.

However, it is possible that certain truncations do lead to the possibility of an action.
In our toy model, an example hereof is provided by the identification
 \begin{align}
  2 \sqrt{7} m_g \varphi = - \tfrac{1}{2} m_\phi \phi \,.
 \label{truncation}
 \end{align}
It can be seen that this truncation is fully consistent with the field equations
\eqref{redfieldeqs} and \eqref{redvectoreqs}. The resulting field equations can be
derived from the Lagrangian
\begin{align}
  \mathcal{L} = \sqrt{-g} [ R - \tfrac{1}{2} c^2 (\partial {\varphi})^2 -
  \tfrac{1}{2} c^2 m_\phi{}^2 e^{4 \varphi/7} ] \,,
\label{trunclagr}
\end{align}
with $c^2 = 1+112 m_g{}^2/m_\phi{}^2$. However, note that this is not the same result as
the insertion of this truncation in the reduced Lagrangian \eqref{redlagr}.

\subsection{General Requirements} \label{sec:no-action-2}

In the above example we have found that in the case of twisted reduction with a trombone
symmetry, one should reduce the field equations and not the action. The lower-dimensional
field equations do not even allow for a corresponding Lagrangian, i.e.~the field
equations can not be interpreted as Euler-Lagrange equations stemming from the
minimalisation of an action. The general rule for twisted reduction seems to be that the
Lagrangian should be invariant under the twist symmetry to allow for a lower-dimensional
Lagrangian. Reduction of the Lagrangian and the field equations are equivalent in such
cases. An example is provided by the twisted reduction with the global symmetry of a
scalar coset, as considered in section~\ref{sec:twist}. Though we know of no general
proof of this statement, it is generally believed to be consistent and no counterexamples
are known.

As for group manifold reductions, one finds a rather similar condition. It turns out
\cite{Sneddon, Pons:2003ka} that only group  manifolds with traceless structure
constants, i.e.~$f_{mn}{}^n = 0$, allow for reduction of the Lagrangian\footnote{Note
that this also proves the correctness of the reduction of the Lagrangian for toroidal
reduction, having $f_{mn}{}^p = 0$.}. Indeed, such manifolds employ a symmetry (stemming
from the higher-dimensional diffeomorphisms) that leaves the higher-dimensional
Lagrangian invariant. In terms of $\intdep^m{}_n$, this corresponds to the
$SL(n,\mathbb{R})$ subgroup of the full $GL(n,\mathbb{R})$. Reduction over such group
manifolds give rise to gauge groups whose adjoint is embedded in the $SL(n,\mathbb{R})$
global symmetry group.

In contrast, group manifolds with traceful structure constants allow for reduction of the
field equations. Indeed, these employ a symmetry that scales the higher-dimensional
Lagrangian. Such symmetries are only embeddable in $GL(n,\mathbb{R})$ and not in
$SL(n,\mathbb{R})$. The corresponding group manifold reduction gauges a subgroup of the
$GL(n,\mathbb{R})$ global symmetry group of the theory, of which only $SL(n,\mathbb{R})$
is a symmetry of the Lagrangian. Examples of such reduction spaces are hyperbolic group
manifolds, which we will encounter in sections \ref{sec:D=9-gaugings} and
\ref{sec:D=8-gaugings}.

This distinction between traceless and traceful structure constants, corresponding to
unimodular and non-unimodular groups respectively, has been the cause of some confusion
in the literature on group manifold reduction. It has been claimed \cite{Scherk:1979zr,
Pons:2003ka} that it is inconsistent to reduce over group manifolds with $f_{mn}{}^n \neq
0$. Another point of view, however, puts emphasis on the consistency of reduction of the
field equations \cite{Alonso-Alberca:2003jq, Cvetic:2003jy, Hull:2003kr}, where
lower-dimensional theories are consistent if every solution uplifts to a
higher-dimensional solution as well. In this article, we will adhere to the latter,
yielding lower-dimensional theories without actions that uplift consistently to the
higher dimension. The same distinction directly carries over to twisted reductions with
symmetries of the Lagrangian and the field equations, respectively.

Indeed, the same situation is encountered in coset reductions, in which one reduces field
equations rather than Lagrangians as well. However, in contrast to the twisted reduction
with a trombone symmetry or over a non-unimodular group manifold, the lower-dimensional
field equations can be obtained from an action. This action can not be derived by
substitution the reduction Ansatz in the higher-dimensional action, though. This is very
much like the truncation \eqref{truncation} in our toy model, leading to field equations
that allow for an action but that do not follow from the reduced action. We will
encounter such situations after reduction over non-unimodular group manifolds in
subsection~\ref{sec:D=8-gaugings-5}.

 \chapter{Gauged Maximal Supergravities} \label{ch:gauged}

\section{Introduction} \label{sec:gauged-intro}

In this chapter we will consider a number of deformations of the massless supergravities with
maximal supersymmetry, as discussed in chapter~\ref{ch:supergravity}. These deformations
are proportional to a parameter $m$ of mass dimension one. Indeed, some fields will
acquire masses proportional to the deformation parameter $m$. Often, another consequence
of the parameter will be the gauging of a global symmetry of the massless theory. For
this reason, such theories will be called gauged supergravities, which comprise the
larger part of this chapter. In cases where the mass parameter does not induce a gauging,
the theory is called a massive supergravity. The only known example of such a deformation
of maximal supergravity is the massive IIA supergravity \cite{Romans:1986tz}. Examples
with sixteen supercharges are the massive iia supergravities in six dimensions
\cite{Haack:2001iz}.

An important property of the massive deformations that we consider is that they do not
break any supersymmetry. The gauged or massive supergravities therefore have the same
number (i.e.~32) of supercharges as the corresponding ungauged or massless supergravity. This
preservation of supersymmetry under the massive deformation is in many cases guaranteed
due to a higher-dimensional origin: if a gauged supergravity can be obtained by any of
the techniques of chapter~\ref{ch:reductions}, it necessarily has the same amount of
supercharges as the higher-dimensional theory. Equivalently stated, reduction with a
twist or over a group or coset manifold does not break supersymmetry\footnote{This can
be contrasted with e.g.~Calabi-Yau compactifications, which break a fraction of the
supersymmetry. Reduction of IIA supergravity over the four-dimensional Calabi-Yau
manifold K3 with fluxes yields the massive iia supergravities with $N=2$ in $D=6$
\cite{Haack:2001iz}, see also \cite{Janssen:2001hy, Behrndt:2001ab}.}. Starting from a
maximal higher-dimensional supergravity, one can apply the different reductions of
chapter~\ref{ch:reductions} to generate many gauged maximal supergravities. We will
perform such reductions to construct gauged maximal supergravities in ten, nine and eight
dimensions. In addition, we will include massive IIA supergravity, which is the only
massive deformation of maximal supergravity without a known higher-dimensional origin.

Throughout this chapter we will reduce supersymmetry variations and field equations
rather than Lagrangians, since some of the rigid symmetries we employ for reduction scale
the Lagrangian. As was explained in detail in section~\ref{sec:no-action}, reduction with
a symmetry that scales the Lagrangian can only be performed on the field equations and
the supersymmetry variations, but not on the Lagrangian.

As a consequence of the non-trivial dimensional reduction, the supersymmetry variations and field
equations receive two types of massive deformations. There are implicit mass terms that
appear via the covariant field strengths, which generally acquire terms that are proportional to the mass parameter.
In addition, the supersymmetry transformations and field equations have explicit mass
terms. The explicit deformations of the massless supersymmetry variations $\delta_0$ are
denoted by $\delta_m$, which are linear in the mass parameter $m$. The fermionic field
equations, symbolically denoted by $X = 0$, also consist of a massless part $X_0$ plus
linear deformations $X_m$. In contrast, the bosonic field equations receive quadratic
massive deformations.

In the cases where it is possible to construct a Lagrangian, the quadratic deformations
of the bosonic field equations can be derived from the explicit mass terms in the
Lagrangian. These are also quadratic in $m$, only depend on the scalars and are therefore
called the scalar potential $V$. In many cases, the scalar potential can be written in
terms of a superpotential $W$, which is linear in the mass parameter:
 \begin{align}
  V =\frac{1}{4} \left( g^{AB} \frac{\delta W}{\delta\Phi^A}
  \frac{\delta W}{\delta \Phi^B}
  - \frac{D-1}{D-2} W^2 \right) \,.
  \label{super-potential}
 \end{align}
Here $g^{AB}$ is the inverse of the scalar metric $g_{AB}$ which occurs as $- g_{AB}
\partial\Phi^A \partial\Phi^B$ in the kinetic scalar terms, where $\Phi^A$ represents the
different scalars of the theory (both dilatons and axions). This expression follows from
the requirement of positive energy \cite{Townsend:1984iu}, as we will show in
subsection~\ref{sec:CSO-DW-2}. In supergravities with a scalar potential of this form,
the explicit deformation $\delta_m$ of the gravitino $\psi_\mu$ and the dilatini
$\lambda$ are proportional to the superpotential $W$ and its derivatives $\delta W /
\delta \Phi^A$, respectively. We will encounter such deformations in all maximal
supergravities except 11D and IIB.

A useful truncation of the full field content of the gauged or massive supergravities
consists of the metric and the scalars only, for which we will derive the bosonic field
equations. This subsector is interesting to us for two reasons. Firstly, it allows for
an investigation of the feasibility of combinations of mass parameters. Suppose one has
a massless theory with two different, separate deformations. One can wonder whether it
is possible to combine these two while preserving all supersymmetry. As we will show, an
investigation of the bosonic field equations for the metric and scalars suffices to
answer this question. Secondly, the vacua of gauged or massive theories are often carried
by the metric and scalars only, as we will see in chapter~\ref{ch:domain-walls}.

In the next section we will review the possible deformations in IIA supergravity. In the
following two sections we will construct different gauged maximal supergravities in nine
and eight dimensions, respectively. In the last section of this chapter we will consider
a general structure of gauged maximal supergravities in various dimensions, which are
obtainable via coset manifold and other reductions. Furthermore, we will discuss the
relation to the gauged supergravities in ten, nine and eight dimensions of the preceding
three sections.

\section{Massive and Gauged IIA Supergravity} \label{sec:massive-gauged-IIA}

In this section we will consider two deformations of IIA supergravity, one of which leads
to a massive version of IIA while the other gives rise to the gauged IIA theory.

\subsection{IIA Supergravity} \label{sec:massive-gauged-IIA-1}

As discussed in subsection~\ref{sec:parent-sugra-4}, toroidal reduction of the
eleven-dimensional theory over a circle yields the massless and ungauged IIA theory in
ten dimensions. The appropriate reduction Ans\"atze given in~\eqref{11Dred} with
$m_{11}=0$. The field content of the $D=10$ IIA supergravity theory is given by
\begin{align}
  \text{D=10 IIA:} \qquad \{ \nephat{e}_{\nephat\mu}{}^{\nephat a},       \nephat{B}_{\nephat\mu\nephat\nu},
  \nephat{\phi}, \nephat \C{1}_{\nephat\mu}, \nephat \C{3}_{\nephat\mu\nephat\nu\nephat\rho} ;
  \nephat{\psi}_{\nephat\mu}, \nephat{\lambda}
  \}\,,
\end{align}
with corresponding Lagrangian~\eqref{IIAaction} and supersymmetry transformations rules
\eqref{IIAsusy}. As discussed in subsection~\ref{sec:parent-sugra-3} and indicated in
table~\ref{tab:IIA-weights}, the IIA theory has two scaling symmetries\footnote{We use a
different basis of these symmetries in this section than in
subsection~\ref{sec:parent-sugra-3}.}. One is called $\nephat\alpha$, which scales the
Lagrangian and is the reduction of the 11D trombone symmetry. The other is
$\nephat\beta$, which leaves the Lagrangian invariant and stems from the internal
coordinate transformations of 11D supergravity.

\begin{table}[ht]
\begin{center}
\begin{tabular}{||c||c|c|c|c|c|c|c|c||c||c||}
\hline \rule[-1mm]{0mm}{6mm}
 $\mathbb{R}^+$ & $\nephat{e}_{\nephat\mu}{}^{\nephat a}$ & $\nephat{B}$ &
$e^{\nephat{\phi}}$ & $\nephat \C{1}$ & $\nephat \C{3}$ & $\nephat{\psi}_{\nephat\mu}$ &
$\nephat{\lambda}$
& $\nephat \epsilon$ & $\nephat{\mathcal{L}}$ & Origin \\
\hline \hline \rule[-2mm]{0mm}{6mm} $\nephat{\alpha}$      & $\ft{9}{8}$  & $3$  &
$\ft{3}{2}$   & $0$ & $3$   & $\ft{9}{16}$   & $-\ft{9}{16}$   & $\ft{9}{16}$   &
$9$   & 11D \\
\hline \rule[-2mm]{0mm}{6mm} $\nephat{\beta}$      & 0   & $\ft12$  & 1  & $-\ft34$ &
$-\ft14$ & $0$   & $0$   & $0$   &
$0$   & \\
\hline
\end{tabular}
\caption{\it The scaling weights of the $D=10$ IIA supergravity fields and action under
the scaling symmetries $\alpha$ and $\beta$ and their origin as higher-dimensional
scaling symmetries.} \label{tab:IIA-weights}
\end{center}
\end{table}

Note that the \RR\ vector $A$ is invariant under $\alpha$ while it scales under $\beta$.
This has important consequences when considering the possible gaugings of IIA
supergravity. Since gauge vectors transform in the adjoint of the gauge group and the
adjoint of $\mathbb{R}^+$ is trivial, only the symmetry $\alpha$ can be gauged while this
is impossible for the symmetry $\nephat\beta$ \cite{Giani:1984wc}. Indeed, we will
encounter the gauging of $\alpha$ below. In addition, the IIA theory allows for another
deformation, which we will first discuss.

\subsection{Massive IIA Supergravity} \label{sec:massive-gauged-IIA-2}

The first massive deformation, with mass parameter $m_{\rm R}$, was already encountered
in subsection~\ref{sec:parent-sugra-3} and was constructed by Romans
\cite{Romans:1986tz}. The explicit deformations of the supersymmetry transformations are
denoted by $\delta_{m_{\rm R}}$ and are given in terms of a superpotential $W$ and its
derivative with respect to the dilaton:
\begin{align}
  \delta_{m_{\rm R}} {\nephat \psi} _{\nephat \mu}
  = - \ft{1}{32} W \nephat{\Gamma}_{\nephat \mu} \nephat \epsilon \,, \qquad
  \delta_{m_{\rm R}} \nephat \lambda = \delta_\phi W \nephat \epsilon \,, \qquad
  W = e^{5 \nephat \phi /4} m_{\rm R} \,,
  \label{Romans-susy}
\end{align}
where $\delta_\phi W = \delta W / \delta \phi$. Furthermore, there are implicit massive
deformations to the original supersymmetry rules $\delta_0$, given in~\eqref{IIAsusy},
due to the fact that one must replace all massless field strengths by the
following massive counterparts:
\begin{align}
  {\nephat \G{2}} & = {\rm d} {\nephat \C{1}} + m_{\rm R} {\nephat B} \,, \quad
  {\nephat H} = {\rm d} {\nephat B} \,, \quad
  {\nephat \G{4}} = {\rm d} {\nephat \C{3}} + {\nephat \C{1}} \wdg {\nephat H}
  + \ft{1}{2} m_{\rm R} {\nephat B} \wdg {\nephat B} \,.
\end{align}

The Lagrangian contains terms linear and quadratic in $m_{\rm R}$. Again there are
implicit deformations, via the massive field strengths, and explicit deformations. The
explicit deformations of the bosonic sector are quadratic in the mass parameter and
define the scalar potential, which can be written in terms of the superpotential $W$ and its
derivative via the general expression~\eqref{super-potential}:
 \begin{align}
   V_{m_{\text{R}}} = \tfrac{1}{2} ( \delta_\phi W )^2 - \tfrac{9}{32} W^2
     = \tfrac{1}{2} e^{5\nephat{\phi}/2} m_{\rm R}^2 \,.
 \label{IIA-scalar-potential}
 \end{align}
Note that this scalar potential can be naturally included in the massless IIA Lagrangian
\eqref{IIAaction} by including the case of $d=-1$ in the summation (and identifying
$G^{(0)} = m_{\text{R}}$).

In the fermionic sector, one finds the following linear deformations of the gravitino and
dilatino field equations in the massive IIA theory:
\begin{align}
  {{X}}_{m_{\text{R}}} (\nephat{\psi}^{\nephat{\mu}}) & =
  m_{\text{R}} e^{5 \nephat{\phi} /4} \nephat{{\Gamma}}^{\nephat{{\mu}}\nephat{{\nu}}}
  (\tfrac{1}{4} \nephat{{\psi}}_{\nephat{{\nu}}} +
  \tfrac{5}{288} \nephat{{\Gamma}}_{\nephat{{\nu}}} \nephat{\lambda}) \,, \notag \\
  {{X}}_{m_{\text{R}}} (\nephat{\lambda}) & =
  m_{\text{R}} e^{5 \nephat{\phi} /4} \nephat{{\Gamma}}^{\nephat{{\nu}}}
  (-\tfrac{5}{4} \nephat{{\psi}}_{\nephat{{\nu}}} -
  \tfrac{21}{160} \nephat{{\Gamma}}_{\nephat{{\nu}}} \nephat{\lambda}) \,.
\end{align}
The undeformed equations, ${{X}}_0 (\nephat{\psi}^{\nephat{\mu}})=0$ and ${{X}}_0
(\nephat{\lambda}) = 0$, are given in \eqref{X0IIA}.

Supersymmetry transforms the fermionic field equations, $X_0 +  {{X}}_{m_{\text{R}}} = 0$,
into the bosonic equations of motion. For later purposes it is
convenient to truncate away all bosonic fields except the metric and the dilaton. After
this truncation we find that the fermionic field equations transform into
\begin{align}
  (\delta_0+\delta_{m_{\text{R}}}) &
  ({{X}}_0 + {{X}}_{m_{\text{R}}}) (\nephat{\psi}^{\nephat{\mu}})
  = \tfrac{1}{2} \nephat{{\Gamma}}^{\nephat{{\nu}}} \nephat{{\epsilon}} \,
  [\nephat{{R}}^{\nephat{{\mu}}}{}_{\nephat{{\nu}}}
  - \tfrac{1}{2} \nephat{{R}} \nephat{{g}}^{\nephat{{\mu}}}{}_{\nephat{{\nu}}}
  - \tfrac{1}{2} (\partial^{\nephat{{\mu}}} \nephat{\phi}) (\partial_{\nephat{{\nu}}} \nephat{\phi})
  + \tfrac{1}{4} (\partial \nephat{\phi})^2
  \nephat{{g}}^{\nephat{{\mu}}}{}_{\nephat{{\nu}}}+
  \notag \\
  & \hspace{4cm}
  + \tfrac{1}{4} m_{\text{R}}^2 e^{5 \nephat{\phi}/2}
  \nephat{{g}}^{\nephat{{\mu}}}{}_{\nephat{{\nu}}}] = 0
  \,, \notag \\
  (\delta_0+\delta_{m_{\text{R}}}) & ({{X}}_0 + {{X}}_{m_{\text{R}}}) (\nephat \lambda) =
  \nephat{\epsilon} \, [ \Box \nephat{\phi} - \tfrac{5}{4} m_{\text{R}}^2 e^{5 \nephat{\phi}/2} ] = 0 \,.
\end{align}
At the right hand side, we thus find the massive IIA bosonic field equations for the
metric and the dilaton. Indeed, these field equations can be derived from the
massless Lagrangian plus the scalar potential~\eqref{IIA-scalar-potential}.

The parameter $m_{\text{R}}$ breaks both symmetries $\alpha$ and $\beta$ of the IIA
theory. This can easily be seen from the scalar potential \eqref{IIA-scalar-potential}:
the former symmetry is broken since the dilaton scales while the Lagrangian is invariant,
while the trombone symmetry is broken since the scalar potential is not a two-derivative
term like the other bosonic terms. However, there is a linear combination that is not
broken by the massive terms: it is given by the linear combination $12 \beta - 5 \alpha$.

As argued in subsection~\ref{sec:parent-sugra-3}, the mass parameter $m_{\text{R}}$
should be seen as a zero-form \RR\ field strength: it appears naturally in the democratic
formulation, including all \RR\ potentials and field strengths. The scalar potential
\eqref{IIA-scalar-potential} then appears as the kinetic term for the zero-form field
strength. The corresponding D-brane is the D8-brane of section~\ref{sec:D8-brane}, which
is magnetically charged with respect to $m_{\text{R}}$ \cite{Polchinski:1995mt}.

The massive IIA theory is different from the other massive deformations
that we will consider in this chapter. Firstly, it is not known to have a
higher-dimensional supergravity origin\footnote{For different approaches to the M-theory
origin of massive IIA supergravity, see \cite{Bergshoeff:1998ak, Hull:1998vy,
West:2004st}.}. Secondly, it is not a gauged supergravity: no global symmetry of
the massless theory has been promoted to a local one. Therefore, this deformation of IIA
gives rise to a massive rather than gauged supergravity.

\subsection{Gauged IIA Supergravity} \label{sec:massive-gauged-IIA-3}

The second massive deformation, with mass parameter $m_{11}$, does give rise to a gauged
IIA supergravity, where the symmetry $\alpha$ has been gauged. It was first obtained in
\cite{Howe:1998qt}, on whose procedure we will comment below. Afterwards, it was shown in
\cite{Lavrinenko:1998qa} that the same theory can also be obtained by a twisted reduction
of $D=11$ supergravity using the trombone symmetry \eqref{11Dtrombone}. The corresponding
twisted reduction Ans\"atze are given in \eqref{11Dred} with $m_{11} \neq 0$.

This leads to the following explicit massive deformations of the $D=10$ IIA supersymmetry rules:
 \begin{align}
  \delta_{m_{11}} {\nephat \psi} _{\nephat \mu}
    = \ft{9}{16} m_{11} e^{-3\nephat \phi/4} \nephat{\Gamma}_{\nephat \mu}
  \Gamma_{11} \nephat \epsilon \,, \qquad
  \delta_{m_{11}} \nephat \lambda = \ft{3}{2} m_{11} e^{-3\nephat \phi/4}
  \Gamma_{11} \nephat \epsilon \,.
 \end{align}
The implicit massive deformations of the original supersymmetry rules $\delta_0$ arise
from the massive bosonic field strengths
\begin{alignat}{2}
  {\rm D}{\nephat\phi} & = {\rm d}{\nephat\phi} + \ft32 m_{11} {\nephat \C{1}} \,, \qquad
  &
  {\nephat \G{2}} & = {\rm d} {\nephat \C{1}} \,, \notag \\
  {\nephat H} & = {\rm d} \nephat B + 3 m_{11} \nephat \C{3}  \,, \qquad &
  {\nephat \G{4}} & = {\rm d} {\nephat \C{3}} + {\nephat \C{1}} \wdg {\nephat H} \,,
\end{alignat}
while the covariant derivative of the supersymmetry parameter is given by
\begin{align}
  D_{\nephat \mu} \nephat \epsilon & = (\partial_{\nephat \mu} +
  \nephat{\omega}_{\nephat \mu} + \ft{9}{16} m_{11} \nephat{\Gamma}_{\nephat \mu}
  \slashed{C}^{(1)} ) \nephat \epsilon \,.
\end{align}

There is no Lagrangian for the IIA gauged supergravity, but there are field equations.
The linear deformations of the fermionic field equations read in this case
\begin{align}
  {{X}}_{m_{11}} (\nephat{\psi}^{\nephat{\mu}}) & =
  m_{11} e^{-3 \nephat{\phi} /4} \Gamma_{11} \nephat{{\Gamma}}^{\nephat{{\mu}}\nephat{{\nu}}}
  (-\tfrac{9}{2} \nephat{{\psi}}_{\nephat{{\nu}}} +
  \tfrac{17}{48} \nephat{{\Gamma}}_{\nephat{{\nu}}} \nephat{\lambda}) \,, \notag \\
  {{X}}_{m_{11}} (\nephat{\lambda}) & =
  m_{11} e^{-3 \nephat{\phi} /4} \Gamma_{11} \nephat{{\Gamma}}^{\nephat{{\nu}}}
  (\tfrac{3}{2} \nephat{{\psi}}_{\nephat{{\nu}}} -
  \tfrac{9}{16} \nephat{{\Gamma}}_{\nephat{{\nu}}} \nephat{\lambda}) \,.
\end{align}
We first consider the truncation with all bosonic fields equal to zero except the metric
and the dilaton. Under supersymmetry the fermionic field equations transform into
\begin{align}
  (\delta_0+\delta_{m_{11}}) ({{X}}_0 + {{X}}_{m_{11}}) (\nephat{\psi}^{\nephat{\mu}})=
  & \, \tfrac{1}{2} \nephat{{\Gamma}}^{\nephat{{\nu}}} \nephat{{\epsilon}} \,
  \big[ \nephat{{R}}^{\nephat{{\mu}}}{}_{\nephat{{\nu}}}
  - \tfrac{1}{2} \nephat{{R}} \nephat{{g}}^{\nephat{{\mu}}}{}_{\nephat{{\nu}}}
  - \tfrac{1}{2} (\partial^{\nephat{{\mu}}} \nephat{\phi}) (\partial_{\nephat{{\nu}}} \nephat{\phi})
  + \tfrac{1}{4} (\partial \nephat{\phi})^2 \nephat{{g}}^{\nephat{{\mu}}}{}_{\nephat{{\nu}}} + \notag \\
  & \hspace{1cm} + 36 m_{11}^2 e^{-3 \nephat{\phi}/2} \nephat{{g}}^{\nephat{{\mu}}}{}_{\nephat{{\nu}}} \big]
  +\notag \\
  & + \Gamma_{11} \nephat{\epsilon}
  [ 3 m_{11} e^{-3 \nephat{\phi}/4} \partial^{\nephat{\mu}} \nephat{\phi} ] =0 \,, \notag \\
  (\delta_0+\delta_{m_{11}}) ({{X}}_0 + {{X}}_{m_{11}}) (\nephat \lambda) =
  & \, \nephat{\epsilon} \, [ \Box \nephat{\phi} ] +
  \nephat{{\Gamma}}^{\nephat{{\nu}}} \Gamma_{11}  \nephat{\epsilon}
  [ 9 m_{11} e^{-3 \nephat{\phi}/4} \partial_{\nephat{\nu}} \nephat{\phi} ] =0 \,.
\end{align}
The terms involving $\Gamma_{11}$ are part of the vector field equation. Therefore, to
obtain a consistent truncation, we must further truncate the dilaton to zero. One is then
left with only the metric satisfying the Einstein equation with a positive cosmological
constant.

The reduced theory is a gauged supergravity, where the scaling symmetry $\nephat \alpha$ of
table~\ref{tab:IIA-weights} has been gauged. In particular, the gauge parameter and
transformation of the Ramond-Ramond potentials read as follows{}\footnote{It is
understood that each field with $w_{\nephat\alpha}\ne 0$ is multiplied by $\Lambda$.
Also, the gauge parameter should not be confused with the dilatino, which is also denoted
by $\lambda$.}:
\begin{align}
  \Lambda = e^{w_{\nephat \alpha} m_{11} \nephat \lambda} \,, \qquad
  {\nephat \C{1}} \rightarrow {\nephat \C{1}} - {\rm d} \nephat \lambda \,, \qquad
  {\nephat \C{3}} \rightarrow e^{3 m_{11} \nephat \lambda} ({\nephat \C{3}} - {\rm d} \nephat \lambda
 \, {\nephat B}) \,,
\end{align}
where $w_{\nephat \alpha}$ are the weights under $\nephat \alpha$. One can
take two different limits of the $\nephat\alpha$ gauge transformations. Firstly, the limit
$m_{11} \rightarrow 0$ leads to the massless gauge transformations of the \RR\ potential.
Secondly, one can take the limit where $\nephat\alpha$ is constant. This leads to the
ungauged scaling symmetry $\nephat\alpha$ of table~\ref{tab:IIA-weights}.

A noteworthy feature of the $D=10$ gauged supergravity is that no Lagrangian can be
defined for it, since the symmetry that is gauged is not a symmetry of the Lagrangian but
only of the equations of motion. This is clear from its higher-dimensional origin, which
involves a twisting with a symmetry of the field equations only. As discussed in
section~\ref{sec:no-action}, this generally gives rise to field equations that can not be
interpreted as Euler-Lagrange equations.

As mentioned above, there exists an alternative way to construct this theory. In
\cite{Howe:1998qt} it was constructed by allowing for a more general solution of the
Bianchi identities of $D=11$ superspace involving a conformal spin connection. This
generalised connection is equivalent to standard $D=11$ supergravity for a topologically trivial
\st\ but leads to a new possibility for a non-trivial \st\ of the form $M_{10} \times
S^1$. The reduction over the circle leads to the $D=10$ gauged supergravity theory. It is
not properly understood why these two procedures give rise to the same lower-dimensional
description.

\subsection{Combinations of Mass Parameters and $\alpha^\prime$ Corrections} \label{sec:massive-gauged-IIA-4}

In the previous subsections we have considered two deformations of IIA supergravity. We
would like to examine the possibility to combine these massive deformations
\cite{Bergshoeff:2002nv}. If possible, the resulting theory will have two mass parameters
characterising the different deformations. However, not all combinations are necessary
consistent with supersymmetry. This complication only appears when investigating the
bosonic field equations: the supersymmetry algebra with a combination of massive
deformations always closes, as can be seen from the following argument.

Suppose one has a supergravity with one massive deformation $m$ and supersymmetry
transformations $\delta_0 + \delta_m$. In all cases discussed in this chapter, only the
supersymmetry variations of the fermions receive explicit massive corrections: $\delta_m
(\text{boson}) =0$. This implies that the issue of the closure of the supersymmetry
algebra is a calculation with $m$-independent parts and parts linear in $m$, but no parts
of higher order\footnote{That is, up to cubic order in fermions. We have not checked the
higher-order fermionic terms, but we do not expect these to affect any of our findings.}
in $m$. On the one hand $[ \delta(\epsilon_1) , \delta(\epsilon_2) ]$ has no terms
quadratic in $m$, since one of the two $\delta$'s acts on a boson. On the other hand the
supersymmetry algebra closes modulo fermionic field equations, which also only have terms
independent of and linear in $m$. Therefore, given the closure of the massless algebra,
the closure of the massive supersymmetry algebra only requires the cancellation of terms
linear in $m$.

The closure of the supersymmetry algebras with a single massive deformation is guaranteed
by their higher-dimensional origin. The argument of linearity then allows one to combine
different massive deformations. Suppose one has two massive supersymmetry algebras with
transformations $\delta_0 + \delta_{m_a}$ and $\delta_0 + \delta_{m_b}$. Both
supersymmetry algebras close modulo fermionic field equations with (different) massive
deformations. Then the combined massive algebra with transformation $\delta_0 +
\delta_{m_a}+ \delta_{m_b}$ also closes modulo fermionic field equations whose massive
deformations are given by the sum of the separate massive deformations linear in $m_a$
and $m_b$. The closure of the combined algebra is guaranteed by the closure of the two
massive algebras, since it requires a cancellation at the linear level.

Under supersymmetry variation of the fermionic field equations, one in general finds
linear and quadratic deformations of the bosonic equations of motion. In addition to
these corrections, we find that there are also algebraic equations posing constraints on
the mass parameters. Solving these equations generically excludes the possibility of
combining massive deformations by requiring mass parameters to vanish. At first sight, it
might seem surprising that the supersymmetry variation of the fermionic equations of
motion leads to constraints other than the bosonic field equations. However, one should
keep in mind that the multiplets involved cannot be linearised around a Minkowski vacuum
solution. Therefore, the usual rules for linearised Minkowski multiplets do not apply
here.

As a first application of this rationale, let us try to combine the two massive
deformations $m_{\text{R}}$ and ${m_{11}}$ of IIA supergravity theory. Based on the
linearity argument presented above, one would expect a closed supersymmetry algebra. The
bosonic field equations (with up to quadratic deformations) can be derived by applying
the supersymmetry transformations (with only linear deformations) to the fermionic field
equations (containing only linear deformations). For simplicity, we truncate all bosonic
fields to zero except the metric and the dilaton, since inclusion of the full field
content will not change the conclusions. We thus find
 \begin{align}
  & (\delta_0+\delta_{m_{\text{R}}}+\delta_{m_{11}})
  ( {{X}}_0 + {{X}}_{m_{\text{R}}} + {{X}}_{m_{11}} )
  (\nephat{\psi}^{\nephat \mu})  = \notag \\
  & \hspace{0cm} = \tfrac{1}{2} \nephat{{\Gamma}}^{\nephat{{\nu}}}
  \nephat{{\epsilon}} \,
  [\nephat{{R}}^{\nephat{{\mu}}}{}_{\nephat{{\nu}}}
  - \tfrac{1}{2} \nephat{{R}} \nephat{{g}}^{\nephat{{\mu}}}{}_{\nephat{{\nu}}}
  - \tfrac{1}{2} (\partial^{\nephat{{\mu}}} \nephat{\phi})
  (\partial_{\nephat{{\nu}}} \nephat{\phi})
  + (\tfrac{1}{4} (\partial \nephat{\phi})^2
  + \tfrac{1}{4} m_{\text{R}}^2 e^{5 \nephat{\phi}/2}
  + 36 m_{11}^2 e^{-3 \nephat{\phi}/2}) \nephat{{g}}^{\nephat{{\mu}}}{}_{\nephat{{\nu}}}] \notag \\
  & \hspace{0.5cm} + \Gamma_{11} \nephat{\epsilon}
  [ 3 m_{11} e^{-3 \nephat{\phi}/4} \partial^{\nephat{\mu}} \nephat{\phi} ]
  + \Gamma_{11} \nephat{{\Gamma}}^{\nephat{{\mu}}} \nephat{\epsilon} \,
  [\tfrac{15}{4} m_{\text{R}} m_{11} e^{\nephat{\phi}/2} ] = 0 \,, \notag \\
  & (\delta_0+\delta_{m_{\text{R}}}+\delta_{m_{11}})
  ( {X}_0 + {X}_{m_{\text{R}}} + {X}_{m_{11}} ) (\nephat \lambda) = \notag \\
  & \hspace{0cm} = \nephat{\epsilon} \,
    [ \Box \nephat{\phi} - \tfrac{5}{4} m_{\text{R}}^2 e^{5 \nephat{\phi}/2} ] +
  \nephat{{\Gamma}}^{\nephat{{\nu}}} \Gamma_{11}  \nephat{\epsilon}
  [ 9 m_{11} e^{-3 \nephat{\phi}/4} \partial_{\nephat{\nu}} \nephat{\phi} ]
  + \Gamma_{11} \nephat{\epsilon} \,
  [\tfrac{33}{2} m_{\text{R}} m_{11} e^{\nephat{\phi}/2} ] = 0 \,.
 \label{10D-combinations}
\end{align}
At the right hand sides we find four different structures. Three of them correspond to
the field equations of the metric, dilaton and \RR\ vector. The vector field equation
corresponds to the term containing $m_{11} \partial_\mu \phi$, which implies that
truncating away the \RR\ vector forces one to set $\phi = c$, provided $m_{11} \neq 0$.
More interesting is the fourth structure which is bilinear in the mass parameters,
leading to the requirement $m_{\text{R}} m_{11} = 0$. This constraint cannot be a remnant
of a higher-rank form field equation due to its lack of Lorentz indices. It could only
fit in the scalar field equation, but the $\Gamma_{11}$ factor prevents this. It is an
extra constraint which does not restrict degrees of freedom but rather restricts mass
parameters.

Independent of this constraint from supersymmetry, one can question whether the mass
parameters $m_{\text{R}}$ and $m_{11}$ are consistent with higher-order corrections of
IIA string theory to supergravity. Starting with the former, it is believed that the
massive IIA deformation is allowed at all orders in $\alpha^\prime$, due to the
connection with the D8-brane. As for the second mass parameter, it arises from the
trombone symmetry of 11D supergravity. However, the higher-order derivative terms which
arise as corrections in M-theory break this symmetry. The twisted reduction of
\cite{Lavrinenko:1998qa} will therefore be prohibited by M-theory corrections to 11D
supergravity. Presumably this also means that the method of \cite{Howe:1998qt} involving
the generalised spin connection does not work in the presence of higher-order
corrections.

Concluding, IIA supergravity allows for two massive deformations with parameters
$m_{\text{R}}$ and $m_{11}$. While the closure of the algebra is a linear calculation and
therefore always works for combinations, the bosonic field equations rule out the
possibility of including both mass parameters \cite{Bergshoeff:2002nv}. Moreover, string
theory corrections to IIA supergravity exclude the $m_{11}$ massive deformations. We
therefore conclude that only Romans' massive IIA supergravity is consistent with
supersymmetry and string theory.

\section{Gauged Maximal Supergravities in $D=9$} \label{sec:D=9-gaugings}

In this section we will consider a number of massive deformations of maximal supergravity
in $D=9$, which all give rise to gauged supergravities and have a higher-dimensional
origin. In addition, we will find relations between these parameters and investigate to
which extent one can combine the different deformations. To end with, we will discuss the
quantisation of a certain class of mass parameters. Many of the results of this section
were first obtained in \cite{Bergshoeff:2002nv}.

\subsection{Maximal Supergravity in $D=9$} \label{sec:D=9-gaugings-1}

Toroidal reduction of both massless IIA and IIB supergravity over a circle yields the
unique $D=9$, $N=2$ massless supergravity theory, as explained in
section~\ref{sec:parent-sugra}. Its field content is given by
\begin{align}
  \text{D=9:} \qquad \{ e_\mu{}^a, \phi, \varphi, \chi, A_\mu, A^i_\mu,
   B^i_{\mu\nu}, C_{\mu\nu\rho}; \psi_\mu, \lambda, \tilde \lambda \}\, ,
\end{align}
with $SL(2,\mathbb{R})$ indices $i=1,2$. These indices are raised and lowered with
$\varepsilon_{ij} = - \varepsilon^{ij}$ with $\varepsilon_{12} = - \varepsilon_{21} = 1$.

The supersymmetry rules $\delta_0$ of the massless or ungauged 9D supergravity are given
in~\eqref{9Dsusy}. The theory inherits several global symmetries from its
higher-dimensional parents. Among these is the $SL(2,\mathbb{R})$ symmetry\footnote{As
can be seen in \eqref{SL2R9D}, the symmetry transformations of both the scalars and the
fermions do not change if we replace $\Omega$ by $-\Omega$; therefore these fields
transform under $PSL(2,\mathbb{R})$. In this section, we will usually only consider group
elements $\Omega$ that are continuously connected to the identity.} from IIB
supergravity. The latter comprises an elliptic $SO(2)$ symmetry $\theta$, a hyperbolic
$SO(1,1)^+ \sim \mathbb{R}^+$ symmetry $\gamma$ and a parabolic $\mathbb{R}$ symmetry
$\zeta$. With a fixed gauge of the local $SO(2)$ symmetry (see
section~\ref{sec:global-symmetries}), the $SL(2,\mathbb{R})$ transformations in 9D read
 \begin{align}
  {\tau} \rightarrow \frac{a {\tau} +b}{c{\tau} +d} \,, & \quad
  {{A}}_i \rightarrow \Omega_i{}^j {A}_j \,, \quad
  {{B}}_i  \rightarrow \Omega_i{}^j {B}_j \,, \quad
  \Omega_i{}^j =
    \left( \begin{array}{cc} a&b\\c&d \end{array} \right) \in SL(2,\mathbb{R}) \,.
    \notag \\
  {\psi}_{{\mu}} & \rightarrow
    \left( \frac{c \, {\tau}^*+d}{c\, {\tau}+d} \right)^{1/4}
    {\psi}_{{\mu}} \,, \qquad
  {\lambda}  \rightarrow \left( \frac{c \, {\tau}^*+d}{c \, {\tau}+d}
    \right)^{3/4} {\lambda} \,, \notag \\
  {\tilde \lambda} & \rightarrow \left( \frac{c \, {\tau}^*+d}{c \, {\tau}+d}
    \right)^{-1/4} {\tilde \lambda} \,, \qquad
  {\epsilon}  \rightarrow \left( \frac{c \, {\tau}^*+d}{c \, {\tau}+d}
    \right)^{1/4} {\epsilon} \,,
 \label{SL2R9D}
 \end{align}
while $\varphi$ and $C$ are invariant.

\begin{table}[ht]
\begin{center}
\begin{tabular}{||c||c|c|c|c|c|c|c|c|c|c|c|c||c||c||}
\hline \rule[-1mm]{0mm}{6mm}
  & $e_\mu{}^a$ & $e^\phi$ & $e^\varphi$ & $\chi$ & $A$ &
  $A^1$ & $A^2$ & $B^1$ & $B^2$ &
  $C$ & $\psi_\mu, \epsilon$ & $\lambda, \tilde \lambda$ & $\mathcal{L}$
  & Orig. \\
\hline \hline \rule[-1mm]{0mm}{6mm}
  $\alpha$ & $\tfrac{9}{7}$ & $0$
   & $\tfrac{6}{\sqrt{7}}$ & $0$ & $3$ & $0$ & $0$ &
  $3$ & $3$ & $3$ & $\tfrac{9}{14}$ & $-\tfrac{9}{14}$ &
  $9$ & 11D \\
\hline \rule[-1mm]{0mm}{6mm}
  $\beta$ & $0$ &
  $\tfrac{3}{4}$ & $\tfrac{\sqrt{7}}{4}$ & -$\tfrac{3}{4}$ & $\tfrac{1}{2}$ & $-\tfrac{3}{4}$ & $0$
  & $-\tfrac{1}{4}$ & $ \tfrac{1}{2}$ & $-\tfrac{1}{4}$ & $0$ & $0$ &
  $0$ & IIA \\
\hline \rule[-1mm]{0mm}{6mm}
  $\gamma$ & $0$ &
  $-2$ & $0$ & $2$ & $0$ & $1$ & $-1$ & $1$ & $-1$ & $0$ & $0$ & $0$ &
  $0$ & IIB \\
\hline \rule[-1mm]{0mm}{6mm}
  $\delta$ & $\tfrac{8}{7}$ &
  $0$ & $-\tfrac{4}{\sqrt{7}}$ & $0$ & $0$ & $2$ & $2$ & $2$ & $2$ & $4$ & $\tfrac{4}{7}$ &
  $-\tfrac{4}{7}$ & $8$ & IIB \\
\hline
\end{tabular}
\caption{\it The scaling weights of the 9D supergravity fields and action under the
scaling symmetries $\alpha, \beta, \gamma$ and $\delta$, subject to the relation
\eqref{rel}, and their origin as higher-dimensional scaling symmetries.
\label{tab:9D-weights}}
\end{center}
\end{table}

In addition to $SL(2,\mathbb{R})$, including the scaling symmetry $\gamma$, the 9D theory
inherits two other scaling symmetries $\alpha$ and $\beta$ from IIA and one
trombone symmetry $\delta$ from IIB. The weights of the different scaling symmetries are
given in table~\ref{tab:9D-weights}. It turns out that only three of the four scaling
symmetries are linearly independent:
\begin{equation}
  8 \alpha - 48 \beta = 18 \gamma + 9 \delta \,. \label{rel}
\end{equation}
This relation gives rise to the following pattern. Using~\eqref{rel} to eliminate one of
the scaling symmetries, we are left with three independent scaling symmetries. Each of
the three gauge fields $A_\mu, A_\mu^{1}, A_\mu^{2}$ has weight zero under two linear
combinations of these three symmetries: one is a symmetry of the action, the other is a
symmetry of the equations of motion only. As we found in 10D, the symmetries that leave a
vector invariant can be gauged. We will now construct the corresponding massive
deformations by performing twisted reductions of IIA and IIB supergravity.

\subsection{Twisted Reduction of IIB using $SL(2,\mathbb{R})$} \label{sec:D=9-gaugings-2}

We will start with the case that has received most attention in the literature: twisted
reductions of $D=10$ IIB supergravity using the $SL(2,\mathbb{R})$ symmetry. It has been
treated in increasing generality by \cite{Bergshoeff:1996ui, Lavrinenko:1998qa,
Meessen:1998qm}.

The reduction Ans\"atze are given in~\eqref{IIBred} with $m_{\text{IIB}}=0$. This yields
three mass parameters $\vec{m} = (m_1,m_2,m_3)$ in 9D, parameterising the algebra element
 \begin{align}
  C_i{}^j = \tfrac{1}{2}
  \left(%
  \begin{array}{cc}
  m_1 & m_2+m_3 \\
  m_2 - m_3 & -m_1 \\
  \end{array}
  \right) \,.
 \label{traceless-mass-matrix}
 \end{align}
The massive deformations will always occur via the superpotential, containing the scalars
via the $SL(2,\mathbb{R}) / SO(2)$ coset $M$:
\begin{align}
  W = e^{2\varphi/\sqrt{7}} \, {\rm Tr}({M} Q) \,, \qquad
  Q^{ij} = \varepsilon^{ik} C_k{}^j = \ft12
    \left(
    \begin{array}{cc}
      -m_2 + m_3 & m_1 \\
      m_1 & m_2 + m_3
    \end{array} \right) \,,
  \label{superpotential-9D}
\end{align}
where $\varepsilon^{12} = - \varepsilon^{21} = -1$, giving rise to the symmetric matrix
$Q$.

The twisted reduction results in explicit deformations of the supersymmetry
transformations, given in \cite{Gheerardyn:2001jj}
\begin{align}
    \delta_{\vec m} \psi_\mu = \ft{1}{28} \gamma_{\mu} W \epsilon \,, \qquad
    \delta_{\vec m} \lambda  = i (\delta_\phi W + i e^{- \phi} \delta_\chi W) \epsilon^* \,,
    \qquad
    \delta_{\vec m} \tilde \lambda  = i \delta_\varphi W \epsilon^* \,,
\label{susyexplmiib}
\end{align}
while the implicit massive deformations read
\begin{align}
  {\rm D} \tau
  & = {\rm d} \tau + e^{-2\varphi/\sqrt{7} -\phi}
    (\delta_\phi W + i e^{-\phi} \delta_\chi W) A \,, \notag \\
  F & = {\rm d} A \,, \qquad
  F_i  = {\rm d} A_i - C_i{}^j B_j \,, \qquad
  H^i = {\rm d} B^i - A F^j \,, \notag \\
  G  & = {\rm d} C + B_i F^i  + \ft12 C_i{}^j B^i B_j \,,
  \label{massive-SL2R-field-strengths}
\end{align}
for the bosons and
\begin{align}
  D_{\mu} \epsilon & = (\partial_{\mu} + {\omega}_{\mu} + \tfrac{i}{4} e^{\phi} \partial_\mu \chi
  - \tfrac{1}{4} i e^{-2 \varphi / \sqrt{7}} W A_\mu ) \epsilon
\end{align}
for the supersymmetry parameter.

The bosonic sector of the field equations is deformed via a scalar potential, that has the
generic form for twisted reductions \eqref{twisted-scalar-potential}:
 \begin{align}
  V_{\vec{m}} & = \tfrac{1}{2} e^{4 \varphi / \sqrt{7}} \text{Tr}[C^2 + C^T M^{-1} C M] \notag \\
    & = \tfrac{1}{2} e^{4 \varphi / \sqrt{7}} \, [ 2 {\rm Tr}({ M}{ Q}{ M}{ Q})
              - ({\rm Tr}({ M}{ Q}))^2 ] \notag \\
  & = \tfrac{1}{2} ( \delta_\phi W )^2
    +\tfrac{1}{2} \, e^{-2\phi} ( \delta_\chi W )^2
    +\tfrac{1}{2} ( \delta_\varphi W )^2 -{\tfrac{2}{7}} W^2 \,,
  \label{9D-potential}
 \end{align}
which we have also written in terms of the mass matrix $Q$ and the form
\eqref{super-potential} with the superpotential $W$ and its derivatives. The field
equations of the 9D fermions receive the following explicit massive corrections:
\begin{align}
  X_{\vec{m}} (\psi^\mu) & = - \tfrac{1}{4} \gamma^{\mu\nu} [ W \psi_\nu - \tfrac{1}{16} i
    ( \delta_\phi W + i e^{-\phi} \delta_\chi W ) \gamma_\nu \lambda^*
    -\tfrac{1}{16} i \delta_\varphi W \gamma_\nu \tilde{\lambda}^* ] \,, \notag \\
  X_{\vec{m}} (\lambda) & = - i \gamma^\mu [
    ( \delta_\phi W + i e^{-\phi} \delta_\chi W ) \psi_\mu^*
    -\tfrac{1}{12} i W \gamma_\mu \lambda
    -\tfrac{2}{9\sqrt{7}} i
    ( \delta_\phi W + i e^{-\phi} \delta_\chi W )
    \gamma_\mu \tilde{\lambda} ] \,, \notag \\
  X_{\vec{m}} (\tilde{\lambda}) & = - i \gamma^\nu [
    \delta_\varphi W \psi_\nu^* - \tfrac{2}{9\sqrt{7}} i
    ( \delta_\phi W - i e^{-\phi} \delta_\chi W )
    \gamma_\nu \lambda
    - \tfrac{1}{28} i W \gamma_\nu \tilde{\lambda} ] \,.
\end{align}

The inclusion of the three mass parameters breaks the $SL(2,\mathbb{R})$ invariance.
Rather than being a symmetry, the transformations now relate theories with different mass parameters:
\begin{align}
  C \rightarrow \Omega^{-1} C \Omega \,.
  \label{SL2R-transformations}
\end{align}
This can always be used to set $m_1 = 0$, yielding an off-diagonal matrix $C$ and a
diagonal matrix $Q$. Due to \eqref{SL2R-transformations}, one says that the massive
theories are covariant under $SL(2,\mathbb{R})$ transformations rather than invariant.
Note that the combination $\det(C) = \det(Q) = \tfrac14 (-m_1{}^2 - m_2{}^2 + m_3{}^2)$
is always invariant under these transformations, which can therefore be used to label the
different massive deformations.

As discussed in subsection~\ref{sec:twist-2}, the mass matrix is only invariant under
\eqref{SL2R-transformations} if
 \begin{align}
  \Omega = \text{exp}(C \lambda) \,, \label{SL2R-gauge-transformations}
 \end{align}
The transformations of this one-dimensional subgroup have special properties;
for example, the superpotential $W$ is invariant under it. In fact, this subgroup of the
global $SL(2,\mathbb{R})$ symmetry has been gauged by the massive deformations $\vec{m}$:
\begin{align}
  \Omega = e^{C \lambda} \,, \qquad
  A \rightarrow A - {\rm d} \lambda \,, \qquad
  {B}_i \rightarrow \Omega_i{}^j
    ( {B}_j - {A}_j \, {\rm d} \lambda ) \,,
\end{align}
with gauge vector $A$ and parameter $\lambda$. We distinguish three distinct cases
depending on the value of $\det(Q)$ \cite{Hull:1998vy, Hull:2002wg, Bergshoeff:2002mb}:
 \begin{itemize}
  \item $\det(Q) = 0$: we gauge the $\mathbb{R}$ subgroup of $SL(2,\mathbb{R})$ with
  parameter $\zeta$,
\item $\det(Q) < 0$: we gauge the $SO(1,1)^+$ subgroup of $SL(2,\mathbb{R})$ with
parameter $\gamma$,
\item $\det(Q) > 0$: we gauge the $SO(2)$ subgroup of $SL(2,\mathbb{R})$ with parameter
$\theta$.
\end{itemize}
All these three cases are one-parameter massive deformations. In
subsection~\ref{sec:D=9-gaugings-6} we will discuss the quantisation of the mass
parameters $m_1$, $m_2$ and $m_3$ in the context of string theory.

\subsection{Toroidal Reduction of Massive IIA} \label{sec:D=9-gaugings-3}

In addition to the twisted reductions, one can also generate mass terms in nine
dimensions by reducing higher-dimensional deformations, i.e.~the massive and gauged IIA
supergravity theories of section~\ref{sec:D=9-gaugings}. We will start with reducing the
first possibility.

Toroidal reduction of the massive IIA supergravity, with reduction Ans\"atze
\eqref{IIAred} with $m_4 = m_{\text{IIA}} = 0$, leads to a gauged nine-dimensional
supergravity. Its deformations coincide with those parameterised by the mass parameters
$\vec{m}$ with the identifications \cite{Bergshoeff:1996ui}
 \begin{align}
  \vec{m} = (0,m_{\text{R}},m_{\text{R}}) \,.
 \label{massive-T-duality}
 \end{align}
Thus the reduction of massive IIA supergravity corresponds to a twisted reduction of IIB
supergravity, employing the $\mathbb{R}$ subgroup of $SL(2,\mathbb{R})$. This
nine-dimensional equivalence is called massive T-duality and can be seen as a deformation of the massless T-duality.

An interesting feature of massive T-duality is that massive IIA becomes a gauged theory
upon reduction. The emergence of this gauging can be seen as a generalisation of the
enhanced gaugings discussed in subsection~\ref{sec:twist-4}, in which the extra gauge
vector comes from a higher-dimensional vector. In the massive IIA case, however, the
gauge vector is $A$, which comes from the \NS\ two-form $B$ in IIA.

\subsection{Overview of Massive Deformations in 9D} \label{sec:D=9-gaugings-4}

In addition to the $SL(2,\mathbb{R})$ twisted reduction of IIB, we can also perform
twisted reductions of both IIA and IIB using the scaling symmetries $\alpha$, $\beta$ and
$\delta$; the corresponding mass parameters are denoted by $m_{\text{IIA}}$, $m_4$ and
$m_{\text{IIB}}$, respectively. The reduction Ans\"atze are given in \eqref{IIAred} and
\eqref{IIBred}. Also, like the massive IIA theory, the gauged version of IIA supergravity
can be toroidally reduced to nine dimensions. The different possibilities are illustrated
in figure~\ref{fig:reductions}, while the resulting implicit and explicit deformations of
the 9D theory are given in appendix~\ref{app:9D}. In total, this amounts to seven
deformations of the unique $D=9$ supergravity, with parameters $m_1, m_2, m_3, m_4,
m_{\text{IIA}}, m_{\text{IIB}}$ and $m_{11}$. As noted before, the parameter
$m_{\text{R}}$ is not independent but yields a subset of the parameters $\vec{m}$.

\begin{figure}[tb]
\centerline{\epsfig{file=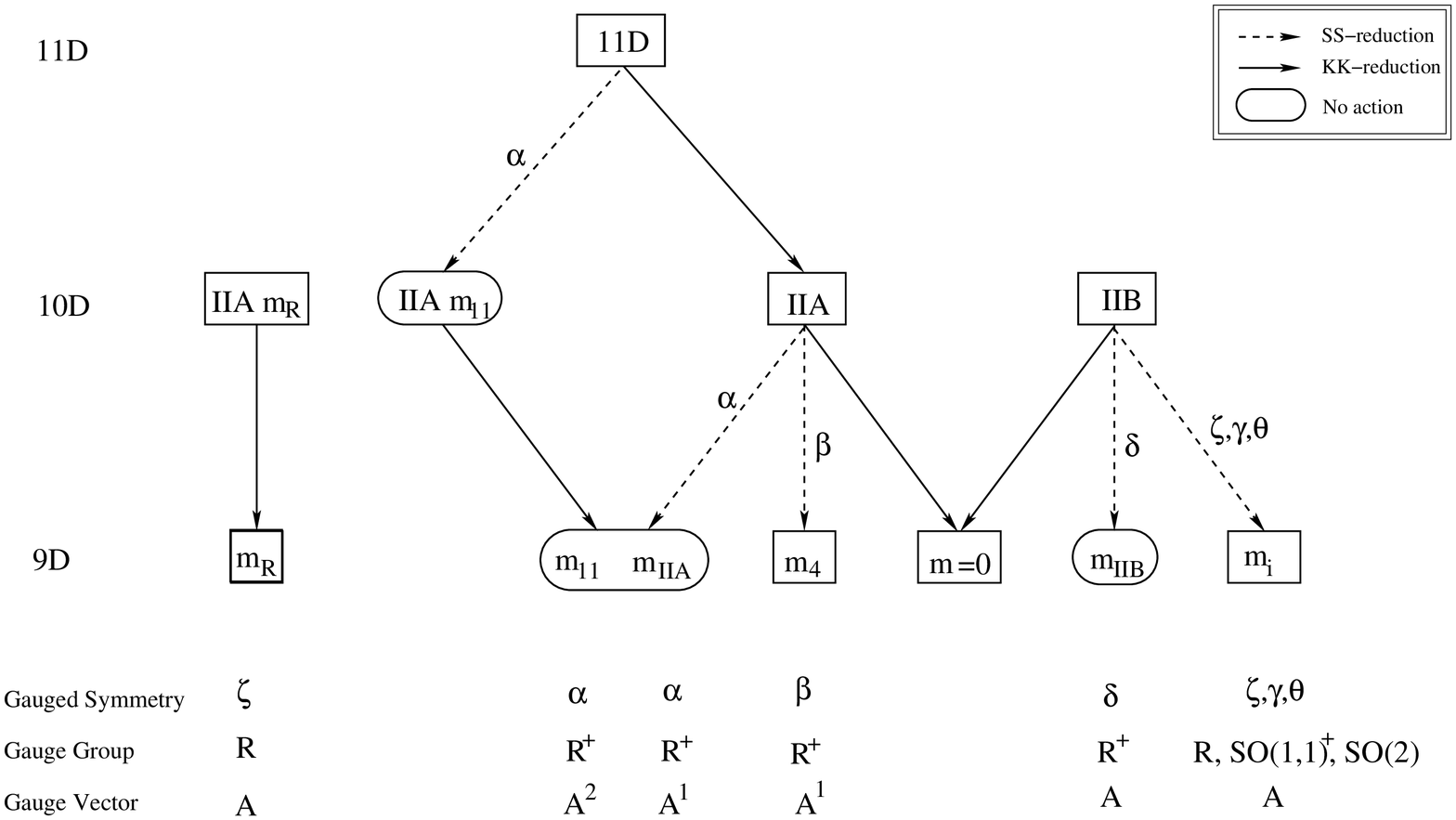,width=1.00\textwidth}} \caption{\it Overview of
all twisted reductions performed in this section with the employed symmetries and
resulting mass parameters. Mass parameters in the same box form a multiplet under
$SL(2,\mathbb{R})$ (see table~\ref{tab:mass-parameters}). We also give the gauged
symmetry and gauge vector in 9D.} \label{fig:reductions}
\end{figure}

However, various massive deformations are related. Symmetries of the massless theory
become field redefinitions in the gauged theory, that only act on the massive
deformations (exactly like in \eqref{SL2R-transformations}). This means that the mass
parameters transform under such transformations: they have a scaling weight under the
different scaling symmetries and fall in multiplets of $SL(2,\mathbb{R})$. In
table~\ref{tab:mass-parameters}, the multiplet structure of the massive deformations
under $SL(2,\mathbb{R})$ is given. The mass parameter $\tilde m_4$ is defined as the
S-dual partner of $m_4$ and can not be obtained by a twisted reduction of IIA
supergravity.

As an example, consider the two mass parameters $(m_{11}$ and $m_{\rm IIA})$, which form
a doublet under $SL(2,\mathbb{R})$ field redefinitions. This can be understood from their
higher-dimensional origin. For $m_{\text{IIA}}$ one first performs an ordinary toroidal
reduction and next a twisted reduction with $\nephat\alpha$, while for $m_{11}$ the order
of these reductions is reversed: one first performs a twisted reduction with $\nephat
{\nephat \alpha}$ and next a toroidal reduction. Since $SL(2,\mathbb{R})$ in 9D comes
from the reparameterisations of the two-torus, it also relates the two mass parameters
$m_{11}$ and $m_{\text{IIA}}$.

\begin{table}[ht]
\begin{center}
\begin{tabular}{||c||c||}
\hline \rule[-2.5mm]{0mm}{8mm}
 Mass parameters & $SL(2,\mathbb{R})$  \\
\hline \hline \rule[-1mm]{0mm}{6mm}
$(m_1,m_2,m_3)$ & adjoint \\
$(m_4,\tilde m_4)$ & doublet \\
$(m_{11},m_{\rm IIA})$ & doublet \\
$m_{\rm IIB}$& singlet\\
\hline
\end{tabular}
\caption{\it The $D=9$ mass parameters of the different reduction schemes (see
figure~\ref{fig:reductions}) form different multiplets under $SL(2,\mathbb{R})$.}
\label{tab:mass-parameters}
\end{center}
\end{table}

All the 9D deformations correspond to a gauging of a global symmetry. As shown in
section~\ref{sec:twist}, it is always the symmetry that is employed in the twisted
reduction Ansatz that becomes gauged upon reduction. The corresponding gauge vector is
provided by the metric, i.e.~it is the Kaluza-Klein vector of the dimensional reduction
(being $A^1$ for IIA and $A$ for IIB). In all cases but one, this is the complete story
and one finds an Abelian gauged supergravity. The exception is the mass parameter $m_4$,
which leads to a non-Abelian symmetry. Indeed, the 10D vector of IIA has a non-trivial
scaling under $\beta$; as discussed in subsection~\ref{sec:twist-4}, this leads to
symmetry enhancement. In the other cases such enhancement is impossible, due to the
absence of gauge vectors with a non-trivial scaling weight.

\subsection{Combining Massive Deformations in 9D and $\alpha^\prime$ Corrections} \label{sec:D=9-gaugings-5}

We would like to consider the feasibility of combinations of massive deformations in nine
dimensions. One might hope that, due to the large amount of mass parameters, the bosonic
field equations do not exclude all possible combinations, as we found in $D=10$.

For the present purposes, we will focus on specific terms in the supersymmetry variations
of the fermionic field equations. In the following, $\delta_m$ and $X_m$ are understood
to mean the supersymmetry variation and fermionic field equation at linear order
containing the sum of all seven possible massive deformations derived in the previous
subsections. Variation of the fermionic field equations gives, amongst other
$\gamma$-structures, the terms
\begin{align}
  (\delta_0 + \delta_m)(X_0 + X_m)(\psi^\mu)
  & \sim i \, \gamma^\mu \epsilon [ \ldots ]
    +  \gamma^\mu \epsilon^* [ \ldots ] + i \, \gamma^\mu \epsilon^* [ \ldots ] \,, \notag \\
  (\delta_0 + \delta_m)(X_0 + X_m)(\lambda)
  & \sim \epsilon [ \ldots ] + i \, \epsilon [ \ldots ] \,, \notag \\
  (\delta_0 + \delta_m)(X_0 + X_m)(\tilde \lambda)
  & \sim \epsilon [ \ldots ] + i \, \epsilon [ \ldots ] + \epsilon^* [ \ldots ]  \,,
\end{align}
where the $[ \ldots ]$ denote different bosonic real expressions of bilinear mass terms
and scalar factors. These are the analogue of the ten-dimensional expression $[
m_{\text{R}} m_{11} e^{\nephat{\phi}/2} ]$ (see \eqref{10D-combinations}) and give rise
to constraints on the mass parameters. Requiring all expressions  $[ \ldots ]$ to vanish,
one is led to the following possible combinations (with the other mass parameters
vanishing):
\begin{itemize}
\item {\bf Case 1} with  $\{ m_{\rm IIA}, m_4 \}$: this combination can also be obtained
by twisted reduction of IIA employing a linear combination of the symmetries $\nephat
\alpha$ and $\nephat \beta$, which guarantees its consistency. It is also a gauging of both
this symmetry and (for $m_4 \neq 0$) the parabolic subgroup of $SL(2,\mathbb{R})$ in 9D,
giving a non-Abelian gauge group.
\item {\bf Case 2,3,4} with $\{ \vec{m}, m_{\rm IIB} \}$: as in the case with $m_{\rm
IIB}=0$ and only $\vec{m}$ this combination contains three different, inequivalent cases
depending on $\det(Q)$ (depending crucially on the fact that $m_{\rm IIB}$ is a singlet
under $SL(2,\mathbb{R})$):
\begin{itemize}
\item {\bf Case 2} with $\{ \vec{m}, m_{\rm IIB} \}$ and $\det(Q) = 0$. \item {\bf Case
3} with $\{ \vec{m}, m_{\rm IIB} \}$ and $\det(Q) > 0$. \item {\bf Case 4} with $\{
\vec{m}, m_{\rm IIB} \}$ and $\det(Q) < 0$.
\end{itemize}
All these combinations can also be obtained by twisted reduction of IIB employing a
linear combination of the symmetries $\nephat \delta$ and (one of the subgroups of)
$SL(2,\mathbb{R})$, implying consistency of the combinations. All cases (assuming that $m_{\rm IIB}
\ne 0$) correspond to the gauging of an Abelian scaling symmetry in 9D.
\item {\bf Case 5} with $\{ 5 m_4 = - 12 m_{\rm IIA}, m_2 = m_3 \}$: this case can be
understood as the twisted reduction of Romans' massive IIA theory, employing the scaling
symmetry that is not broken by the $m_{\rm R}$ deformations: it is given by the
combination $12 \nephat \beta - 5 \nephat \alpha$ of table~\ref{tab:IIA-weights}. This
deformation gauges both the linear combination of scaling symmetries and the parabolic
subgroup of $SL(2,\mathbb{R})$ in 9D, which form a non-Abelian gauge group.
\end{itemize}
Another solution to the quadratic constraints has parameters $\{ m_{\rm IIA}, m_{11} \}$,
but this combination does not represent a new case: it can be obtained from only $m_{\rm
IIA}$ (and thus a truncation of case 1) via an $SL(2,\mathbb{R})$ field redefinition
(since they form a doublet). Thus the most general deformations are the five cases given
above, all containing two mass parameters. All of these are gauged theories and have
a higher-dimensional origin. Both case 1 and case 5 have a non-Abelian gauge group
provided $m_4 \neq 0$.

We will now consider the viability of the different mass parameters in string theory
rather than supergravity. The massive deformations that are based on a symmetry that is
broken by $\alpha^\prime$ corrections do not correspond to a sector of compactified
string theory. Only the symmetries that are preserved by the higher-order string
corrections to supergravity give rise to gauged supergravities that are embeddable in
string theory. We have two such symmetries:
\begin{itemize}
 \item
The $SL(2,\mathbb{R})$ (or rather its $SL(2,\mathbb{Z})$ subgroup) symmetry of IIB. Thus
the $\vec{m}= (m_1, m_2, m_3)$ deformations correspond to the low-energy limits of three
different sectors of compactified IIB string theory (depending on $\det(Q) = \tfrac{1}{4}
(- m_1{}^2 - m_2{}^2 + m_3{}^2)$).
 \item
The linear combination $\alpha + 12 \beta$ of scaling symmetries of IIA. Thus one can
define a massive deformation $m_s$ within case 1 with $\{ m_{\rm IIA} = m_s, m_4= 12 m_s
\}$ which corresponds to the low-energy limit of a sector of compactified IIA string
theory.
\end{itemize}

One gains a better understanding of the $m_s$ massive deformation and the $\alpha + 12
\beta$ symmetry of IIA from the following point of view. This combination of scaling
symmetries of IIA can be understood from its 11D origin as the general coordinate
transformation $x^{11} \rightarrow \lambda \, x^{11}$. This explains why all
$\alpha^\prime$ corrections transform covariantly under this specific scaling symmetry:
the higher-order corrections in 11D are invariant under general coordinate
transformations and upon reduction they must transform covariantly under the reduced
coordinate transformations, among which is the $\alpha + 12 \beta$ scaling symmetry.

In fact, the twisted reduction from IIA to 9D using the transformation $x^{11}
\rightarrow \lambda \, x^{11}$ is equivalent to the unique group manifold reduction from
11D to 9D: upon relating the components of $f_{ab}{}^c$ (of which only one is independent
for 2D groups) to $m_s$, the deformations from the twisted and group reductions coincide.
Indeed, this explains why the $m_s$ deformations correspond to a gauging of the 2D
non-Abelian group rather than only the scaling symmetry $\alpha + 12 \beta$. This is an
example of the relation between the different methods of dimensional reduction, as
indicated in subsection~\ref{sec:group-5}.

\subsection{Quantisation Conditions on $SL(2,\mathbb{R})$ Mass Parameters} \label{sec:D=9-gaugings-6}

The classical $SL(2,\mathbb{R})$ symmetry of IIB supergravity is broken to
$SL(2,\mathbb{Z})$ by string theory. We would like to consider the effect of this on the
twisted reductions of IIB with the $SL(2,\mathbb{R})$ symmetry of
subsection~\ref{sec:D=9-gaugings-2}. In particular, it implies that the monodromy matrix
must be an element of $SL(2,\mathbb{Z})$, the arithmetic subgroup of $SL(2,\mathbb{R})$:
\begin{align}
  \nephat{M} (x+2\pi R) = \Lambda \, \nephat{M}(x) \Lambda^T
  \qquad \text{with~~} \Lambda = e^{2 \pi R \, C} \in SL(2,\mathbb{Z}) \,,
\label{quantcond}
\end{align}
where $C$ is given by \eqref{traceless-mass-matrix}. This will imply a quantisation of
the mass parameters $\vec{m}$.

We will apply the following procedure. The mass parameters will be parameterised by
$\vec{m}=\tilde m \, (p,q,r)$. Then, given the radius of compactification $R$ and the
relative coefficients $(p,q,r)$ of the mass parameters, one should choose the overall
coefficient $\tilde m$ such that the monodromy lies in $SL(2,\mathbb{Z})$. This is not
always possible; a necessary requirement in all cases but one will be that $(p,q,r)$ are
integers and satisfy a so-called diophantic equation, i.e.~an equation for integer
numbers. Furthermore we must require $q$ and $r$ to be either both even or both odd. Thus
we get all $SL(2,\mathbb{Z})$ monodromies that can be expressed as products of the
elements
 \begin{align}
  S = \left( \begin{array}{cc}
                       0 & 1 \\
                       -1 & 0
                      \end{array} \right) \,, \qquad
  T = \left( \begin{array}{cc}
                       0 & 1 \\
                       0 & 0
                      \end{array} \right) \,,
 \end{align}
and their inverses. The conjugacy classes of $SL(2,\mathbb{Z})$ have been classified in
\cite{DeWolfe:1998eu, DeWolfe:1998pr}. We will discuss the results for the different
possibilities of $\det(Q)$ \cite{Hull:2002wg, Bergshoeff:2002mb}.

The case $\det(Q) < 0$ gives rise to a monodromy $\Lambda \in SL(2,\mathbb{Z})$ provided
we have
\begin{align}
  \tilde m = \frac{\text{arccosh}(n/2)}{\pi R\sqrt{n^2-4}}
  \text{~~and~~} p^2+q^2-r^2=n^2-4 \,,
\end{align}
for some integer $n \geq 3$. One set of solutions to this diophantic equation is
$(p,q,r)=(\pm n,0,\pm 2)$ with monodromy $\Lambda=(S \, T^{-n})^{\pm 1}$. There are other
conjugacy classes, however: not all other solutions are related to it by
$SL(2,\mathbb{Z})$.

For $\det(Q) = 0$, we find that $\Lambda$ is an element of $SL(2,\mathbb{Z})$ provided we
have
\begin{align}
  \tilde m = \frac{1}{2\pi R}
  \text{~~and~~} p^2+q^2-r^2=0 \,.
  \label{R-quant}
\end{align}
All the solutions of the diophantic equation are related via $SL(2,\mathbb{Z})$ to the
solution $(p,q,r)=(0,n,n)$ with $n$ an arbitrary integer. This gives rise to the
monodromy $\Lambda=T^n$. The quantisation on $\tilde m$ is the same charge quantisation
condition as found in \cite{Bergshoeff:1996ui}.

For the remaining case, $\det(Q) > 0$, we find that there are three distinct
possibilities for $\Lambda$ to be an element of $SL(2,\mathbb{Z})$. For the first
possibility we must have
\begin{align}
  \tilde m = \frac{1}{4 R}
  \text{~~and~~} p^2+q^2-r^2=-4 \,. \label{quant1}
\end{align}
One solution to this diophantic equation is $(p,q,r)=(0,0,\pm 2)$, yielding
$\Lambda=S^{\pm 1}$. All other solutions to the diophantic equation are related by
$SL(2,\mathbb{Z})$. For the second possibility one must require
\begin{align}
  \tilde m = \frac{1}{3 \sqrt{3} R}
  \text{~~and~~} p^2+q^2-r^2=-3 \,, \label{quant2}
\end{align}
which is solved by $(p,q,r)=(\pm1,0,\pm2)$ with monodromy $\Lambda=(T^{-1} \, S)^{\pm1}$.
Again all other solutions are related by $SL(2,\mathbb{Z})$. The third possibility is of
a different sort: it requires
\begin{align}
  \tilde m = \frac{1}{R} \text{~~and~~} p^2+q^2-r^2=-4 \,, \label{quant3}
\end{align}
but $(p,q,r)$ are not necessarily integer-valued. This gives rise to trivial monodromy
$\Lambda=\mathbb{I}$ and thus corresponds to a truncation of the untwisted Kaluza-Klein
tower to a set of massive rather than massless modes, see subsection~\ref{sec:twist-2}
and \cite{Dabholkar:2002sy}.

\section{Gauged Maximal Supergravities in $D=8$} \label{sec:D=8-gaugings}

In this section we will perform all possible 3D group manifold reductions of 11D
supergravity, resulting in different 8D gauged maximal supergravities. These results were
first obtained in \cite{Alonso-Alberca:2003jq, Bergshoeff:2003ri}.

\subsection{The Bianchi Classification of 3D Groups} \label{sec:D=8-gaugings-1}

We will first review the Bianchi classification\footnote{Actually, the classification
method used nowadays and presented here is not Bianchi's original one, but it is due to
Sch\" ucking and Behr (see Kundt's paper based on the notes taken in a seminar given by
Sch\" ucking \cite{Kundt} and the editorial notes \cite{Krasinski}), and the earliest
publications in which this method is followed are \cite{Estabrook, Ellis:1969vb}. The
history of the classification of three- and four-dimensional real Lie algebras is also
reviewed in \cite{MacCallum}. We will adhere to the common use of Bianchi classification,
however.} \cite{Bianchi} of three-dimensional Lie groups. The generators of the group
satisfy the commutation relations ($m,n,p=(1,2,3)$)
\begin{align}
  [ T_m , T_n ] = f_{mn}{}^p T_p \,,
\end{align}
with constant structure coefficients $f_{mn}{}^p$ subject to the Jacobi identity $
f_{[mn}{}^q f_{p]q}{}^r = 0$. For three-dimensional Lie groups, the structure constants
have nine components, which can be conveniently parameterised by
 \begin{align}
  f_{mn}{}^p = \varepsilon_{mnq} Q^{pq} + 2 \delta_{[m}{}^p a_{n]} \,, \qquad
  Q^{pq} a_q =0 \,.
 \label{3D-structure-constants}
 \end{align}
Here $Q^{pq}$ is a symmetric matrix with six components, and $a_m$ is a vector with three
components. The constraint on their product follows from the Jacobi identity. Having
$a_q=0$ corresponds to an algebra with traceless structure constants: $f_{mn}{}^n =0$.
The Bianchi classification distinguishes between class A and B algebras which have
vanishing and non-vanishing trace, respectively.

Of course Lie algebras are only defined up to changes of basis: $T_m \rightarrow R_m{}^n
\, T_n$ with $R_m{}^n \in GL(3,\mathbb{R})$. The corresponding transformation of the
structure constants and its components reads
\begin{align}
  f_{mn}{}^p \rightarrow f'_{mn}{}^p = R_m{}^q R_n{}^r
  (R^{-1})_s{}^p f_{qr}{}^s \,: \quad
  \begin{cases}
    Q^{mn} \rightarrow \det(R) ((R^{-1})^T Q R^{-1}))^{mn}
\,, \\
    a_m \rightarrow R_m{}^n a_n \,.
  \end{cases}
\label{structtransf}
\end{align}
These transformations are naturally divided into two complementary sets. First there is
the group of automorphism transformations with $f_{mn}{}^p = f'_{mn}{}^p$, whose
dimension is given in table~\ref{tab:3Dalgebras} for the different algebras
\cite{Schirmer:1995dy}. Then there are the transformations that change the structure
constants, and these can always be used \cite{Wald:1984rg, Schirmer:1995dy} to transform
$Q^{pq}$ into a diagonal form and $a_q$ to have only one component. We will explicitly go
through the argument.

Consider an arbitrary symmetric matrix $Q^{mn}$ with eigenvalues $\lambda_m$ and
orthogonal eigenvectors $\vec{u}_m$. Taking
\begin{align}
  R^T = (\sqrt{d_2 d_3} \, \vec{u}_1, \sqrt{d_1 d_3} \, \vec{u}_2, \sqrt{d_1 d_2} \, \vec{u}_3) \,,
\label{diagonalisation}
\end{align}
with $d_m \neq 0$ and ${\rm sgn}(d_1) = {\rm sgn}(d_2) = {\rm sgn}(d_3)$ we find that
\begin{equation}
 Q^{mn} \rightarrow {\rm diag}(d_1 \lambda_1, d_2 \lambda_2, d_3 \lambda_3)\, .
\end{equation}
We now distinguish between four cases, depending on the rank of $Q^{mn}$:
\begin{itemize}
\item Rank$(Q^{mn})=3$: in this case all components of $a_m$ necessarily vanish (due to
the Jacobi identity), and we can take $d_m = \pm 1 / |\lambda_m|$ to obtain
\begin{align}
  Q^{mn} = \pm {\rm diag}({\rm sgn}(\lambda_1), {\rm sgn}(\lambda_2),
{\rm sgn}(\lambda_3)) \,,
  \qquad a_m=(0,0,0) \,.
\end{align}
\item Rank$(Q^{mn})=2$: in this case one eigenvalue vanishes which we take to be
$\lambda_1$. Then we set  $d_i = \pm 1 / |\lambda_i|$, with $i=2,3$, to obtain $Q^{mn} =
\pm {\rm diag}(0, {\rm sgn}(\lambda_2), {\rm sgn}(\lambda_3))$. From the Jacobi identity,
it then follows that $a_m = (a,0,0)$. We distinguish between vanishing and non-vanishing
vector. In the case $a \neq 0$, one might think that one can use $d_1$ to set $a=1$, but
from the transformation rule of $a_m$~\eqref{structtransf} and the form of $R$
\eqref{diagonalisation} it can be seen that $a \sim \sqrt{d_2 d_3}$, and therefore $a$
can not be fixed by $d_1$. In this case we thus have a one-parameter family of Lie
algebras:
\begin{align}
  Q^{mn} = \pm {\rm diag}(0, {\rm sgn}(\lambda_2), {\rm sgn}(\lambda_3))
\,, \qquad
  \begin{cases}
    a_m = (0,0,0) \,, \\
    a_m = (a,0,0) \,.
  \end{cases}
\label{oneparameter}
\end{align}
\item Rank$(Q^{mn})=1$: in this case two eigenvalues vanish, e.g.~$\lambda_1 = \lambda_2
= 0$. We set $d_3 = \pm 1 / |\lambda_3|$ to obtain $Q^{mn} = \pm {\rm diag}(0, 0, {\rm
sgn}(\lambda_3))$. Again one distinguishes between $a_m = 0$ and $a_m \neq 0$. In the
latter case one is left with a vector $a_m = (a_1,a_2,0)$, of which $a_1 \sim \sqrt{d_2
d_3}$ and $a_2 \sim \sqrt{d_1 d_3}$. Thus, one can use $d_1$ and $d_2$ to adjust the
length of $\vec{a}$ to $1$, after which an $O(3)$ transformation in the $(1,2)$-subspace
gives the final result:
\begin{align}
  Q^{mn} = \pm {\rm diag}(0, 0, {\rm sgn}(\lambda_3)) \,, \qquad
  \begin{cases}
    a_m = (0,0,0) \,, \\
    a_m = (1,0,0) \,.
  \end{cases}
\end{align}
\item Rank$(Q^{mn})=0$: in this case all three eigenvalues vanish and therefore $Q^{mn} =
0$. Thus, the transformation with matrix~\eqref{diagonalisation} is irrelevant. For $a_m
\neq 0$, it follows from~\eqref{structtransf} that one can first do a scaling to get
$|\vec{a}|=1$ and then an $O(3)$ transformation to obtain:
\begin{align}
  Q^{mn} = {\rm diag}(0,0,0) \,, \qquad
  \begin{cases}
    a_m = (0,0,0) \,, \\
    a_m = (1,0,0) \,.
  \end{cases}
\end{align}
\end{itemize}
Thus, we find that the most general three-dimensional Lie algebra can be described by
 \begin{align}
  Q^{mn} = \text{diag}(q_1,q_2,q_3) \,, \qquad a_m=(a,0,0) \,.
 \label{Q-and-a}
 \end{align}
In this basis the commutation relations take the form
\begin{equation}
   [T_1 , T_2] = q_3 T_3 -a T_2 \,, \qquad
   [T_2 , T_3] = q_1 T_1 \,, \qquad
   [T_3 , T_1] = q_2 T_2 +a T_3 \,.
\label{commutations}
\end{equation}
The different three-dimensional Lie algebras are obtained by taking different signatures
of $Q^{mn}$ and are given in table~\ref{tab:3Dalgebras}. Na\"\i vely one might conclude
that the classification as given above leads to ten different algebras. However, it turns
out that one has to treat the subcase $a=1/2$ of~\eqref{oneparameter} as a separate
case\footnote{The distinction between $a=1/2$ and $a \neq 1/2$ arises when considering
the isometries on the group manifold, see also \cite{Bergshoeff:2003ri}.}. Thus, the
total number of inequivalent three-dimensional Lie algebras is eleven, two of which are
one-parameter families.

\begin{table}[ht]
\begin{center}
\begin{tabular}{||c||c|c||c|c|c|c||}
\hline \rule[-3mm]{0mm}{8mm}
  Bianchi & $a$ & $(q_1,q_2,q_3)$ & Class & Algebra & Dim(Aut) \\
\hline \hline \rule[-3mm]{0mm}{8mm}
  I & 0 & $(0,0,0)$ & A & $u(1)^3$ & $9$ \\
\hline \rule[-3mm]{0mm}{8mm}
  II & 0 & $(0,0,1)$ & A & $heis_3$ & $6$ \\
\hline \rule[-3mm]{0mm}{8mm}
  III & $1$ & $(0,-1,1)$ & B & & $4$ \\
\hline \rule[-3mm]{0mm}{8mm}
  IV & 1 & $(0,0,1)$ & B & & $4$ \\
\hline \rule[-3mm]{0mm}{8mm}
  V & 1 & $(0,0,0)$ & B & & $6$ \\
\hline \rule[-3mm]{0mm}{8mm}
  VI$_0$ & 0 & $(0,-1,1)$ & A & $iso(1,1)$ & $4$ \\
\hline \rule[-3mm]{0mm}{8mm}
  VI$_a$ & $a$ & $(0,-1,1)$ & B & & $4$ \\
\hline \rule[-3mm]{0mm}{8mm}
  VII$_0$ & 0 & $(0,1,1)$ & A & $iso(2)$ & $4$ \\
\hline \rule[-3mm]{0mm}{8mm}
  VII$_a$ & $a$ & $(0,1,1)$ & B & & $4$ \\
\hline \rule[-3mm]{0mm}{8mm}
  VIII & 0 & $(1,-1,1)$ & A & $so(2,1)$ & $3$ \\
\hline \rule[-3mm]{0mm}{8mm}
  IX & 0 & $(1,1,1)$ & A & $so(3)$ & $3$ \\
\hline
\end{tabular}
\caption{\label{tab:3Dalgebras}\it The Bianchi classification of three-dimensional Lie
algebras in terms of the components $a$ and $q_1,q_2,q_3$ of their structure constants.
Note that there are two one-parameter families VI$_a$ and VII$_a$ with special cases
VI$_0$, VII$_0$ and VI$_{a=1/2}$=III. The algebra $heis_3$ denotes the three-dimensional
Heisenberg algebra. The table also gives the dimensions of the automorphism groups.}
\end{center}
\end{table}

Of the eleven Lie algebras, only $SO(3)$ and $SO(2,1)$ are simple while the rest are all
non-semi-simple \cite{Schirmer:1995dy,Hamermesh}. In the non-semi-simple cases, we can
always choose $q_1=0$. In this case, the Abelian invariant subgroup consists of $T_2$ and
$T_3$, since $T_1$ does not appear on the right-hand side in~\eqref{commutations}. The
algebras of class B with non-vanishing trace $f_{mn}{}^n$ always give rise to non-compact
groups \cite{Ashtekar:1991wa}. In contrast, the algebras of class A correspond to both
compact and non-compact groups; an example is the algebra of type IX, which always gives
rise to the compact $SO(3)$ group. All algebras of class A can be seen as group
contractions and analytic continuations of $so(3)$, see
subsection~\ref{sec:CSO-gaugings-1}.

\subsection{Reduction over a 3D Group Manifold} \label{sec:D=8-gaugings-2}

In this subsection we perform the reduction of $D=11$ supergravity over a
three-dimensional group manifold to $D=8$ dimensions. The prime example is the reduction
over the three-sphere $S^3$, which gives rise to the $SO(3)$ gauged supergravity of Salam
and Sezgin \cite{Salam:1985ft}. By choosing other structure constants, corresponding to
other three-dimensional Lie algebras, one employs other group manifolds, some of which
give rise to non-compact gaugings. Since these algebras are ordered via the Bianchi
classification, the different group manifold reductions give rise to a Bianchi
classification of 8D gauged maximal supergravities \cite{Bergshoeff:2003ri}.

To perform the dimensional reduction, it is convenient to make an $8+3$ split of the
eleven-dimensional space-time: $x^{\hat{\mu}} = (x^\mu, z^m)$ with $\mu=(0,1,\ldots7)$ and
$m=(1,2,3)$. Eleven-dimensional fields will be hatted while unhatted quantities are 8D.
Using a particular Lorentz frame the reduction Ansatz for the eleven-dimensional fields is
 \begin{align}
   \hat{e}_{\hat{\mu}}{}^{\hat{a}} =
   \left(
   \begin{array}{cr}
    e^{-\varphi/6} e_{\mu}{}^{a} &
    e^{\varphi/3} L_{m}{}^{i}A^{m}{}_{\mu} \\
    0 & e^{\varphi/3}L_{n}{}^{i}\,U^{n}{}_{m} \\
   \end{array} \right) \,,
 \end{align}
and
 \begin{alignat}{2}
  \hat{C}_{abc} & =  e^{\varphi/2} \, C_{abc} \,, \qquad &
  \hat{C}_{abi} & = L_{i}{}^{m}B_{m\, ab} \,, \notag \\
  \hat{C}_{aij} & = e^{-\varphi/2}\, \varepsilon_{mnp} L_{i}{}^{m} L_{j}{}^{n}\, V_a{}^p
  \,, \qquad &
  \hat{C}_{ijk} & = e^{-\varphi}\varepsilon_{ijk} \ell \,,
 \end{alignat}
for the bosonic fields and
 \begin{align}
  {\hat \psi}_{\nephat a} = e^{\varphi/12} ( \psi_{a} - \tfrac{1}{6} \Gamma_{a}
  \Gamma^{i} \lambda_i ) \,, \qquad {\hat\psi}_{i} =  e^{\varphi/12} \lambda_i \,,
  \qquad \hat \epsilon  = e^{-\varphi/12} \epsilon \,,
 \end{align}
for the fermions.  Thus the full eight-dimensional field content consists of the following
$128 + 128$ field components (omitting space-time indices on the potentials):
 \begin{align}
  {\rm{8D:~~~}}
  \{ e_\mu{}^a, L_m{}^i, \varphi,
  \ell, A^m, V^m, B_m, C; \psi_\mu, \lambda_i \} \,.
 \end{align}
We will now describe the quantities appearing in this reduction Ansatz.

The matrix $L_m{}^i$ describes the five-dimensional $SL(3,\mathbb{R}) / SO(3)$ scalar
coset of the internal space. It transforms under a global $SL(3,\mathbb{R})$ acting from
the left and a local $SO(3)$ symmetry acting from the right. We take the following
explicit representative~\eqref{L-representative}, thus fixing the gauge of the local $SO(3)$ symmetry:
 \begin{align}
  L_m{}^i = \left(
  \begin{array}{ccc}
    e^{-\sigma/\sqrt{3}} &
    e^{-\phi/2+\sigma/2\sqrt{3}} \chi_1 & e^{\phi/2+\sigma/2\sqrt{3}} \chi_2 \\
    0 & e^{-\phi/2+\sigma/2\sqrt{3}} & e^{\phi/2+\sigma/2\sqrt{3}} \chi_3 \\
    0 & 0 & e^{\phi/2+\sigma/2\sqrt{3}}
  \end{array} \right) \,,
 \label{Lscalar}
 \end{align}
which contains two dilatons $\phi,\sigma$ and three axions $\chi_1,\chi_2,\chi_3$. It is
useful to define the $SO(3)$ invariant scalar matrix
\begin{equation}
  M_{mn} = L_m{}^i L_n{}^j \eta_{ij} \, ,
\label{Mscalar}
\end{equation}
where $\eta_{ij} = \mathbb{I}_3$ is the internal flat metric. Similarly, the
two-dimensional $SL(2,\mathbb{R}) / SO(2)$ scalar coset is parameterised by the dilaton
$\varphi$ and the axion $\ell$ via the $SO(2)$ invariant scalar matrix
 \begin{align}
  {W}_{IJ} = e^{\varphi} \left( \begin{array}{cc} \ell^2 + e^{-2 \varphi} & \ell  \\ \ell & 1 \\
  \end{array} \right)\, .
 \label{Wscalar}
 \end{align}

The only dependence on the internal coordinates $z^m$ comes in via the $GL(3,\mathbb{R})$
matrices $U^{m}{}_{n}$.  These can be interpreted as the components of the three
Maurer-Cartan one-forms $\sigma^{m}= U^{m}{}_{n}dz^{n}$ of some three-dimensional Lie
group. By definition they satisfy the Maurer-Cartan equations~\eqref{MC}, giving rise to
the structure constants $f_{mn}{}^{p}$ of the group, which are independent of $z^m$.
Using a particular frame in the internal directions, the explicit coordinate dependence
of the Maurer-Cartan one-forms is given by
\begin{align}
  U^m{}_n = \left(
  \begin{array}{ccc}
    1 & 0 &-s_{1,3,2} \\
    0 & e^{a z^1}\,c_{2,3,1} & e^{a z^1}\,c_{1,3,2} \, s_{2,3,1} \\
    0 & -e^{a z^1}\,s_{3,2,1} & e^{a z^1}\,c_{1,3,2} \, c_{2,3,1}
  \end{array} \right) \,,
\label{explicitU}
\end{align}
where we have used the following abbreviations
 \begin{align}
  c_{m,n,p} = \cos(\sqrt{q_m} \sqrt{q_n} \, z^p) \,, \qquad
  s_{m,n,p} = \sqrt{q_m} \sin(\sqrt{q_m}\sqrt{q_n} \, z^p) /\sqrt{q_n} \,,
  \label{cmnp-definition}
 \end{align}
This gives rise to structure constants~\eqref{3D-structure-constants} with
\eqref{Q-and-a}. It is understood that the structure constants satisfy the Jacobi
identity, amounting to $q_1 a =0$.

A subtlety which is not obvious from the analysis by Scherk and Schwarz
\cite{Scherk:1979zr} is that one only can reduce the action for traceless structure
constants ($f_{mn}{}^n = 0 $). These cases lead to the class A gauged supergravities. For
structure constants with non-vanishing trace ($f_{mn}{}^m \neq 0 $), one has to resort to
a reduction of the field equations, see section~\ref{sec:no-action}. These cases lead to
the class B gauged supergravities. Note that the adjoint of the gauge group ${\cal G}$ in
embedded in the fundamental of $GL(3,\mathbb{R})$:
\begin{equation}
  g_n{}^m = e^{\lambda^k f_{kn}{}^m}\, , \label{8D-gauge-transformations}
\end{equation}
where $\lambda^k$ are the parameters of the gauge transformations. Therefore, in the case
of a non-vanishing trace, the gauge group ${\cal G}$ is a subgroup of $GL(3,\mathbb{R})$
and not of $SL(3,\mathbb{R})$.

The relation between the Maurer-Cartan one-forms $\sigma^m$ and the three-dimensional
isometry groups is as follows. The metric on the group manifold reads
\begin{align}
  ds_G^2 = e^{2 \varphi /3} M_{mn} \sigma^m \sigma^n \,,
\label{bs}
\end{align}
where the scalars $\varphi$ and $M$ are constants from the three-dimensional point of
view. A vector field $L$ defines an isometry if it leaves the metric invariant
\begin{align}
{ L}_{{\scriptscriptstyle L}}g_{mn}=0\,.
\end{align}
For all values of the scalars, the group manifold has three isometries generated by the
left invariant Killing vector fields, as explained in section~\ref{sec:group}. These
fulfill the stronger requirement
\begin{align}
  { L}_{{\scriptscriptstyle L}_m} \sigma^n= 0 \label{lsigma}
\end{align}
for all three Maurer-Cartan forms on the group manifold and generate the algebra as given
in~\eqref{MC}. In the class A case, i.e.~$a=0$, the left-invariant Killing vectors
generating the three isometries are given by
 \begin{align}\label{isometries}
  L_1 & =\frac{c_{1,2,3}}{c_{1,3,2}}\,\frac{\partial}{\partial z^1}
      - s_{2,1,3}\,\frac{\partial}{\partial z^2}
      +\frac{c_{1,2,3}\,s_{3,1,2}}{c_{1,3,2}}\,\frac{\partial}{\partial z^3}\,,\nonumber\\
  L_2 & =\frac{s_{1,2,3}}{c_{1,3,2}}\,\frac{\partial}{\partial z^1}
      +c_{1,2,3}\,\frac{\partial}{\partial z^2}
      -\frac{s_{1,2,3}\,s_{1,3,2}}{c_{1,3,2}}\,\frac{\partial}{\partial z^3}\,,\\\nonumber
  L_3 &=\frac{\partial}{\partial z^3}\,,
 \end{align}
whereas in the class B case, i.e.~$q_1 = 0$ and $a \ne 0$, they are given by
 \begin{align}\label{isometries2}
  L_1 &=\frac{\partial}{\partial z^1}-(a z^2 + q_2 z^3) \,\frac{\partial}{\partial z^2}
     +(q_3 z^2-a z^3) \,\frac{\partial}{\partial z^3}\,,\nonumber\\
  L_2 &=\frac{\partial}{\partial z^2}\,,\hskip 2truecm
  L_3 =\frac{\partial}{\partial z^3}\,.
 \end{align}
Here, $\partial / \partial z^2$ and $\partial / \partial z^3$ are manifest isometries.
This follows from the fact that the matrix $U^n{}_m$ is independent of $z^2$ and $z^3$.

In this section, we have not heeded any global issues concerning the group manifold
reductions. This amounts to taking the universal cover of the group manifold. For this
reason, the manifolds of types I-VIII are non-compact and have the topology of
$\mathbb{R}^3$, while the type IX manifold has the topology of $S^3$. The latter case
therefore does not raise any issues when compactifying. In the case of non-compact
groups, there are two approaches:
 \begin{itemize}
 \item
One reduces over a non-compact group manifold. Supersymmetry is preserved, but the
non-compact internal manifold leads to a continuous spectrum in the lower-dimensional
theory; this spectrum can be consistently truncated to an 8D gauged maximal supergravity,
however. This is the so-called non-compactification scheme.
 \item
The group manifold is compactified by dividing out by discrete symmetries
\cite{Thurston:1979}. For all Bianchi types except types IV and VI$_a$, it is possible to
construct compact manifolds in this way \cite{Barrow:2000ka}. Sometimes, supersymmetry is
preserved under this operation, like for the three-torus. In other cases, in particular
for class B group manifolds, we do not know whether any supersymmetry is preserved under
such an identification.
 \end{itemize}
In this article, we will concentrate on local aspects, and therefore not take sides
regarding this issue.

\subsection{Supersymmetry Transformations and Global Symmetries} \label{sec:D=8-gaugings-3}

With the Ansatz above, all class A and B gauged supergravities can be obtained. We will
first consider the supersymmetry transformations of these theories. Reduction of the 11D
supersymmetry rules \eqref{11Dsusy} yields
\begin{align}
\delta e_{\mu}{}^{a} = &
-\frac{i}{2}\overline{\epsilon} \Gamma^{a} {\psi}_{\mu} \, \notag \\
\delta \psi_\mu  = & 2 \partial_\mu \epsilon
 - \frac{1}{2} {\slashed \omega}_\mu \epsilon
  +\frac{1}{2}L_{[i|}{}^{m} {\cal D}_{\mu} L_{m|j]}\Gamma^{ij} \epsilon \,
+\frac{1}{24}e^{-\varphi/2}f_{ijk}\Gamma^{ijk} \Gamma_\mu \epsilon
-\frac{1}{6}\,e^{-\varphi/2} f_{ij}{}^{j}\Gamma_\mu \Gamma^i\epsilon
\, \notag \\
& +\frac{1}{24} e^{\varphi/2}\Gamma^iL_{i}^{\ m} ( \Gamma_{\mu}^{\ \nu
\rho}-10\delta_\mu^{\ \nu}\Gamma^\rho )
 F_{m\nu \rho}\epsilon
-\frac{i}{12}e^{-\varphi}\Gamma^{ijk}L_i{}^m L_j{}^n L_k{}^p \G{1}_{\mu mnp}
\epsilon \notag \\
& +\frac{i}{96}e^{\varphi/2}(\Gamma_{\mu}^{\ \nu \rho \delta \epsilon} -4 \delta^{\
\nu}_{\mu} \Gamma^{\rho \delta \epsilon}) G_{\nu \rho \delta \epsilon} \epsilon +
\frac{i}{36}\Gamma^i L_i^{\ m}(\Gamma_{\mu}^{\ \nu \rho \delta} -6 \delta^{\ \nu}_\mu
\Gamma^{\rho \delta}) H_{\nu \rho \delta m}\epsilon
\notag \\
& +\frac{i}{48}e^{-\varphi/2}\Gamma^i \Gamma^j L_i^{\ m} L_j^{\ n} (\Gamma_\mu^{\ \nu
\rho}-10\delta_\mu^{\ \nu}\Gamma^\rho)
F_{\nu \rho mn} \epsilon \,, \notag \displaybreak[2] \\
\delta \lambda_i  = &
 \frac{1}{2}L_i^{\ m} L^{jn}{\slashed {\cal D}}{ M}_{mn}
 \Gamma_j \epsilon -\frac{1}{3} {\slashed \partial} \varphi
\Gamma_i \epsilon
-\frac{1}{4}e^{-\varphi/2} (2f_{ijk}-f_{jki})\Gamma^{jk}\epsilon \, \notag \\
& + \frac{1}{4}e^{\varphi/2}L_i^{\ m}{ M}_{mn}{\slashed F}^n \epsilon +
\frac{i}{144}e^{\varphi/2}\Gamma_i {\slashed G}\epsilon +\frac{i}{36}(2\delta_i^{\
j}-\Gamma_{i}^{\ j})L_j^{\ m}{\slashed H}_m
\epsilon \notag \\
& +\frac{i}{24}e^{-\varphi/2}\Gamma^j L_j^{\ m}L_k^{\ n} (3\delta_i^{\ k}-\Gamma_{i}^{\
k}){\slashed F}_{mn} \epsilon +\frac{i}{6}e^{-\varphi}\Gamma^{jk} L_i{}^m L_j{}^n L_k{}^p
{\slashed G}^{(1)}_{mnp} \epsilon \,,
\notag \displaybreak[2] \\
\delta A^m{}_{\mu} = & -\frac{i}{2}e^{-\varphi/2} L_{i}^{\ m} \overline{\epsilon}  (
   \Gamma^{i} \psi_{\mu} -{\Gamma}_{\mu}
(\eta^{ij}- \frac{1}{6}\Gamma^{i}\Gamma^{j})\lambda_j )  \,, \notag \\
\delta V_{\mu\, mn} = & \varepsilon_{mnp} [-\frac{i}{2}e^{\varphi/2} L_i^{\ p} \bar
\epsilon ( \Gamma^i \psi_\mu +\Gamma_\mu (\eta^{ij}-\frac{5}{6}\Gamma^i \Gamma^j)
\lambda_j )-
\ell\, \delta A^p{}_\mu ]   \,, \notag \displaybreak[2] \\
\delta B_{\mu \nu\, m} = & L_m^{\ \ i} \bar \epsilon  (\Gamma_{i[\mu} \psi_{\nu ]}
+\frac{1}{6} \Gamma_{\mu \nu}
(3\delta_i^{\ j}-\Gamma_i \Gamma^j)\lambda_j ) -2\, \delta A^n{}_{[\mu} V_{\nu]\, mn} \,, \notag \displaybreak[2] \\
\delta C_{\mu \nu \rho} = & \frac{3}{2}e^{-\varphi/2} \bar \epsilon \Gamma_{[\mu \nu}(
\psi_{\rho]} -\frac{1}{6}\Gamma_{\rho]} \Gamma^i \lambda_i )
-3 \delta A^m{}_{[\mu} B_{ \nu \rho]\,m} \,, \notag \displaybreak[2] \\
L^{\ n}_{i}\delta L_n{}_j = & \frac{i}{4}e^{\varphi/2}
  \overline\epsilon  (\Gamma_i \delta_j^{\ k} + \Gamma_j \delta^{\ k}_i-
  \frac{2}{3}\eta_{ij}\Gamma^k )\lambda_k \,,  \notag \displaybreak[2] \\
\delta \varphi
   = & -\frac{i}{2} \overline{\epsilon} \Gamma^{i} \lambda_i \,, \notag \\
\delta \ell = & -\frac{i}{2}e^{\varphi} \bar \epsilon \Gamma^i \lambda_i \,.
 \label{8D-susy}
 \end{align}
where reduction of the 11D field strength $\hat G$ gives rise to the 8D field strengths
\begin{alignat}{2} \label{curvatures}
 G & = d C + F^{m} \wdg B_m \,, \qquad &
 F_{mn} & = {\cal D} V_{mn} - f_{mn}{}^{p} B_{p} + \ell \varepsilon_{mnp} F^p \,, \nonumber \\
 H_{m} & = {\cal D} B_m + F^{n} \wdg V_{mn} \,, \qquad &
 \G{1}_{mnp} & = \varepsilon_{mnp} d \ell + 3 \left( V_{r[ m}+\ell A^q \varepsilon_{qr[m} \right) f_{np]}{}^{r}\, ,
\end{alignat}
and where the field strengths of the Kaluza-Klein vectors are given by
 \begin{equation}
  F^m  = d A^{m} - \tfrac{1}{2} f_{np}{}^{m} A^{n} \wdg A^{p} \,,
 \end{equation}
which are the non-Abelian gauge field strengths.

The ungauged theory has a global symmetry group (see table~\ref{tab:bosons})
 \begin{align}
  SL(3,\mathbb{R}) \times SL(2,\mathbb{R}) \,.
 \end{align}
The first group acts on the indices $m,n,p$ of the bosonic sector in the obvious way. For
$SL(2,\mathbb{R})$ covariance, one needs to construct the $SL(2,\mathbb{R}) / SO(2)$
scalar coset $W_{IJ}$ given in \eqref{Wscalar} and the doublet of vector field strengths
$F^{I \, m} = (\epsilon^{mnp} F_{np}, F^m)$, with $I=1,2$. The $SO(1,1)^+ \sim
\mathbb{R}^+$ subgroup of $SL(2,\mathbb{R})$ can be combined with the $SL(3,\mathbb{R})$
group to yield the full $GL(3,\mathbb{R})$, that one would expect from the 11D origin.

In the gauged theory, this $GL(3,\mathbb{R})$ is in general no longer a symmetry, since
it does not preserve the structure constants. The unbroken part is exactly given by the
automorphism group of the structure constants as given in table~\ref{tab:3Dalgebras}. Of
course, this always includes the gauge group, which is embedded in $GL(3,\mathbb{R})$ via
\eqref{8D-gauge-transformations}. However, the full automorphism group can be bigger. For
instance, it is nine-dimensional in the $U(1)^3$ case; this amounts to the fact that the
ungauged $D=8$ theory has a $GL(3,\mathbb{R})$ symmetry. Note that all other cases have
Dim(Aut) $< 9$ and thus break the $GL(3,\mathbb{R})$ symmetry to some extent. The scaling
symmetry that corresponds to the determinant of the $GL(3,\mathbb{R})$ element  (or,
equivalently, to the $SO(1,1)^+$ subgroup of $SL(2,\mathbb{R})$), is broken by all
non-vanishing structure constants. To understand the fate of the other subgroups of
$SL(2,\mathbb{R})$, one needs to define the doublet $f^I_{mn}{}^p = ( f_{mn}{}^p, 0)$.
Under a global $SL(2,\mathbb{R})$ transformation the full theory is invariant up to a
transformation of the structure constants:
\begin{align}
  f^I_{mn}{}^p \rightarrow \Omega^I{}_J f^I_{mn}{}^p \,, \qquad
  \Omega^I{}_J \in SL(2,\mathbb{R}) \,.
  \label{doublet}
\end{align}
From this transformation, one can see that the $SO(2)$ and $\mathbb{R}^+$ subgroups of
$SL(2,\mathbb{R})$ are broken by any non-zero structure constants and thus in all
theories except the Bianchi type I. In contrast, the doublet of structure
constants~\eqref{doublet} is invariant under an $\mathbb{R}$ subgroup of the
$SL(2,\mathbb{R})$ symmetry.

\subsection{Lagrangian for Class A Theories} \label{sec:D=8-gaugings-4}

The bosonic part of the eight-dimensional action for class A theories reads
 \begin{align}
  \mathcal{L} = & \, \sqrt{-g} \, \left[ R + \tfrac{1}{4}{\rm Tr} ({\cal D} { M} {\cal D}{
  M}^{-1} ) + \tfrac{1}{4}{\rm Tr} (\partial { W}
  \partial { W}^{-1}) -\tfrac{1}{4} F^{I\, m}{ M}_{mn}{ W}_{IJ} F^{J\, n}
  + \right. \notag \\
  & \hspace{1cm} \left. -\tfrac{1}{2\cdot 3!} H_{m}{ M}^{mn} H_{n} -\tfrac{1}{2\cdot 4!}
  e^{\varphi} G^{2} - { V} - \tfrac{1}{6} \star (CS) \right] \,, \label{action-8DA}
  \end{align}
with Chern-Simons term
 \begin{align}
   CS = & \ell \, G \wdg G  + 2 \epsilon^{mnp} G \wdg H_{m} \wdg V_{np}
   - 2 G \wdg (\tilde{F}^{m}+\ell F^m) \wdg B_{m} + 2 G \wdg \partial \ell \wdg C + \notag \\
   & +\epsilon^{mnp} H_{m} \wdg H_{n}\wdg B_{p}
   + 2 H_{m} \wdg (\tilde{F}^{m}+\ell F^m) \wdg C \,,
 \end{align}
where we have defined $\tilde{F}^m = \epsilon^{mnp} G_{np}$. The scalar potential ${V}$
reads
\begin{align}
 {V} & = {\textstyle\frac{1}{4}} e^{-\varphi}\,
             [ 2{ M}^{nq}f_{mn}{}^{p} f_{pq}{}^{m}
                  + { M}^{mq}{ M}^{nr}{ M}_{ps}
                  f_{mn}{}^{p}f_{qr}{}^{s} ]\,  \notag \\
          & = - \tfrac{1}{2} \, e^{-\varphi}\, [ ({\rm Tr}({ M}{ Q}))^2
              - 2 {\rm Tr}({ M}{ Q}{ M}{ Q}) ] \,,
 \label{8D-potential}
\end{align}
where we have used the relation \eqref{3D-structure-constants} between the structure
constants and the mass matrix.

The massive deformations of class A can be written in terms of a superpotential $W$,
which is given by
\begin{align}
  W = e^{-\varphi/2} {\rm Tr}({ M}{ Q}) \,.
\end{align}
The deformations of the supersymmetry transformation of the gravitino can be written in
terms of $W$, while the dilatino variations contain terms with $\delta_\Phi W$, where
$\Phi$ denotes a generic scalar. The scalar potential \eqref{8D-potential} can also be
written in terms of the superpotential and its derivatives via the general formula
\eqref{super-potential}. We will come back to this in
subsection~\ref{sec:CSO-gaugings-3}.

\subsection{Lagrangians for Truncations of Class B Theories} \label{sec:D=8-gaugings-5}

The class B gaugings and group manifolds are parameterised by three parameters $a \neq 0$
and $(q_2,q_3)$ while $q_1=0$. The full set of field equations for class B gaugings
cannot be derived from an action. However, for specific truncations this is possible, as
discussed in section~\ref{sec:no-action}. We know of three such cases, leading to a
Lagrangian with a single exponential potential \cite{Bergshoeff:2003vb}:
\begin{itemize}
 \item Type III with the truncation\footnote{The off-diagonal components of $M$ (corresponding
 to non-zero axions) are consequences of our basis choice for the structure constants. An
 $SO(2)$ rotation renders $M$ diagonal but introduces off-diagonal components in ${Q}$.}
  \begin{align}
  { M} = \left( \begin{array}{ccc}
  e^{-\sigma/\sqrt{3}} & 0 & 0 \\
  0 & e^{\sigma / 2 \sqrt{3}} \text{cosh}(\tfrac{1}{2} \sqrt{3} \sigma)
  & -e^{\sigma / 2 \sqrt{3}} \text{sinh}(\tfrac{1}{2} \sqrt{3} \sigma) \\
  0 & -e^{\sigma / 2 \sqrt{3}} \text{sinh}(\tfrac{1}{2} \sqrt{3} \sigma)
  & e^{\sigma / 2 \sqrt{3}} \text{cosh}(\tfrac{1}{2} \sqrt{3} \sigma)
  \end{array} \right)
  \label{typeIIItrunc}
  \end{align}
  which corresponds to the manifold $S^1 \times \mathbb{H}^2$. It leads to the Lagrangian
 \begin{align}
  {\cal L} = \sqrt{-g} [R - \tfrac{1}{2} (\partial \varphi)^2
  - \tfrac{1}{2} (\partial \sigma)^2 - \tfrac{3}{2} e^{- \varphi - \sigma / \sqrt{3}}] \,,
 \label{typeIII}
 \end{align}
 which has $\Delta = -1$.
 \item Type V with ${ M} = \mathbb{I}_3$, corresponding to the manifold $\mathbb{H}^3$:
 \begin{align}
  {\cal L} = \sqrt{-g} [R - \tfrac{1}{2} (\partial \varphi)^2 - \tfrac{3}{2} e^{- \varphi }]
  \,,
  \label{LagrangianH3}
 \end{align}
 with a dilaton coupling giving rise to $\Delta = -4/3$.
 \item Type VII$_{a}$ with ${ M} = \mathbb{I}_3$, also corresponding to the manifold
 $\mathbb{H}^3$ and leading to the same Lagrangian \eqref{LagrangianH3}.
\end{itemize}
Note that in all three cases the group manifold (partly) reduces to a hyperbolic
manifold, i.e.~the maximally symmetric space of constant negative curvature with enhanced
isometry and isotropy groups.

\subsection{Nine-dimensional Origin} \label{sec:D=8-gaugings-6}

In this subsection, we will discuss how all $D=8$ gauged supergravities, except those
whose gauge group is simple (i.e.~$SO(3)$ or $SO(2,1)$), can be obtained by a twisted
reduction of maximal $D=9$ ungauged supergravity using its global symmetry group
$\mathbb{R}^+ \times \SLTR$. This is possible since all these theories follow from the
reduction over a non-semi-simple group manifold, which has two commuting isometries.
These can always be arranged to be manifest, as in~\eqref{explicitU} with $q_1=0$. In
these cases, one first can perform a toroidal reduction over $T^2$ to nine dimensions,
followed by a twisted reduction to eight dimensions.

Restricting ourselves to symmetries that are not broken by $\alpha^\prime$-corrections,
the $D=9$ global symmetry group is given by
 \begin{equation}\label{dualitygroup}
   SL(2,\mathbb{R}) \times \mathbb{R}^+\,.
 \end{equation}
Here the duality group $SL(2,\mathbb{R})$ is a symmetry of the action and is not broken
by $\alpha^\prime$-corrections, since it descends from the duality group
$SL(2,\mathbb{R})$ of type IIB string theory. We denote its elements by $\Omega$. The
explicit $\mathbb{R}^+$ symmetry with elements $\Lambda$ is given by\footnote{The
symmetry $\alpha+12 \beta$ considered in subsection~\ref{sec:D=9-gaugings-5} is a linear
combination of the explicit $\mathbb{R}^+$ and the $SO(1,1)^+ \sim \mathbb{R}^+$ symmetry
of $SL(2,\mathbb{R})$.} the combination $4 \alpha - 3 \delta$ of
table~\ref{tab:9D-weights} and is valid on the equations of motion only. Since it has an
M-theory origin as the scaling symmetry $x^{\nephat \mu} \rightarrow \Lambda \,
x^{\nephat \mu}$ for ${\nephat \mu}=10,11$, this symmetry is not broken by
$\alpha^\prime$-corrections either. This scaling symmetry is precisely the transformation
with parameter $\Lambda = \exp(a z^1)$, generated by the matrix $U^m{}_n$,
see~\eqref{explicitU}, for $q_1=q_2=q_3=0$. Note that this scaling symmetry scales the
volume-element of the two-torus, which explains why it is only a symmetry of the $D=9$
equations of motion.

\begin{table}[ht]
\begin{center}
\hspace{-1cm}
\begin{tabular}{||c||c|c||}
\hline \rule[-1mm]{0mm}{6mm}
  $D=9 \Rightarrow D=8$ & $\Lambda = 1$ & $\Lambda \neq 1$ \\
  Reduction Ansatz & ($\Rightarrow$ class A) & ($\Rightarrow$ class B) \\
\hline \hline \rule[-1mm]{0mm}{6mm}
  $\Omega = \mathbb{I}_2$ & I $= U(1)^3$ & V \\
\hline \rule[-1mm]{0mm}{6mm}
  $\Omega \in \mathbb{R}$ & II $= {\rm Heis}_3$ & VI \\
\hline \rule[-1mm]{0mm}{6mm}
  $\Omega \in \mathbb{R}^+$ &  VI$_0$ $= ISO(1,1)$ & III = VI$_{a=1/2}$, VI$_a$ \\
\hline \rule[-1mm]{0mm}{6mm}
  $\Omega \in SO(2)$ & VII$_0$ $= ISO(2)$ & VII$_a$ \\
\hline
\end{tabular}
\caption{\it The $D=8$ non-semi-simple gauged maximal supergravities, resulting from
reduction of $D=9$ ungauged maximal supergravity by using the different global symmetries
in $D=9$. Here $\Omega$ and $\Lambda$ denote elements of $SL(2,\mathbb{R})$ and
$\mathbb{R}^+$, respectively. \label{ninedimred}}
\end{center}
\end{table}

When performing the $D=9$ to $D=8$ twisted reduction \cite{Scherk:1979ta}, we distinguish
between the cases where $\Lambda = 1\ (a=0)$ and where $\Lambda \neq 1\ (a\ne 0)$.
Furthermore, we allow $\Omega$ to be either the identity or an element of the three
subgroups of $SL(2,\mathbb{R})$. Reduction to $D=8$ thus gives rise to eight different
possibilities, one of which has to be split in two. These correspond to the nine $D=8$
gauged maximal supergravities with non-semi-simple gauge groups, i.e.~all Bianchi types
except type VIII with gauge group $SO(2,1)$ and type IX with gauge group $SO(3)$. The
result is given in table~\ref{ninedimred}.

It can be seen that class A gauged supergravities are obtained by using only a subgroup
of $SL(2,\mathbb{R})$, which is a reduction that can be performed on the $D=9$ ungauged
action. Class B gauged supergravities, however, require the use of the extra scaling
symmetry which indeed can only be performed at the level of the field equations.

An alternative to the twisted reduction of 9D ungauged theories is the trivial reduction
of the gauged theories of section~\ref{sec:D=9-gaugings}. When restricting to gauge
groups that are embeddable in string theory, we have four possibilities in nine
dimensions: the three subgroups $\zeta$, $\gamma$ and $\theta$ of $SL(2,\mathbb{R})$ and
the scaling symmetry $\alpha + 12 \beta$. Upon reduction, we find that these theories are
related to Bianchi types up to $SO(2) \subset \SLTR$ rotation of $90$ degrees. The
specific types are II, VI$_0$ and VII$_0$ (of class A) and III (of class B),
respectively.

\section{$CSO$ Gaugings of Maximal Supergravities} \label{sec:CSO-gaugings}

In this section we will discuss $CSO$ gauged maximal supergravities, appearing in diverse
dimensions, and describe the relation to the previously constructed theories. We will
conclude by mentioning some other possibilities of gauged maximal supergravities.

\subsection{$CSO$ Algebras and Groups} \label{sec:CSO-gaugings-1}

An important role in gauged maximal supergravity is played by the so-called $CSO$ groups,
see e.g.~\cite{Hull:1985rt, Andrianopoli:2000fi, Hull:2002cv}. These groups can be seen
as analytic continuations and group contractions of $SO$ groups, as is demonstrated
below.

We start with the algebra $so(n)$ with generators in the fundamental representation (with
$i,j,\ldots = 1, \ldots, n$)
 \begin{align}
  (g_{ij})^k{}_l = \delta_{[i}^k Q_{j] l} \,,
 \label{generators}
 \end{align}
with $Q$ equal to the identity matrix for $so(n)$. The generators are labelled by an
antisymmetric pair of indices, giving rise to $\tfrac{1}{2} n (n-1)$ different
generators. These satisfy the commutation relations
 \begin{align}
  [ g_{ij} , g_{kl} ] = f_{ij,kl}{}^{mn} g_{mn} \,, \qquad
  f_{ij,kl}{}^{mn} = 2 \delta_{[i}^{[m} Q_{j][k} \delta_{l]}^{n]} \,.
 \end{align}
The corresponding group elements leave the matrix $Q$ invariant:
 \begin{align}
   \text{exp} (\lambda^{ij} g_{ij}) Q \text{exp} (\lambda^{ij} g^T_{ij}) = Q \,,
 \label{Q-invariant}
 \end{align}
where $\lambda^{ij}$ are the (real) parameters of the group elements. The above
properties hold for an arbitrary matrix $Q$, which equals $\mathbb{I}_n$ for the $SO(n)$
group.

Consider the following scaling of the $so(n)$ algebra, where $i,j=1,\ldots,n-1$:
 \begin{align}
  g_{ij} \rightarrow g_{ij} \,, \qquad g_{in} \rightarrow \lambda g_{in}
 \label{scale-generators}
 \end{align}
A straightforward calculation shows that the only effect on the above algebra is a
scaling of the matrix $Q$:
 \begin{align}
  Q = \mathbb{I}_{n} \rightarrow
  \left( \begin{array}{cc}
                       \mathbb{I}_{n-1} & 0 \\
                       0 & \lambda^{-2}
                      \end{array} \right) \,.
 \end{align}
Therefore, different choices for $\lambda$ result in different algebras:
 \begin{itemize}
  \item $\lambda \rightarrow 1$ is the trivial case, retaining the $so(n)$ algebra,
  \item $\lambda \rightarrow i$ is an analytic continuation, yielding the $so(n-1,1)$ algebra and
  \item $\lambda \rightarrow \infty$ corresponds to a group contraction, giving the $iso(n-1)$ algebra,
 \end{itemize}
as can be seen from the defining equation \eqref{Q-invariant}. Thus, the (imaginary or
infinite) rescaling of the generators \eqref{scale-generators} takes one from the $so(n)$
algebra with $Q = \mathbb{I}_{n}$ to the algebras $so(n-1,1)$ or $iso(n-1)$.

One can perform the operation \eqref{scale-generators} a number of times with different
generators, leading to the algebra \eqref{generators} with the matrix
 \begin{align}
  Q = \left( \begin{array}{ccc}
                       \mathbb{I}_p & 0 & 0 \\
                       0 & -\mathbb{I}_q & 0 \\
                       0 & 0 & 0_r
                      \end{array} \right) \,,
 \label{general-Q}
 \end{align}
with $p+q+r = n$. The corresponding algebra is called the $cso(p,q,r)$ algebra,
satisfying the equations \eqref{generators}-\eqref{Q-invariant}. Therefore, the
$cso(p,q,r)$ algebras with $p+q+r = n$ are analytic continuations and group contractions
of the prime example $so(n)$. This generalises the $so(n)$ algebra to $[ n^2 / 4 + n ]$
different possible algebras.

Note that a generator $g_{ij}$ vanishes if and only if $Q_{ii} = Q_{jj} = 0$. For this
reason, the matrix \eqref{general-Q} gives rise to $\tfrac{1}{2} r (r-1)$ vanishing
generators. The number of non-trivial generators of a $cso(p,q,r)$ algebra therefore
equals
 \begin{align}
  \tfrac{1}{2} (p+q+r)(p+q+r-1) - \tfrac{1}{2} r (r-1) = \tfrac{1}{2} (p+q)(p+q+2r-1) \,.
 \end{align}
Also note that $cso(p,q,r)$ and $cso(q,p,r)$ are isomorphic, while $cso(p,q,0) = so(p,q)$
and $cso(p,q,1) = iso(p,q)$.

\begin{figure}[th]
  \centerline{\epsfig{file=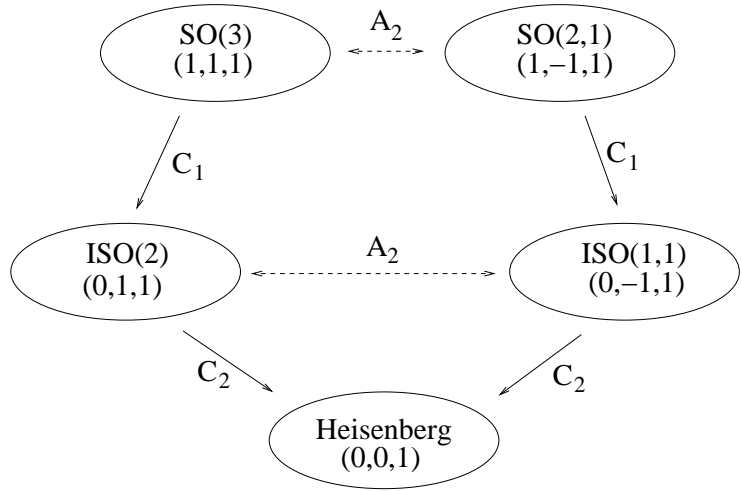,width=.6\textwidth}}
  \caption{\it Relations between the different $CSO$ groups with $n=3$ under analytic continuations {\rm A} and
  group contractions {\rm C}. The boxes give the groups and the diagonal components of $Q$.}
  \label{fig:3Drelations}
\end{figure}

The corresponding $CSO$ group elements satisfy \eqref{Q-invariant}. The simplest examples
are
 \begin{itemize}
  \item $n = 2$: $SO(2)$, $SO(1,1)$, $ISO(1) \sim \mathbb{R}$,
  \item $n = 3$: $SO(3)$, $SO(2,1)$, $ISO(2)$, $ISO(1,1)$, $CSO(1,0,2) \sim Heis_3$.
 \end{itemize}
The $n=2$ case are the one-dimensional subgroups of $SL(2,\mathbb{R})$, while the $n=3$
case exactly comprises the class A groups of the Bianchi classification (see
table~\ref{tab:3Dalgebras}). The relations under analytic continuations and group
contractions are illustrated in figure~\ref{fig:3Drelations} for $n=3$.

\subsection{Gauged Maximal Supergravity} \label{sec:CSO-gaugings-2}

One might have noticed a certain familiarity with the $CSO(p,q,r)$ groups with $p+q+r=n$
for $n=2$ and $n=3$. Indeed, these are exactly the gauge groups for a subset of the
gaugings considered in sections~\ref{sec:D=9-gaugings} and~\ref{sec:D=8-gaugings}. Such
$CSO$ groups also emerge in lower-dimensional gauged maximal supergravities\footnote{For
the purposes of uniformity, we will restrict ourselves to $D \geq 4$. Gauged maximal
supergravities in $D=3$ have a number of remarkable properties, see
e.g.~\cite{Nicolai:2001sv, deWit:2004yr}.}, as we will now discuss.

It has been known for long that certain gauged maximal supergravities with global
symmetry groups $SL(n,\mathbb{R})$ allow for the gauging of the $SO(n)$ subgroup of the
global symmetry. An example is the $SO(8)$ gauging in four dimensions
\cite{deWit:1982ig}. Subsequently, it was realised that such gauged supergravities could
be obtained by the reduction of a higher-dimensional supergravity over a sphere, with a
flux of some field strength through the sphere. An example is the reduction of 11D
supergravity over $S^7$, with magnetic flux of the four-form field strength through the
seven-sphere, yielding the $SO(8)$ theory \cite{deWit:1987iy}. Other examples are given
in table\footnote{We have included massive IIA supergravity in table~\ref{tab:gaugings},
even though it is not a gauged theory and its higher-dimensional origin is unknown, for
reasons that will be discussed in the next subsection.}$^,$\footnote{The $S^5$ reduction
of IIB has not (yet) been proven in full generality. The linearised result was obtained
by \cite{Kim:1985ez} while the full reduction of the $SL(2,\mathbb{R})$ invariant part of
IIB supergravity was performed by \cite{Cvetic:2000nc}.}~\ref{tab:gaugings}.

\begin{table}[ht]
\begin{center}
\hspace{-1cm}
\begin{tabular}{||c|c||c||c||}
\hline \rule[-1mm]{0mm}{6mm}
  $D$ & $n$ & $\phi$ & Origin \\
\hline \hline \rule[-1mm]{0mm}{6mm}
  $10$ & $1$ & $\surd$ & Massive IIA \cite{Romans:1986tz}\\
\hline \rule[-1mm]{0mm}{6mm}
  $9$ & $2$ & $\surd$ & IIB with $SO(2)$ twist \cite{Meessen:1998qm} \\
\hline \rule[-1mm]{0mm}{6mm}
  $8$ & $3$ & $\surd$ & IIA on $S^2$ \cite{Salam:1985ft} \\
\hline \rule[-1mm]{0mm}{6mm}
  $7$ & $5$ & $-$ & 11D on $S^4$ \cite{Nastase:1999cb, Nastase:1999kf} \\
\hline \rule[-1mm]{0mm}{6mm}
  $6$ & $5$ & $\surd$ & IIA on $S^4$ \cite{Cvetic:2000ah} \\
\hline \rule[-1mm]{0mm}{6mm}
  $5$ & $6$ & $-$ & IIB on $S^5$ \cite{Kim:1985ez, Cvetic:2000nc} \\
\hline \rule[-1mm]{0mm}{6mm}
  $4$ & $8$ & $-$ & 11D on $S^7$ \cite{deWit:1987iy} \\
\hline
\end{tabular}
\caption{\it The different gauged maximal supergravities in $D$ dimensions with $n$ mass
parameters. The relevant scalar subsector consists of the coset $SL(n,\mathbb{R})/SO(n)$
plus, for the cases with a $\surd$ in the third column, an extra dilaton $\phi$. We also
give the higher-dimensional origin of the $SO(n)$ prime examples. \label{tab:gaugings}}
\end{center}
\end{table}

In addition to $SO(n)$, the global symmetry group $SL(n,\mathbb{R})$ has more subgroups
that can be gauged. It was found that many more gaugings could be obtained from the
$SO(n)$ prime examples by analytic continuation or group contraction of the gauge group
\cite{Hull:1984vg, Hull:1984qz}. This leads one from $SO(n)$ to the group $CSO(p,q,r)$
with $p+q+r = n$, as we have seen in the previous subsection.

At first the generalisation of $SO(p,q)$ to $CSO(p,q,r)$ was thought to be possible only
for even-dimensional gauged supergravities, due to problems with the number of degrees of
freedom of gauge potentials in odd dimensions. The resolution lies in the role played by
the massive self-dual gauge potentials in odd dimensions\footnote{See
\cite{Alonso-Alberca:2002tb} for an alternative proposal based on the St\"uckelberg
mechanism.} \cite{Townsend:1984xs}. For example, the resulting field content in $D=5$
contains $15+r$ gauge vectors and $12-r$ massive self-dual two-form potentials
\cite{Andrianopoli:2000fi}. In $D=7$ one would expect $r$ massless two-forms and $5-r$
massive self-dual three-forms, of which the case $r=1$ is confirmed in
\cite{Cvetic:2000ah}. Surprisingly, this phenomenon does not occur in $D=9$, where one
has one massless three-form potential for all values of $r$ \cite{Meessen:1998qm}. This
is related to the fact that the 9D potential is a singlet, while the lower-dimensional
potentials transform non-trivially under the gauge group, see table~\ref{tab:bosons}. In
this section, we will be concerned with the scalar subsector of these theories and
therefore not mind the subtleties associated with the gauge potentials.

The question of the higher-dimensional origin\footnote{For discussions of the
higher-dimensional origin of self-duality relations, see \cite{Townsend:1984xs,
Nastase:1999cb, Hull:2003kr}.} of the $CSO(p,q,r)$ gaugings was clarified in
\cite{Hull:1988jw}, where the same operations of analytic continuations and group
contractions were applied to the internal manifold. The resulting manifolds are
hypersurfaces defined by
 \begin{align}
  \sum_{i=1}^n q_i \mu_i{}^2 = 1 \,,
 \label{hypersurface}
 \end{align}
with $n$ parameters\footnote{Another approach to the introduction of these parameters in
the lower dimension is the inclusion of $n$ Killing vectors in 11D supergravity
\cite{Bergshoeff:1998ak, Meessen:1998qm, Alonso-Alberca:2000gh, Alonso-Alberca:2002tb}.}
$q_i$ of which $p$ are positive, $q$ are negative and $r$ are vanishing; hence $p+q+r =
n$. The manifold corresponding to \eqref{hypersurface} is denoted by $H^{p,q} \times T^r$
\cite{Hull:1988jw}. The hyperbolic manifold $H^{p,q}$ can be endowed with a
positive-definite metric, which generically is inhomogeneous \cite{Cvetic:2004km}; the
exceptions are the (maximally symmetric) coset spaces
 \begin{align}
  S^n = H^{n+1,0} \simeq \frac{SO(n+1)}{SO(n)} \,, \qquad
  H^n = H^{1,n} \simeq \frac{SO(1,n)}{SO(n)} \,,
 \end{align}
i.e.~the sphere and the hyperboloid. Generically the spaces $H^{p,q}$ are non-compact;
the only exception is the sphere with $q=0$.

Thus non-compact gauge groups $CSO(p,q,r)$ with $q \neq 0$ are obtained from reduction
over non-compact manifolds, as first suggested in \cite{Pernici:1985nw}. It can be argued
that the corresponding reduction is consistent provided the compact case, with reduction
over $S^{n-1}$, has been proven consistent \cite{Hull:1988jw}.

A special case of this reduction is provided by $p+q = 1$ or $2$. In such cases,
$H^{p,q}$ corresponds to a one-dimensional manifold, over which one performs a twisted
reduction (see section~\ref{sec:twist}). The difference between $(p,q,r) = (2,0,0)$,
$(1,1,0)$ and $(1,0,1)$ is the flux of the scalars: the different values correspond to
twisting with the subgroups $SO(2)$, $SO(1,1)$ and $\mathbb{R}$ of a global symmetry
group $SL(2,\mathbb{R})$, respectively.

Examples of these cases are provided by the reduction of IIB with an $SL(2,\mathbb{R})$
twist, giving rise to $CSO$ gauged supergravity in 9D with $n=2$ (see
subsection~\ref{sec:D=9-gaugings-2}). This requires the identification
 \begin{align}
 Q = \ft12
    \left(
    \begin{array}{cc}
      -m_2 + m_3       & m_1 \\
      m_1 & m_2 + m_3
    \end{array} \right)
  = \left(%
  \begin{array}{cc}
  q_1 & 0 \\
  0 & q_2 \\
  \end{array}%
  \right)\,,
 \label{qs}
 \end{align}
between the parameters $\vec{m} = (m_1,m_2,m_3)$ of the $SL(2,\mathbb{R})$ twisted
reduction \eqref{IIBred} and the parameters $(q_1,q_2)$ of the reduction over the
hypersurface \eqref{hypersurface}. The choice of diagonal $Q$ corresponds to vanishing
$m_1$, which can always be obtained by $SL(2,\mathbb{R})$ field redefinitions (as
explained in subsection~\ref{sec:D=9-gaugings-2}). Note that generic twisted reductions
\eqref{twisted-reduction} give rise to a traceless matrix $C$, which only for $n=2$ can
be related to a symmetric matrix $Q$, see \eqref{superpotential-9D}.  The explicit
relation between the twisted reduction coordinate $y$ and the Cartesian coordinates
$\mu_i$ reads
 \begin{align}
  \mu_1 = \text{sin}(\sqrt{q_1 q_2} y) / \sqrt{q_1} \,, \qquad
  \mu_2 = \text{cos}(\sqrt{q_1 q_2} y) / \sqrt{q_2} \,. \label{2D-mu-coordinates}
 \end{align}
This explains the relation between twisted reduction and the case $p+q \leq 2$ of
\eqref{hypersurface}.

Another noteworthy remark concerns the next case, $p+q = 3$. This defines two-dimensional
spaces, e.g.~$S^2$ and $H^2$, over which one can perform coset reductions. Alternatively,
these cases can be viewed as group manifold reductions over three-dimensional group
manifolds, e.g.~$SO(3)$ and $SO(2,1)$. For example, one can either perform a
two-dimensional coset reduction of IIA or a three-dimensional group manifold reduction of
11D to obtain the class A gauged supergravities in 8D \cite{Alonso-Alberca:2003jq}. The
structure constants of these class A group manifolds are given by
 \begin{align}
  f_{mn}{}^p = \varepsilon_{mnq} Q^{pq} \,, \qquad
  Q^{mn} = \text{diag} (q_1,q_2,q_3) \,,
 \end{align}
which relates the parameters of the group manifold reduction and the reduction over the
hypersurface. Note that the structure constants only contain a symmetric matrix $Q$ for
the case $n=3$, confirming the relation between 3D group manifolds and
\eqref{hypersurface} with $n=3$. Explicitly, the relations between the three-dimensional
group manifold reductions and the reductions over the two-dimensional hypersurface
\eqref{hypersurface} are
 \begin{align} \nonumber
 \mu_1&=\sin(\sqrt{q_2 q_3}\,y^2)/\sqrt{q_1}\,,\\
 \mu_2&=\sin(\sqrt{q_1 q_3}\,y^1)\cos(\sqrt{q_2 q_3}\,y^2)/\sqrt{q_2}\,,\\\nonumber
 \mu_3&=\cos(\sqrt{q_1 q_3}\,y^1)\cos(\sqrt{q_2 q_3}\,y^2)/\sqrt{q_3}\,,
 \label{3D-mu-coordinates}
 \end{align}
where $y^{1,2}$ are the two coordinates of the 3D group manifold that remain after
reduction over the manifest isometry direction $y^3$.

We expect the following relations between the different maximal supergravities with $CSO$
gauge groups upon toroidal reduction. Consider the mass parameters in dimensions $D$ and
$d < D$, denoted by $n_D$ and $n_d \geq n_D$, respectively. Then the $n_D$ mass
parameters in $D$ dimensions reduce to the $n_d$ mass parameters in $d$ dimensions with
$n_d - n_D$ vanishing entries:
 \begin{align}
  Q_D \qquad
  \begin{array}{c} T^{D-d} \\ \Longrightarrow \\ ~~ \end{array} \qquad
  Q_d = \left(%
 \begin{array}{cc}
  Q_D & 0 \\
  0 & 0_{n_d - n_D} \\
 \end{array}%
 \right) \,.
 \end{align}
Therefore, the set of all $CSO$ gaugings in $D$ dimensions reduces to (generically) a
subset of all $CSO$ gaugings in $d$ dimensions. In the reduction Ansatz from 11D or 10D
to $d$ dimensions, the $n_d - n_D$ vanishing mass parameters correspond to a torus
$T^{D-d}$ over which one can reduce first, as can be seen from \eqref{hypersurface}. This
conjecture relating the different $CSO$ gauged supergravities will be proven below for
the scalar subsector of the theories.

\subsection{Scalar Potential} \label{sec:CSO-gaugings-3}

In addition to the gauging of the group $CSO(p,q,r)$, the non-trivial reduction over the
spaces $H^{p,q} \times T^r$ gives rise to a scalar potential. To this end, we consider
the scalar subsector of these theories.

In all cases, it contains a scalar coset $SL(n,\mathbb{R}) / SO(n)$, which is
parameterised by a symmetric matrix $M$. We will restrict ourselves to a diagonal matrix,
for reasons that will be explained in section~\ref{sec:CSO-DW}. The diagonal part of the
scalar is given by
 \begin{align}
  M = \text{diag} (e^{\vec{\alpha}_1 \cdot \vec{\phi}},\ldots,e^{\vec{\alpha}_n \cdot \vec{\phi}}) \,,
 \end{align}
where the $n$ vectors $\vec{\alpha}_i=\{\alpha_{iI}\}$ are weights of $SL(n,\mathbb{R})$
fulfilling the following relations
 \begin{align}
  \sum_i \alpha_{iI}=0 \,, \qquad
  \sum_i \alpha_{iI}\,\alpha_{iJ}=2\,\delta_{IJ} \,, \qquad
  \vec{\alpha}_{i}\cdot\vec{\alpha}_{j}=2\,\delta_{ij}-\frac{2}{n} \,.
  \label{weights}
 \end{align}
In addition, the scalar coset can contain an extra scalar $\phi$, as indicated in
table~\ref{tab:gaugings}. Note that $M$ and $\phi$ generically do not correspond to the
full scalar coset, as can be inferred from table~\ref{tab:bosons}; however, they do
constitute the part that is relevant to the $CSO$ gauging and scalar potential.
Similarly, the full global symmetry will often be larger than $SL(n,\mathbb{R})$; it is
for example given by $SO(5,5)$ in 6D. Its $SL(n,\mathbb{R})$ subgroup will generically be
the largest symmetry of the Lagrangian, however, and is the only part of the symmetry
group that is relevant for the present discussion.

The scalar potential of all $CSO$ gaugings has the universal form
 \begin{align}
  V = - \tfrac{1}{2} e^{a \phi} ((\text{Tr}[QM])^2 - 2 \text{Tr}[QMQM]) \,, \qquad
  Q = \text{diag} (q_1,\ldots,q_n) \,,
 \label{CSO-potential}
 \end{align}
in terms of the mass parameters $q_i$ of the hypersurface \eqref{hypersurface}. The
dilaton coupling $a$ is given by
 \begin{align}
  a^2 = \frac{8}{n} - 2 \, \frac{D-3}{D-2} \,, \label{DNA}
 \end{align}
for the different cases. This scalar potential can be written in terms of the
superpotential
 \begin{align}
  W = e^{a \phi/2} \text{Tr}[QM] \,,
  \label{CSO-superpotential}
 \end{align}
via the general formula~\eqref{super-potential} for the scalar potential:
 \begin{align}
  V = \tfrac{1}{2} (\delta_\phi W)^2 + \tfrac{1}{2}(\delta_{\vec{\phi}} W)^2
    -  \frac{D-1}{4(D-2)} W^2\,.
 \end{align}
This superpotential also parameterises the explicit deformations of the supersymmetry
transformations: the gravitino variation will be proportional to $W$ while the dilatini
variations will be proportional to $\delta W / \delta \vec{\phi}$ and $\delta W / \delta
{\phi}$.

In accordance with table~\ref{tab:gaugings}, $a$ vanishes for $(D,n) = (7,5)$, $(5,6)$
and $(4,8)$, for which the extra dilaton $\phi$ is absent. The $SL(2,\mathbb{R})$ twisted
reduction of IIB and class A group manifold reduction of 11D yield scalar potentials
\eqref{9D-potential} and \eqref{8D-potential} that coincide with \eqref{CSO-potential}
for $(D,n) = (9,2)$ and $(8,3)$, respectively. In addition, the scalar potential
\eqref{IIA-scalar-potential} of massive IIA also is of exactly this form with
$(D,n)=(10,1)$ and is therefore included in table~\ref{tab:gaugings}.

For the $SO(n)$ cases, i.e.~all $q_i = 1$, the scalar subsector can be truncated by
setting $M = \mathbb{I}$. In this truncation, the scalar potential reduces to a single
exponential potential
 \begin{align}
  V = - \tfrac{1}{2} n (n-2) e^{a \phi} \,,
 \end{align}
Note the dependence of the sign of the potential on $n$: it is positive for $n=1$,
vanishing for $n=2$ and negative for $n \geq 3$. If $a=0$ (which necessarily implies $n
\geq 3$ in $D \geq 4$), the scalar potential becomes a cosmological constant and allows
for a fully supersymmetric AdS solution; for this reason, such theories are called AdS
supergravities. Theories with $a \neq 0$ are called DW supergravities since the natural
vacuum is a domain wall solution, see section~\ref{sec:CSO-DW}.

\subsection{Group Contraction and Dimensional Reduction} \label{sec:CSO-gaugings-4}

We would like to consider two operations on the scalar sector of the $CSO$ gauged
supergravity. The first operation corresponds to a contraction of the $CSO$ gauge group
and corresponds to setting one mass parameter equal to zero, as explained above. For
concreteness, it is taken to be the last one: $q_i = (q_p , 0)$, where we have split up
$i=(p,n)$ and $p = 1,\ldots,n-1$. The superpotential now reads
 \begin{align}
  W = e^{a \phi/2} \sum_{p} q_p e^{\vec{\alpha}_{p} \cdot \vec{\phi}}
   = e^{a \phi/2 + \vec{\beta} \cdot \vec{\phi}} \sum_p q_p e^{\vec{\beta}_p \cdot \vec{\phi}}
   \,,
  \label{contracted-superpotential}
 \end{align}
where we have chosen to extract an overall part $\vec{\beta} \cdot \vec{\phi}$ according
to $\vec{\alpha}_p = \vec{\beta} + \vec{\beta}_p$. A convenient choice for $\vec{\beta}$
is
 \begin{align}
  \vec{\beta} = - \frac{1}{n-1} \vec{\alpha}_n = (0,\ldots,0,\frac{1}{\sqrt{n(n-1)/2}})
  \,.
 \end{align}
This corresponds to the scalar coset split
 \begin{align}
  M = \left(%
 \begin{array}{cc}
  e^{\vec{\beta} \cdot \vec{\phi}} \tilde{M} & 0 \\
  0 & e^{-(n-1) \vec{\beta} \cdot \vec{\phi}} \\
 \end{array}%
 \right) \,, \qquad
  \tilde{M} = \text{diag}(e^{\vec{\beta}_1 \cdot \vec{\phi}} , \ldots , e^{\vec{\beta}_{n-1} \cdot
  \vec{\phi}})\,,
 \end{align}
where the weight vectors $\vec{\beta}_p$ are subject to the reduction of \eqref{weights}:
 \begin{align}
  \sum_p \beta_{pI}=0 \,, \qquad
  \sum_p \beta_{pI}\,\beta_{pJ} = 2 \, \delta_{IJ} \,, \qquad
  \vec{\beta}_{p} \cdot \vec{\beta}_{q} = 2 \, \delta_{pq}-\frac{2}{n-1} \,,
  \label{weights-beta}
 \end{align}
while the last component of all vectors $\vec{\beta}_p$ vanishes: $\beta_{pn} = 0$. .
Therefore, the contracted superpotential \eqref{contracted-superpotential} only depends
on the smaller coset $SL(n-1,\mathbb{R}) / SO(n-1)$. Also note that the overall dilaton
coupling has changed due to the contraction. For the scalar potential, this will amount
to $a \phi + 2 \vec{\beta} \cdot \vec{\phi}$ instead of $a \phi$. After a change of
basis, corresponding to an $SO(n+1)$ rotation in $(\phi,\vec{\phi})$-space, this takes
the form $\tilde{a} \tilde{\phi}$ with
 \begin{align}
  \tilde{a}^2 = a^2 + 4 \vec{\beta} \cdot \vec{\beta} = \frac{8}{n-1} - 2 \, \frac{D-3}{D-2} > a \,,
  \label{DNA-contracted}
 \end{align}
which is exactly the original relation \eqref{DNA} with $n$ decreased by one. It should
be clear that this contraction can be employed several times, each time reducing $n$ by
one.

The second operation we wish to perform corresponds to dimensionally reducing the scalar
sector. We take trivial Ans\"{a}tze for the scalars, $\hat{M} = M$ and $\hat{\phi} =
\phi$, and the usual Ansatz \eqref{Ansatz-gravity-circle} for the metric (obtaining
Einstein frame with a canonically normalised Kaluza-Klein scalar $\varphi$ in the lower
dimension):
 \begin{align}
  \hat{ds}_D{}^2 = e^{2 \gamma \varphi} ds_{D-1}{}^2 + e^{- 2 (D-3) \gamma \varphi} dz^2 \,, \qquad
  \gamma^2 = \frac{1}{2(D-2)(D-3)} \,,
 \end{align}
where we have truncated the Kaluza-Klein vector away. The resulting scalar potential is
of the same form \eqref{CSO-potential}, but again the dilaton coupling has changed: the
factor $a \phi$ is replaced by $a \phi + 2 \gamma \varphi$. After a field redefinition,
this corresponds to $\tilde{a} \tilde{\phi}$ with
 \begin{align}
  \tilde{a}^2 = a^2 + 4 \gamma^2 = \frac{8}{n} - 2 \, \frac{D-4}{D-3} > a \,,
  \label{DNA-reduced}
 \end{align}
which is exactly the original relation \eqref{DNA} with $D$ decreased by one. Again,
dimensional reduction can be performed any number of times, reducing $D$ by one at each
step.

Concluding, after any number of group contractions or dimensional reductions, the scalar
subsector will always have a scalar potential \eqref{CSO-potential} with dilaton coupling
\eqref{DNA}. The only effect of these operations is to decrease $D$ or $n$ by one,
respectively: the resulting system still satisfies all equations with the new values of
the parameters $D$ and $n$. This proves that the scalar subsectors of different gauged
supergravities reduce onto each other upon matching $D$ and $n$ by dimensional reductions
and/or group contractions. We expect this to hold for the full theories as well.

\subsection{Other Gauged Maximal Supergravities} \label{sec:CSO-gaugings-5}

The $CSO$ gaugings generalise the gaugings of subgroups of $SL(2,\mathbb{R})$ and
$SL(3,\mathbb{R})$ in nine and eight dimensions, respectively. These are not the only
possibilities in lower dimensions, however. Other examples were constructed in
e.g.~\cite{Andrianopoli:2002mf, Hull:2002cv}.

An interesting approach was taken in \cite{Nicolai:2001sv, deWit:2002vt}, where possible
gaugings were classified by a purely group-theoretical analysis. For example, different
gaugings were found in 4D, depending on the global symmetry group of the
Lagrangian\footnote{Hodge duality relates electric and magnetic vectors in 4D. While this
does not affect the symmetry group of the field equations, the different choices give
rise to different global symmetries of the Lagrangian.} \cite{deWit:2002vt}. The
Lagrangian with $SL(8,\mathbb{R})$ invariance allows for the $CSO(p,q,r)$ gauging with
$p+q+r = n$, as found above, but other gaugings in $D=4$ and $D=5$ were also found. For
example, after a number of Hodge duality transformations can bring one to an equivalent
Lagrangian with $SL(6,\mathbb{R}) \times SL(2,\mathbb{R}) \times SO(1,1)$ invariance,
which allows for other gaugings. These gauged theories are obtainable from dimensional
reduction of IIB supergravity with fluxes\footnote{Interestingly, when truncating from
$N=8$ to $N=4$ gauged supergravities, the higher-dimensional origin becomes IIB on an
orientifold with fluxes and branes \cite{D'Auria:2003jk, Angelantonj:2003rq,
deWit:2003hq}.} \cite{deWit:2003hq}. Indeed, the global symmetry group has a natural
origin from the IIB point of view: the $SL(6,\mathbb{R})$ stems from the six internal
coordinates, while the $SL(2,\mathbb{R})$ is already present in ten dimensions.

In addition to theories with a Lagrangian, it was found in
sections~\ref{sec:D=9-gaugings} and \ref{sec:D=8-gaugings} that M-theory allows for other
gauged supergravities, that do not have an action but only field equations. In nine
dimensions, there was one such theory with parameter $m_s$. In eight dimensions, there
were five theories, with parameters $q_2,q_3$ and $a$. Clearly, one can expect such
theories also in the lower dimensions. It is not clear to us what the general
pattern\footnote{Note that the number of mass parameters in nine and eight dimensions
coincides with the number of antisymmetric components of the matrices $Q^{mn}$ in these
dimensions (i.e.~one and three, respectively). It would be interesting to investigate
whether the gauged theories without an action are somehow related to antisymmetric mass
matrices.} will be, however.

 \chapter{Domain Walls} \label{ch:domain-walls}

In this chapter, we will construct half-supersymmetric domain wall solutions to the
massive and gauged supergravities of the previous chapter and we will discuss their
physical interpretation in terms of branes. In the last section we will consider 1/4
supersymmetric intersections of domain walls with strings.

\section{D8-brane in Massive IIA} \label{sec:D8-brane}

\subsection{D8-brane Solution} \label{sec:D8-brane-1}

In section~\ref{sec:susy-solutions} we have discussed the different supersymmetric
solutions of massless IIA supergravity. The situation for massive IIA supergravity is
radically different: there is no maximally supersymmetric solution
\cite{Figueroa-O'Farrill:2002ft} and only one half-supersymmetric solution, the D8-brane
solution \cite{Polchinski:1996df}. It is carried by the metric and the dilaton, which
read
 \begin{align}
  ds^2 & \, =  H^{1/8} dx_9{}^2 + H^{9/8} dy^2 \,, \qquad
  e^{\phi} = H^{-5/4} \,.
 \label{D8-brane}
 \end{align}
Note that this is of the form of the generic $p$-brane solutions \eqref{p-brane-metric}
with $d=9$, $\tilde{d}=-1$ and $\Delta = 4$, and has $\sqrt{-g} g^{yy} = 1$. It is
expressed in terms of one harmonic function $H = c + m_{\text{R}} y$, where we take
$m_{\text{R}}$ positive and $c$ is an arbitrary integration constant. This solution
preserves half of supersymmetry under the supersymmetry rules \eqref{IIAsusy} with
explicit massive deformations \eqref{Romans-susy} with Killing spinor
\begin{align}
  {\epsilon} & \, = H^{1/32}{\epsilon}_{0} \,,
  \text{~~with~}
  (1 + {\Gamma}^{\underline{y}}) \, {\epsilon}_{0} =0 \,,
\end{align}
where ${\epsilon}_{0}$ is a constant spinor that satisfies the above linear constraint.
Thus the D8-brane has 16 unbroken supersymmetries.

For later use we would also like to present the D8-brane in a different coordinate
system, which is related via
 \begin{align}
  \tilde{H} = 2 m_{\text{R}} \tilde{y} + \tilde{c} = H(y)^2 \,.
  \label{D8-coordinate-transformation}
 \end{align}
In the new transverse coordinate $\tilde{y}$ the solution reads
 \begin{align}
  ds^2 & \, =  \tilde{H}^{1/16} dx_9{}^2 + \tilde{H}^{-7/16} d\tilde{y}^2 \,, \qquad
  e^{\phi} = \tilde{H}^{-5/8} \,.
 \label{D8-brane-2}
 \end{align}
Note that we now have $\sqrt{-g} g^{tt} = -1$. For the present section, we will use the
first parameterisation \eqref{D8-brane}, however.

As discussed in subsection~\ref{sec:susy-solutions-2}, a domain wall with harmonic
function $H = c + m_{\text{R}} y$ is not well-defined. The zeroes in $H$ induce
singularities in the solution. To avoid these, one has to include source terms
(corresponding to a thin domain wall) to modify the behaviour of the harmonic function.
We will discuss such source terms for the D8-brane solution in the next subsection.

\subsection{Source Terms and Piecewise Constant Parameters} \label{sec:D8-brane-2}

To this end we introduce a number of source terms, corresponding to eight-branes. Since
these couple to a nine-form potential it is necessary to dualise the mass parameter of
massive IIA to a ten-form field strength:
 \begin{align}
  m_{\text{R}} = e^{-5 \phi/2} \star G^{(10)} \,, \qquad \G{10} = d C^{(9)} \,,
 \end{align}
as discussed in subsection~\ref{sec:parent-sugra-3}. In the absence of sources, the field
equation for $\C{9}$ implies $m_{\text{R}}$ to be constant. When sources are present,
however, the parameter $m_{\text{R}}$ is required to be piecewise constant, i.e.~it can
take different (constant) values in different regions of the transverse space. This
property is the reason why the corresponding solution is called a domain wall; the
eight-brane sources separate physically different regions.

The eight-brane source terms are given by
 \begin{align}
  S_{8} = - \frac{2 \pi}{(2\pi \ell_s)^9} \int d^9 x \{ e^{-\phi} \sqrt{-g_{(9)}} + \tfrac{1}{9!} \varepsilon^{(9)} C^{(9)} \} \,,
 \label{8braneaction}
 \end{align}
with $\varepsilon^{(9) \; {\mu_0} \cdots {\mu_8}} = \varepsilon^{(10) \; {\mu_0} \cdots
{\mu_8} y}$ and we use the ranges $\mu, \nu = (0, \ldots, 8)$ in this section. Depending
on the coefficients of $S_8$ in the total action, the source terms have a different
interpretation in string theory:
\begin{itemize}
 \item
Objects with positive coefficients correspond to D8-branes. Passing through such a domain
wall leads to a decrease of the slope of the harmonic function \cite{Polchinski:1996df,
Chamblin:1999ea, Bergshoeff:2001pv}. The prime example is
 \begin{align}
  H = \begin{cases}
  c - m_{\text{R}} y \,, \qquad y > 0 \,, \\
  c + m_{\text{R}} y \,, \qquad y < 0 \,,
  \end{cases}
 \end{align}
with $c$ and $m_{\text{R}}$ positive. This can be written as $H = c - m_{\text{R}} |y|$,
where the absolute value of the transverse coordinate $y$ can be seen as a consequence of
the piecewise constant parameter $m_{\text{R}}$. It follows that $H$ will vanish for some
critical value of $y$.
 \item
Objects with negative coefficients correspond to so-called O8-planes. These are
orientifold planes\footnote{Orientifold planes arise when modding out with a discrete
symmetry that involves $\Omega$, the string world sheet parity operation; see
\cite{Dabholkar:1997zd} for a nice introduction.}, which arise by dividing out by a
specific $\mathbb{Z}_2$ symmetry. In this case the relevant symmetry is $I_y \Omega$,
where $I_y$ is a reflection in the transverse space and $\Omega$ is the world-sheet
parity operation. Its effect on the IIA supergravity fields reads
\begin{align}
  y & \rightarrow - y \, , \notag \\
  \big\{ \phi, g_{\mu \nu },
     B_{\mu \nu} \big\}
  & \rightarrow \big\{ \phi, g_{\mu \nu},
    -B_{\mu \nu} \big\}\, , \notag \\
  \big\{ \C{2n-1}_{{\mu_1} \cdots {\mu_{2n-1}}} \big\}
  & \rightarrow (-)^{n+1} \big\{
    \C{2n-1}_{{\mu_1} \cdots {\mu_{2n-1}}} \big\}
    \,, \notag \\
  \big\{ \psi_{\mu} , \lambda, \epsilon \big\}
  & \rightarrow \Gamma^{\underline{y}} \big\{ \psi_{\mu} ,
   -\lambda, \epsilon \big\} \,,
 \label{I9Omega}
\end{align}
and the parity of the fields with one or more indices in the $y$-direction is given by
the rule that every index in the $y$-direction gives an extra minus sign compared to the
above rules.

Due to the inclusion of such source terms, the harmonic function will be e.g.~$H = c +
m_{\text{R}} |y|$ with $c$ and $m_{\text{R}}$ positive \cite{Chamblin:1999ea,
Bergshoeff:2001pv}. Thus $H$ is positive for all values of $y$ and has a minimum at the
O8-plane.
\end{itemize}
One thus finds that the introduction of D8-branes leads to zeroes in $H$ and thus to a
'critical distance'. It forces one to include O8-planes at a smaller distance, such that
the zero in $H$ is avoided. If the transverse space is $\mathbb{R}/\mathbb{Z}_2$, we can
take one O8-plane with \RR\ charge $-16$ (in units where a D8-brane has charge $+1$) and
$n$ D8-branes and their images with $n \le 8$. For $n>8$ the total tension is positive
and a zero in the harmonic function will occur. On the other hand, if the transverse
space is $S^1/\mathbb{Z}_2$ (i.e.~the range of $y$ is compact), the total tension has to
vanish and one is led to type~I${}^\prime$ string theory with two O8-planes at the two
fixed points and 16 D8-branes and their images in between \cite{Polchinski:1996df}.

\subsection{Type I$^\prime$ String Theory and Supergravity} \label{sec:D8-brane-3}

We will consider an example of the latter situation in full detail. First we choose our
space-time to be $M^9 \times S^1$. All fields satisfy $\Phi(y)=\Phi(y+2 \pi R)$ with $R$
the radius of $S^1$. Furthermore, the fields are either even or odd under $I_9 \Omega$.
Modding out this $\mathbb{Z}_2$ symmetry, the odd fields vanish on the fixed points $y=0$
and $y= \pi R$ of the orientifold, where we will put the brane sources. However, the
type~IIA theory would be inconsistent under orientifold truncation unless extra gauge
degrees of freedom appear in the theory. It turns out we have to place 32 D8-branes
between these O8-planes \cite{Polchinski:1996df}, leading to type~${\rm I^\prime}$ string
theory. It is T-dual to type~I string theory, which is obtained by modding the IIB theory
with the $\mathbb{Z}_2$ symmetry $\Omega$. This also explains the origin of the 32
D8-branes: the type~I gauge group $SO(32)$ can be seen to come from 32 unoriented
D9-branes (filling all of space-time) and performing T-duality yields the 32
D8-branes~\cite{Horava:1996qa}.

We will consider the special situation where all D-branes coincide with either one of the
O-planes. In addition, we assume that there is no matter on the branes. Thus, we are
describing the vacuum solution of the D-brane system, switching off the excitations on
the branes. Therefore, our total effective action is given by
\begin{align}
 S = 2(n-8) S_{8} \delta(y) + 2(8-n) S_{8} \delta(y-\pi R)
 - \frac{2 \pi}{(2 \pi \ell_s)^9} \int d^9 x
    \mathcal{L}_{\text{bulk}} \,,
  \label{bulk+brane}
\end{align}
which is given by the bulk action and an O8-plane and $2n$ D8-branes at $y=0$ and an
O8-plane and $32-2n$ D8-branes at $\pi R$. For definiteness we will take $8 < n \leq 16$,
i.e.~the D8-branes dominate the O8-plane at $y=0$ while the latter dominates at $y=\pi
R$.

The D8-brane solution is given by \eqref{D8-brane} with harmonic function
\cite{Bergshoeff:2001pv}
 \begin{align}
  H = c + \frac{(8-n)}{2\pi\ell_{s}} |y| \,.
 \label{harmonic-kink}
 \end{align}
Thus we may identify the mass parameter as follows:
\begin{equation}
  m_{\text{R}} = \begin{cases}
   \vspace{.1cm} \displaystyle{\frac{8-n}{2 \pi \ell_{s}}} \,, \qquad y > 0 \,, \\
   \displaystyle{\frac{n-8}{2 \pi \ell_{s}}} \,, \qquad  y < 0 \,.
  \end{cases}
 \label{quant}
\end{equation}
The harmonic function \eqref{harmonic-kink} with piecewise constant mass parameter
$m_{\text{R}}$ will have a zero if the range of $y$ is too large; the distance between
the branes must be small enough to prevent the harmonic function from vanishing. The
radius of the circle and distance between the O-planes is thus restricted to
\begin{equation}
  R < {2 c \ell_s \over (n-8)} \,.
\end{equation}
The saturating case is called the critical distance $R_c$. Thus it seems that type~${\rm
I^\prime}$ \ST\ is consistent only on $M^9 \times (S^1 / \mathbb{Z}_2)$ with a circle of
restricted radius.

Of course we have only considered a special case of the type~${\rm I^\prime}$ theory with
all D-branes on one of the fixed points. However, also with D-branes in between the
O-planes we expect the vacuum solution to imply a critical distance: each O8-plane
necessarily has $16$ D8-branes in its vicinity. The same phenomenon of type~${\rm
I^\prime}$ was found in \cite{Polchinski:1996df} in the context of the duality between
the heterotic and type~I theories. Note that the maximal distance depends on the
distribution of the D-branes. In the most asymmetric case ($n=16$) it is smallest while
in the most symmetric case ($n=8$) there is no restriction on $R$.

Note that the identification \eqref{quant} implies a quantisation of the mass parameter
of massive IIA supergravity. Upon dimensional reduction, this should coincide with the
special case of the $SL(2,\mathbb{R})$ mass parameters
 \begin{align}
  \vec{m} = (0, \frac{\tilde{n}}{2 \pi R} , \frac{\tilde{n}}{2 \pi R}) \,, \qquad
  \tilde{n} \in \mathbb{Z} \,, \label{quantB}
 \end{align}
as can be seen from \eqref{massive-T-duality} and \eqref{R-quant}. At first sight, these
quantisation conditions do not seem to match. The resolution can be found in
\cite{Green:1996bh}, where factors of $g_s$ are properly taken into account. Being
related to the mass of D-branes, both quantised masses \eqref{quant} and \eqref{quantB}
are inversely proportional to $g_s$ of IIA and IIB, respectively (see the discussion
below \eqref{gs-scaling}):
 \begin{align}
  m_A = \frac{\pm (n-8)}{2 \pi \ell_{s} g_{A}} \,, \qquad
  m_B = \frac{\tilde{n}}{2 \pi R_B g_{B}} \,,
 \end{align}
where we have included the A and B labels and omitted the $s$ subscript of $g_s$. The
T-duality relations between the IIA and IIB parameters
 \begin{align}
  R_A R_B = \ell_s{}^2 \,, \qquad g_A \ell_s = g_B R_B \,,
 \end{align}
then exactly relate the two expressions for the quantised mass with $\tilde{n} = \pm
(n-8)$.

It is clear that the D8-O8 system can be generalized further. To start with, placing
D-branes in any compact transverse space requires the presence of oppositely charged
branes that need to have opposite tensions in order to be in supersymmetric equilibrium
\cite{Bergshoeff:2001pv}. If all the negative-tension branes are identified with
orientifold planes, as we have suggested here, then the compact transverse spaces must be
orbifolds with the orientifold planes placed at the orbifold points. The $\mathbb{Z}_{2}$
reflection symmetries associated to the orientifold planes can be part of more general
orbifold groups ($\mathbb{Z}_{n}$ etc.).  It would be interesting to realize these bulk
\& brane configurations explicitly.

\section{Domain Walls in $CSO$ Gaugings and their Uplift} \label{sec:CSO-DW}

\subsection{The DW/QFT Correspondence} \label{sec:CSO-DW-1}

Due to the AdS/CFT correspondence \cite{Maldacena:1998re}, it has been realized that
there is an intimate relationship between certain branes of string or M-theory and
corresponding lower-dimensional $SO(n)$ gauged supergravities. The relation is
established via a maximally supersymmetric vacuum configuration of string or M-theory,
which is the direct product of an AdS space and a sphere (see
subsection~\ref{sec:susy-solutions-3}): for a $p$-brane with $n$ transverse directions,
we are dealing with an $AdS_{p+2} \times S^{n-1}$ vacuum configuration. On the one hand,
this vacuum configuration arises as the near-horizon limit of an M2-, D3- or M5-brane; on
the other hand, the coset reduction over the spherical part leads to the related $SO(n)$
gauged supergravity in $p+2$ dimensions, which allows for a maximally supersymmetric
$AdS_{p+2}$ vacuum configuration (see section~\ref{sec:CSO-gaugings}). The gauge theory
of the AdS/CFT correspondence can be taken at the boundary of this $AdS_{p+2}$ space. All
dilatons are constant for this vacuum configuration (with no extra dilaton present,
i.e.~$a=0$). This is related to the conformal invariance of the gauge theory.

There are two ways to depart from conformal invariance, which both involve exciting some
of the dilatons in the vacuum configuration. The first deformation can be introduced via
the $n-1$ dilatons of the AdS supergravities. By exciting some of these dilatons one
obtains a deformed Anti-de Sitter configuration. In the AdS/CFT correspondence this
corresponds to considering the (non-conformal) Coulomb branch of the gauge theory
\cite{Maldacena:1998re}.

Alternatively, one can obtain a non-conformal theory by considering the other branes of
string and M-theory, for which there is an extra dilaton present in the scalar potential
of the gauged supergravities (corresponding to the $\surd$ in table~\ref{tab:branes}).
This leads to DW supergravities, where the maximally supersymmetric AdS vacuum is
replaced by a non-conformal and half-supersymmetric domain wall solution. This situation
is encountered when one generalises the AdS/CFT correspondence to a DW/QFT correspondence
\cite{Itzhaki:1998dd, Boonstra:1998mp}.

\begin{table}[ht]
\begin{center}
\hspace{-1cm}
\begin{tabular}{||c|c||c||c||}
\hline \rule[-1mm]{0mm}{6mm}
  $D$ & $n$ & $\phi$ & Brane \\
\hline \hline \rule[-1mm]{0mm}{6mm}
  $10$ & $1$ & $\surd $ & D8 \\
\hline \rule[-1mm]{0mm}{6mm}
  $9$ & $2$ & $\surd$ & D7 \\
\hline \rule[-1mm]{0mm}{6mm}
  $8$ & $3$ & $\surd$ & D6 \\
\hline \rule[-1mm]{0mm}{6mm}
  $7$ & $5$ / $4$ & $-$ / $\surd$ & M5 / NS5A \\
\hline \rule[-1mm]{0mm}{6mm}
  $6$ & $5$ & $\surd$ & D4 \\
\hline \rule[-1mm]{0mm}{6mm}
  $5$ & $6$ & $-$ & D3 \\
\hline \rule[-1mm]{0mm}{6mm}
  $4$ & $8$ / $7$ & $-$ / $\surd$ & M2 / D2 \\
\hline
\end{tabular}
\caption{\it The domain walls and gauged supergravities in $D$ dimensions with $n$ mass
parameters are related to M- or D-branes with $n$ transverse directions.}
\label{tab:branes}
\end{center}
\end{table}

A natural generalisation is to excite some of the $n-1$ dilatons describing the Coulomb
branch of the CFT and the extra dilaton that leads to a non-conformal QFT at the same
time. This leads to domain wall solutions of $SO(n)$ gauged DW supergravities
\cite{Cvetic:2000zu} that describe the Coulomb branch of the (non-conformal) QFT. The
uplift of these multiple domain walls leads to (the near-horizon-limit of) brane
distributions in string or M-theory, as we will see in the next subsections. This is
based on results from \cite{Bergshoeff:2004nq}.

A $p$-brane can be reduced in two ways: via a double dimensional reduction (leading to a
($p-1$)-brane in one dimension lower) or a direct dimensional reduction (leading to a
$p$-brane in one dimension lower). It has been pointed out \cite{Boonstra:1998mp} that
direct dimensional reduction leads from $SO(n)$ gauged supergravities to the generalised
$CSO$ gauged supergravities of \cite{Hull:1984vg, Hull:1984qz}. Thus, direct dimensional
reduction corresponds to a group contraction of the gauged supergravity (see
subsection~\ref{sec:CSO-gaugings-4}). In contrast, double dimensional reduction of a
brane corresponds to a dimensional reduction of the gauged supergravity (as was discussed
in the same subsection). Recapitulating, we have the following correspondences between
the brane and gauged supergravity points of views:
\begin{center}
\begin{tabular}{ccc}
${\underline {\rm Brane}}$ && ${\underline {\rm Gauged\ supergravity}}$\\
\rule[-1mm]{0mm}{8mm}
direct\ dimensional\ reduction & $\Leftrightarrow$ & group contraction\\
double\ dimensional\ reduction & $\Leftrightarrow$ & toroidal reduction
\end{tabular}
\end{center}

Note that not all branes of string or M-theory are present in table~\ref{tab:branes}. The
missing cases of the D0 and F1A can rather easily be included, as has been done in
\cite{Bergshoeff:2004nq}. The corresponding $D=1,2$ gauged supergravities have $n=9,8$,
respectively \cite{Nicolai:2000zt, Fischbacher:2003yw}. The remaining cases are the IIB
doublets of NS5B/D5 and F1B/D1 branes. The associated theories are the reduction of IIB
over $S^3$ or $S^7$ with an electric or magnetic flux of the NS-NS/R-R three form field
strength \cite{Cvetic:2000dm, Cvetic:2003jy}. For the $D=3$ $SO(8)$ theories
corresponding to the IIB strings (which are different from the F1A result), see
\cite{Fischbacher:2003yw}. The five-brane cases are supposed to lead to new $D=7$ $SO(4)$
gauged supergravities, which might be related to the theories constructed in
\cite{Alonso-Alberca:2002tb}.

\subsection{Domain Walls} \label{sec:CSO-DW-2}

In this subsection, we give a unified description of a class of domain wall solutions for
the $CSO$ gauged supergravities in various dimensions, which is of particular relevance
to the DW/QFT correspondence.

We consider the following Ansatz for the domain wall with $D-1$ world-volume coordinates
$\vec{x}$ and one transverse coordinate $y$:
 \begin{align}
   ds^2 & = g(y)^2 d\vec{x}^2 + f(y)^2 dy^2 \,, \qquad
   M = M(y) \,, \qquad \phi = \phi (y) \,.
 \label{Ansatz}
 \end{align}
The idea is to substitute this Ansatz into the action, consisting of the Einstein-Hilbert
term, scalar kinetic terms and the scalar potential \eqref{CSO-potential}, and write this
as a sum of squares \cite{Bakas:1999fa}. Using \eqref{DNA}, the reduced one-dimensional
action can be written as
\begin{align}\nonumber
  S = \int dy \,
    g^{D-1} f \Big[ & \frac{D-1}{4(D-2)}\Big(\frac{2(D-2)}{fg}\frac{dg}{dy}- W\Big)^2
    -\frac{1}{2}\Big(\frac{1}{f}\frac{d\vec{\phi}}{dy}+\vec{\partial} W\Big)^2+\\
& -\frac{1}{2} \Big(\frac{1}{f}\frac{d\phi}{dy}+\partial_\phi W\Big)^2
+\frac{1}{f}\,\frac{dW}{dy}+(D-1)\frac{1}{fg}\frac{dg}{dy}W\Big]\,, \label{Bogomolnyi}
\end{align}
which is a sum of squares, up to a boundary term. Minimalisation of this action therefore
corresponds to the vanishing of the squared terms. This gives rise to the first-order
Bogomol'nyi equations
\begin{align}
  \frac{1}{f}\frac{d\vec{\phi}}{dy}=-\vec{\partial}W\,,\quad\frac{1}{f}\frac{d\phi}{dy}=-\partial_\phi
  W\,,\quad \frac{2(D-2)}{fg}\frac{dg}{dy}= W \,, \label{bog}
\end{align}
Note that one should not expect a Bogomol'nyi equation associated to $f$ since it can be
absorbed in a reparameterisation of the transverse coordinate $y$.

The Bogomol'nyi equations can be solved by the elegant domain wall solution, generalising
\cite{Cvetic:1999xx,Cvetic:2000zu},
 \begin{align}
   ds^2 & = h^{1/(2D-4)} d\vec{x}^2 + h^{(3-D)/(2D-4)} dy^2 \,, \notag \\
   M & = h^{1/n} \text{diag}(1/h_1,\ldots,1/h_n) \,, \qquad e^\phi = h^{-a/4}
   \,,
 \label{domain-wall}
 \end{align}
written in terms of $n$ harmonic functions $h_i$ and their product $h$:
 \begin{align}
  h_i = 2 q_i y +l_i^2 \,, \qquad h = h_1 \cdots h_n \,.
 \end{align}
Note that this transverse coordinate basis\footnote{For $D=10$ and $n=1$, the solution
\eqref{domain-wall} coincides with the D8-brane in the $\tilde{y}$-coordinate
\eqref{D8-brane-2}.} has $\sqrt{-g} g^{tt} = -1$. The functions $h_ i$ are necessarily
positive since the entries of $M$ are positive. For all $q_i \geq 0$, this implies that
$y$ can range from $0$ to $\infty$; if there is at least one $q_i<0$, the range of $y$ is
bounded from above.

The solution is parameterised by $n$ integration constants\footnote{Strictly speaking, it
is $l_i{}^2$ rather than $l_i$ that appears as integration constant, allowing for
positive and negative $l_i{}^2$. However, one can always  take these positive by shifting
$y$, in which case the crucial distinction between $l_i$ and $l_i{}^2$ disappears.}
$l_i$. However, if a charge $q_i$ happens to be vanishing, the corresponding $l_i$ can
always be set equal to one (by $SL(n,\mathbb{R})$ transformations that leave the scalar
potential invariant). In addition, one can eliminate one of the remaining $l_i$'s by a
redefinition of the variable $y$. Therefore we effectively end up with $p+q-1$
independent constants, parameterising the $p+q$ harmonics. We define $m$ to be the number
of linearly independent harmonics $h_i$ with $q_i \neq 0$ and call the corresponding
solution \eqref{domain-wall} an $m$-tuple domain wall. For different values of the
constants $l_i$, one finds different numbers $m$ of linearly independent harmonics. For
examples of truncations to single domain walls, see tables~\ref{tab:9D-single-DW} and
\ref{tab:8D-single-DW}.

It should not be a surprise that all scalar potentials of table~\ref{tab:gaugings}
satisfy the relation \eqref{DNA} since these are embedded in a supergravity theory, whose
Lagrangian ``is the sum of the supersymmetry transformations'' and therefore always
yields first-order differential equations. For this reason, domain wall solutions to the
separate terms in \eqref{Bogomolnyi} will always preserve half of supersymmetry. The
corresponding Killing spinor is given by
 \begin{align}
  \epsilon = h^{1/(8D-16)} \epsilon_0 \,, \qquad (1 + \Gamma_{\underline{y}}) \epsilon_0 = 0 \,,
 \end{align}
where the projection constraint eliminates half of the components of $\epsilon_0$. An
exception is $a=0$, $q_i=1$ and $l_i = 0$, in which case the domain wall solution
\eqref{domain-wall} becomes a maximally (super-)symmetric Anti-De Sitter space-time in
horospherical coordinates. Then the singularity at $y=0$ is a coordinate artifact and
there is an extra Killing spinor, yielding fully unbroken supersymmetry.

\subsection{Higher-dimensional Origin and Harmonics} \label{sec:CSO-DW-3}

Upon uplifting these domain walls, one obtains higher-dimensional solutions, which are
related to the 1/2 supersymmetric brane solutions of 11D, IIA and IIB supergravity, as
given in table~\ref{tab:branes}. Note that the number of mass parameters (and therefore
the number of harmonic functions $h_i$ of the transverse coordinate) always equals the
transverse dimension of the brane. Thus, in $D$ dimensions, the number $n$ of mass
parameters is given by the co-dimension of the half-supersymmetric $(D-2)$-brane of IIA,
IIB or M-theory.

The metric of the uplifted solution can in all cases be written in the form
 \begin{align}\label{brane}
  ds^2=H_n^{(2-n)/(D+n-3)}\,dx_{{\scriptscriptstyle{D-1}}}^2+H_n^{(D-1)/(D-n-3)}\,ds_{{\scriptscriptstyle{n}}}^2\,,
 \end{align}
where $H_n$ is a harmonic function on the transverse space, whose powers are appropriate
for the corresponding D-brane solution in ten dimensions or M-brane solution in eleven
dimensions (as can be checked from section~\ref{sec:susy-solutions}). From the form of
the metric one infers that the solution corresponds to some brane distribution. For all
$q_i=1$, these solutions were found in \cite{Bakas:1999ax,Cvetic:1999xx,Bakas:1999fa} for
the D3-, M2- and M5-branes and in \cite{Cvetic:2000zu,Bakas:2000nt} for other branes.

The harmonic function takes the form
 \begin{align}
  H_n (y, \mu_i)=h^{-1/2} \Big( \sum_{i=1}^n\frac{q_i^2\mu_i^2}{h_i} \Big)^{-1} \,,
  \label{harm}
 \end{align}
where $\mu_i$ are Cartesian coordinates, fulfilling \eqref{hypersurface}. The transverse
part of the metric is given by \cite{Cvetic:1999xx}
 \begin{align}\label{transverse}
  ds_{{\scriptscriptstyle{n}}}^2 = H_n^{-1} h^{-1/2}dy^2+\sum_{i=1}^n h_i\,d\mu_i^2\,,
 \end{align}
With a change of coordinates, it can be seen that the $n$-dimensional transverse space is
flat\footnote{For $D=10$ and $n=1$, this coordinate transformation coincides with
\eqref{D8-coordinate-transformation}. Indeed, the brane solution \eqref{brane} is
identical to the D8-brane with transverse $y$-coordinate \eqref{D8-brane}.}
\cite{Russo:1998mm,Cvetic:1999xx}
 \begin{align}\label{mutoz}
  z_i=\sqrt{h_i}\mu_i\,,\quad
  ds_{{\scriptscriptstyle{n}}}^2=\sum_{i=1}^n dz_i dz_i\,.
 \end{align}
The above is easily verified
\begin{align}
dz_i=h_i^{-1/2} q_i\mu_i\, dy+h_i^{1/2}\,d\mu_i\,,\quad \sum_{i=1}^ndz_i
dz_i=\sum_{i=1}^n \frac{q_i^2\mu_i^2}{h_i}\,dy^2+\sum_{i=1}^n h_i\,d\mu_i^2\,,
\end{align}
where we have used $\sum_{i=1}^n q_i\,\mu_i\,d\mu_i=0$, which follows from
\eqref{hypersurface}. Note that one has $\sqrt{-g} g^{ii}=1$ after the coordinate change
to $z_i$.

The harmonic function $H_n$ specifies the dependence on the $n$ transverse coordinates
$z_i$. The constants $l_i$ parameterise the possible harmonics that are consistent with
the reduction Ansatz. The mass parameters $q_i$ specify this reduction Ansatz. Thus,
changing a mass parameter $q_i$ changes both the reduction Ansatz and the harmonic
function that is compatible with that Ansatz. Sending a mass parameter to zero, e.g.~$q_n
\rightarrow 0$, corresponds to truncating the harmonic function on $n$-dimensional flat
space to
 \begin{align}
  H_n (q_n =0, l_n = 1) = H_{n-1} \,,
  \label{trunc-harm}
 \end{align}
i.e.~a harmonic function on $(n-1)$-dimensional flat space.

It is difficult to obtain the explicit expression for the harmonic function $H_n$ in
terms of the Cartesian coordinates $z_i$ (the example of $n=2$ will be given in the next
section). Nevertheless, one can show that $H_n$ is indeed harmonic on $\mathbb{R}^n$ for
all values of $q_i$, thus extending the analysis of \cite{Bakas:1999ax} where $q_i = 1$.
The calculation is facilitated by the following definitions
\begin{align}
A_m=\sum_{i=1}^n\frac{q_i^m z_i^2}{h_i^m}\,,\quad B_m=\sum_{i=1}^n\frac{q_i^m}{h_i^m}\,.
\end{align}
In terms of $A_m$ and $B_m$ we calculate
\begin{align}
\partial_i H_n=h^{-1/2}\Big(-\frac{q_i z_i}{h_i}\frac{B_1}{A_2^2}+4\,\frac{q_i
z_i}{h_i}\frac{A_3}{A_2^3}-2\frac{q_i^2 z_i}{h_i^2}\frac{1}{A_2^2}\Big)\,,
\end{align}
from which we finally get
\begin{align}
 \sum_{i=1}^n\partial_i\partial_i H_n
 & =h^{-1/2}\Big(2 \frac{B_2}{A_2^2}-2\frac{B_1 A_3}{A_2^3}-16 \frac{A_4}{A_2^3}+16\frac{A_3^2}{A_2^4}-2
  \frac{B_2}{A_2^2}+16\frac{A_4}{A_2^3}+2\frac{B_1 A_3}{A_2^3}-16\frac{A_3^2}{A_2^4}\Big) \notag \\
 & = 0 \,,
\end{align}
which proves the harmonicity of $H_n$ on $\mathbb{R}^n$.

\subsection{Brane Distributions for $SO(n)$ Harmonics} \label{sec:CSO-DW-4}

Since the harmonic function $H_n$ depends on the angular variables in addition to the
radial, the uplifted solution will in general correspond to a distribution of branes
rather than a single brane. For $D<9$ and all $q_i=1$ (i.e.~the $SO(n)$
cases\footnote{Note that we can also include the cases where some $q_i=0$, using
\eqref{trunc-harm}.} with $n \geq 3$) this means that the harmonic function can be
written in terms of a charge distribution $\sigma$ as follows
\cite{Cvetic:1999xx,Cvetic:2000zu}
 \begin{align}
  H_n(\vec{z}) = \int d^n z' \frac{\sigma(\vec{z}\,')}{\vert \vec{z}-\vec{z}\,'\vert^{n-2}}\,,
 \end{align}
and since $H_n$ appears without an integration constant, the distributions will actually
be a near-horizon limit of the brane distribution.

\begin{figure}[tb]
\centerline{\epsfig{file=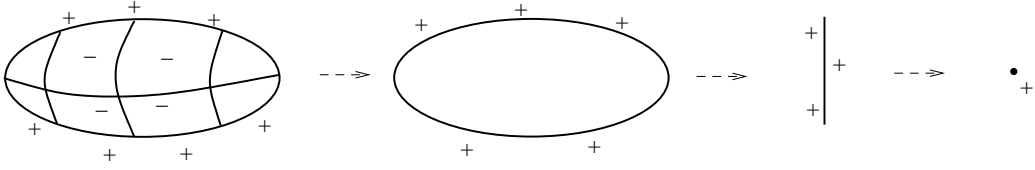,width=.85\textwidth}} \caption{\it The
distributions of NS5A-branes corresponding to the uplift of the 7D $ISO(4)$ domain walls
with three, two, one and zero non-vanishing $l_i$'s, respectively.}
\label{fig:D5-distrib}
\end{figure}

It turns out that the distributions are given in terms of higher dimensional ellipsoids
\cite{Kraus:1998hv,Cvetic:1999xx}. The dimensions of these ellipsoids are given in terms
of the number $m$ of independent harmonics $h_i$ or, equivalently, the number $m-1$ of
non-vanishing constants $l_i$. It is convenient to define
 \begin{align}
  x_{m-1} = 1 - \sum_{i=1}^{m-1} \frac{z_i^2}{l_i^2} \,, \qquad \vec{l} = ( l_1 , \ldots l_{m-1} , 0, \ldots , 0)
  \,,
 \end{align}
where the last $n-m+1$ constants $l_i$ are vanishing. Starting with the case $m=n$,
i.e.~all harmonics $h_i$ independent and only $l_n$ equal to zero, we have a negative
charge distributed inside the ellipsoid and a positive charge distributed on the
boundary:
 \begin{align}
  \sigma_{n} \sim \frac{1}{l_1 \cdots l_{n-1}}\, \Big(-x_{{\scriptscriptstyle{n-1}}}^{-3/2}
  \Theta(x_{{\scriptscriptstyle{n-1}}})+ 2 \,
  x_{{\scriptscriptstyle{n-1}}}^{-1/2}
  \delta(x_{{\scriptscriptstyle{n-1}}})\Big)\,\delta^{(1)}(z_{{\scriptscriptstyle{n}}})
  \,,
  \label{double-distribution}
 \end{align}
where $\Theta(x)$ is the Heaviside step function: $\Theta(x<0) = 0$ and $\Theta(x>0) =
1$. Upon sending $l_{n-1}$ to zero, the charges in the interior of the ellipsoid cancel,
leaving one with a positive charge on the boundary of a lower dimensional ellipsoid:
 \begin{align}
  \sigma_{n-1} \sim \frac{1}{l_1 \cdots l_{n-2}}\,
  \delta(x_{{\scriptscriptstyle{n-2}}})\,\delta^{(2)}(z_{{\scriptscriptstyle{n-1}}},z_{{\scriptscriptstyle{n}}})
  \,,
 \end{align}
which is the brane distribution corresponding to $n-1$ independent harmonics since
$h_{n-1}$ and $h_n$ are linearly dependent. Next, the contraction of more constants will
yield brane distributions over the inside of an ellipsoid. The distribution $\sigma(z_i)$
is then a product of a delta-function and a theta-function and the branes are localised
along $n-m+1$ coordinates and distributed within an $m-1$-dimensional ellipsoid, defined
by $x_{m-1}=0$. For $2 \leq m \leq n-2$ non-zero constants, one has
 \begin{align}
  \sigma_{m} \sim \frac{1}{l_1 \cdots l_{m-1}}\, x_ {{\scriptscriptstyle{m-1}}}^{(n-m-3)/2} \,
  \Theta(x_{{\scriptscriptstyle{m-1}}}) \,
  \delta^{(n-m+1)}(z_{{\scriptscriptstyle{m}}},\ldots,z_{{\scriptscriptstyle{n}}})
  \,.
 \end{align}
Finally, one is left with all constant $l_i$ vanishing, in which case the distribution
has collapsed to a point and generically reads
 \begin{align}
  \sigma_1 =
  \delta^{(n)}(z_{{\scriptscriptstyle{1}}},\ldots,z_{{\scriptscriptstyle{n}}})\,,
 \label{point-distribution}
 \end{align}
i.e.~we are left with a single brane with all harmonics $h_i$ linearly dependent. All
these distributions satisfy
 \begin{align}
  \sigma_{m-1} = \delta(z_{m-1}) \int \sigma_m \,,
 \end{align}
consistent with the picture of distributions that collapse the $z_{m-1}$-coordinate upon
sending $l_{m-1}$ to 0. The case of NS5A-branes is illustrated in figure
\ref{fig:D5-distrib}.

The general lesson to be drawn from this section is that the domain wall solutions uplift
to branes with harmonic functions given by \eqref{harm}. For the cases with all $q_i \geq
0$, these harmonic functions correspond to the near-horizon limit of the brane
distributions \eqref{double-distribution}. In the simplest case, with all relevant $l_i =
0$ and therefore $m=1$, this distribution collapses to a point \eqref{point-distribution}
and the harmonic function stems from the near-horizon limit of a single brane. In the
next sections we will see whether these findings also hold for the special cases $n=2$
and $n=3$.

\section{Domain Walls in 9D and Uplift to IIB} \label{sec:9D-DW}

In this section, we will consider domain wall solutions to the 9D gauged supergravities
that were constructed in section~\ref{sec:D=9-gaugings}. We will first focus on the
$SL(2,\mathbb{R})$ gauged theories and later comment on the possibility of domain walls
in the other 9D gauged theories.

\subsection{Domain Walls in $SL(2,\mathbb{R})$ Gauged Supergravities} \label{sec:9D-DW-1}

We first consider domain walls in the $SL(2,\mathbb{R})$ gauged supergravities in 9D,
which are specified by three mass parameters. We will take $m_1 = 0$, which can be
obtained by an $SL(2,\mathbb{R})$ transformation. Thus, we are left with a symmetric mass
matrix $Q$ with diagonal entries $q_1$ and $q_2$, see \eqref{qs}. By choosing appropriate
values for $q_1$ and $q_2$, one can still cover each of the three conjugacy classes of
$SL(2,\mathbb{R})$, corresponding to $q_1 q_2$ positive, negative or vanishing.

For the present purpose of domain walls, it suffices to consider a truncation to gravity
and the scalars. The supersymmetry transformations of the fermions, which are given in
\eqref{9Dsusy} in full generality, then reduce to (see \eqref{9Dsusy} and
\eqref{susyexplmiib})
 \begin{align}
  \delta \psi_\mu
    & = (\partial_\mu + \omega_\mu +\ft{i}{4} e^\phi \partial_\mu \chi
    + \tfrac{1}{28} \gamma_\mu W) \epsilon \,, \notag \\
  \delta_0 \lambda
    & = i (\slashed{\partial} \phi + \delta_\phi W) \, \epsilon^*
    - e^\phi (\slashed{\partial} \chi + \delta_\chi W) \, \epsilon^* \,, \notag \\
  \delta_0 \tilde\lambda & = i (\slashed \partial \varphi + \delta_\varphi W) \,
  \epsilon^* \,,
 \label{Killing-9D}
\end{align}
with the superpotential $W$ given by \eqref{superpotential-9D}. The projector
corresponding to a domain wall Ansatz is
\begin{align}
  \Pi_{\text{DW}} = \tfrac{1}{2} (1+\gamma^{\underline{y}}) \,,
 \label{projection-9D}
\end{align}
where $\underline{y}$ indicates a tangent space direction, see
appendix~\ref{app:conventions}.

Half-supersymmetric domain walls correspond to configurations satisfying the Killing
spinor equations, which are obtained by requiring \eqref{Killing-9D} subject to the
projection \eqref{projection-9D} to vanish. The most general solutions were first
classified in \cite{Bergshoeff:2002mb} (for other discussions of 9D domain walls, see
\cite{Cowdall:2000sq,Nishino:2002zi}) and read\footnote{In \cite{Bergshoeff:2002mb} a
different transverse coordinate $\tilde{y}$ was used, which is related via $h(y) =
\tilde{h}(\tilde{y})^2$, where the function $\tilde{h}(\tilde{y})$ appears in the metrics
of \cite{Bergshoeff:2002mb} and is not necessarily harmonic. Each different conjugacy
class has a different function $\tilde{h}(\tilde{y})$ and therefore requires a different
coordinate transformation. \label{coordinate-transformation}}
\begin{align}
  ds^2 & = h^{1/14} (-dt^2+dx_7^2) + h^{-3/7}dy^2\, ,\notag \\
  e^\phi & = h^{-1/2} h_1 \,,\quad e^{\sqrt{7}\varphi} = h^{-1}\, , \qquad
  \chi = c_1 h_1{}^{-1} \,,
  \label{9D-DW}
\end{align}
with the functions
\begin{align}
  h = h_1 h_2 - c_1^2 \,, \qquad
  h_1 = 2 q_1 y + l_1{}^2 \,, \qquad
  h_2 = 2 q_2 y + l_2{}^2 \,.
\end{align}
This is the most general half-supersymmetric domain wall solution.

The general 9D domain walls are parameterised by three constants. However, as also
explained for the general case in subsection~\ref{sec:CSO-DW-2}, one can always do a
coordinate transformation $y \rightarrow y + c$ to shift either $l_1$ or $l_2$ to zero.
The third parameter $c_1$ can be understood as corresponding to the gauge symmetry with
constant parameters: by performing $SL(2,\mathbb{R})$ transformations of the form
\eqref{SL2R-gauge-transformations} one shifts $c_1$. For this reason, one can always
choose a gauge in which it vanishes. In this case the Killing spinor reads
 \begin{align}
  \epsilon = h^{1/56} \, \epsilon_0 \,,
 \end{align}
while in general it depends on $c_1$. Since the transformation to shift $c_1$ to zero is
a gauge transformation with constant parameter, it does not affect the gauge potentials.
Note that the most general domain walls therefore are $SL(2,\mathbb{R})$ orbits of the
prime example \eqref{domain-wall} and is expressed in terms of two harmonic functions and
one constant. In the $SL(2,\mathbb{R})$ frame with $c_1=0$, it can be seen as a harmonic
superposition of the domain walls with harmonics $h_1$ and $h_2$. Due to the two
independent harmonic functions, we call this the double domain wall.

In certain truncations, the general solution \eqref{9D-DW} becomes a single domain wall
with only one independent harmonic function. This can happen either due to the vanishing
of a mass parameter $q_i$ or due to special values of the constants $l_i$. In
table~\ref{tab:9D-single-DW} we give the two possible truncations leading to single
domain walls and the corresponding value of $\Delta$ as defined in \cite{Lu:1995cs}. Note
that the $SO(2)$ case cannot be assigned a $\Delta$-value since it has vanishing
potential, as already noted in \cite{Bergshoeff:2002mb}. The domain wall is carried by
the non-vanishing massive contributions to the BPS equations. In other words, the
potential is zero but there is a non-vanishing superpotential.

\begin{table}[h]
\begin{center}
\begin{tabular}{||c||c|c|c|c||}
\hline \rule[-3mm]{0mm}{8mm}
  Gauge group & $(q_1,q_2)$ & $h_1$ & $h_2$ & $\Delta$ \\
\hline \hline \rule[-3mm]{0mm}{8mm}
  $\mathbb{R}$ & $(0,q)$ & $1$ & $2 q y $ & $4$ \\
\hline \rule[-3mm]{0mm}{8mm}
  $SO(2)$ & $(q,q)$ & $2 q y $ & $2 q y$ & $\times$ \\
\hline
\end{tabular}
\caption{\it The single domain walls as truncations of the 9D double domain wall
solution. We give the two possible truncations and the corresponding value of $\Delta$.
Note that there does not exist a $\Delta$-value in the $SO(2)$ case due to the vanishing
of the potential.\label{tab:9D-single-DW}}
\end{center}
\end{table}

\subsection{Seven-branes and Orientifold Planes} \label{sec:9D-DW-2}

As discussed in the section~\ref{sec:D8-brane}, the occurrence of domain walls with
positive tension leads to a harmonic function that vanishes at a point in the transverse
space. To avoid this, one has to include orientifold planes with negative tension as
well, which can be introduced by modding out the theory with a
$\mathbb{Z}_2$-transformation. In 10D IIA the relevant symmetry is $I_y \Omega$
\eqref{I9Omega} which introduces (in the case of $y$ compact) 16 D8-branes and their
images and two O8-planes. In 9D the relevant $\mathbb{Z}_2$-symmetry can be obtained from
the IIA transformation $I_y \Omega$ \eqref{I9Omega} by the reduction in a direction other
than $y$. Alternatively, one could reduce the IIB transformation $(-)^{F_L} I_{xy}
\Omega$ in the $x$ direction. Upon reduction these give the same transformation and
therefore are T-dual \cite{Bergshoeff:2001pv}. In particular, the 9D
$\mathbb{Z}_2$-symmetry acts on the mass parameters as $Q \rightarrow - Q$. Thus all
three mass parameters flip sign. However, one can always use an
$SL(2,\mathbb{R})$-transformation to set $m_1=0$. Then one is left with $q_1$ and $q_2$
and since both mass parameters flip sign, one introduces orientifold planes which carry a
charge of $-16$ with respect to both $q_1$ and $q_2$. Taking $y$ compact (for a
non-compact transverse space the discussion is analogous), one also has to introduce a
number of positive tension branes to cancel the total charges. For the $q_2$-charge this
correspond to 32 D7-branes. The cancellation of $q_1$-charge requires 32 Q7-branes, which
are defined as S-duals of the D7-branes. Thus the following picture seems to emerge:
\begin{itemize}
\item Two orientifold planes, one at each of the fixed points of the $S^1$, each carrying
a charge of $(-16,-16)$ with respect to the two mass parameters $(q_1,q_2)$. \item
Sixteen D7-branes and their images, located at arbitrary points between the two O7-planes
and each carrying a charge of $(0,1)$. \item Sixteen Q7-branes and their images, defined
as S-duals of the D7-branes, also distributed between the two O7-planes and each carrying
a charge of $(1,0)$.
\end{itemize}
Depending on the positioning of the various 7-branes, the mass parameters can take
different values. Note that the gauge group can change when passing through a 7-brane,
since it can affect only $q_1$ or $q_2$ and thus $\det(Q)$ need not be invariant. The
reduction of the type~I${}^\prime$ theory would correspond to a special case of this
general set-up, in which eight of the Q7-branes and their images are positioned at each
O7-plane, thereby cancelling the $(-16,0)$ charge and inducing $q_1=0$ everywhere in the
bulk\footnote{Toroidal compactifications of type~I${}^\prime$ string theory have been
considered in \cite{Chaudhuri:2000aa} from a somewhat different point of view. It would
be interesting to link its results to our analysis here.}.

\subsection{Uplift to IIB and D7-branes} \label{sec:9D-DW-3}

Instead of the nine-dimensional discussion of source terms above, one can also uplift the
domain walls to solutions of IIB supergravity. The general formula \eqref{harm} applied
to the 9D case yields the harmonic function
 \begin{align}
  H_2 = \Big( \sqrt{h_2} \mu_1{}^2 + \sqrt{h_1} \mu_2{}^2 \Big)^{-1} \,,
 \end{align}
with the identifications \eqref{2D-mu-coordinates} to make contact with the explicit
twisted reduction Ansatz \eqref{IIBred} with reduction coordinate $y$. In this case, it
is straightforward (though perhaps not very insightful) to perform the coordinate
transformation to $z_i$, which yields:
 \begin{align}
  H_2 = \Big( \frac{\alpha z_1{}^2 + \beta z_2{}^2 + \gamma(z_1{}^2 + z_2{}^2)}{2 \gamma^2} \Big)^{1/2} \,,
 \label{D7-harmonic}
 \end{align}
with the definitions
 \begin{align}
  \begin{array}{cc}
  \alpha = q_1 q_2 (z_1{}^2 + z_2{}^2) + q_1 l_2^2 - q_2 l_1^2 \,, \\
  \beta = q_1 q_2 (z_1{}^2 + z_2{}^2) - q_1 l_2^2 + q_2 l_1^2 \,,
  \end{array} \qquad
  \gamma = \sqrt{\tfrac{1}{2}(\alpha^2 + \beta^2) + q_1 q_2 (z_1{}^2 -z_2{}^2)(\alpha-\beta)} \,.
 \end{align}
Indeed, it can be checked that this function is harmonic with respect to flat
$(z_1,z_2)$-space for all values of $q_i$ and $l_i$.

The harmonic function $H_2$ generically depends on both $z_1{}^2 + z_2{}^2$ and $z_1{}^2
- z_2{}^2$. Note that the dependence on the latter only disappears if
$\alpha=\beta=\gamma$, in which case the harmonic function reads
 \begin{align}
  H_2 = (q_1 q_2)^{-1/2}\,.
 \end{align}
This requires the relation $q_1 l_2^2 = q_2 l_1^2$. This cannot be satisfied with the
charges $(q_1,q_2) = (1,-1)$ and $(1,0)$ while keeping both $h_i > 0$. Therefore, the
only possibility is charges $(1,1)$, implying that the two constants $l_i$ need to be
equal, yielding a harmonic function given by $H_2 = 1$. Another case with a manifest
isometry is provided by the charges $(1,0)$, where the harmonic function becomes
 \begin{align}
  H_2 (q_2 = 0) = |z_1| / l_2 \,,
 \end{align}
which is a harmonic function in a one-dimensional transverse space, in agreement with
\eqref{trunc-harm}.

For the $SO(2)$ case, in which we take $q_1 = q_2 = 1$, the IIB solution can be
understood as a distribution of D7-branes. Without loss of generality we take $l_2 = 0$.
Then one has
 \begin{align}
  H = 1 + \int dz_1' dz_2' \sigma(z_1', z_2'; l_1) \log((z_1-z_1')^2+(z_2-z_2')^2) \,,
  \label{D7-SO2-harmonic}
 \end{align}
with the D7-brane distribution
 \begin{align}
  \sigma(z_1', z_2'; _1) = \frac{1}{2 \pi l_1} \Big[
   - \big( 1- \frac{z_1'{}^2}{l_1{}^2} \big)^{-3/2} \Theta \big( 1- \frac{z_1'{}^2}{l_1{}^2} \big)
   + 2 \big( 1- \frac{z_1'{}^2}{l_1{}^2} \big)^{-1/2} \delta \big( 1- \frac{z_1'{}^2}{l_1{}^2} \big) \Big] \,,
 \label{D7-distribution}
 \end{align}
for the case $m=2$ (implying that $l_1 \neq 0$). Note that this distribution consists of
a line interval of negative D7-brane density with a positive contribution at both ends of
the interval. Both positive and negative contributions diverge but these cancel exactly:
 \begin{align}
  \int dz_1' dz_2' \sigma(z_1', z_2') = 0 \,,
 \end{align}
i.e.~the total charge in the distribution \eqref{D7-distribution} vanishes.

\begin{figure}[tb]
\centerline{\epsfig{file=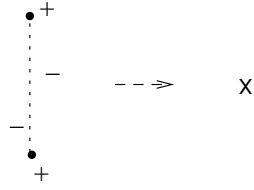,width=.2\textwidth}} \caption{\it The
distributions of D7-branes corresponding to the uplift of the 9D $SO(2)$ domain walls
with one and zero non-vanishing $l_i$'s, respectively. The cross indicates the conical
singularity of the locally flat space-time.} \label{fig:D7-distrib}
\end{figure}

The parameter $l_1$ of the general $SO(2)$ solution can be set to zero, truncating to
only one independent harmonic function: $m=1$. This corresponds to a collapse of the line
interval to a point, as can be seen from \eqref{D7-distribution}. However, due to the
fact that the total charge vanishes, this leaves us without any D7-brane density:
 \begin{align}
  \sigma (z_1', z_2'; l_1 = 0) = 0 \,,
 \end{align}
Indeed, the general harmonic function \eqref{D7-harmonic} equals one for the $SO(2)$ case
with $l_1 = l_2 = 0$. The D7-brane distributions are shown in figure
\ref{fig:D7-distrib}.

Therefore, the two-dimensional $SO(2)$ harmonic function \eqref{D7-SO2-harmonic} of the
D7-brane differs in two important ways from the generic $SO(n)$ harmonic function with
$n>2$. Firstly, the total charge distribution of D7-branes vanishes, while it adds up to
a finite and positive number in the other cases. Secondly, but not unrelated, one needs
to include a constant in the harmonic function \eqref{D7-SO2-harmonic} in terms of the
distribution. In the generic cases this constant was absent, corresponding to the
near-horizon limit of these branes. In this respect, the D7-brane is special, as can also
be seen from the following observation.

As discussed in section~\ref{sec:susy-solutions}, the near-horizon limit of D-branes
\eqref{confAdSxS} yields a metric that is conformally\footnote{Except for the case $p =
5$, which has Minkowski$_7$ rather than AdS$_7$ \cite{Gibbons:1993sv}.} AdS $\times$ S.
To absorb the conformal factor, one needs to go to the so-called dual frame, in which the
tension of the brane is independent of the dilaton:
 \begin{align}
  g_{\mu \nu}^{\text{dual}} = \exp \left( \frac{(3-p)}{2(p-7)} \, \phi  \right) g_{\mu \nu}^{\text{Einstein}}
  \,.
 \end{align}
In the dual frame, the near-horizon geometry of all D-branes with $p \leq 6$ reads
$\text{AdS}_{p+2} \times S^{n-1}$. Clearly, this formula does not hold for the D7-brane;
a related complication is the fact that the dual object is the D-instanton, which lives
on a Euclidean space.

Having found the general $SL(2,\mathbb{R})$ domain walls, we would like to impose the
different quantisation conditions on $\vec{m}$ of subsection~\ref{sec:D=9-gaugings-6}.
For the $SO(2)$ case with $q_1 = q_2 = q$, these translate in a deficit angle: since the
argument of the trigonometric functions of $\mu_i$ \eqref{2D-mu-coordinates} is
$\sqrt{q_1 q_2} y$ and the variable $y$ is identified up to $2 \pi R$, our $SO(2)$
angular variable has a range of $2 \pi q R$. For this reason, the locally flat
space-times with $H_2=1$ are conical space-times with a deficit angle $2 \pi (1- q R)$.
Let us go through the three quantisation possibilities for $SO(2)$
\cite{Bergshoeff:2002mb}, giving the result for $l_1 = l_2$:
 \begin{itemize}
 \item
The first quantisation condition \eqref{quant1} has $q = 1 / (4R)$ and thus yields a
deficit angle of $3\pi/2$. In other words, this is a half-supersymmetric $\text{Mink}_8
\times \mathbb{C} / \mathbb{Z}_4$ space-time with non-trivial monodromy, the bosonic part
of which was also mentioned in \cite{Greene:1990ya}.
 \item
The second quantisation condition \eqref{quant2} cannot be applied to $q$ but only to an
$SL(2,\mathbb{R})$ related partner of our uplifted domain wall, since it requires an
off-diagonal matrix $Q$. It gives rise to a deficit angle of $5\pi/3$, leading to a
half-supersymmetric $\text{Mink}_8 \times \mathbb{C} / \mathbb{Z}_6$ space-time with
non-trivial monodromy.
 \item
The third quantisation condition \eqref{quant3} has $q = 1/R$ and thus leads to fully
supersymmetric $\text{Mink}_{10}$ space-time. The monodromy is trivial and there is a
second Killing spinor with opposite chirality. For the previous two quantisation
conditions this second Killing spinor had a different monodromy and was therefore not a
globally consistent solution of the Killing spinor equations.
 \end{itemize}

\subsection{Domain Walls for other Gauged Supergravities} \label{sec:9D-DW-4}

In the previous subsections, we have constructed and discussed the most general domain
wall solution to the three $SL(2,\mathbb{R})$ gauged supergravities. In this subsection
we would like to address the possibility of domain walls for the other 9D gauged
supergravities of section~\ref{sec:D=9-gaugings}.

Since we are looking for 1/2 BPS solutions, we have to solve the Killing spinor
equations. These are obtained by setting the supersymmetry variation of the gravitino and
dilatini to zero, while the supersymmetry parameter is subject to a certain projection.
The projector for a domain wall is given by $\frac{1}{2}(1 \pm \gamma^{\underline{y}})$,
where $y$ denotes the transverse direction.

In this way we solve first order equations instead of second order equations, which we
would encounter if we would solve the field equations directly. For static
configurations, a solution to the Killing spinor equation is also a solution to the field
equations. From an analysis of the massive supersymmetry transformations $\delta_0 +
\delta_m$ of the gravitino and the dilatino, it was found \cite{Bergshoeff:2002nv} that
to solve the Killing spinor equations, one has to set all mass parameters to zero except
for $\vec m$. Therefore, there are not more half-supersymmetric domain wall solutions
than the ones given in \eqref{9D-DW} with mass parameters $q_1$ and $q_2$.

By analysing the possibilities for other projectors in nine dimensions (i.e.~demanding
that the projector squares to itself and that its trace is half of the spinor dimension),
we find that there is another projector given by $\frac{1}{2}(1 \pm i
\gamma^{\underline{t}})$. This projector would give a time-dependent solution, which can
be seen as a Euclidean domain wall having time as a transverse direction. See
\cite{Kerimo:2004qx} for an example.

\section{Domain Walls in 8D and Uplift to IIA and 11D} \label{sec:8D-DW}

In this section, we will construct the most general half-supersymmetric domain wall
solution to the 8D gauged maximal supergravities of section~\ref{sec:D=8-gaugings}. We
will start with the class A theories and only comment on class B at the end.

\subsection{Domain Walls of Class A Theories} \label{sec:8D-DW-1}

In section~\ref{sec:D=8-gaugings} we have obtained the bosonic action \eqref{action-8DA}
and supersymmetry transformations \eqref{8D-susy} of the $D=8$ gauged maximal
supergravities with gauge groups of class A. We now look for domain wall solutions that
preserve half of the supersymmetry\footnote{For a nice review of wrapped domain walls
with less supersymmetry in the $SO(3)$ case, see \cite{Paredes:2004xw}. These uplift to
purely gravitational solutions in 11D involving manifolds of special holonomy.},
following \cite{Alonso-Alberca:2003jq}. For an earlier discussion of a subset of these
solutions, see \cite{Cowdall:1997tw}.

We consider the following domain wall Ansatz:
\begin{align}
  ds^2 & = g(y)^2 dx_7{}^2 + f(y)^2 dy^2 \,, \qquad
  {\cal M} = {\cal M}(y)\, ,\hskip .6truecm \varphi = \varphi (y)\,, \qquad
  \epsilon = \epsilon(y) \,.
\label{8D-Ansatz}
\end{align}
Our Ansatz only includes the metric and the scalars. All other fields are vanishing
except the $SL(2,\mathbb{R}) / SO(2)$ scalar $\ell$ which we have set constant. It turns
out that there are no half-supersymmetric domain walls for non-constant $\ell$. We need
to satisfy the Killing spinor equations (which are a truncation of \eqref{8D-susy} to the
fields of the Ansatz \eqref{8D-Ansatz})
\begin{align}
 \delta \psi_\mu & = 2 \partial_\mu \epsilon -\tfrac{1}{2} \slashed{\omega}_\mu \epsilon
   +\tfrac{1}{2} \slashed{Q}_\mu \epsilon
   + \tfrac{1}{24} e^{- \varphi /2} f_{ijk} \Gamma^{ijk}
\Gamma_\mu \epsilon = 0 \, , \notag \\
\delta \lambda_i & = - \slashed P_{ij}\Gamma^{j}\epsilon
  - \tfrac{1}{3} \slashed{\partial} \varphi \Gamma_i \epsilon
  - \tfrac{1}{4} e^{-\varphi/2} (2 f_{ijk}
-f_{jki}) \Gamma^{jk} \epsilon = 0 \, , \notag
\end{align}
where the Killing spinor satisfies the condition
\begin{align}
  (1+ \Gamma_{\underline{y123}}) \epsilon = 0 \, . \label{8D-projection}
\end{align}
The indices $1,2,3$ refer to the internal group manifold directions.

The most general class A domain wall solution reads
\begin{align}\label{triple}
  ds^2 & = h^{\frac{1}{12}}dx_7^2 + h^{-\frac{5}{12}}dy^2\,, \nonumber \\
  e^\varphi & = h^{\frac{1}{4}}, \hspace{0.5cm}
  e^\sigma = h^{-\frac{1}{2\sqrt{3}}}h_1^{\frac{\sqrt{3}}{2}}, \hspace{0.5cm}
  e^\phi= h^{-\frac{1}{2}}h_1^{-\frac{1}{2}}(h_1h_2-c_1^2)\,, \notag \\
  \chi_1 & = c_1h_1^{-1}, \hspace{0.5cm}
  \chi_2=\chi_1\chi_3+c_2h_1^{-1},\hspace{0.5cm}
  \chi_3=(c_1c_2+c_3 h_1)\left(h_1h_2-c_1^2 \right )^{-1},
\end{align}
where the dependence on the transverse coordinate $y$ is governed by
\begin{align}
& h(y) = h_1 h_2 h_3 - c_3^2 h_1-c_2^2 h_2-c_1^2 h_3-2c_1 c_2 c_3 \,, \notag \\
& h_1 = 2 q_1y+l_1{}^2, \hspace{0.5cm} h_2 = 2 q_2y+l_2{}^2, \hspace{0.5cm} h_3 = 2
  q_3y+l_3{}^2\,. \label{harmonics}
\end{align}
The corresponding Killing spinor is quite intricate so we will not give it here. Note
that the solution is given by three harmonic function $h_i$. For this reason we call the
general solution a triple domain wall.

The general solution has six integration constants $c_i$ and $l_i$. As before, one can
eliminate one of the constants $l_i$ by a redefinition of the variable $y$. The other
three constants $c_1$, $c_2$ and $c_3$ can be understood to come from the following
symmetry. The mass deformations do not break the full global $SL(3,\mathbb{R})$; indeed,
they gauge the three-dimensional subgroup of $SL(3,\mathbb{R})$ that leaves the mass
matrix $Q$ invariant. Thus one can use the unbroken global subgroup to transform any
solution\footnote{Note that one cannot use the unbroken local subgroup of
$SL(3,\mathbb{R})$ (the gauge transformations) since this would induce non-vanishing
gauge vectors and thus would be inconsistent with our Ansatz \eqref{8D-Ansatz}.},
introducing three constants. In our solution these correspond to $c_1$, $c_2$ and $c_3$,
which can therefore be set to zero by fixing the $SL(3,\mathbb{R})$ frame. From now on we
will always assume the frame choice $c_1 = c_2 = c_3 = 0$ unless explicitly stated
otherwise. This results in
\begin{align}\label{basis}
  \chi_1 = \chi_2 = \chi_3 = 0 \,, \qquad
  \mathcal{M} = h^{-2/3} \text{diag}( h_2 h_3, h_1 h_3, h_1 h_2 ) \,, \qquad
  h = h_1 h_2 h_3 \,.
\end{align}
In this $SL(3,\mathbb{R})$ frame the expression for the Killing spinor simplifies
considerably and reads $\epsilon = h^{1/48} \epsilon_0$. Thus, analogously to 9D, we find
that the most general domain wall solution to these gauged supergravities is given by the
$SL(3,\mathbb{R})$ orbits of the generic solution \eqref{domain-wall}.

The triple domain wall can be truncated to double or single domain walls when restricting
the constants $l_1, l_2$ and $l_3$. In table~\ref{tab:8D-single-DW} we give the three
possible truncations leading to single domain walls and the corresponding value of
$\Delta$ as defined in \cite{Lu:1995cs}. The Bianchi II case was given in
\cite{Cowdall:1997tw} and the Bianchi IX case in \cite{Boonstra:1998mp} (up to coordinate
transformations). Note that the Bianchi VII$_0$ case cannot be assigned a $\Delta$-value
since it has vanishing potential. The domain wall is carried by the non-vanishing massive
contributions to the BPS equations. The same mechanism occurs in $SO(2)$ gauged $D=9$
supergravity \cite{Bergshoeff:2002mb}, see table~\ref{tab:9D-single-DW}.

\begin{table}[ht]
\begin{center}
\begin{tabular}{||c||c|c|c|c||c||c||}
\hline \rule[-3mm]{0mm}{8mm}
  Bianchi & $Q=\text{diag}$ & $h_1$ & $h_2$ & $h_3$ & $\Delta$ & Uplift \\
\hline \hline \rule[-3mm]{0mm}{8mm}
  II & $(0,0,q)$ & $1$ & $1$ & $2 q y$ & $4$ & \eqref{4EH-II-Heis} \\
\hline \rule[-3mm]{0mm}{8mm}
  VII$_0$ & $(0,q,q)$ & $1$ & $2 q y$ & $2 q y$ & $\times$ & \eqref{4EH-I-ISO}  \\
\hline \rule[-3mm]{0mm}{8mm}
  IX & $(q,q,q)$ & $2 q y$ & $2 q y$ & $2 q y$ & $-\tfrac{4}{3}$ & \eqref {4flat} \\
\hline
\end{tabular}
\caption{\it The single domain walls as truncations of the 8D triple domain wall
solution. Note that there exists no $\Delta$-value in the Bianchi VII$_0$ case due to the
vanishing of the potential. In the last column we indicate where the uplifted solution to
11D is given. \label{tab:8D-single-DW}}
\end{center}
\end{table}

\subsection{Uplift to IIA and D6-branes} \label{sec:8D-DW-2}

\begin{figure}[tb]
\centerline{\epsfig{file=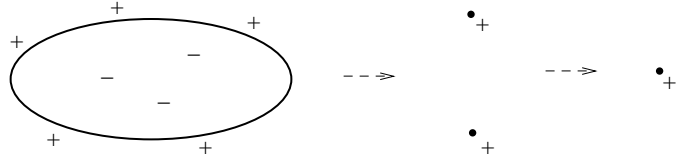,width=.55\textwidth}} \caption{\it The
distributions of D6-branes corresponding to the uplift of the 8D $SO(3)$ domain walls
with two, one and zero non-vanishing $l_i$'s, respectively.} \label{fig:D6-distrib}
\end{figure}

The special case of the $SO(3)$ D6-brane distributions (with all $q_i=1$) was first
discussed in \cite{Cvetic:2000zu}. This splits up in three separate possibilities, with
$m=3, 2$ or $1$. The first distribution $\sigma_3$ consists of positive and negative
densities and is given by the general formula \eqref{double-distribution}. Upon sending
$l_2$ to zero, this collapses to
 \begin{align}
  \sigma_2 \sim \frac{1}{l_1} \,
  \delta \big( 1 - \frac{z_1{}^2}{l_1{}^2} \big) \, \delta^{(2)}(z_2 , z_3) \,.
 \end{align}
This is a distribution at the boundary of a one-dimensional ellipse, i.e.~it is localised
at the points $z_1 = \pm l_1$. For this reason, the corresponding harmonic function is
given by
 \begin{align}
  H_3 ({\vec z},l_1) = \frac{1}{2((z_1 - l_1)^2 + z_2{}^2 + z_3{}^2)^{1/2}} + \frac{1}{2((z_1 + l_1)^2 + z_2{}^2 +
  z_3{}^2)^{1/2}} \,, \label{double-center-harmonic}
 \end{align}
i.e.~the near-horizon limit of the double-centered harmonic. Upon sending $l_1$ to zero,
the brane distribution $\sigma_1$ collapses to a point, as given in
\eqref{point-distribution}. Indeed, the harmonic function becomes
 \begin{align}
  H_3 = \frac{1}{|\vec{z}|} \,, \label{single-center-harmonic}
 \end{align}
i.e.~the near-horizon limit of the single-centered harmonic with $SO(3)$ isometry. The
different distributions of D6-branes are shown in figure \ref{fig:D6-distrib}.

\subsection{Uplift to 11D and KK-monopoles} \label{sec:8D-DW-3}

The D6-brane solution is different from the other branes in table~\ref{tab:branes} in the
following sense: they can be uplifted to a purely gravitational solution in 11D, the
Kaluza-Klein monopole (see also subsection~\ref{sec:susy-solutions-4}). In the $z_i$
coordinates, the higher-dimensional metric reads
 \begin{align}
   ds^2 = dx_{7}^2 + H^{-1}(dy^3 + \sum_{i=1}^3 A_i dz_i)^2 + \sum_{i=1}^3 H dz_i{}^2 \,,
  \label{KK-monopole}
 \end{align}
where $y^3$ is the isometry direction of the KK-monopole. The function $H=H(z_i)$ is
given implicitly in \eqref{harm} with $n=3$ and $A_i=A_i(z_j)$ is subject to the
condition \eqref{KK-monopole-vector}.

However, since we do not have the harmonic function $H=H(z_i)$ explicitly, we will rather
use the coordinates $y$ and $\mu_i$. The $\mu_i$ are related to the coordinates $y_1$ and
$y_2$ of the group manifold via \eqref{3D-mu-coordinates}. In these coordinates and using
the $SL(3,\mathbb{R})$ frame of (\ref{basis}), the triple domain wall solutions becomes a
purely gravitational solutions with a metric of the form $\hat{ds}{}^2 = dx_7{}^2 +
ds_4{}^2$, where
\begin{align}
  ds_4{}^2 = h^{-\frac{1}{2}}dy^2+h^{\frac{1}{2}}
\left ( \frac{\sigma_1^2}{h_1}+\frac{\sigma_2^2}{h_2}+\frac{\sigma_3^2}{h_3} \right) \,.
\label{4dmetric}
\end{align}
Here $\sigma_1$, $\sigma_2$ and $\sigma_3$ are the Maurer-Cartan 1-forms defined in
\eqref{MC} (with $y_i$ instead of $z_i$), $h=h_1h_2h_3$ and the three harmonics $h_i$ are
defined in \eqref{harmonics}. The uplifted solutions are all half-supersymmetric.

The solutions (\ref{4dmetric}) are cohomogeneity one solutions of different Bianchi
types. The $SO(3)$ expression of this four-dimensional metric was obtained previously in
the context of gravitational instanton solutions as self-dual metrics of Bianchi type IX
with all directions unequal \cite{Belinskii:1978}. More recently, the Heisenberg,
$ISO(1,1)$ and $ISO(2)$ cases and their relations to domain wall solutions were
considered in \cite{Gibbons:1998ie, Lavrinenko:1998qa}, whose results are related to ours
via coordinate transformations. In the following discussion we will focus on the
four-dimensional part of the eleven-dimensional metric since it characterises the
uplifted domain walls.

Without loss of generality, we take $q_3=2$ in this subsection. The coordinate
transformation $2 y = \tfrac{1}{2} r^4- l_3{}^2$ then eliminates the constant $l_3$ and
results in the metric
\begin{align}\label{2pardmetric}
  ds_4{}^2 & = (k_1 k_2k_3)^{-1/2} \left[ dr^2+
  r^2  ( k_2 k_3 \sigma_1^2 + k_1k_3 \sigma_2^2 + k_1 k_2 \sigma_3^2 )
  \right] \, ,
\end{align}
where $k_j = q_j/2 + s_j r^{-4}$ with $s_j=l_j{}^2 - q_j l_3{}^2$ for $j=1,2$, and $k_3=
1$. As anticipated, the metric (\ref{2pardmetric}) depends only on two constant
parameters $s_1$ and $s_2$, which are restricted by the (gauge group dependent) condition
 \begin{align}
 s_j > - q_j r^4 / 2 \,, \label{sj-condition}
 \end{align}
in order to satisfy the requirements $h_j > 0$.

In general the metrics have curvatures that both go to zero as $r^{-6}$ for large $r$ and
diverge at $r=0$ and $k_j = 0$, producing incomplete metrics \cite{Eguchi:1978,
Belinskii:1978}. There are two exceptions to this behaviour, which coincide with the
special cases of the D6-brane discussion with $SO(3)$ isometry:
 \begin{itemize}
 \item
The first one corresponds to the case with $s_1=s_2=0$, which can only be obtained for
the $SO(3)$ case due to \eqref{sj-condition}. Taking $q_1=q_2=2$, the metric is locally
flat space:
\begin{align}\label{4flat}
  ds_4{}^2 & = dr^2+ r^2 (\sigma_1^2+\sigma_2^2+ \sigma_3^2) \, ,
\end{align}
where $r$ is the radius of the three-dimensional spheres. This corresponds to the uplift
of the 9D Bianchi type IX single domain wall or the D6-brane with harmonic function
\eqref{single-center-harmonic}.
 \item
The second exception corresponds to the $SO(3)$ gauging (taking $q_1 = q_2 =2$) with $s_1
= s_2 = s < 0$. It is known as the Eguchi-Hanson (EH), or Eguchi-Hanson II, metric
\cite{Eguchi:1978}:
\begin{align}\label{4EH-II}
  ds_4{}^2 & = \left( 1 + \frac{s}{r^4} \right)^{-1}dr^2
+ r^2 (\sigma_1^2+\sigma_2^2) +\left( 1 +\frac{s}{r^4} \right)\sigma_3^2 \, .
\end{align}
This metric corresponds to the uplift of the D6-brane with harmonic function
\eqref{double-center-harmonic}.
 \end{itemize}
Another case that we want to discuss, although it is singular, is obtained in the $SO(3)$
gauging (with $q_1=q_2=2$) by choosing $s_1= s \neq 0$ and $s_2 = 0$:
\begin{align}\label{4EH-I}
  ds_4{}^2 & = \left( 1+\frac{s}{r^4} \right)^{-1}(dr^2
+ r^2 \sigma_1^2) +\left( 1 +\frac{s}{r^4} \right)(\sigma_2^2+\sigma_3^2) \, .
\end{align}
This metric is called the Eguchi-Hanson I (EH-I) metric \cite{Eguchi:1978}.

The uplifted metrics for the singular cases with $\det(Q) = 0$ are also given in
(\ref{2pardmetric}). Among them are two special metrics, which can be obtained by
contraction of the EH metrics (\ref{4EH-II}) and (\ref{4EH-I}). This contraction is
possible because the solutions contain at least one non-zero constant $s_i$, and must be
performed before identifying the charges $q_1$ and $q_2$ and the constants $s_1$ and
$s_2$. We will take $q_1=0$ (which implies $s_1 > 0$ due to \eqref{sj-condition}) and
$q_2 \neq 0$ to get the uplifted metrics for the $ISO(2)$ gaugings.

As an example, let us perform such contractions on the contracted EH metrics, leading
from $SO(3)$ isometries to $ISO(2)$ isometries. After contraction, the expression for the
EH-I metric is
\begin{align}\label{4EH-I-ISO}
  ds_4{}^2 & = \left(\frac{s}{r^4} \right)^{-1/2}(dr^2
+ r^2 \sigma_1^2) +\left(\frac{s}{r^4} \right)^{1/2}(\sigma_2^2+\sigma_3^2) \,,
\end{align}
and the EH-II metric reads
\begin{align}\label{4EH-II-ISO}
  ds_4{}^2 = & \left(\frac{s}{r^4}(1+\frac{s}{r^4}) \right)^{-1/2}dr^2
+ \left( \frac{s}{r^4}(1 +\frac{s}{r^4}) \right)^{1/2}\sigma_3^2 +\nonumber \\
& + r^2 \Big( ( 1+\frac{r^4}{s} )^{1/2}\sigma_1^2+ ( 1+\frac{r^4}{s} )^{-1/2}\sigma_2^2
\Big) \, ,
\end{align}
In the EH-I contracted metric (\ref{4EH-I-ISO}) we have taken $s_1=s$ whereas in the EH
contracted metric (\ref{4EH-II-ISO}) we have set $s_1=s_2= s$, while $q_2 = q_3 = 2$ in
both cases. Notice that the contracted EH-I metric with $ISO(2)$ isometry is precisely
the four-dimensional part of the uplifted metric for the Bianchi type VII$_0$ single
domain wall.

The metrics with Heisenberg isometry are obtained by a further contraction $q_2=0$ in the
metric (\ref{2pardmetric}). Among these metrics, there is again one special case that can
also be obtained by a contraction of the contracted EH metric with isometry $ISO(2)$.
Notice that it is not possible to have a contracted EH-I metric with Heisenberg isometry
since this would require $s_2=0$ and the metrics with Heisenberg isometry have $s_2 \neq
0$. The expression for the contracted EH metric with Heisenberg isometry is
\begin{align}\label{4EH-II-Heis}
  ds_4{}^2 & = \left(\frac{s}{r^4} \right)^{-1}dr^2
+ r^2 (\sigma_1^2+\sigma_2^2) +\left(\frac{s}{r^4} \right)\sigma_3^2 \,, \notag \\
  & = \left(\frac{s}{r^4} \right)^{-1}dr^2
+ r^2 (dz_1{}^2+dz_2{}^2) +\left(\frac{s}{r^4} \right)(dz_3+ 2 z^1 d z_2)^2 \,,
\end{align}
where $s_2 = s$ and $q_3=2$. This is the four-dimensional part of the uplifted metric for
the Bianchi type II single domain wall. This contraction was considered previously in
\cite{Gibbons:2002}.

It is interesting to note that the uplifting of the triple domain wall solution
(\ref{triple}) does not lead to the most general four-metrics. For example, there are
three complete and non-singular hyper-K\"ahler metrics with $SO(3)$ isometry in four
dimensions: the Eguchi-Hanson, Taub-NUT and Atiyah-Hitchin metric (for a useful
discussion of these metrics, see \cite{Bakas:1997gf}). The absence of the Taub-NUT and
Atiyah-Hitchin metrics in our analysis stems from the fact that only the Eguchi-Hanson
metric allows for a covariantly constant spinor that is independent of the $SO(3)$
isometry directions \cite{Gauntlett:1997pk}. In performing the group manifold reductions,
we have assumed that our spinors are independent of the internal coordinates. This is not
compatible with the Taub-NUT and the Atiyah-Hitchin metrics, which therefore reduce to
non-supersymmetric domain walls in 8D. It would be interesting to see whether one can
alter the procedure of group manifold reductions, to allow for the Taub-NUT and
Atiyah-Hitchin metrics to reduce to half-supersymmetric domain walls in $D=8$ dimensions.

\subsection{Isometries of the 3D Group Manifold} \label{sec:8D-DW-4}

The internal three-dimensional manifolds are by definition invariant under the
three-dimensional group of isometries given in \eqref{isometries} and
\eqref{isometries2}. This holds for arbitrary values of the scalars in \eqref{bs}.
However, there can be more isometries, which rotate two of the Maurer-Cartan one-forms
$\sigma^i$ and $\sigma^j$ into each other. This is an isometry of the metric in two
cases:
\begin{itemize}
\item $q_i = q_j = 0$: In this case one can use the automorphism group to set $l_i = l_j
= 1$. Equation \eqref{4dmetric} shows that a rotation between $\sigma^i$ and $\sigma^j$
is an isometry for all solutions of this class.
\item $q_i = q_j \neq 0$: In this case one must set $l_i = l_j$ by hand, after which a
rotation between $\sigma^i$ and $\sigma^j$ is an isometry. Thus, this only holds for a
truncation of the solutions of this class and since $h_i=h_j$ corresponds to decreasing
$m$ by one.
\end{itemize}
This leads to the different possibilities summarised in
table~\ref{tab:isometryenhancement}. These exhaust all possible number of isometries on
three-dimensional class A group manifolds \cite{Bergshoeff:2003ri}.

\begin{table}[ht]
\begin{center}
\begin{tabular}{||c|c||c|c|c|c||}
\hline \rule[-3mm]{0mm}{8mm}
  Bianchi & $(q_1,q_2,q_3)$ & $m=0$ & $m=1$ & $m=2$ & $m=3$ \\
\hline \hline \rule[-3mm]{0mm}{8mm}
  I & $(0,0,0)$ & $6$ & - & - & - \\
\hline \rule[-3mm]{0mm}{8mm}
  II & $(0,0,1)$ & - & $4$ & - & - \\
\hline \rule[-3mm]{0mm}{8mm}
  VI${}_0$ & $(0,-1,1)$ & - & - & $3$ & - \\
\hline \rule[-3mm]{0mm}{8mm}
  VII${}_0$ & $(0,1,1)$ & - & $6$ & $3$ & - \\
\hline \rule[-3mm]{0mm}{8mm}
  VIII & $(1,-1,1)$ & - & - & $4$ & $3$ \\
\hline \rule[-3mm]{0mm}{8mm}
  IX & $(1,1,1)$ & - & $6$ & $4$ & $3$ \\
\hline
\end{tabular}
\caption{\label{tab:isometryenhancement}\it The numbers of isometries of the
three-dimensional group manifold for the different multiple domain wall solutions with
$m$ independent harmonic functions $h_i$. For a given type one finds isometry enhancement
by decreasing $m$, i.e.~upon identifying two harmonic functions.}
\end{center}
\end{table}

The extra fourth isometry was constructed by Bianchi \cite{Bianchi} for the types II,
VIII and IX. He claimed that type VII${}_0$ did not allow for isometry enhancement but
the existence of three extra Killing vectors\footnote{We thank Sigbj\o rn Hervik for a
valuable discussion on this point.} was later shown in \cite{Szafron}. These three extra
isometries appear upon identifying the two $y$-dependent harmonics. Note that the extra
isometries may not be isometries of the full manifold in which the group submanifold is
embedded. Indeed, this is what happens for type VII$_0$ where two of the extra isometries
are $y$-dependent and therefore do not leave the full metric invariant \cite{Szafron}.
The extra Killing vectors of the group manifold for the uplifted domain wall solutions
(\ref{triple}) are explicitly given by \cite{Bergshoeff:2003ri}
\begin{itemize}
\item { Type I} with $Q={\rm diag}(0,0,0)$: \ \
\begin{align}
L_4&=-z^2\,\frac{\partial}{\partial z^1}+z^1\,\frac{\partial} {\partial z^2}\,, \qquad
L_5=-z^3\,\frac{\partial}{\partial z^1}+z^1\,\frac{\partial} {\partial z^3}\,, \notag \\
L_6& =-z^3\,\frac{\partial} {\partial z^2}+z^2\,\frac{\partial} {\partial z^3}\,.
\end{align}
\item { Type II} with $Q={\rm diag}(0,0,1)$: \ \
\begin{align}
L_4&=-z^2\,\frac{\partial}{\partial z^1}+z^1\,\frac{\partial} {\partial
z^2}+\tfrac{1}{4}\,((z^1)^2-(z^2)^2)\,\frac{\partial} {\partial z^3}\,.
\end{align}
\item { Type VII$_0$} with $Q={\rm diag}(0,1,1)$ with $h(y)=h_2=h_3$: \ \
\begin{align}
L_4&=-h^{-1/2}z^2\,\frac{\partial} {\partial z^1}+h^{1/2}z^1\,\frac{\partial}
{\partial z^2}\,, \nonumber\\
L_5&=-h^{-1/2}z^3\,\frac{\partial} {\partial z^1}+h^{1/2}z^1\,\frac{\partial}
{\partial z^3}\,,\\
L_6&=-z^3\,\frac{\partial} {\partial z^2}+z^2\,\frac{\partial} {\partial z^3}\,.\nonumber
\end{align}
\item { Type VIII} with $Q={\rm diag}(1,-1,1)$: \ \
\begin{align}
L_5&=\frac{s_{3,2,1}s_{1,3,2}}{c_{1,3,2}}\,\frac{\partial}{\partial
z^1}+c_{3,2,1}\,\frac{\partial} {\partial
z^2}+\frac{s_{3,2,1}}{c_{1,3,2}}\,\frac{\partial} {\partial z^3}\,.
\end{align}
\item { Type IX} with $Q={\rm diag}(1,1,1)$: \ \
\begin{align}
L_4&=-\frac{c_{3,2,1}s_{1,3,2}}{c_{1,3,2}}\,\frac{\partial}{\partial
z^1}+s_{2,3,1}\,\frac{\partial} {\partial
z^2}-\frac{c_{3,2,1}}{c_{1,3,2}}\,\frac{\partial} {\partial z^3}\,,\nonumber\\
L_5&=\frac{s_{3,2,1}s_{1,3,2}}{c_{1,3,2}}\,\frac{\partial}{\partial
z^1}+c_{3,2,1}\,\frac{\partial} {\partial
z^2}+\frac{s_{3,2,1}}{c_{1,3,2}}\,\frac{\partial} {\partial z^3}\,,\\
L_6&=-\,\frac{\partial}{\partial z^1}\,, \nonumber
\end{align}
\end{itemize}
where we have used the definitions \eqref{cmnp-definition}. The extra Killing vectors
$L_4$, $L_5$ and $L_6$ correspond to rotations between $\sigma^1$ and $\sigma^2$,
$\sigma^1$ and $\sigma^3$ and $\sigma^2$ and $\sigma^3$, respectively.

As we have mentioned above, two of the class A solutions uplift to flat space-time in
$D=11$: the Bianchi type IX solutions with $m=1$ and all Bianchi type I solutions (having
$m=0$). In view of the discussion above, we can now understand why this happens. One can
check that the only way to embed three-dimensional submanifolds of zero (for type I) or
constant positive (for type IX) curvature in four Euclidean Ricci-flat dimensions is to
embed them in four-dimensional flat space. Indeed, this is exactly what we find: the two
solutions both have a maximally symmetric group manifold with six isometries and hence
constant curvature and uplift to flat $D=11$ space-time.

The type VII$_0$ group manifold can also have six isometries and zero curvature. For the
domain wall solutions above, this cannot be embedded in four-dimensional flat space due
to the $y$-dependence of two of its isometries. Note, however, that there is another type
VII$_0$ solution with flat geometry and vanishing scalars that coincides with the type I
solution \eqref{triple} given above\footnote{
  This solution coincides, after an $SO(2)$ rotation of 90 degrees,
  with the Kaluza-Klein reduction of the Mink$_9$ solution \cite{Gheerardyn:2001jj,Bergshoeff:2002nv}
  of the $SO(2)$ gauged supergravity in $D=9$.}.
The corresponding group manifold can be embedded in four-dimensional flat space and
indeed this solution uplifts to 11-dimensional Minkowski just as the type I solution.
However, unlike its type I counterpart, the eight-dimensional type VII$_0$ solution with
flat geometry and vanishing scalars breaks all supersymmetry.

\subsection{Domain Walls for Class B Theories} \label{sec:8D-DW-5}

We would like to see whether there are also supersymmetric domain wall solutions to the
class B supergravities of section~\ref{sec:D=8-gaugings}. It turns out that for this case
there are no domain wall solutions preserving any fraction of supersymmetry, much like
the situation in nine dimensions. This can be seen as follows.

The structure of the BPS equations requires the projector for the Killing spinor of a
half-supersymmetric domain wall solution to be the same as \eqref{8D-projection}. The
presence of the extra term in $\delta \psi_\mu$ (see \eqref{8D-susy}), depending on the
trace of the structure constants, implies that there are no domain wall solutions with
this type of Killing spinor, since the structure of $\Gamma$-matrices of this term cannot
be combined with other terms. To get a solution, one is forced to put $f_{ij}{}^j=0$,
thus leading back to the class A case. This also follows from $\delta\lambda_i$, since
the resulting equation is symmetric in two indices, except for a single antisymmetric
term, containing $f_{ij}{}^j$.

\section{Domain Walls with Strings Attached} \label{sec:DW-string-particle}

In this section, we will consider an extension of the domain walls of the previous
sections where strings and particles are included. The resulting solutions will be
1/4-supersymmetric. The analysis will only be performed in massive IIA supergravity and
$SL(2,\mathbb{R})$ gauged supergravity in $D=9$, but we expect that some generalisation
exists for all $CSO$ gauged supergravities of section~\ref{sec:CSO-gaugings}.

\subsection{D8-F1-D0 Solution} \label{sec:DW-string-particle-1}

In the mid 1990's, it was found that D-branes can be understood as hyperplanes on which a
fundamental string, or F-string, can end \cite{Polchinski:1995mt}. The endpoint of an
F-string appears as an electrically charged particle on the world-volume of the D-brane.
An exception to this generic phenomenon is the D-particle, on which a single F-string
cannot end due to charge conservation\footnote{Indeed, the Born-Infeld vector, which
carries the corresponding degrees of freedom on the D-brane \wv, is not present on the
world-line of the D-particle.} \cite{Strominger:1996ac, Townsend:1997em}.

The situation changes in the presence of a domain wall in which case charge conservation
no longer forbids an F-string to end on a D-particle \cite{Polchinski:1996sm}. In fact,
when a D-particle crosses a D8-brane, a stretched fundamental string with endpoints on
the D0- and D8-brane is created\footnote{The same phenomenon is found for the T-dual
configuration of two crossing D4-branes, which can be at angles \cite{Kitao:1998vn,
Ohta:1997ir}.} \cite{Danielsson:1997wq, Danielsson:1998gi, Bergman:1997gf}. This process
is, via duality, related to the Hanany-Witten effect in which a stretched D3-brane is
created if a D5-brane crosses an NS5-brane \cite{Hanany:1997ie}. The intersecting
configuration for this case is given by{}\footnote{Each horizontal entry indicates one of
the 10 directions $0,1,\ldots, 9$ in space-time.  A $\times (-)$ means that the
corresponding direction is in the world-volume or (transverse to) the brane.}
\begin{equation}
\label{hw1}
 \begin{array}{c|c}
{\rm D5}:\ \ \          \times & \times    \times   -   -   -  \times  \times   \times   - \\
{\rm NS5}:\          \times & \times   \times  \times  \times  \times  -  -   -   -    \\
{\rm D3}: \ \ \       \times & \times \times - - - - - - \times
                         \end{array}
\end{equation}
The intersecting configuration of \cite{Danielsson:1997wq, Danielsson:1998gi,
Bergman:1997gf} is obtained by first applying T-duality in the directions 1 and 2, next
applying an S-duality and, finally, applying a T-duality in the directions 6,7 and 8:
\begin{equation}
\label{hw2}
 \begin{array}{c|c}
{\rm D0}:\ \ \          \times & -    -   -   -   -  -  - -    - \\
{\rm D8}:\ \ \          \times & \times   \times  \times  \times  \times  \times \times \times   -    \\
{\rm F1}: \ \ \       \times & - - - - - - - - \times
                         \end{array}
\end{equation}

In this subsection, we consider the massive IIA supergravity background of the F1-string
that is stretched between the D8-domain wall and the D0-particle. It is given by the
following solution\footnote{There are also other string-like solutions to massive IIA
supergravity \cite{Janssen:1999sa, Imamura:2001cr}, which we will not consider.}
\cite{Massar:1999sb} to the equations of motion of the $D=10$ Romans' massive IIA
supergravity theory \cite{Romans:1986tz}:
\begin{align}
\nephat{ds}^2 &= -H^{1/8}\tilde{H}^{-13/8}dt^2+H^{9/8}\tilde{H}^{-5/8}dy^2+H^{1/8}\tilde{H}^{3/8}dx_8^2 \,, \notag \\
{B}_{ty}&= - \tilde{H}^{-1} \,, \qquad \C{1}_t=H \tilde{H}^{-1} \,, \qquad e^{\nephat
\phi} =H^{-5/4}\tilde{H}^{1/4} \,, \label{D8-F1-D0}
\end{align}
where the harmonic functions $H$ and $\tilde{H}$ are defined as
\begin{align}
H=c+m_{\text{R}}y\,,\quad \tilde{H}=1+\frac{Q}{r^6} \,,
\end{align}
and the radial coordinate is given in terms of the coordinates longitudinal to the
D8-brane, $r^2=x_1^2+\ldots+x_8^2$. The solution preserves 1/4 of supersymmetry and the
Killing spinor is annihilated by the following projectors
\begin{align}
  \nephat{\Pi}_{\text{D0}} = \tfrac{1}{2} (1+\Gamma^{\underline{0}}\Gamma_{11}) \,, \qquad
  \nephat{\Pi}_{\text{F1}} = \tfrac{1}{2} (1+\Gamma^{\underline{0y}}) \,, \qquad
  \nephat{\Pi}_{\text{D8}} = \tfrac{1}{2} (1+\Gamma^{\underline{y}}) \,,
\end{align}
where any of the three projectors can be obtained from the other two. The solution is a
harmonic superposition of two elements, which can be obtained by taking different limits:
\begin{itemize}
\item The limit $Q \rightarrow 0$ leads to the single D8-brane solution \eqref{D8-brane}
which preserves 1/2 supersymmetry. \item The limit $m_{\text{R}} \rightarrow 0$ leads to
an (infinite) F-string with D-particles smeared in the string direction, preserving 1/4
supersymmetry. The F1- and D0-brane charges are related and therefore it is not possible
to obtain these as single objects from the above solution.
\end{itemize}
More precisely, the flux distributions of the F-string and D-particle described by the
solution \eqref{D8-F1-D0} are given by
\begin{align}\label{Q}
  \nephat{\mathcal{Q}}_1 & = e^{-\nephat{\phi}} \star (d \nephat{B}) = - Q H \, d \Omega_7 \,, \notag \\
  \nephat{\mathcal{Q}}_0 & = e^{3\nephat{\phi}/2} \star (d \nephat{\C{1}} + m_{\text{R}} \nephat{B}) = Q \, dy \wedge d \Omega_7 \,,
\end{align}
with $d\Omega_7$ the volume form of $S^7$. To obtain the corresponding charges these are
to be integrated over $S^7$ and $S^7 \times \mathbb{R}$, respectively, where  $S^7$
together with the 8D radius $r$ spans $\mathbb{R}^8$  and $\mathbb{R}$ covers the
$y$-direction transverse to the domain wall. The flux distributions are related by
\begin{align}
  d \nephat{\mathcal{Q}}_1 & = - m_{\text{R}} \nephat{\mathcal{Q}}_0 \,,
\end{align}
as required by the field equation for $\nephat{B}$. This relation shows that in the
presence of a domain wall ($m_{\text{R}} \ne 0$), the D-particle
($\nephat{\mathcal{Q}}_0\ne 0$) leads to the creation of a fundamental string ($ d
\nephat{\mathcal{Q}}_1\ne 0$). A similar result was obtained in \cite{Imamura:2001cr} for
the NS5-D6-D8 system, i.e.~when a NS5-brane passes through a D8-brane a D6-brane,
stretched between the NS5-brane and the D8-brane, is created. Both processes are related
to the Hanany-Witten effect via duality.

We now return to the distribution of D-particles and F-strings described by
\eqref{D8-F1-D0}. First of all we note that all non-zero tensor components of
\eqref{D8-F1-D0} are even under the $\mathbb{Z}_2$ orientifold symmetry $I_y \Omega$
\eqref{I9Omega}. If this were not the case, one would be forced to include a source term,
corresponding to the non-zero tensor components, which is smeared over a 9D hyperplane.
The only odd field that we allow for is the mass parameter and the corresponding source
terms are the domain walls. The supergravity solution \eqref{D8-F1-D0} that we consider
only has even non-zero tensor components and the inclusion of source terms for this
solution was discussed in \cite{Kallosh:2001zc}\footnote{The particle and strings source
terms of \cite{Kallosh:2001zc} are smeared in the $y$-direction and directly relate to
the charge distributions \eqref{Q}.}, resulting in a globally well-defined solution on
$S^1 / \mathbb{Z}_2$. We will now discuss its physical implications.

\begin{figure}[ht]
  \centerline{\epsfig{file=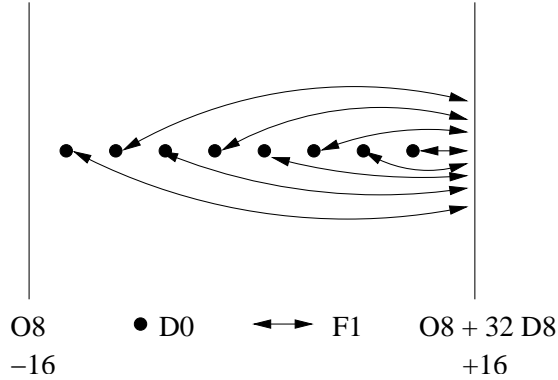,width=.45\textwidth}}
  \caption{\it The creation of strings in type~I${}^\prime$: a (continuous) distribution of D-particles with
  a monotonically increasing distribution of unoriented F-strings ending on the D8-branes.
  The distribution of these F-strings has a maximum at the position of the D8-branes.} \label{HWDO}
\end{figure}

We note that the distribution of F-strings is linear in $H$, see \eqref{Q}. When we are
dealing with a D8-brane, we have $H= c- m_{\text{R}} |y|$ which is a linearly increasing
function when going towards the domain wall. This is in agreement with the idea of
creation of strings when passing through a D8-brane \cite{Danielsson:1997wq,
Danielsson:1998gi, Bergman:1997gf}. It is pictorially given in figure \ref{HWDO}, where
we have taken all D8-branes to coincide with one of the orientifolds. The strings are
unoriented due to the identification $y \sim - y$ which superposes two strings of
opposite orientation, see e.g.~\cite{Bergman:1997gf}. It should be noted that the linear
behaviour of $\mathcal{Q}_1$ is an artifact of the coordinate frame for the transverse
coordinate $y$. The important feature is that it is monotonically increasing when
approaching the domain wall.

\subsection{Domain Walls with $SL(2,\mathbb{R})$ Strings in $D=9$} \label{sec:DW-string-particle-2}

In \cite{Bergshoeff:2003sy} the D8-F1-D0 solution of massive IIA has been generalised to
the 9D gauged supergravities with gauge group $CSO(p,q,r)$ with $p+q+r=2$.

We start from a general Ansatz, respecting $SO(7)$ symmetry. The fields are thus allowed
to depend on $r=(x_1{}^2 + \ldots + x_7{}^2)^{1/2}$ and the transverse direction $y$. Our
strategy will be to solve the BPS-equations obtained from the supersymmetry variations of
the fermions. In analogy with the solution \eqref{D8-F1-D0} for the Romans' mass
parameter, we will assume that the dependence on $r$ and $y$ coordinates can be separated
in a product, i.e.~$f(y,r)=f(r) \, f(y)$. This assumption will simplify the equations
drastically.

The BPS-equations are obtained by requiring the spinor $\epsilon$ to be annihilated by
the projection operators for the relevant branes. We search for solutions, which include
domain walls, strings and particles. Since we search for 1/4 BPS solutions the 3
projection operators corresponding to the domain walls, strings and particles should not
be independent. In other words,  once we have two of the operators, the third should
follow as a combination of these. In contrast to the IIA solution \eqref{D8-F1-D0}, we
have the possibility of $SL(2,\mathbb{R})$ doublets of both the particles and the strings
in 9D. By analysing the supersymmetry variations in type~IIB in $D=10$, it can be seen
that the projectors for the F1- and the D1-strings are actually different, and this will
therefore also be the case for the strings and particles in $D=9$. Choosing a specific
string projector corresponds to choosing an $SL(2,\mathbb{R})$ frame. We take the
following projectors
 \begin{align}
  \Pi_{\text{D0}} = \tfrac{1}{2} (1+\gamma^{\underline{0}}*) \,, \qquad
  \Pi_{\text{F1}} = \tfrac{1}{2} (1+\gamma^{\underline{0y}}*) \,, \qquad
  \Pi_{\text{DW}} = \tfrac{1}{2} (1+\gamma^{\underline{y}})\,,
 \end{align}
where $*$ is seen as an operator, i.e.~$*\epsilon=\epsilon^\ast$. Any third projector is
implied by the other two:
 \begin{align}
  \Pi_{\text{DW}} = \Pi_{\text{D0}} + \Pi_{\text{F1}} - 4 \, \Pi_{\text{D0}} \Pi_{\text{F1}}\, ,
 \end{align}
and cyclic. Since $\epsilon$ transforms under $SL(2,\mathbb{R})$, the choice of
$SL(2,\mathbb{R})$ frame can be seen as a choice of $\epsilon$. To get the most general
solution, we should keep the mass parameters as general as possible. We can, however,
still perform $SL(2,\mathbb{R})$ transformations, which are upper triangular, without
changing $\epsilon$. This can easily be seen by noting that $\epsilon$ transforms as
 \begin{align}
  \epsilon\rightarrow\Big(\frac{c\tau^\ast+d}{c\tau+d}\Big)^{\frac{1}{4}}\epsilon
  \end{align} under the $SL(2,\mathbb{R})$ transformation \begin{align}
  \Lambda=\left(\begin{array}{cc} a&b\\c&d
  \end{array}\right)\, .
 \end{align}
We see that $\epsilon$ is invariant for $c=0$. The mass matrix transforms under $\Lambda$
as well. Even with $c=0$ we can always use $\Lambda$ to put $m_1$ to zero.

Analysing the BPS equations we find that, in order to make up the relevant projection
operators, the following components must be put to zero:
\begin{align}
  & F_{ty}=F_{tm}=F^1_{ty}=F^1_{tm}=H^i_{tmn}=H^2_{tym} + \chi H^1_{tym}=0
\label{hchi} \,,
\end{align}
where $m,n$ are indices of the spatial coordinates $x^m \neq y$. The Bianchi identity for
$F^1$ reads $dF^1= - q_1 H^2$. Since $F^1=0$, this will lead to further restrictions when
$q_1$ is non-vanishing. We find that $H^2=0$ and, using \eqref{hchi}, also $\chi
H^1_{tyi} = 0$. We require $H^1 \neq 0$, since otherwise no F-strings would be present
and we conclude that $\chi = 0$ if $q_1 \ne 0$. If $q_1=0$, one can draw the same
conclusion but from a different argument. In this case, the BPS equations directly imply
$\partial_\mu \chi = 0$ and therefore $\chi$ is a constant. The only non-zero mass
parameter $q_2$ gauges the $\mathbb{R}$ subgroup of $SL(2,\mathbb{R})$, which shifts the
axion. Thus one can always use a global gauge transformation to set $\chi = 0$. Then
\eqref{hchi} implies $H^2=0$. On top of this we take $F^2_{ty}=0$ since a non-zero value
requires D0-brane sources smeared on the domain-wall world-volume and we want to avoid
such 'walls' of D0-branes. We thus find that, for all values of the mass parameters, we
are left with just two non-vanishing tensor components, $F^2_{tr}$ and $H^1_{tyr}$.

We now substitute our Ansatz in the supersymmetry variations of the fermions, which are
given in \eqref{9Dsusy} and \eqref{susyexplmiib}. This leads to two undetermined
functions, one depending on $r$ and one depending on $y$. The latter can be fixed
arbitrarily by using a general coordinate transformations in $y$. To determine the
function of $r$, we need at least one field equation, e.g.~the one for $\varphi$. We have
computed this field equation, and the result is that the $r$-dependent function can be
expressed in terms of a harmonic function. The resulting particle-string-domain wall
solution can be expressed in a unified way, i.e.~including all cases $\det(Q) = 0$,
$\det(Q) > 0$ and $\det(Q) <0$, as follows
 \begin{align} \notag \label{9Dsolution}
  ds^2&=-(h_1 h_2)^{1/14} \tilde{H}^{-11/7}dt^2+(h_1 h_2)^{-3/7} \tilde{H}^{-4/7} dy^2+(h_1 h_2)^{1/14}
  \tilde{H}^{3/7}dx_7^2\, ,\\\notag
  e^\phi&=h_1^{1/2} h_2^{-1/2}\,,\quad e^{\sqrt{7}\varphi}=(h_1 h_2)^{-1} \tilde{H}\, ,\\
  A^2_t&=-h_2^{1/2} h^{-1}\,,\quad B^1_{ty}= - h_2^{-1/2} h^{-1}\, ,\\\notag \epsilon&=(h_1
  h_2)^{1/56} \tilde{H}^{-11/28}\,\epsilon_0\, .
 \end{align}
The solution is given in terms of three harmonic functions
\begin{align}
  h_1 = 2 q_1 y + l_1{}^2 \,, \qquad h_2 = 2 q_2 y + l_2{}^2 \,, \qquad \tilde{H}=1+\frac{Q}{r^5} \,.
\end{align}
The $q$'s are given in terms of the mass parameters in \eqref{qs} and $l_1$ and $l_2$ are
integration constants. Just as in $D=10$, the solution is a harmonic superposition of
D-particles, F-strings and domain walls with string and particle fluxes
\begin{align}
  \mathcal{Q}_1 & = e^{-\phi-\varphi/\sqrt{7}} \star (d B^1) = - Q \, h_2{}^{1/2} \, d \Omega_6 \,, \notag \\
  \mathcal{Q}_0 & = e^{\phi + 3 \varphi/\sqrt{7}} \star (d A^2 + q_2 B^1) = - Q \, h_2{}^{-1/2} \, dy \wedge d \Omega_6 \,,
\label{D0F1}
\end{align}
with $d\Omega_6$ the volume form of the $S^6$. The charges are obtained by integrating
the fluxes over $S^6$ and $S^6 \times \mathbb{R}$, respectively, where the $S^6$,
together with the 7D radius $r$, spans $\mathbb{R}^7$ and $\mathbb{R}$ covers the
$y$-range. The flux distributions are related by
\begin{align}
  d \mathcal{Q}_1 = q_2 \mathcal{Q}_0 \,,
\end{align}
as required by the $B^2$ equation of motion.

Of course one can perform an $SL(2,\mathbb{R})$ transformation on the solution
\eqref{9Dsolution} and obtain intersections with more general strings and particles. The
$SL(2,\mathbb{R})$ generalised flux distributions are given by
\begin{align}
  {\mathcal{Q}}_{i \,1} & = e^{-\varphi/\sqrt{7}} \mathcal{M}_{ij} \star (d {B}^j)
    = q_{i \, 1} \mathcal{Q}_1 \,, \notag \\
  {\mathcal{Q}}_{i \,_0} & = e^{3 \varphi/\sqrt{7}} \mathcal{M}_{ij} \star (d A^j - Q^{jk} B_k)
    = q_{i \, 0} \mathcal{Q}_0 \,,
\end{align}
where the massive field strengths are taken from \eqref{massive-SL2R-field-strengths}. In
this notation the F-strings and D-particles \eqref{D0F1} have charges ${q}_{i \,1} =
(1,0)$ and ${q}_{i \,0} = (0,1)$. A transformation with parameter
\begin{align} \label{sl2r}
  \Lambda = \left(\begin{array}{cc} r & p \\ s & q \end{array}\right) \in SL(2,\mathbb{R}) \,,
\end{align}
would take the distributions of F-strings and D-particles \eqref{D0F1} to ${q}_{i \,1} =
(r,s)$ and ${q}_{j \,0} = (p,q)$. This corresponds to $(p,q)$-particles and
$(r,s)$-strings subject to the condition $qr - ps = 1$. Furthermore, the
$SL(2,\mathbb{R})$ transformation \eqref{sl2r} rotates the diagonal background \eqref{qs}
into
\begin{align}
  Q^{ij} = \tfrac{1}{2} \left(\begin{array}{cc} -m'_2 + m'_3 & m'_1 \\
   m'_1  & m'_2 + m'_3 \end{array}\right) =
  \left(\begin{array}{cc} q^2 q_1 + s^2 q_2 & -pq q_1 - rs q_2 \\
    -pq q_1 - rs q_2 & p^2 q_1 + r^2 q_2 \end{array}\right) \,.
\label{mass}
\end{align}
From now on we will omit the primes on the mass parameters. Thus we find that the most
general intersection of $(p,q)$-particles, $(r,s)$-strings and an $(m_1,m_2,m_3)$-domain
wall is subject to two conditions:
\begin{itemize}
 \item
The $SL(2,\mathbb{R})$ condition $qr - ps = 1$ should be satisfied. This condition
requires orthogonality of the strings and particle charges. It can be expressed as
$\varepsilon^{ij} q_{i \, 0} q_{i \, 1} = 1$.
 \item
The form of the mass matrix is given in \eqref{mass}. This mass matrix contains only two
independent parameters $q_1$ and $q_2$ rather than three for an arbitrary but symmetric
mass matrix. This restriction corresponds to $Q^{ij} q_{i \, 0} q_{i \, 1} = 0$.
\end{itemize}
The two orthogonality conditions are manifestly $SL(2,\mathbb{R})$ invariant and the
parameters $q_1$ and $q_2$ specify the only $SL(2,\mathbb{R})$ orbits that solves the BPS
equations.

\begin{figure}[h]
  \centerline{\epsfig{file=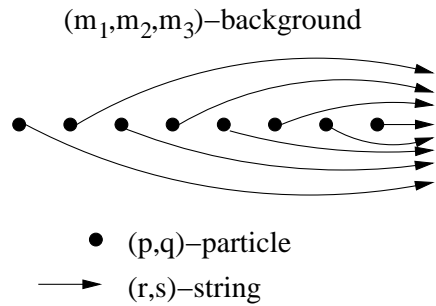,width=.35\textwidth}}
  \caption{\it The creation of strings in 9D: a (continuous) distribution of {\rm (p,q)}-particles with
  a monotonically increasing distribution of emanating oriented {\rm (r,s)}-strings  in an {\rm (m$_1$,m$_2$,m$_3$)}-background.
  There are two orthogonality conditions on the charges.} \label{HWpq}
\end{figure}

The physical picture consists of a distribution of particles from which strings are
emanating towards the domain wall, like in the IIA case. However, we now have an
$SL(2,\mathbb{R})$ generalisation of $(r,s)$-strings stretching between $(p,q)$-particles
in an $(m_1,m_2,m_3)$ background with two orthogonality conditions. The two conditions
reduce the seven parameters to five, three of which correspond to the $SL(2,\mathbb{R})$
freedom while the two remaining parameters are $q_1$ and $q_2$. In addition the charge
$Q$ is the unit string charge. The general solution is illustrated in figure \ref{HWpq}.
The interval in this case is Max$(-q_1/l_1{}^2, -q_2/l_2{}^2) < y < 0$ with all $q_i$
positive. Note that the charge distribution of the strings is not linear, as opposed to
the massive IIA solution in 10D. This is due to the freedom of reparameterisation of the
$y$-coordinate; the important feature is that $\mathcal{Q}_0$ is continuous and positive,
implying $\mathcal{Q}_1$ to be monotonically increasing on this interval.

One can take different limits of the general solution \eqref{9Dsolution}. First of all,
one can set the parameter $q_1=0$. This case corresponds to the reduction of the massive
IIA solution of \cite{Massar:1999sb} and indeed the Kaluza-Klein reduction \eqref{IIAred}
of \eqref{D8-F1-D0} along one of the world-volume directions of the D8-brane gives
(changing $y$ to $\tilde{y}$ for reasons that will become clear shortly)
 \begin{align}\notag
 ds^2 &= -H^{1/7}\tilde{H}^{-11/7}dt^2+H^{8/7}\tilde{H}^{-4/7}d\tilde{y}^2+
 H^{1/7}\tilde{H}^{3/7}dx_7^2\, ,\\
 e^\phi&=H^{-1}\,,\quad e^{\sqrt{7}\varphi}=H^{-2}\tilde{H}\, ,\\\notag
 B^{1}_{t\tilde{y}}&=-\tilde{H}^{-1}\,,\quad A^{2}_t=-H \tilde{H}^{-1}\, ,
 \end{align}
where the harmonic functions are defined as
 \begin{align}
  H= c + 2 m_{\text{R}} \tilde{y}\,,\quad
  \tilde{H}= 1 + \frac{Q}{r^5}\, .
 \end{align}
The above is a special solution to the nine-dimensional gauged supergravity where the
mass parameters obey $q_1 = 0$ and $q_2 = m_{\text{R}}$. Exactly the same identifications
were found in the case of the reduced massive IIA supergravity, see
\eqref{massive-T-duality}. It is related to the generic solution \eqref{9Dsolution} by a
coordinate transformation $y=y(\tilde{y})$ defined by $h_2(y) = H(y)^2$, which is a
special case of the coordinate transformation of footnote~\ref{coordinate-transformation}
of this chapter.

Another possible truncation of the general solution \eqref{9Dsolution} is obtained by
setting both mass parameters $q_1$ and $q_2$ equal to zero. This yields a harmonic
superposition of the F-string solution with a distribution of D-particles on it. The two
charge distributions are related (both are linear in $Q$) and therefore it is impossible
to obtain either one separately.

\begin{figure}[h]
  \centerline{\epsfig{file=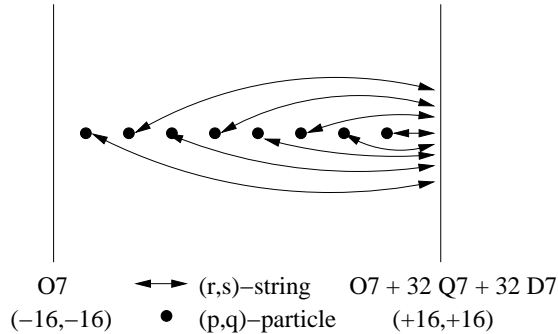,width=.45\textwidth}}
  \caption{\it The creation of strings in 9D on $S^1 /\mathbb{Z}_2$: a (continuous) distribution of
  {\rm (p,q)}-particles with a monotonically increasing distribution of unoriented {\rm (r,s)}-strings
  ending on the D7- and Q7-branes. } \label{fig:HWDQO}
\end{figure}

The 9D particle-string-domain wall solution \eqref{9Dsolution} corresponds to a region
between two domain walls on the $S^1 /\mathbb{Z}_2$, as illustrated in figure \ref{HWpq}.
One might wonder about the possibility of extending this to a globally well-defined
solution by including source terms for the domain walls and the particle-string
intersection. Note that all tensor components of \eqref{9Dsolution} are even under the
relevant 9D $\mathbb{Z}_2$-symmetry. In fact, the reason to discard the possibility of
non-zero $F^1_{ty}$ was its odd transformation under this $\mathbb{Z}_2$-symmetry. Thus
one is led to think that it is possible to embed the solution \eqref{9Dsolution} in a
globally well-defined solution on $S^1 /\mathbb{Z}_2$, as is illustrated in
figure~\ref{fig:HWDQO}. It would be interesting to investigate the boundary conditions in
a manner analogous to the IIA analysis of \cite{Kallosh:2001zc}.

Recapitulating, the 9D solution \eqref{9Dsolution} consists of a smeared distribution of
$(p,q)$-particles, from which $(r,s)$-strings are emanating and ending on the
$(m_1,m_2,m_3)$-domain wall. There are two orthogonality conditions on the seven
parameters, $\varepsilon^{ij} q_{1 \, 0} q_{i \, 1} = 1$ and $Q^{ij} q_{1 \, 0} q_{i \,
1} = 0$, which are manifestly $SL(2,\mathbb{R})$ invariant. This is the natural
generalisation of the 10D IIA solution \eqref{D8-F1-D0}. These solutions suggest new
possibilities of string creation in nine dimensions that are not the result of the
reduced type~I${}^\prime$ mechanism.

In the light of the general domain wall solution \eqref{domain-wall} of
section~\ref{sec:CSO-DW}, it would be very interesting to extend our 10D and 9D analysis
here and consider intersections of domain walls and strings in all dimensions. It is not
obvious to us, however, what the general structure in $D \leq 7$ will be. Consider as
example the $SO(5)$ gauged theory in $D=7$. Due to the lack of one- and two-forms in the
fundamental representation of $SO(5)$, our construction does not trivially carry over. We
do expect solutions like \eqref{9Dsolution} in the $ISO(4)$ gauged theory in $D=7$,
however. It would be very interesting to investigate such solutions in lower-dimensional
gauged supergravities and their uplift to the higher-dimensional theories.

 \mychap{Acknowledgements}

This review article is based on my Ph.D. thesis\footnote{For the unabridged version,
including a historical introduction to high-energy physics and a crash course on
perturbative string theory, see http://www.ub.rug.nl/eldoc/dis/science/d.roest/. The
other chapters are virtually identical to the material presented here. If you are
interested in a hard copy version, please contact me.}, and many results of the research
during my Ph.D. period have been used in chapters \ref{ch:gauged}
and~\ref{ch:domain-walls} and appendix~\ref{appendix:Appendix}. For this reason, I would
very much like to thank my supervisor Eric Bergshoeff for his guidance and enthousiasm
and my other collaborators, Natxo Alonso-Alberca, Klaus Behrndt, Andres Collinucci, Ulf
Gran, Renata Kallosh, Rom\'{a}n Linares, Mikkel Nielsen, Tom\'{a}s Ort\'{\i}n, Antoine
van Proeyen, Per Sundell, Stefan Vandoren and Tim de Wit, for pleasant and fruitful
collaborations. In addition, I would like to thank the members of my Ph.D.~committee,
Professors Jan Louis, Tom\'{a}s Ort\'{\i}n and Bernard de Wit, for their efforts.

I am grateful to the Centre for Theoretical Physics of the University of Groningen for
the opportunity to perform this research during the last four years. Additionally, this
work has been supported in part by the European Community's Human Potential Programme
under contract HPRN-CT-2000-00131 Quantum Spacetime, in which I am associated to Utrecht
University.

 \appendix

 \chapter{Conventions} \label{app:conventions}

\section{Generalities}

We use mostly plus signature $(-+\cdots +)$. Greek indices $\mu,\nu,\rho\ldots$ denote
world coordinates and Latin indices $a,b,c\ldots$ represent tangent space-time. The
different indices are related by the Vielbeins $e_{a}{}^{\mu}$ and inverse Vielbeins
$e_{\mu}{}^{a}$, that satisfy
\begin{equation}
 e_{a}{}^{\mu}e_{b}{}^{\nu}g_{\mu\nu}=\eta_{ab}\, , \hspace{1cm}
 e_{\mu}{}^{a}e_{\nu}{}^{b}\eta_{ab}=g_{\mu\nu}\, .
\end{equation}
Here $\eta_{ab}$ is the Minkowski space-time metric and the space-time metric is $g_{\mu
\nu}$. Underlined explicit indices $0, \ldots, D-1$ refer to the tangent space-time
coordinates.

The covariant derivative on fermions is given by $D_{{\mu}}=\partial_{{\mu}}+ {\omega}_{
\mu}$ with the spin connection ${\omega}_{ \mu} = {\textstyle{1\over 4}}
{\omega}_{{\mu}}{}^{{{a}} {{b}}} {\Gamma}_{{{a}}{{b}}}$, where
\begin{align}
\label{eq:spincon} \omega_{abc} = -\Omega_{abc}+\Omega_{bca} -\Omega_{cab}\, , \qquad
\Omega_{ab}{}^{c} = e_{a}{}^{\mu}e_{b}{}^{\nu} \partial_{[\mu}e^{c}{}_{\nu]}\, .
\end{align}
The Riemann curvature tensor is given in terms of the spin connection by
\begin{align}
R_{\mu\nu a}{}^{b} = 2\partial_{[\mu}\, \omega_{\nu] a}{}^{b} -2\omega_{[\mu|
a}{}^{c}\,\omega_{|\nu]c}{}^{b}\, .
\end{align}
We symmetrise and anti-symmetrise with weight one.

Gauge potentials of rank $d$ are denoted by $\C{d}$ with field strength $\G{d+1}$. For
notational compactness, we sometimes omit the superscript label and denote gauge
potentials of rank $0$ up to $3$ by $\chi$, $A$ or $V$, $B$ and $C$ respectively. The
corresponding field strengths are given by the symbol $\G{1}, F, H$ and $G$,
respectively.

Our conventions in form notation in $D$ dimensions are as follows:
\begin{align}
  P^{(p)} = & \, \frac{1}{p!} P^{(p)}_{\mu_1 \cdots \mu_p} \rmd x^{\mu_1} \wdg \cdots \wdg
     \rmd x^{\mu_p} \,, \notag \\
  P^{(p)}  \cdot Q^{(p)} = & \, \frac{1}{p!}
  P^{(p)}_{\mu_1 \cdots \mu_p} Q^{(p)\,\mu_1 \cdots \mu_p}\,,
  \notag \displaybreak[2] \\
  P^{(p)} \wdg Q^{(q)} =
  & \, \frac{1}{p!q!} P^{(p)}_{\mu_1 \cdots \mu_p}
    Q^{(q)}_{\mu_{p+1} \cdots \mu_{p+q}} \rmd x^{\mu_1} \wdg \cdots \wdg
    \rmd x^{\mu_{p+q}} \,, \displaybreak[2] \notag \\
  \star \, P^{(p)} = & \, \frac{1}{(D-p)!p!} \sqrt{-g}
    \varepsilon^{(D)}_{\mu_0 \cdots \mu_{D-1}} P^{(p)\,\mu_{D-p} \cdots \mu_{D-1}}
    \rmd x^{\mu_0} \wdg \cdots \wdg \rmd x^{\mu_{D-p-1}} \,, \notag \\
  & \text{with~~} \, \varepsilon^{(D)}_{0123\cdots D-1}=-\varepsilon ^{0123\ldots D-1}=1 \,, \notag \\
  \star \star \, P^{(p)} = & \, (-)^{p(D-p)+1} P^{(p)} \,, \notag \\
  d =
  & \, \partial_\mu \rmd x^\mu \,,
\label{formconv}
\end{align}
where the last line is the exterior derivative, acting from the left.

In the case of dimensional reduction, we will always be reducing a
$\nephat{D}$-dimensional theory to a $D$-dimensional one, over an internal manifold of
$n=\nephat{D}-D$ dimensions. The higher-dimensional fields will be hatted and the
lower-dimensional ones unhatted. The corresponding split in the coordinates reads
$x^{\nephat{\mu}} = (x^\mu , z^m)$, with indices $\nephat{\mu}$ and $\mu$ ranging from
$0$ to $\nephat{D}-1$ and $D-1$, respectively, while $m = 1,\ldots,n$.

\section{Spinors and $\Gamma$-matrices in Various Dimensions}

We will denote the $\Gamma$-matrices by $\Gamma_\mu$ (of dimensions 32) in eleven and ten
dimensions and by $\gamma_\mu$ (of dimensions 16) in nine dimensions. They can be chosen
to satisfy
\begin{equation}
 \Gamma_{\nephat\mu}^\dagger=\eta_{\nephat\mu\nephat\mu}\Gamma_{\nephat\mu}
 \qquad {\rm and} \qquad
 \gamma_\mu^\dagger=\eta_{\mu\mu}\gamma_\mu\,,
\end{equation}
respectively. We can also choose the $\Gamma$-matrices to be real, while in nine
dimensions they will be purely imaginary, which implies that
\begin{equation}
 \Gamma_{\nephat\mu}^T=\eta_{\nephat\mu\nephat\mu}\Gamma_{\nephat\mu}
 \qquad {\rm and} \qquad
 \gamma_\mu^T=-\eta_{\mu\mu}\gamma_\mu\,.
\end{equation}
The following notation is used to denote the antisymmetric product of $n$
$\Gamma$-matrices:
 \begin{align}
  \Gamma_{\mu_1 \cdots \mu_n} = \Gamma_{[\mu_1} \cdots \Gamma_{\mu_n]} \,.
 \end{align}
Slashes are used to contract $\Gamma$-matrices and field strengths in the following
sense:
 \begin{align}
  \not\!\!H = H^{\mu \nu \rho }\Gamma _{\mu \nu \rho } \,, \qquad
  \not\!\!H_\mu =H_{\mu \nu \rho}\Gamma ^{\nu \rho } \,,
 \end{align}
with similar formulae for other field strengths. In nine dimensions the same notation is
used with $\Gamma$ replaced by $\gamma$.

In eleven and ten dimensions we use the 32-dimensional spinor representation, with
$\Gamma$-matrices $\Gamma_\mu$ (and $\Gamma_{11}$ in 10D). Upon reduction to nine
dimensions we will split this into 16-dimensional representations, with $\Gamma$-matrices
$\gamma_\mu$. This will be discussed below. In contrast, upon reduction to eight
dimensions we will use the corresponding spinor representation; rather, we preserve the
32-dimensional representation, with $\Gamma$-matrices $\Gamma_\mu$ and $\Gamma_i$ with
$i=1,2,3$.

In ten dimensions the minimal spinor is a 32-component Majorana-Weyl spinor with 16
(real) degrees of freedom. With the choice
\begin{equation}
 \Gamma_{11} = -\Gamma_{\underline{0} \cdots \underline{9}} \,,\qquad\qquad
 \Gamma_{11}=\left(\begin{matrix} \id & \;\;\,0 \\ 0 & -\id \end{matrix}\right),
\end{equation}
we can write a ten-dimensional Majorana-Weyl spinor as being composed of
nine-dimensional, 16 component, Majorana-Weyl spinors according to
\begin{equation}
\psi^{MW}_+=\left(\begin{array}{c}\psi_1\\0\end{array}\right)\,,\qquad\psi^{MW}_-=\left(\begin{array}{c}0\\\psi_2\end{array}\right)\,,
\end{equation}
where $\psi_i$ are nine-dimensional Majorana-Weyl spinors and $+$ or $-$ denotes the
chirality of the ten-dimensional spinor. The split of an arbitrary ten-dimensional spinor
into two Majorana-Weyl spinors of opposite chirality can of course be done without
reference to nine dimensions (through the specific choice of $\Gamma_{11}$), but each
ten-dimensional Majorana-Weyl spinor will then in general have 32 non-zero components
even though it only has 16 degrees of freedom. In order to reduce to nine dimensions we
use
\begin{equation}
\Gamma_{11}=\sigma_3\otimes\id\,,\qquad
\Gamma_{\underline{z}}=\sigma_1\otimes\id\,,\qquad \Gamma_{a}=\sigma_2\otimes\gamma_a\,,
\end{equation}
where $z$ is the reduction coordinate and the Pauli matrices are defined as
\begin{equation}
\sigma_1=\left(\begin{matrix}0 & 1 \\ 1 & 0\end{matrix}\right)\,,\qquad
\sigma_2=\left(\begin{matrix}0 & -i \\ i & 0\end{matrix}\right)\,,\qquad
\sigma_3=\left(\begin{matrix}1 & 0 \\ 0 & -1\end{matrix}\right)\,.
\end{equation}
As mentioned above the nine dimensional $\gamma$-matrices are purely imaginary. If we
work with a reduction of type IIB, where the two spinors have the same chirality, it may
be convenient to introduce complex, nine-dimensional, Weyl spinors according to
\begin{alignat}{2}
\psi_c&= \psi_1+i\psi_2\,,\qquad &\lambda_c=\lambda_2+i \lambda_1\,,\\
\epsilon_c&=\epsilon_1+i\epsilon_2\,,\qquad
&\tilde\lambda_c=\tilde\lambda_2+i\tilde\lambda_1\,,
\end{alignat}
which in ten-dimensional notation can be written as, e.g.,
\begin{equation}
\psi^{W}_+=\left(\begin{array}{c}\psi_1\\0\end{array}\right)+i\left(\begin{array}{c}\psi_2\\0\end{array}\right)\,.
\end{equation}
If we instead work with a reduction of type IIA the two spinors will have opposite
chirality, and can thus be composed into a ten-dimensional Majorana spinor according to
\begin{equation}
\psi^{M}=\left(\begin{array}{c}\psi_1\\0\end{array}\right)+\left(\begin{array}{c}0\\\psi_2\end{array}\right)\,.
\end{equation}

\begin{figure}[tb]
\begin{center}
\epsfig{file=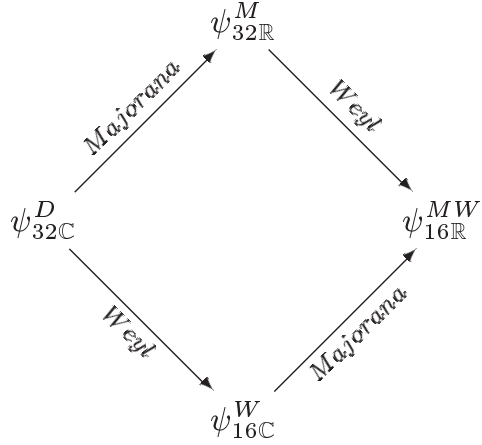, width=.4\textwidth} \caption{Schematic view of how a
ten-dimensional Dirac spinor can be projected down to a Majorana-Weyl spinor along two
different routes. The number of real or complex degrees of freedom for each spinor is
also indicated. The relation between the spinors at the intermediate stage (in nine
dimensions) is given by (\ref{MWrelation}).}\label{spinordiagram}
\end{center}
\end{figure}

When working with these non-minimal spinors, which are either just Majorana
($\psi_\mu^M$) or just Weyl ($\psi_\mu^W$)~\cite{Bergshoeff:2002mb}, the two formulations
are (in nine dimensions) related via
\begin{equation}
\begin{aligned}
  \tfrac{1}{2} (1+\Gamma_{11}) \psi_\mu^M & = \text{Re} (\psi_\mu^W)\, , \\
  \tfrac{1}{2} (1+\Gamma_{11}) \lambda^M & = \text{Im} (\Gamma_{\underline{z}} \lambda^W)\,, \\
  \tfrac{1}{2} (1+\Gamma_{11}) \tilde\lambda^M & = \text{Im} (\Gamma_{\underline{z}} \tilde\lambda^W)\,, \\
  \tfrac{1}{2} (1+\Gamma_{11}) \epsilon^M & = \text{Re} (\epsilon^W)\,, \qquad
\end{aligned}
\begin{aligned}
  \tfrac{1}{2} (1-\Gamma_{11}) \psi_\mu^M & = \text{Im} (\Gamma_{\underline{z}} \psi_\mu^W)\, , \\
  \tfrac{1}{2} (1-\Gamma_{11}) \lambda^M & = \text{Re} (\lambda^W)\, , \\
  \tfrac{1}{2} (1-\Gamma_{11}) \tilde\lambda^M & = \text{Re}
    (\tilde\lambda^W)\, , \\
  \tfrac{1}{2} (1-\Gamma_{11}) \epsilon^M & = \text{Im} (\Gamma_{\underline{z}} \epsilon^W)\, ,
\end{aligned}\label{MWrelation}
\end{equation}
for positive ($\psi_\mu^W, \epsilon^W$) and negative ($\lambda^W,\tilde\lambda^W$)
chirality Weyl fermions. With the above mentioned decomposition into nine-dimensional
Majorana-Weyl spinors we can write
\begin{equation}
\psi_\mu^M=\left(\begin{array}{c}\psi_1\\\psi_2\end{array}\right)\,,\quad
\epsilon^M=\left(\begin{array}{c}\epsilon_1\\\epsilon_2\end{array}\right)\,,\quad
\lambda^M=\left(\begin{array}{c}\lambda_1\\\lambda_2\end{array}\right)\,,\quad
\tilde\lambda^M=\left(\begin{array}{c}\tilde\lambda_1\\ \tilde\lambda_2\end{array}\right)
\end{equation}
and
\begin{align}
\psi_\mu^W&=\left(\begin{array}{c}\psi_1+i\psi_2\\0\end{array}\right)\,,\quad
\epsilon_\mu^W=\left(\begin{array}{c}\epsilon_1+i\epsilon_2\\0\end{array}\right)\,,\quad\\
\lambda^W&=\left(\begin{array}{c}0\\ \lambda_2+i\lambda_1\end{array}\right)\,,\quad
\tilde\lambda^W=\left(\begin{array}{c}0\\
\tilde\lambda_2+i\tilde\lambda_1\end{array}\right)\,,\quad
\end{align}
where the spinors without an $M$ or $W$ superscript are Majorana-Weyl spinors. The two
different routes to obtain Majorana-Weyl spinors are illustrated in figure
\ref{spinordiagram}. Note also that it follows from the Clifford algebra and the choice
of $\Gamma_{11}$ that $\Gamma_{\underline{z}}$ is off-diagonal, which is crucial for this
construction.

 \chapter{Supergravity and Reductions} \label{appendix:Appendix}

\section{11D Supergravity}

\subsection{11D Supersymmetry Transformations and Field Equations}

The supersymmetry transformation rules of $N=1$ eleven-dimensional supergravity read
\begin{align}
  \delta { e}_{{\mu}}{}^{{ a}}
  &= \bar{{ \epsilon}} {\Gamma}^{{ a}} {\psi}_{{\mu}}\, , \notag \\
  \delta {{ C}}_{{\mu}{\nu}{\rho}}
  &= -3\, \bar{{{\epsilon}}} {\Gamma}_{[{\mu}{\nu}} {\psi}_{{\rho}]}\, , \notag \\
  \delta {\psi}_{{\mu}}
  &= D_{{ \mu}} {\epsilon} + \tfrac{1}{192}(\slashed{G} {\Gamma}_{{\mu}}
  - \tfrac{1}{3} {\Gamma}_{{\mu}} \slashed{G})  { \epsilon}\, ,
\label{11Dsusy}
\end{align}
with the field strengths ${G} = \text{d} { C}$ and $D_{{ \mu}} {\epsilon} = (\partial_{{
\mu}} + { \omega}_{{ \mu}}) {\epsilon}$. The 11D fermionic field content consists solely
of a 32--component gravitino, whose field equation reads
\begin{align}
  X_0 (\nephat{\nephat{\psi}}^{\nephat{\nephat{\mu}}}) =
  \nephat{\nephat{\Gamma}}^{\nephat{\nephat{\mu}}\nephat{\nephat{\nu}}\nephat{\nephat{\rho}}}
  \nephat{\nephat{D}}_{\nephat{\nephat{\nu}}}
  \nephat{\nephat{\psi}}_{\nephat{\nephat{\rho}}} = 0 \,,
\end{align}
with $\nephat{\nephat{D}}_{\nephat{\nephat{\nu}}} =  \partial_{\nephat{\nephat{\nu}}} +
\nephat{\nephat{\omega}}_{\nephat{\nephat{\nu}}}$ and where we have set the three-form
equal to zero. Under supersymmetry this fermionic field equations transforms into
\begin{align}
  \delta_0 X_0 (\nephat{\nephat{\psi}}^{\nephat{\nephat{\mu}}})  =
  \tfrac{1}{2} \nephat{\nephat{\Gamma}}^{\nephat{\nephat{\nu}}} \nephat{\nephat{\epsilon}} \,
  [\nephat{\nephat{R}}^{\nephat{\nephat{\mu}}}{}_{\nephat{\nephat{\nu}}}
   - \tfrac{1}{2} \nephat{\nephat{R}} \nephat{\nephat{g}}^{\nephat{\nephat{\mu}}}{}_{\nephat{\nephat{\nu}}} ] \,,
\end{align}
which implies the bosonic Einstein equation for the metric.

\subsection{11D Reduction Ans\"atze to IIA}

We use the following reduction Ans\"atze (where hatted quantities are 11D and unhatted
are IIA)
\begin{align}
  {{\hat e}}_{{\hat\mu}}{}^{{\hat a}} & = e^{m_{11} z} \left(
    \begin{array}{cc} e^{-\phi/12}  e_{\mu}{}^{ a}
      & - e^{2\phi/3}  \C{1}_{\mu} \notag \\
      0 & e^{2\phi/3}
    \end{array} \right)\,, \notag \\
{\hat\psi}_{ a}
& = e^{-m_{11} z/2} e^{\phi/24} [\psi_{ a} - \tfrac{1}{24} \Gamma_{ a} \lambda]\,, \notag \\
{\hat\psi}_{\underline z}
& = \tfrac13 e^{-m_{11} z/2} e^{\phi/24} \Gamma_{\underline z} \lambda\,, \notag \\
{\hat\epsilon}
& = e^{m_{11} z/2} e^{-\phi/24} \epsilon\,, \notag \\
{{\hat C}}_{ \mu  \nu  \rho} & = e^{3 m_{11} z}  \C{3}_{ \mu  \nu  \rho}\,, \notag \\
{\hat C}_{ \mu  \nu z} & = - e^{3 m_{11} z}  B_{ \mu  \nu}\,, \label{11Dred}
\end{align}
to arrive at the IIA supersymmetry transformations in ten dimensions, where
\begin{itemize}
 \item $m_{11} = 0$ for toroidal reduction and
 \item $m_{11} \neq 0$ for twisted reduction using the trombone symmetry of 11D
 supergravity.
\end{itemize}

\section{IIA Supergravity}

\subsection{IIA Supersymmetry Transformations and Field Equations}

The supersymmetry transformation rules of ten-dimensional massless or ungauged IIA
supergravity read
\begin{align}
  \delta_0 \nephat e_{\nephat \mu}{}^{\nephat a}
  & = \overline{\nephat \epsilon} \Gamma^{\nephat a}
    {\nephat \psi}_{\nephat \mu} \,, \notag \\
  \delta_0 {\nephat \psi}_{\nephat \mu}
  & = \big( D_{\nephat \mu}
    +\tfrac{1}{48} e^{-\nephat \phi/2}( \slashed{\nephat H} \nephat{\Gamma}_{\nephat \mu}
    +\tfrac{1}{2} \nephat{\Gamma}_{\nephat \mu} \slashed{\nephat H}) \Gamma_{\text{11}}
    +\tfrac{1}{16} e^{3\nephat \phi/4}( \slashed{G}^{(2)} \nephat{\Gamma}_{\nephat \mu}
    -\tfrac{3}{4} \nephat{\Gamma}_{\nephat \mu} \slashed{G}^{(2)}) \Gamma_{\text{11}}+ \notag \\
  & \; \; \; +\tfrac{1}{192} e^{\nephat \phi/4}( \slashed{G}^{(4)} \nephat{\Gamma}_{\nephat \mu}
    -\tfrac{1}{4} \nephat{\Gamma}_{\nephat \mu} \slashed{G}^{(4)})
    \big) \nephat \epsilon \,, \notag \\
  \delta_0 {\nephat B}_{\nephat \mu \nephat \nu}
  & = 2 e^{\nephat \phi/2} \overline{\nephat \epsilon} \Gamma_{\text{11}}
    {\nephat \Gamma}_{[\nephat \mu} ( {\nephat \psi}_{\nephat \nu]}
    +\tfrac18 {\nephat \Gamma}_{\nephat \nu]} \nephat \lambda) \,, \notag \\
  \delta_0 \nephat \C{1}_{\nephat \mu}
  & = - e^{-3\nephat \phi/4} \overline{\nephat \epsilon} \Gamma_{\text{11}}
    ( {\nephat \psi}_{\nephat \mu} -\tfrac{3}{8} {\nephat \Gamma}_{\nephat \mu}
    \nephat \lambda) \,, \notag \\
  \delta_0 \nephat \C{3}_{\nephat \mu \nephat \nu \nephat \rho}
  & = - 3 e^{-\nephat \phi/4} \overline{\nephat \epsilon}
    {\nephat \Gamma}_{[\nephat \mu \nephat \nu}
    ( {\nephat \psi}_{\nephat \rho]} -\tfrac{1}{24} {\nephat \Gamma}_{\nephat \rho]}
    \nephat \lambda)
    +3 \nephat \C{1}_{[\nephat \mu} \delta_0 {\nephat B}_{\nephat \nu \nephat \rho]} \,, \notag \\
  \delta_0 \nephat \lambda
  & = \big( \slashed \partial \nephat \phi
    + \tfrac{1}{12} e^{-\nephat \phi/2} \slashed{\nephat H} \Gamma_{\text{11}}
    + \tfrac{3}{8} e^{3\nephat \phi/4} \slashed{G}^{(2)} \Gamma_{\text{11}}
    + \tfrac{1}{96} e^{\nephat \phi/4} \slashed{G}^{(4)} \big) \nephat \epsilon \,,
    \notag \\
  \delta_0 \nephat \phi
  & = \tfrac{1}{2} \overline{\nephat \epsilon} \nephat \lambda \,,
\label{IIAsusy}
\end{align}
with the following field strengths:
\begin{align}
  {\nephat \G{2}} & = {\rm d} {\nephat \C{1}} \,, \qquad
  {\nephat H} = {\rm d} {\nephat B} \,, \qquad
  {\nephat \G{4}} = {\rm d} {\nephat \C{3}} + {\nephat \C{1}} \wdg {\nephat H} \,,
\end{align}
and $D_{\nephat{\mu}} \nephat{\epsilon} = (\partial_{\nephat{\mu}} +
\nephat{\omega}_{\nephat{\mu}}) \nephat{\epsilon}$. Upon (massless) reduction with our
Ans\"atze the 11D field equation splits up in two field equations for the 10D IIA
fermionic field content, a gravitino and a dilatino:
\begin{align}
  X_0 ({\nephat{\psi}}^{{\nephat{\mu}}}) & = \nephat{{\Gamma}}^{\nephat{{\mu}}
\nephat{{\nu}}\nephat{{\rho}}}
  {\nephat{D}}_{\nephat{{\nu}}} \nephat{{\psi}}_{\nephat{{\rho}}}
  - \tfrac{1}{8} (\slashed{\partial} \nephat{\phi})
  \nephat{{\Gamma}}^{\nephat{{\mu}}} \nephat{\lambda} = 0 \,, \qquad
  X_0 (\nephat{\lambda}) = \nephat{{\Gamma}}^{\nephat{{\nu}}}
 {\nephat{D}}_{\nephat{{\nu}}} \nephat{\lambda}
  - \nephat{{\Gamma}}^{\nephat{{\nu}}} (\slashed{\partial} \nephat{\phi})
\nephat{{\psi}}_{\nephat{{\nu}}} =0 \,, \label{X0IIA}
\end{align}
with ${\nephat{D}}_{\nephat{{\nu}}} = (\partial_{\nephat{{\nu}}} +
\nephat{{\omega}}_{\nephat{{\nu}}})$ and where we have set the vector, two- and
three-form equal to zero. Under supersymmetry these fermionic field equations transform
into
\begin{align}
  \delta_0 X_0 ({\nephat{\psi}}^{{\nephat{\mu}}})  =
  & \tfrac{1}{2} \nephat{{\Gamma}}^{\nephat{{\nu}}} \nephat{{\epsilon}} \,
  [\nephat{{R}}^{\nephat{{\mu}}}{}_{\nephat{{\nu}}}
  - \tfrac{1}{2} \nephat{{R}} \nephat{{g}}^{\nephat{{\mu}}}{}_{\nephat{{\nu}}}
  - \tfrac{1}{2} (\partial^{\nephat{{\mu}}} \nephat{\phi}) (\partial_{\nephat{{\nu}}} \nephat{\phi})
  + \tfrac{1}{4} (\partial \nephat{\phi})^2 \nephat{{g}}^{\nephat{{\mu}}}{}_{\nephat{{\nu}}}] \,, \notag \\
  \delta_0 X_0 (\nephat{\lambda}) =
  & \nephat{\epsilon} \, [ \Box \nephat{\phi} ]\,,
\end{align}
which imply the usual graviton-dilaton field equations.

\subsection{IIA Reduction Ans\"atze to 9D}

We use the following reduction Ansatz with $z$-dependence implied by the
$SO(1,1)$-symmetries (where hatted quantities are IIA and unhatted are 9D):
\begin{align}
  {\hat e}_{\hat \mu}{}^{\hat a} & = e^{9 m_{\text{IIA}}z/8}\left(
    \begin{array}{cc} e^{\phi/16-3\varphi/16\sqrt{7}} e_\mu{}^a
      & e^{-7\phi/16+3\sqrt{7}\varphi/16} A_{\mu}^{1} \\
      0 & e^{-7\phi/16+3\sqrt{7}\varphi/16}
    \end{array} \right) \,, \notag \\
  {\hat \psi}_a & = e^{-9 m_{\text{IIA}}z/16} e^{-\phi/32 + 3\varphi/32\sqrt{7}}
    [\psi_a +\tfrac{1}{32} \Gamma_a (\lambda-\tfrac{3}{\sqrt{7}} {\tilde \lambda})] \,, \notag \\
  {\hat \psi}_{\underline z} & =
    -\tfrac{7}{32} e^{-9 m_{\text{IIA}}z/16} e^{-\phi/32 + 3\varphi/32\sqrt{7}} \Gamma_{\underline{z}}
    (\lambda - \tfrac{3}{\sqrt{7}} \tilde\lambda ) \,, \notag \\
  {\hat B}_{\mu \nu} & = - e^{3 m_{\text{IIA}}z+m_4 z/2}
    (B_{\mu \nu}^{1} - 2 A^1_{[\mu} A_{\nu]}) \,, \notag \\
  {\hat B}_{\mu z} & = - e^{3 m_{\text{IIA}}z+m_4 z/2} A_\mu \,, \notag
 \displaybreak[2] \\
 {\hat{C}}^{(1)}_\mu & = -e^{-3m_4 z/4}
    (A_\mu^{2} + \chi A^1_\mu ) \,, \notag \\
  {\hat{C}}^{(1)}_z & = - e^{-3m_4z/4} \chi \,, \notag \\
  {\hat{C}}^{(3)}_{\mu \nu \rho} & = e^{3 m_{\text{IIA}}z-m_4 z/4}
    (C_{\mu \nu \rho} -3 A_{i \, [\mu} B^i_{\nu\rho]}
    + 6 A^1_{[\mu} A^2_{\nu} A_{\rho]} ) \,, \notag \\
  {\hat C}^{(3)}_{\mu \nu z} & = - e^{3 m_{\text{IIA}}z-m_4 z/4}
    (B_{\mu \nu}^{2} - 2 A^2_{[\mu} A_{\nu]} ) \,, \notag \\
  {\hat \lambda} & = \tfrac14 e^{-9m_{\text{IIA}}z/16} e^{-\phi/32 + 3\varphi/32\sqrt{7}}
    (3 \lambda + \sqrt{7} \tilde \lambda ) \,, \notag \\
  \hat \epsilon & = e^{9m_{\text{IIA}}z/16} e^{\phi/32 - 3\varphi/32\sqrt{7}} \epsilon \,, \notag \\
  \hat \phi & = \tfrac14 (3 \phi + \sqrt{7} \varphi) + \left( \tfrac{3}{2}m_{\text{IIA}}+m_4 \right) z
  \,,
\label{IIAred}
\end{align}
where the mass parameters are given by
\begin{itemize}
 \item $m_{\text{IIA}} = 0$ and $m_4 =0$ for toroidal reduction,
 \item $m_{\text{IIA}} = 0$ and $m_4 \neq 0$ for twisted reduction using the scale symmetry
 $\alpha$,
 \item $m_{\text{IIA}} = 0$ and $m_4 \neq 0$ for twisted reduction using the trombone symmetry
 $\beta$ and
 \item $m_{\text{IIA}} \neq 0$ and $m_4 \neq 0$ for a combination of the latter two.
\end{itemize}

\section{IIB Supergravity}

\subsection{IIB Supersymmetry Transformations and Field Equations}

The supersymmetry transformation rules of ten-dimensional IIB supergravity read (in
complex notation)
\begin{align}
\delta { e}_{ \mu}\,^{ a}
&= \tfrac12  {\overline{\epsilon}}\,{ \Gamma}^{ a}{ \psi}_{ \mu} + \text{h.c.} \,, \notag\\
\delta  \psi_{ \mu}
&= D_{ \mu}  \epsilon - \tfrac{i}{16 \cdot 5!} \slashed{ G}^{\text{(5)}} { \Gamma}_{ \mu}  \epsilon\, \notag\\
& \quad + \tfrac{i}{16 \cdot 3!} e^{ \phi/2} \left({ \Gamma}_{ \mu}{ \Gamma}^{(3)} + 2 { \Gamma}^{(3)}{ \Gamma}_{ \mu}\right) \left({ H}^{2} + \tau { H}^{1}\right)_{(3)}{ \epsilon}^* \,, \notag\\
\delta  \lambda
&= - e^{\phi} \slashed \partial  \tau {\epsilon}^* - \tfrac{1}{2 \cdot 3!} e^{\phi/2} {\Gamma}^{(3)} \left( { H}^{2} + \tau { H}^{1} \right)_{(3)}  \epsilon \,, \notag\\
\delta { B}^{1}_{\mu \nu}
&= e^{\phi/2}\left({ {\overline{\epsilon}}}^*\,{\Gamma}_{[\mu} {\psi}_{\nu]} - \tfrac{i}{8}  {\overline{\epsilon}}\,{\Gamma}_{\mu \nu} \lambda \right) + \text{h.c.} \,, \notag\\
\delta { B}^{2}_{\mu \nu}
&= - e^{\phi/2} { \tau}^* \left({\overline{\epsilon}}^* {\Gamma}_{[\mu} { \psi}_{\nu]} - \tfrac{i}{8} {\overline{\epsilon}}\,{\Gamma}_{\mu \nu} \lambda \right) + \text{h.c.} \,, \notag\\
\delta \C{4}_{\mu \nu \lambda \rho}
&= 2 i \,  {\overline{\epsilon}}\,{\Gamma}_{[\mu \nu \lambda} {\psi}_{\rho]} - \tfrac32 \, { B}_{i \, [\mu \nu}\delta { B}^{i}_{ \lambda \rho]} +\text{h.c.} \,, \notag\\
\delta \chi
&= -\tfrac14 e^{-\phi}  {\overline {\epsilon}} {\lambda}^* + \text{h.c.} \,, \notag\\
\delta \phi &= \tfrac{i}{4} {\overline {\epsilon}} {\lambda}^* + \text{h.c.} \,, \label{IIBsusy}
\end{align}
with the complex scalar $\tau = \chi + i e^{-\phi}$, the axion $\chi = \C{0}$, the
doublet $B^i = (-B, \C{2})$ and the field strengths\footnote{Note that $\G{5}$ is not of
the canonical form \eqref{RR-duality}; the difference amounts to a field redefinition.}
\begin{align}
 H^i &= {\rm d} B^i \,, \qquad
 \G{5} = {\rm d} \C{4} + \tfrac12 B_i \wdg H^i \,.
\end{align}
Indices $i,j$ of $SL(2,\mathbb{R})$ are contracted with $\varepsilon_{ij} = -
\varepsilon^{ij}$ with $\varepsilon_{12} = -\varepsilon_{21} = 1$. The covariant
derivative of the IIB Killing spinor reads
\begin{align}
 D_{ \mu}  \epsilon &= (\partial_{ \mu} + { \omega}_{ \mu}
  + \tfrac{i}{4} e^{ \phi} \partial_{ \mu}  \chi )  \epsilon \,.
\end{align}
When truncating to the metric, scalars and fermions, the massless 9D fermionic field equations read
\begin{align}
  X_0 ({\nephat \psi}^{\nephat \mu}) & =
  {\nephat \Gamma}^{{\nephat \mu}{\nephat \nu}{\nephat \rho}} (\partial_{\nephat \nu} + {\nephat{\omega}}_{{{{\nephat \nu}}}}+
  \tfrac{1}{4} i e^{{\nephat \phi}} \partial_{\nephat \nu} {\nephat \chi}) {\nephat \psi}_{\nephat \rho}
  + \tfrac{1}{8} e^{\nephat \phi} (\slashed{\partial} {\nephat \tau}) {\nephat \Gamma}^{\nephat \mu} {\nephat \lambda}^* =0 \,, \notag \\
  X_0 ({\nephat \lambda}) & =
  {\nephat \Gamma}^{\nephat \mu} (\partial_{\nephat \mu} + {\nephat{\omega}}_{{{{\nephat \mu}}}}+ \tfrac{3}{4} i e^{{\nephat \phi}} \partial_{\nephat \mu} {\nephat \chi}) {\nephat \lambda}
  + e^{\nephat \phi} {\nephat \Gamma}^{\nephat \mu} (\slashed{\partial} {\nephat \tau}) {\nephat \psi}_{\nephat \mu}^* = 0 \,.
\end{align}

\subsection{IIB Reduction Ans\"atze to 9D}

The reduction Ans\"atze we used for reducing the above rules are (where hatted quantities
are IIB and unhatted are 9D)
\begin{align}
 \hat{e}_{\hat{\mu}}\,^{\hat{a}} & =
  e^{m_{\text{IIB}} z} \left(\begin{array}{cc}
    e^{\sqrt{7}\varphi/28}e_{\mu}\,^{a} & - e^{-\sqrt{7}\varphi/4}A_{\mu} \\
    0 & e^{-\sqrt{7}\varphi/4}
  \end{array} \right) \,, \notag \\
  \hat{\psi}_{a} & =
   e^{-m_{\text{IIB}} z/2} e^{-\sqrt{7}\varphi/56}\left(\frac{c\tau^*+d}{c\tau+d}\right)^{1/4}(\psi_{a}+
   \tfrac{1}{8\sqrt{7}}\Gamma_{a}\tilde{\lambda}^*) \,, \notag \\
 \hat{\psi}_{\underline{z}} & =
   -\tfrac{\sqrt{7}}{8} e^{-m_{\text{IIB}} z/2} e^{-\sqrt{7}\varphi/56}\left(\frac{c\tau^*+d}{c\tau+d}\right)^{1/4}
   \Gamma_{\underline{z}}\tilde{\lambda}^* \,, \notag \\
 \hat{\lambda} & = i e^{-m_{\text{IIB}} z/2} e^{-\sqrt{7}\varphi/56}\left(\frac{c\tau^*+d}{c\tau+d}\right)^{3/4}\lambda \,, \notag \\
 \hat{\epsilon} & = e^{m_{\text{IIB}} z/2} e^{\sqrt{7}\varphi/56}\left(\frac{c\tau^*+d}{c\tau+d}\right)^{1/4}\epsilon
  \displaybreak[2] \,, \notag \\
  \hat{\tau} & = \frac{a\tau+b}{c\tau+d} \,, \notag \\
 \hat{B}^i_{\mu\nu} & = e^{2 m_{\text{IIB}} z} (\Omega(z)^{-1})_j{}^i \, B^j_{\mu\nu}
 \,,\hspace{1cm} \hat{B}^i_{\mu z} = -e^{2 m_{\text{IIB}} z} (\Omega(z)^{-1})_j{}^i \, A^j_{\mu} \,, \notag \\
 \hat{C}^{(4)}_{\mu \nu \lambda \rho} & = e^{4 m_{\text{IIB}} z} D_{\mu \nu \lambda \rho} \,,\hspace{1.5cm}
 \hat{C}^{(4)}_{\mu \nu \lambda z}= e^{4 m_{\text{IIB}} z} (- C_{\mu \nu \lambda}
  + \tfrac32 A_{i \, [\mu} B^i_{\nu \rho]}) \,,
\label{IIBred}
\end{align}
where we take the $\Omega$ to be $z$-dependent:
\begin{align}
  \Omega(z)_i{}^j & = \text{exp} \left(
    \begin{array}{cc}
      \tfrac{1}{2} m_1 z &
      \tfrac{1}{2}(m_2+m_3)z \\
      \tfrac{1}{2}(m_2-m_3)z &
      - \tfrac{1}{2} m_1 z
    \end{array} \right) \,, \notag \\
    & = \left(
    \begin{array}{cc}
      \cosh(\alpha z) + \tfrac{m_1}{2\alpha}\sinh(\alpha z) &
      \tfrac{1}{2\alpha}(m_2+m_3)\sinh(\alpha z) \\
      \tfrac{1}{2\alpha}(m_2-m_3)\sinh(\alpha z) &
      \cosh(\alpha z) - \tfrac{m_1}{2\alpha}\sinh(\alpha z)
    \end{array} \right)\, ,
\end{align}
where $\alpha^2 = \tfrac{1}{4} (m_1{}^2 + m_2{}^2 - m_3{}^2)$. The mass parameters
$\vec{m} = (m_1,m_2,m_3)$ and $m_{\text{IIB}}$ take the following values in the different
reduction schemes
\begin{itemize}
 \item $\vec{m} = 0$ and $m_{\text{IIB}} = 0$ for toroidal reduction,
 \item $\vec{m} \neq$ and $m_{\text{IIB}} = 0$ for twisted reduction with the $SL(2,\mathbb{R})$ symmetry,
 \item $\vec{m} = 0$ and $m_{\text{IIB}} \neq 0$ for twisted reduction with the trombone
 symmetry and
 \item $\vec{m} \neq 0$ and $m_{\text{IIB}} \neq 0$ for a combination of the latter two.
\end{itemize}

\section{9D Maximal Supergravity} \label{app:9D}

\subsection{9D Supersymmetry Transformations and Field Equations}

The unique nine-dimensional $N=2$ supergravity theory has the following supersymmetry
transformations:
\begin{align}
\delta_0 e_{\mu}\,^{a}
&= \ft12 \bar{\epsilon} \gamma^{a} \psi_{\mu} + \text{h.c.} \,,\notag\\
\delta_0 \psi_\mu
&= D_\mu \epsilon + \ft{i}{16} e^{-2 \varphi / \sqrt7} \left( \ft57 \gamma_{\mu} \gamma^{(2)} - \gamma^{(2)} \gamma_{\mu} \right) F_{(2)} \epsilon \notag\\
& \quad - \ft{1}{8\cdot 2!} e^{3 \varphi/ {2\sqrt7}} \left( \ft57 \gamma_{\mu}\gamma^{(2)} - \gamma^{(2)}\gamma_{\mu} \right) \,  e^{\phi/2} \left( F^{2} + \tau F^{1} \right)_{(2)} \epsilon^* \notag \notag\\
& \quad + \ft{i}{8\cdot 3!}e^{- \varphi / {2 \sqrt7}} \left( \ft37 \gamma_{\mu}\gamma^{(3)} + \gamma^{(3)}\gamma_\mu \right)  e^{\phi/2} \left( H^{2} + \tau H^{1} \right)_{(3)}\epsilon^* \notag \notag\\
& \quad - \ft{1}{8\cdot 4!}e^{\varphi / \sqrt7} \left( \ft{1}{7} \gamma_{\mu} \gamma^{(4)} - \gamma^{(4)} \gamma_{\mu} \right) G_{(4)} \epsilon \,,\notag\\[2mm]
\delta_0 \tilde\lambda & = i \slashed \partial \varphi \, \epsilon^* - \ft{1}{\sqrt{7}} e^{-2 \varphi/ \sqrt7} \slashed F \epsilon^* - \ft{3i}{2\cdot 2!\sqrt{7}}e^{3\varphi/{2\sqrt7}} e^{\phi/2}\gamma^{(2)}\left( F^{2} + \tau^* F^{1} \right)_{(2)} \epsilon \notag \\
& \quad + \ft{1}{2\cdot 3!\sqrt{7}}e^{-\varphi/ {2 \sqrt7}}e^{\phi/2}\gamma^{(3)}\left(H^{2} + \tau^* H^{1} \right)_{(3)}\epsilon \notag\\
& \quad + \ft{i}{4!\sqrt{7}} e^{\varphi/ \sqrt7} \slashed G \epsilon^* \,,\notag\\[2mm]
\delta_0 \lambda
&= i \slashed{\partial} \phi \, \epsilon^* - e^\phi \slashed{\partial} \chi \, \epsilon^* - \ft{i}{2\cdot 2!} e^{3\sqrt7\varphi/14} e^{\phi/2}\gamma^{(2)}\left( F^{2} + \tau F^{1} \right)_{(2)}\epsilon \notag\\
& \quad - \ft{1}{2\cdot 3!}e^{-\sqrt{7}\varphi/14}e^{\phi/2}\gamma^{(3)}\left(H^{2} + \tau H^{1} \right)_{(3)}\epsilon \displaybreak[2] \,,\notag\\[2mm]
\delta_0 A_{\mu} &= \ft{i}{2} e^{2 \varphi / \sqrt7} \bar{\epsilon}(\psi_{\mu}
- \ft{i}{\sqrt{7}}\gamma_{\mu} \tilde \lambda^*) + \mbox{h.c.} \,,\notag\\
\delta_0 A^1_{\mu}
&= \ft{i}{2} e^{\phi/2} e^{-3\varphi/2\sqrt7} \left( \overline{\epsilon}^*  \psi_{\mu} + \ft{i}{4} \overline \epsilon \gamma_{\mu} \lambda + \ft{3i}{4\sqrt7} \overline \epsilon^* \gamma_{\mu} {\tilde \lambda}^* \right) + \text{h.c.} \displaybreak[2] \,,\notag\\[2mm]
\delta_0 A^2_{\mu} &=
-\ft{i}{2} e^{\phi/2} \tau^* e^{-3\varphi /2\sqrt7} \left( \overline{\epsilon}^* \psi_{\mu} + \ft{i}{4} \overline \epsilon \gamma_{\mu } \lambda + \ft{3i}{4\sqrt7} \overline \epsilon^* \gamma_{\mu} {\tilde \lambda}^* \right) + \text{h.c.} \,,\notag\\
\delta_0 B^1_{\mu \nu}
&= - e^{\phi/2} e^{\varphi/2\sqrt7} \left( \overline{\epsilon}^* \gamma_{[\mu} \psi_{\nu]} + \ft{i}{8} {\overline{\epsilon}} \gamma_{\mu \nu} \lambda - \ft{i}{8\sqrt7} {\overline{\epsilon}}^* \gamma_{\mu \nu} {\tilde\lambda}^* \right) - A_{[\mu}^{\phantom{()}} \delta_0 A^1_{\nu]} + \text{h.c.} \,,\notag\\
\delta_0 B^2_{\mu \nu}
&= 
e^{\phi/2} \tau^* e^{\varphi/2\sqrt7} \left( \overline{\epsilon}^* \gamma_{[\mu} \psi_{\nu]} + \ft{i}{8} {\overline{\epsilon}} \gamma_{\mu \nu} \lambda - \ft{i}{8\sqrt7} {\overline{\epsilon}}^* \gamma_{\mu \nu} {\tilde\lambda}^* \right) - A_{[\mu}^{\phantom{()}} \delta_0 A^2_{\nu]} + \text{h.c.} \,,\notag\\
\delta_0 C_{\mu \nu \rho} &= \tfrac{3}{2} e^{-\varphi/\sqrt7} \bar{\epsilon}
\gamma_{[\mu\nu} \left( \psi_{\rho]} + \tfrac{i}{6\sqrt{7}} \gamma_{\rho]}
\tilde\lambda^* \right) - \tfrac{3}{2} B_{i \, [\mu\nu} \, \delta_0 A^i_{\rho]} +
\mbox{h.c.} \displaybreak[2] \,, \notag \\ 
\delta_0\varphi
&= -\ft{i}{4} \bar{\epsilon} \tilde \lambda^* + \text{h.c.} \,,\notag\\
\delta_0\chi
&= \ft{1}{4} e^{-\phi}\overline{\epsilon}\lambda^* + \text{h.c.} \,,\notag\\
\delta_0\phi &= -\ft{i}{4} \overline{\epsilon}\lambda^* + \text{h.c.} \,, 
\label{9Dsusy}
\end{align}
with the complex scalar $\tau = \chi + i e^{-\phi}$. The field strengths read
\begin{align}
  \G{1} = {\rm d} \chi \,, \quad
  F = {\rm d} A \,, \quad
  F^i = {\rm d} A^i \,, \quad
  H^i = {\rm d} B^i - A \wdg F^i \,, \quad
  G = {\rm d} C + B_i \wdg F^i \,.
\label{fieldstrengths}
\end{align}
The covariant derivative of the Killing spinor reads
\begin{align}
  D_\mu \epsilon & = (\partial_\mu + \omega_\mu +\ft{i}{4} e^\phi \partial_\mu \chi ) \epsilon \,.
\end{align}
When truncating to the metric, scalars and fermions, the massless 9D fermionic field
equations read
\begin{align}
  X_0 (\psi^\mu) & =
  \gamma^{\mu\nu\rho} (\partial_\nu + {{\omega}}_{{{\nu}}}+ \tfrac{1}{4} i e^{\phi} \partial_\nu \chi) \psi_\rho
  - \tfrac{1}{8} e^\phi (\slashed{\partial} \tau) \gamma^\mu \lambda^*
  + \tfrac{1}{8} i (\slashed{\partial} \varphi) \gamma^\mu \tilde{\lambda}^* = 0 \,, \notag \\
  X_0 (\lambda) & =
  \gamma^\mu (\partial_\mu + {{\omega}}_{{{\mu}}}+ \tfrac{3}{4} i e^{\phi} \partial_\mu \chi) \lambda
  + e^\phi \gamma^\mu (\slashed{\partial} \tau) \psi_\mu^* = 0 \,, \notag \\
  X_0 (\tilde{\lambda}) & =
  \gamma^\mu (\partial_\mu + {{\omega}}_{{{\mu}}} - \tfrac{1}{4} i
 e^\phi \partial_\mu \chi) \tilde{\lambda}
  - i  \gamma^\mu (\slashed{\partial} \varphi) \psi_\mu^* = 0 \,.
\end{align}
Under supersymmetry these yield the variation
\begin{align}
  \delta_0 X_0 (\psi^\mu) & = \tfrac{1}{2} {{\gamma}}^{{{\nu}}} {{\epsilon}}
  \, [ {{R}}^{{{\mu}}}{}_{{{\nu}}}
  - \tfrac{1}{2} {{R}} {{g}}^{{{\mu}}}{}_{{{\nu}}}
  - \tfrac{1}{2} ((\partial^{{{\mu}}} {\phi}) (\partial_{{{\nu}}} {\phi})
  - \tfrac{1}{2} (\partial {\phi})^2 {{g}}^{{{\mu}}}{}_{{{\nu}}}) + \notag \\
  & \hspace{1.5cm} - \tfrac{1}{2} e^{2\phi} ((\partial^{{{\mu}}} {\chi}) (\partial_{{{\nu}}} {\chi})
  - \tfrac{1}{2} (\partial {\chi})^2 {{g}}^{{{\mu}}}{}_{{{\nu}}})
  - \tfrac{1}{2} ((\partial^{{{\mu}}} {\varphi}) (\partial_{{{\nu}}} {\varphi})
  - \tfrac{1}{2} (\partial {\varphi})^2 {{g}}^{{{\mu}}}{}_{{{\nu}}})] \,, \notag \\
  \delta_0 X_0 (\lambda) & = \epsilon^* [ - e^\phi (\Box \chi + 2 (\partial_\mu \phi)(\partial^\mu \chi)) ] +
  i \epsilon^* [ \Box \phi - e^{2\phi}(\partial \chi)^2 ] \,, \notag \\
  \delta_0 X_0 (\tilde{\lambda}) & = i \epsilon^* [ \Box \varphi ] \,,
\end{align}
which are the massless bosonic field equations for the metric and the scalars.

\subsection{Twisted Reduction of IIA using $\nephat \beta$}

The reduction of massless IIA supergravity using the scale symmetry $\nephat \beta$ of
table~\ref{tab:IIA-weights} for twisting, with reduction Ans\"atze~\eqref{IIAred} with
$m_{\text{IIA}}=0$, leads to a massive deformation with mass parameter $m_4$. Only the
supersymmetry variations of the dilatini receive explicit massive deformations:
\begin{align}
    \delta_{m_4} \lambda =
      \tfrac{3}{4} m_4 e^{\phi/2 - 3\varphi/2\sqrt{7}} \epsilon \,, \qquad
    \delta_{m_4} \tilde \lambda =
      \tfrac{\sqrt{7}}{4} m_4 e^{\phi/2 - 3\varphi/2\sqrt{7}} \epsilon \,.
\label{susyexplmiia}
\end{align}
The implicit massive deformations read:
\begin{align}
  {\rm D} \phi & = e^{-\phi} {\rm D} e^{\phi} \,, \qquad
  {\rm D} \varphi = e^{-\varphi} {\rm D} e^{\varphi} \,, \qquad
  \G{1} = {\rm D} \chi + \tfrac{3}{4} m_4 A^2 \,, \qquad
  G = {\rm D} C + B_i \wdg F^i \,, \notag \\
  F & = {\rm D} A + \tfrac{1}{2} m_4 B^1 \,, \qquad
  F^{1} = {\rm d} A^1 \,, \qquad
  F^{2} = {\rm D} A^2 \,, \notag \\
  H^{1} & = {\rm D} B^1 - A \wdg F^{1} \,, \qquad
  H^{2} = {\rm D} B^2 - A \wdg F^{2} -\tfrac{1}{4} m_4 (C + 3 A^2 \wdg B^1) \,.
\label{fieldstrengthsmiia}
\end{align}
The $\mathbb{R}^+$ covariant derivative is defined by ${\rm D}={\rm d} - w_\beta \, m_4
A^1$ with $w_\beta$ the $\beta$ scale weight of the field it acts on, as given in the
table~\ref{tab:9D-weights}, and ${\rm DD} = - w_\beta \, m_4 F^{1}$. The covariant
derivative of the supersymmetry parameter has no massive deformation.

As for the field equations, the explicit deformations in the bosonic sector are given by
the scalar potential
 \begin{align}
  V = \tfrac{1}{2} e^{\phi -3 \varphi / \sqrt{7} } m_4{}^2 \,,
 \end{align}
which can not be written in terms of a superpotential as~\eqref{super-potential}. The
explicit deformations of the fermionic field equations read
\begin{align}
  X_{m_4} (\psi^\mu) & = i m_4 e^{\phi/2-3\varphi/2\sqrt{7}} \gamma^{\mu \nu}
    [-i\tfrac{3}{256} \gamma_\nu \lambda
     -i \tfrac{\sqrt{7}}{256} \gamma_\nu \tilde{\lambda} ] \,, \notag \\
  X_{m_4} (\lambda) & = - m_4 e^{\phi/2-3\varphi/2\sqrt{7}} \gamma^\nu
    [\tfrac{3}{4}\psi_\nu
    +\tfrac{2}{9\sqrt{7}} i \gamma_\nu \tilde{\lambda}^* ] \,, \notag \\
  X_{m_4} (\tilde{\lambda}) & = - m_4 e^{\phi/2-3\varphi/2\sqrt{7}} \gamma^\nu
    [\tfrac{\sqrt{7}}{4}\psi_\nu
    -\tfrac{2}{9\sqrt{7}} i \gamma_\nu \lambda^* ] \,.
\end{align}

These massive deformations can be seen as a gauging of the scale symmetry $\beta$ with
gauge field transformation
\begin{align}
  \qquad \Lambda = e^{- w_\beta m_4 \lambda^1} \,, \qquad
  A^1 & \rightarrow A^1 - {\rm d} \lambda^1 \,,
\label{nonAbtr1}
\end{align}
with gauge vector $A^1$ and parameter $\lambda^1$. In addition, we find that the
parabolic $\mathbb{R}$ subgroup of $SL(2,\mathbb{R})$ is gauged:
 \begin{align}
  \chi \rightarrow \chi + \tfrac{3}{4} m_4 \lambda^2 \,, \quad
  B^2 \rightarrow B^2 - \tfrac{3}{4} m_4 \lambda^2 B^1 \,, \quad
  A^2 \rightarrow A^2 - d \lambda^2 - \tfrac{3}{4} m_4 \lambda^2 A^1 \,,
\label{nonAbtr2}
\end{align}
with gauge vector $A^2$ and parameter $\lambda^2$. These two symmetries do not commute
but rather form the two--dimensional non-Abelian Lie group, consisting of scalings and
translations in one dimension (so-called collinear transformations \cite{Gilmore}). The
algebra reads
\begin{align}
  [ T_1 , T_2 ] = T_2 \,,
\end{align}
which is non--semi--simple. The emergence of this non-Abelian group is an example of the
enhanced gaugings of subsection~\ref{sec:twist-4} and can be understood by the scaling of
the 10D vector $A_\mu$ under $\beta$, see table~\ref{tab:IIA-weights}.

\subsection{Twisted Reduction of IIA using $\nephat  \alpha$}

The twisted reduction of massless IIA supergravity based on the $\nephat  \alpha$
symmetry of table~\ref{tab:IIA-weights}, with reduction Ans\"atze~\eqref{IIAred} with
$m_4=0$, leads to a gauged supergravity with mass parameter $m_{\rm IIA}$. The explicit
massive deformations in this case appear in the variation of the gravitino and one of the
dilatini:
\begin{align}
    \delta_{m_{\text{IIA}}} \psi_\mu = - \tfrac{9}{14} i m_{\text{IIA}}
      e^{\phi/2 - 3\varphi/2\sqrt{7}} \gamma_\mu \epsilon^* \,, \qquad
    \delta_{m_{\text{IIA}}} \tilde \lambda =
    \tfrac{6}{\sqrt{7}} m_{\text{IIA}} e^{\phi/2 - 3\varphi/2\sqrt{7}} \epsilon \,.
\end{align}
The implicit massive deformations are given by
\begin{alignat}{2}
  {\rm D} \phi & = e^{-\phi} {\rm d} e^{\phi} \,,  \qquad
  {\rm D} \varphi  = e^{-\varphi} {\rm D} e^{\varphi} \,, \qquad
  \G{1}  = {\rm d} \chi  \,, \qquad
  G  = {\rm D} C + B_i \wdg F^i \,, \notag \\
  F & = {\rm D} A + 3 m_{\text{IIA}} B^1 \,, \qquad
  F^{1}  = {\rm d} A^1 \,, \qquad
  F^{2}  = {\rm d} A^2 \,,  \notag \\
  H^{1} & = {\rm D} B^1 - A \wdg F^{1} \,, \qquad
  H^{2}   = {\rm D} B^2 - A \wdg F^{2} + 3 m_{\text{IIA}} C \,.
\end{alignat}
The $\mathbb{R}^+$ covariant derivative is defined by ${\rm D}={\rm d} - w_\alpha \,
m_{\text{IIA}} A^1$ with $w_\alpha$ the scale weight under $\alpha$ of the field it acts
on, as given in the table~\ref{tab:9D-weights}, and ${\rm DD} = - w_\alpha \,
m_{\text{IIA}} F^{1}$. The covariant derivative of the supersymmetry parameter is given
by
\begin{align}
  D_{\mu} \epsilon & = (\partial_{\mu} + {\omega}_{\mu} + \tfrac{i}{4} e^{\phi} \partial_\mu \chi
  - \ft{9}{14} m_{\text{IIA}} \Gamma_{\mu} \slashed{A}^{1} ) \epsilon \,.
\end{align}
The 9D fermionic field equations have the following explicit massive deformations:
\begin{align}
  X_{m_{\text{IIA}}} (\psi^\mu) & = i m_{\text{IIA}} e^{\phi/2-3\varphi/2\sqrt{7}} \gamma^{\mu \nu}
    [\tfrac{9}{2}\psi_\nu^* -i\tfrac{9}{32} \gamma_\nu \lambda
    + i \tfrac{3}{4\sqrt{7}} \gamma_\nu \tilde{\lambda} ] \,, \notag \\
  X_{m_{\text{IIA}}} (\lambda) & = - m_{\text{IIA}} e^{\phi/2-3\varphi/2\sqrt{7}}
    \gamma^\nu [ - i \tfrac{\sqrt{7}}{6} \gamma_\nu \tilde{\lambda}^* ] \,, \notag \\
  X_{m_{\text{IIA}}} (\tilde{\lambda}) & = - m_{\text{IIA}} e^{\phi/2-3\varphi/2\sqrt{7}} \gamma^\nu
    [\tfrac{6}{\sqrt{7}}\psi_\nu -\tfrac{11}{6\sqrt{7}} i \gamma_\nu \lambda^*
    +\tfrac{1}{7} i \gamma_\nu \tilde{\lambda}^* ] \,.
\end{align}
This massive deformation is a gauging of the scale symmetry $\alpha$ with transformation:
\begin{align}
   \Lambda = e^{- w_\alpha m_{\text{IIA}} \lambda^1} \,, \qquad
   A^1 \rightarrow A^1 - {\rm d} \lambda^1 \, ,
\end{align}
with gauge vector $A^1$ and parameter $\lambda^1$.

\subsection{Twisted Reduction of IIB using $\nephat \delta$}

The other possibility for twisted reduction of $D=10$ IIB supergravity involves the
trombone symmetry of IIB supergravity. We use the reduction Ans\"atze given in
\eqref{IIBred} with $m_1=m_2=m_3=0$, yielding a massive deformation with parameter
$m_{\text{IIB}}$. The explicit deformations of the supersymmetry rules read
\begin{align}
    \delta_{m_{\text{IIB}}}\psi_\mu = - \ft47 i m_{\text{IIB}} e^{2\varphi/\sqrt7}
      \gamma_\mu \epsilon \,, \qquad
    \delta_{m_{\text{IIB}}}\tilde\lambda = -\ft{4}{\sqrt{7}} m_{\text{IIB}} e^{2\varphi/\sqrt7} \epsilon^* \,.
\end{align}
The implicit deformations read
\begin{align}
F &= {\rm d}A \,, \qquad {F}^i = {\rm d}{A}^i - 2 m_{\text{IIB}} {B}^i \,, \qquad
H^i = {\rm d} B^i - A \wdg F^i \,, \notag \\
G &= {\rm d}C + B_i \wdg F^i + m_{\text{IIB}} B_i \wdg B^i \,, \qquad D\varphi = {\rm
d}\varphi - \ft{4}{\sqrt7} m_{\text{IIB}} A \,,
\end{align}
for the bosons and
\begin{align}
  D_{\mu} \epsilon & = (\partial_{\mu} + {\omega}_{\mu} + \tfrac{i}{4} e^{\phi} \partial_\mu \chi
  + \ft{4}{7} m_{\text{IIB}} \Gamma_{\mu} \slashed{A} ) \epsilon
\end{align}
for the supersymmetry parameter. The explicit deformations of the fermionic field
equations read
\begin{align}
  X_{m_{\text{IIB}}} (\psi^\mu) & = i m_{\text{IIB}} e^{2\varphi/\sqrt{7}} \gamma^{\mu\nu}
    [4 \psi_\nu-\tfrac{15}{16\sqrt{7}} i \gamma_\nu \tilde{\lambda}^*] \,, \notag \\
  X_{m_{\text{IIB}}} (\lambda) & = m_{\text{IIB}} e^{2\varphi/\sqrt{7}} \gamma^{\nu}
    [ \tfrac{4}{9} i \gamma_\nu \lambda ] \,, \notag \\
  X_{m_{\text{IIB}}} (\tilde{\lambda}) & = m_{\text{IIB}} e^{2\varphi/\sqrt{7}} \gamma^{\nu}
    [\tfrac{4}{\sqrt{7}} \psi_\nu^* -i \tfrac{4}{7}
\gamma_\nu \tilde{\lambda}] \, .
\end{align}
This is a supergravity where the scale symmetry $\delta$ has been gauged, whose action
reads
\begin{align}
  \Lambda = e^{w_\delta m_{\text{IIB}} \lambda} \,, \qquad
  A \rightarrow A - {\rm d} \lambda \,, \qquad
  {B}^i \rightarrow e^{2 m_{\text{IIB}}\lambda} ({B}^i - {A}^i \, {\rm d} \lambda)
  \,,
\end{align}
with gauge vector $A$ and parameter $\lambda$.

\subsection{Toroidal Reduction of Gauged IIA}

Finally, one can also consider the toroidal reduction of the $D=10$ IIA gauged
supergravity of subsection~\ref{sec:massive-gauged-IIA-3}, again with reduction Ans\"atze
\eqref{IIAred} with $m_4 = m_{\text{IIA}} = 0$. This leads to a $D=9$ gauged supergravity
with the following explicit deformations
\begin{align}
    \delta_{m_{11}} {\psi}_{\mu}
     = \ft{9}{14} i m_{11} e^{\phi/2-3\varphi/2\sqrt7} \tau \gamma_\mu \epsilon^* \,,
     \qquad
    \delta_{m_{11}} \tilde \lambda =
      - \ft{6}{\sqrt7} m_{11} e^{\phi/2-3\varphi/2\sqrt{7}} \tau^* \epsilon \,.
\end{align}
The bosonic implicit deformations read
\begin{align}
 {\rm D}\varphi & = {\rm d}\varphi -\ft{6}{\sqrt7} m_{11} A^2 \,, \qquad
 F       = {\rm D}A + 3 m_{11} B^2 \,, \qquad
 G      = {\rm D}C + + B_i \wdg F^i \,, \notag \\
 F^i & = d A^i \,, \qquad
 H^{1}  = {\rm D}B^1 - A \wdg F^{(1)} - 3 m_{11} C \,, \qquad
 H^{2}  = {\rm D}B^2 - A \wdg F^{(2)} \,,
\end{align}
with the scale covariant derivative of a field with weight $w$ defined by ${\rm D}={\rm
d}-w_\alpha m_{11}A^2$. For the supersymmetry parameter we find
\begin{align}
  D_{\mu} \epsilon & = (\partial_{\mu} + {\omega}_{\mu} + \tfrac{i}{4} e^{\phi} \partial_\mu \chi
  + \ft{9}{14} m_{11} \Gamma_{\mu} \slashed{A}^{2} )
\epsilon \,.
\end{align}
The fermionic field equations are deformed by the massive contributions
\begin{align}
  X_{m_{11}} (\psi^\mu) & = - i m_{11} e^{\phi/2-3\varphi/2\sqrt{7}} \gamma^{\mu \nu}
    [\tfrac{9}{2} \tau \psi_\nu^* -i\tfrac{9}{32} \tau^* \gamma_\nu \lambda
    + i \tfrac{3}{4\sqrt{7}} \tau \gamma_\nu \tilde{\lambda} ] \,, \notag \\
  X_{m_{11}} (\lambda) & = m_{11} e^{\phi/2-3\varphi/2\sqrt{7}}
    \gamma^\nu [ - i \tau \tfrac{\sqrt{7}}{6} \gamma_\nu \tilde{\lambda}^* ] \,, \notag \\
  X_{m_{11}} (\tilde{\lambda}) & = m_{11} e^{\phi/2-3\varphi/2\sqrt{7}} \gamma^\nu
    [\tfrac{6}{\sqrt{7}} \tau^* \psi_\nu -\tfrac{11}{6\sqrt{7}} i \tau \gamma_\nu \lambda^*
    +\tfrac{1}{7} i \tau^* \gamma_\nu \tilde{\lambda}^* ] \,.
\end{align}
This massive deformation is a gauging of the scale symmetry $\alpha$, reading
\begin{align}
 \Lambda = e^{- w_\alpha m_{11} \lambda^2} \,, \qquad A^2 \rightarrow A^2 - {\rm d} \lambda^2\,,
\end{align}
with gauge vector $A^2$ and parameter $\lambda^2$.

 \addcontentsline{toc}{chapter}{Bibliography}
 \footnotesize
 \bibliographystyle{utphysmodb}
 \bibliography{article}
 \normalsize

\end{document}